\documentclass [11pt] {report}

\usepackage{amssymb,latexsym,revsymb}
\usepackage{amsmath,amsbsy}
\usepackage{epsfig,bm}
\usepackage{graphicx,comment}
\usepackage{cite}
\usepackage{alltt}  %

\DeclareMathOperator{\sign}{\text{sign}}
\DeclareMathOperator{\tr}{\text{Tr}}
\DeclareMathOperator{\res}{\text{Res}}
\DeclareMathOperator{\Res}{\text{Res}}

\begin{document}

\sloppy

\def\a{{\alpha}}
\def\b{{\beta}}
\def\d{{\delta}}
\def\D{{\Delta}}
\def\e{{\varepsilon}}
\def\g{{\gamma}}
\def\G{{\Gamma}}
\def\k{{\kappa}}
\def\l{{\lambda}}
\def\L{{\Lambda}}
\def\m{{\mu}}
\def\n{{\nu}}
\def\o{{\omega}}
\def\O{{\Omega}}
\def\S{{\Sigma}}
\def\s{{\sigma}}
\def\th{{\theta}}
\def\x{{\xi}}
\def\cO{\mathcal{O}}
\def\eps{{\varepsilon}}
\def\epsp{{\varepsilon^\prime}}
\def\epspp{{\varepsilon^{\prime\prime}}}
\def\lp{{\lambda^\prime}}
\def\kppperp{{\mathbf{k}^{\prime \prime \perp}}}
\def\gp{{\gamma^+}}
\def\lamp{{\lambda^\prime}}
\def\kpr{{\mathbf{k}^\perp_{\text{rel}}}}
\def\g5{{\gamma^5}}
\def\kpperp{{\mathbf{k}^{\prime \perp}}}
\def\xt{{\tilde{x}}}
\def\kpp{{\mathbf{k}^{\prime\perp}}}
\def\kppp{{\mathbf{k}^{\prime\prime\perp}}}
\def\ppp{{\mathbf{p}^{\prime\perp}}}
\def\dperp{{\mathbf{\Delta}^\perp}}
\def\koneperp{{\mathbf{k}^{\perp}_{1}}}
\def\ktwoperp{{\mathbf{k}^{\perp}_{2}}}
\def\pperp{{\mathbf{p}^{\perp}}}
\def\Pperp{{\mathbf{P}^{\perp}}}
\def\koneplus{{k^{+}_{1}}}
\def\ktwoplus{{k^{+}_{2}}}
\def\Pplus{{P^{+}}}
\def\pplus{{p^{+}}}
\def\qperp{{\mathbf{q}^{\perp}}}
\def\pmag{{p^{\perp}}}
\def\kperp{{\mathbf{k}^{\perp}}}
\def\kplus{{k^{+}}}
\def\kaperp{{\mathbf{\kappa}^{\perp}}}
\def\kaplus{{\kappa^{+}}}
\def\Pee{{P^{\mu}}}
\def\Pp{{{P}^{\prime}{}^{\mu}}}
\def\Pbar{{\bar{P}^{\mu}}}
\def\Pbarplus{{\bar{P}^{+}}} 
\def\ym{{y^{-}}}
\def\Ppbra{{\langle \; P^{\prime} \; |}}
\def\Pket{{ | \; P \; \rangle}}
\def\phid{{\hat{\phi}^\dagger}}
\def\phiud{{\hat{\phi}}}
\def\dplus{{\partial^{+}}}
\def\meask{{\frac{d\kplus  d\kperp}{\sqrt{ 2 \kplus} (2\pi)^3}}}
\def\measpj{{\frac{d\pplus_{j}  d\pperp_{j}}{\sqrt{ 2 p^+_{j}} (2\pi)^3}}}
\def\measka{{\frac{d\kaplus  d\kaperp}{\sqrt{ 2 \kaplus} (2\pi)^3}}}
\def\ak{{\hat{a}(\kplus,\kperp)}}
\def\adk{{\hat{a}^\dagger(\kplus,\kperp)}}
\def\aka{{\hat{a}(\kaplus,\kaperp)}}
\def\adka{{\hat{a}^\dagger(\kaplus,\kaperp)}}
\def\pket{{ | \; p_{1}, p_{2} \; \rangle}}
\def\qbra{{\langle \; p_{3}, p_{4}  \; |}}
\def\ndpd{{\mathcal{F}(X, \zeta, t)}}
\def\ie{{i \epsilon}}
\def\ofpd{{F(x,\xi,t)}}
\def\dw{{D_{\text{W}}}}
\def\bigdw{{\Delta_{\text{W}}}}
\def\mn{{\langle \mu^2 \rangle}}
\def\pp{{p^{\prime}}}
\def\lam{{\Lambda}}
\def\kminus{{k^{-}}}
\def\pminus{{p^{-}}}
\def\qminus{{q^{-}}}
\def\qplus{{q^+}}
\def\Gt{{\tilde{G}}}
\def\pe{{\mathcal{P}}}
\def\qu{{\mathcal{Q}}}
\def\psiket{{|\psi\rangle}}
\def\psitwoket{{|\psi_{2}\rangle}}
\def\psiquket{{|\psi_{\qu}\rangle}}
\def\psibra{{\langle\psi|}}
\def\psitwobra{{\langle\psi_{2}|}}
\def\psiqubra{{\langle\psi_{\qu}|}}
\def\xpp{{x^{\prime\prime}}}
\def\yp{{y^\prime}}
\def\xp{{x^\prime}}
\def\zp{{z^\prime}}
\def\zpp{{z^{\prime\prime}}}
\def\ypp{{y^{\prime\prime}}}

\def\ol#1{{\overline{#1}}}

\title{{\bf{\Huge Light-Front Dynamics \\ 
and\\
Generalized Parton Distributions\\
}}}
\author{{\LARGE Brian C.~Tiburzi}\\
\\
\\
\\
\\
A dissertation submitted in partial fulfillment of\\
the requirements for the degree of\\
\\
Doctor of Philosophy\\
\\
\\
Advisor\\
Professor Gerald A.~Miller\\
Department of Physics\\
University of Washington
}
\date{July  2004}

\maketitle

\thispagestyle{empty}
\cleardoublepage{\ }

\abstract{

High energy reactions necessarily involve large momentum transfer to 
the target's constituents, which thus rebound near light speed. 
The experimentally observed correlation between initial and 
final states measures the system's response along the light cone. 
A convenient theoretical description of such reactions is light-front 
dynamics where the properties of physical states are described along 
the advance of a wavefront of light.

In this work we report on using light-front dynamics to describe 
generalized parton distributions, which are correlation functions
encountered in virtual Compton scattering at large momentum transfer.
This two photon process requires pair annihilation contributions which 
are subtle to handle in light-front dynamics. We are careful to derive
such contributions in the light-front Bethe-Salpeter formalism. This 
derivation highlights the connection to the Fock space picture as well 
as provides a toy model for generalized parton distributions.

Ultimately we use light-front dynamics to build a phenomenological 
parametrization for the proton's generalized parton distributions. 
This task is non-trivial due to the field theoretic constraints required 
of these distributions. To meet the constraints imposed by Lorentz covariance, 
we approach generalized parton distributions from the perspective of 
double distributions. Despite avid phenomenological use, surprisingly little
work has been done to calculate double distributions. 
We first focus on a number of simple pedagogical examples and demonstrate
how double distributions can be correctly determined. The double
distributions we derive, including those for the proton, 
satisfy known constraints and are adequate for phenomenological estimates 
of cross sections. The models used allow one to study the interplay between 
the light-front formulation and Lorentz covariance.

}

\thispagestyle{empty}
\cleardoublepage{\ }

\thispagestyle{empty}

\pagenumbering{roman}
\setcounter{page}{0}
\setcounter{tocdepth}{3}

\tableofcontents

\chapter*{Glossary}      
\thispagestyle{plain}

\noindent {\bf BSE}:  Bethe-Salpeter equation. 

\bigskip
\noindent {\bf Covariant}: Lorentz covariant, i.e., transforming properly under the Lorentz group.

\bigskip
\noindent {\bf DD}: Double distribution.

\bigskip
\noindent {\bf DIS}: Deep-inelastic scattering.

\bigskip
\noindent {\bf DSE}: Dyson-Schwinger equation, a coupled field theory equation of motion for a Green's function. 

\bigskip
\noindent {\bf DVCS}: Deeply virtual Compton scattering.

\bigskip
\noindent {\bf GDA}: Generalized distribution amplitude, or two pion distribution amplitude.

\bigskip
\noindent {\bf GPD}: Generalized parton distribution.

\bigskip
\noindent {\bf IMF}: Infinite momentum frame.

\bigskip
\noindent {\bf LFTOPT}: Light-front time-ordered perturbation theory.

\bigskip
\noindent {\bf Light-cone coordinates}: The coordinates that describe the world sheets of light, namely
in our conventions $x^\pm = (x^0 \pm x^3)/\sqrt{2}$.

\bigskip
\noindent {\bf Light-front energy}: The Fourier conjugate to light-front time.

\bigskip
\noindent {\bf Light-front time}: The hypersurface defined by $x^+ = \text{constant}$.

\bigskip
\noindent {\bf PDF}: Parton distribution function, or quark distribution.

\bigskip
\noindent {\bf Polynomiality}: Property of the moments of generalized parton distributions required by Lorentz invariance.

\bigskip
\noindent {\bf Positivity}: Bounds that are required of generalized parton distributions due to the positivity of the norm on Hilbert space, properly 
referred to as positivity bounds. 

\bigskip
\noindent {\bf QCD}: Quantum chromodynamics, the quantum field theory of colored quarks and gluons.

\bigskip
\noindent {\bf QED}: Quantum electrodynamics, the quantum field theory of electrically charged particles and photons.

\cleardoublepage{\ }

\thispagestyle{empty}

\chapter{Introduction} \label{chap:intro}

\pagenumbering{arabic}
\setcounter{page}{1}

The description of the bound states of strongly interacting particles calls for a synthesis of relativity and quantum mechanics. While
this synthesis is afforded by quantum field theory, the direct investigation of bound states from a covariant framework has been 
limited. Other complimentary methods are desirable for both varied intuition and differing calculational strategies.

In a Hamiltonian approach, one attempts to calculate wavefunctions for the physical states of the theory. 
More than a half century ago, Dirac's paper on the forms of relativistic dynamics \cite{Dirac:1949cp} introduced the front-form
Hamiltonian approach, which differs from conventional Hamiltonian dynamics by a relativistic boost to infinite momentum.  
Applications of this form of dynamics to quantum mechanics and field theory were overlooked at the time due to the 
appearance of covariant perturbation theory.
The reemergence of front-form dynamics was largely motivated by simplicity as well as physicality. The 
light-front approach has the largest kinematic subgroup of operators \cite{Leutwyler:1978vy} of any Hamiltonian theory.
Today the physical utility of  light-front dynamics is transparent: hard scattering processes probe 
a light-cone correlation of bound states. Not surprisingly, then, many perturbative QCD applications can be treated on the light front, see e.g. 
\cite{Lepage:1980fj}. Outside this realm, physics on the light cone has been extensively developed for non-perturbative QCD 
\cite{Zhang:1994ti,Harindranath:1996hq,Brodsky:1998de,Heinzl:2000ht} as well as applied to nuclear physics \cite{Carbonell:1998rj,Miller:2000kv}.

In this thesis we consider applications of light-front dynamics to generalized parton distributions (GPDs). As GPDs enter into amplitudes 
for hard exclusive reactions, light-cone quantized fields make a natural appearance. GPDs are theoretically challenging objects to calculate. Even simple phenomenological 
parameterizations cannot be written down at will [as is the case of conventional parton distributions functions (PDFs)] due to the field theoretic
constraints GPDs must satisfy. Ultimately we will generate a phenomenological parametrization for the proton's GPDs. Before doing so, however, 
we investigate from a few perspectives the difficulties inherent in describing these new distribution functions. We first motivate the physicality 
of the light-cone framework by considering deep-inelastic scattering and then discuss bound states in field theory 
and the front form of Hamiltonian dynamics

\section{Deep-inelastic scattering} \label{DIS}

Deep-inelastic scattering (DIS) is the canonical hard scattering process. We shall therefore review DIS in some detail in order to introduce
the relevance of light-front dynamics in processes amenable to treatment in perturbative QCD. 
Our treatment parallels the analysis of \cite{Jaffe:1992ra,Jaffe:1996zw}.
We show how light-cone coordinates and the 
collinear approximation  enable
simplification of the DIS amplitude into a product of hard scattering coefficients and universal functions. The latter are the conventional 
parton distribution functions (PDFs). These distribution functions are momentum probabilities with respect to the quark wavefunction
of the target defined along the advance of a wavefront of light. 
The analysis presented below will be extended later in Chapter \ref{chap:GPDs}
to deeply virtual Compton scattering (DVCS).  
Nonetheless, this Section gives practical motivation for the need to investigate bound states on the light cone.

\subsection{Bjorken limit}

In DIS off the proton, the process under consideration is electron\footnote{%
One can also consider muon, positron, neutrino, \emph{etc}. scattering off 
the proton (or other) targets.} 
scattering at large momentum transfer, namely
\begin{equation} \label{eqn:DIS}
e(k) + p(P) \to e(k') + X
.\end{equation}
Here $k$, $k'$ label the momenta of the initial and final electrons respectively, while $P$ is the momentum of the proton. 
The reaction happens at such large momentum transfer that the proton breaks up, as it were, and subsequently forms new hadrons. 
We have used $X$ to label any of the myriad of possible final states.

\begin{figure}
\begin{center}
\epsfig{file=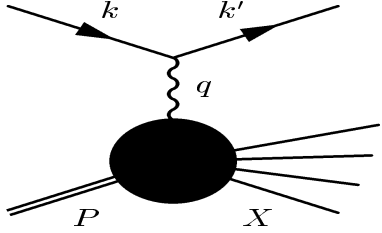,width=2in,height=1.2in}
\end{center}
\caption[Diagrammatic depiction of the DIS process.]{Diagrammatic depiction of the DIS process. An electron of momentum $k$ is incident on a proton of momentum $P$. 
The leading-order electromagnetic process is virtual photon exchange between the electron and proton. As a result, the electron's momentum is altered to $k'$
and due to the high virtuality of the photon, the proton breaks up into various reaction products collectively denoted by $X$.}
\label{f:DIS}
\end{figure}

We define the four-momentum transfered to the proton as $q = k - k'$. In the language of QED, this momentum is transfered via the virtual exchange of a 
photon between the electron and proton, see Figure \ref{f:DIS}. 
The offshellness of the photon is given by $q^2 = - Q^2$, where $Q^2$ is assumed to be very large and positive in the
deep-inelastic limit. In accordance with the Heisenberg uncertainty principle, the larger the momentum transfer, the smaller
the region probed. Hence one can view the electromagnetic process in DIS as resolving the structure deep within the proton. From the electron's perspective, 
however, the proton is severely Lorentz contracted and thus the virtual photon is only capable of resolving small sizes
transverse to the beam direction.

In evaluating the cross section, 
we have the freedom to choose both the beam direction and the overall frame in which to view the scattering reaction.
With a suitable orientation of the beam direction and a convenient choice of reference frame, 
we can work simultaneously with $\mathbf{P}^\perp \equiv  (P^1, P^2) = 0$ and $\mathbf{q}^\perp = 0$. 
Now as is conventional, we introduce two vectors $p^\mu$ and $n^\mu$ into which we decompose the remaining momenta. Explicitly
these vectors are
\begin{equation}
p^\mu  = \frac{\Lambda}{\sqrt{2}} (1,0,0,1) 
\end{equation}
and
\begin{equation}
n^\mu = \frac{1}{\sqrt{2} \Lambda} (1, 0, 0, -1)
,\end{equation}
where $\Lambda$ is an unfixed parameter associated with the unexploited freedom to boost along the $\hat{z}$-direction. 
Notice both vectors are lightlike: $p^2 = n^2 = 0$. Additionally their product is $p \cdot n = 1$.

In terms of these $p$ and $n$ vectors, the proton's momentum appears as
\begin{equation} \label{eqn:P}
P^\mu = p^\mu + \frac{M^2}{2} n^\mu
,\end{equation}
so that $P^2 = M^2$. In the proton's rest frame, $\Lambda = M / \sqrt{2}$. In this frame we define the energy of the virtual 
photon $q^0 = \nu$ and hence $P\cdot q = M \nu$. The photon's momentum in the proton rest frame is thus
\begin{equation}
q^\mu = ( \nu, 0, 0, - \sqrt{Q^2 + \nu^2})
\end{equation}
because the virtuality is $q^2 = - Q^2$.

\begin{figure}
\begin{center}
\epsfig{file=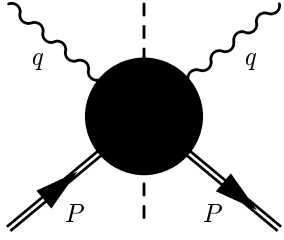}
\end{center}
\caption[The forward virtual Compton amplitude.]{The forward virtual Compton amplitude $T^{\mu \nu}(q,P)$. The dashed line denotes the cut that
yields the imaginary part needed for the DIS cross section.}
\label{f:T}
\end{figure}

Lastly we define the Lorentz invariant Bjorken variable $x$
\begin{equation} \label{eqn:Bjorken}
x =  - \frac{q^2}{2 P \cdot q}  
.\end{equation}
In the proton rest frame, we have $x = Q^2 / 2 M \nu$. In the limit of  elastic proton-electron scattering $x$ is fixed by momentum conservation,  $x = 1$.
For a general inelastic process, $x$ is kinematically restricted to $ \in (0,1]$. Notice in particular this implies $\nu^2 \gg Q^2$. 
In DIS, one refers to the Bjorken limit
in which $Q^2 \to \infty$ while $x$ remains fixed by the kinematics. 
In this limit the doubly differential cross section for the reaction in Eq.~\eqref{eqn:DIS}
has the form
\begin{equation}
\frac{d^2 \sigma}{dE' d\Omega'} = \frac{\alpha_E^2}{M Q^4} \frac{E'}{E} L^{\mu \nu}(k,q) W_{\mu \nu}(q,P)
,\end{equation}
where we have included only the leading-order electromagnetic contribution from one photon exchange.
Here $E'$ is the final electron energy and $d \Omega'$ its differential solid angle, while $E$ is the initial electron energy. 
$L^{\mu \nu}(k,q)$ is the leptonic tensor representing the QED amplitude squared for virtual photon emission off the electron line and
$\alpha_E$ is the electromagnetic fine-structure constant. The unknown quantity in the cross section 
is the hadronic tensor $W^{\mu \nu}(q,P)$ which denotes the amplitude squared for the virtual photon-proton interaction. The hadronic reaction products $X$, however, 
are not observed in \emph{inclusive} DIS and are accordingly summed over in $W^{\mu \nu}(q,P)$. Thus we utilize the optical theorem
\begin{equation}
W^{\mu \nu}(q,P) = - \frac{1}{2 \pi} \Im \text{m} \left[ T^{\mu \nu}(q,P) \right]
\end{equation}  
to relate $W^{\mu \nu}(q,P)$ to the forward virtual Compton amplitude $T^{\mu \nu}(q,P)$ that is depicted in 
Figure \ref{f:T}. In terms of the Heisenberg current operator $J^\mu(y)$, $T^{\mu \nu}(q,P)$ can be written as the diagonal 
proton matrix element of the time-ordered product of two currents, namely
\begin{equation} \label{eqn:T}
T^{\mu \nu}(q,P) = i \int d^4 y \, e^{i q \cdot y} \langle P | T \Big\{ J^\mu(y) J^\nu(0) \Big\} | P \rangle
.\end{equation}

\subsection{Collinear approximation}

Above we have defined the kinematics of DIS and written the experimentally observed cross-section in terms
of the forward virtual Compton amplitude $T^{\mu \nu}(q,P)$. We now proceed to isolate the leading contribution
to $T^{\mu \nu}(q,P)$ in perturbative QCD.\footnote{%
Throughout this work, we shall work only to leading order in the strong coupling. Important points concerning
$\cO(\a_s)$ corrections are contained in footnotes such as this. For a comprehensive discussion of perturbative
QCD corrections to DIS, see e.g.~\cite{Jaffe:1996zw}.
} 
Large momentum $q$ flows through the diagram depicted in Figure \ref{f:T}. Since $T^{\mu \nu}(q,P)$ is 
a Fourier transform, most contributions are averaged out as $q^2 \to - \infty$. Indeed the only non-vanishing contributions 
that remain arise from singularities in the product of current operators as $y^2 \to 0$.

\begin{figure}
\begin{center}
\epsfig{file=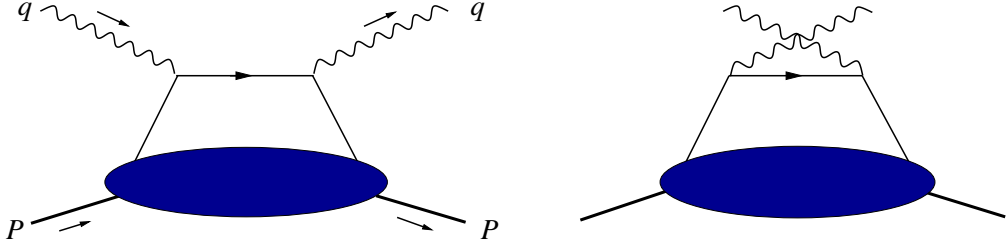,width=5.25in}
\end{center}
\caption[Graphical depiction of the handbag mechanism in DIS.]{Graphical depiction of the handbag mechanism in DIS. Due to asymptotic freedom, the forward virtual Compton
amplitude is dominated by free quark scattering. There are two contributions: the handbag shown on the left, and the crossed
bag shown on the right. Figure adapted from \cite{Diehl:2003ny}.}
\label{f:hand}
\end{figure}

Due to the nature of the QCD running coupling, at large momentum transfer the highly virtual photon couples to asymptotically free quarks. 
Asymptotic freedom justifies the use of the impulse approximation for the product of currents and the so called handbag dominance emerges
(see Figure \ref{f:hand}). More rigorously, the handbag 
mechanism picks out the leading singularity as $y^2 \to 0$ in Eq.~\eqref{eqn:T}. The handbag contribution to $W^{\mu \nu}(q,P)$ is thus\footnote{%
Quark masses have been neglected in the propagator with large momentum flow. The restriction $m_q^2 / Q^2 \ll 1$ automatically follows
from $M^2 / Q^2 \ll 1$, which, as we shall see, is the natural scale separation condition in DIS. 
}
\begin{equation}
W^{\mu \nu}(q,P) = \frac{1}{2\pi} \Im \text{m}
\left\{ 
i \int \frac{d^4 k}{(2\pi)^4} \tr
\left[
\gamma^\mu \frac{i (\rlap\slash k + \rlap\slash q)}{(k+ q)^2 + i \varepsilon}
\gamma^\nu M(k,P)
\right]  
\right\}
\end{equation} 
along with an additive similar contribution from the crossed handbag. The matrix element $M(k,P)$
in the above expression is the forward free-quark proton scattering amplitude given by
\begin{equation} \label{eqn:M}
M_{\a\b} (k,P) = \int d^4 y \, e^{i k \cdot y} \langle P | T \Big\{\ol \psi_\b(0) \psi_\a(y) \Big\}
| P \rangle
,\end{equation}
where $\a$ and $\b$ are Dirac indices.

Lastly we need to isolate the dominant contribution from the matrix element $M(k,P)$ in the Bjorken limit. 
To do this we decompose 
the active quark momentum $k^\mu$ in terms of the light-cone vectors $p^\mu$ and $n^\mu$, \emph{viz}.
\begin{equation}
k^\mu = (k\cdot n) \, p^\mu + (k \cdot p) \, n^\mu + \mathbf{k}_\perp^\mu
,\end{equation}
and analyze the situation in each momentum channel. The situation is the simplest with respect to the 
transverse component. There, the photon carries no transverse momentum and so the active quark's transverse momentum 
is unchanged after interaction. In the $p^\mu$ direction, the photon adds $q \cdot n = - x$ to the active quark's momentum
forcing the final momentum to be $k \cdot n - x$. Lastly in the $n^\mu$ direction, the photon adds 
$q \cdot p \approx \infty$ and the rebounding quark has nearly infinite momentum in the $n^\mu$ channel.
Thus relative to this direction the struck quark has effectively zero momentum, i.e.
$\mathbf{k}^\perp \approx 0$ and $k \cdot n - x  \approx 0$. The latter we enforce
by inserting $1 = \int dx \, \delta( x - k \cdot n)$. 
Our analysis leads to the collinear approximation to the active quark's momentum
\begin{equation} \label{eqn:col}
k^\mu \approx ( k \cdot n ) \, p^\mu
,\end{equation}
because the transverse momentum is effectively zero and after interaction with the photon the $n^\mu$ component is infinite, 
which leads to suppression of the matrix element. In the collinear approximation, we have the resulting contribution to $W^{\mu \nu}(q,P)$
\begin{equation} \label{factorization}
W^{\mu \nu}(q,P) = - \frac{1}{2 \pi} \Im \text{m} \, 
\left\{
\int dx \, \tr 
\left[
\gamma^\mu \frac{x \rlap \slash p  + \rlap \slash q}{(x p + q)^2 + i \varepsilon} \gamma^\nu M_{\a\b}(x)
\right]
\right\}
.\end{equation}

Above we have reduced the free-quark proton Green's function $M(k,P)$ into a function of only\footnote{%
This statement and the factorization in Eq.~\eqref{factorization} are modified by $\cO(\a_s)$ corrections. These arise
from perturbative gluon interactions which lead to renormalization scale dependence, i.e.~$M_{\a \b} = M_{\a \b}(x,\mu^2)$. 
Physically $\mu$ is the scale at which the partons are resolved. Consequently the notion of a parton
is renormalization scale and scheme dependent. The Compton amplitude $W^{\mu \nu}(q,P)$ remains scale independent
because the  $\cO(\a_s)$ corrections to the hard scattering kernel in Eq.~\eqref{factorization} exactly compensate 
for the $\mu$ dependence in the quark-proton Green's function.
}
Bjorken $x$
\begin{equation}
M_{\a\b}(x) \equiv \int \frac{d^4 k}{(2 \pi )^4} \delta( x - k \cdot n) M_{\a\b}(k,P)
.\end{equation} 
In this form, the integral over $y$ in Eq.~\eqref{eqn:M} can be performed trivially. 
We arrive at
\begin{equation}
M_{\a\b}(x) = \int \frac{d \l}{2 \pi} e^{i \l x} 
\langle P | \ol \psi_\b(0) \psi_\a(\l n) | P \rangle
,\end{equation}
here we have used $\l n$ to denote evaluation at the position $y^\mu = \l n^\mu$. 
Thus the DIS scattering cross-section depends upon a correlation function containing 
two quark fields with a lightlike separation (above the resulting separation is 
$y^- = \l / \sqrt{2} \Lambda$, $y^+ = 0$, and $\mathbf{y}^\perp = 0$). This means that the relevant part of the
quark propagator in DIS is only that along the light cone. Consequently the amplitude depends upon a 
correlation of proton states along the light cone and inherent to DIS is the system's response
along the advance of a wavefront of light. Furthermore, as a result of the collinear approximation
Eq.~\eqref{eqn:col}, the active quark's plus momentum is a fraction of the proton's, i.e., $k^+ = x P^+$.

For unpolarized DIS, we can decompose the Dirac structure of the light-cone correlator $M(x)$ into $p^\mu$ and $n^\mu$ vectors
\begin{equation}
M_{\a\b}(x) = \frac{1}{2} \rlap \slash p_{\a\b} \, f_1(x) + \frac{1}{2} M^2 \rlap \slash n_{\a\b} \, f_4(x)
.\end{equation}
Taking the trace of $M(x)$ with $\rlap \slash p$ and $\rlap \slash n$ separately, we deduce
\begin{equation} \label{eqn:1and4}
\int \frac{d \l}{2 \pi} e^{i \l x} \langle P | \ol \psi(0) \gamma^\mu \psi(\l n) | P \rangle
= 2 \left[
f_1(x) \, p^\mu + M^2 f_4(x) \, n^\mu
\right]
.\end{equation}
Now since both $n^2$ and $p^2$ are zero, and $n \cdot p = 1$, contributions to the DIS cross section from the $f_4(x)$ structure function 
must come out proportional to $\frac{M^2}{Q^2} f_4(x)$ at the end of the day. One can neglect this structure due to 
the power-law suppression in the Bjorken limit.\footnote{%
This statement is not altered by perturbative QCD corrections. Indeed Eq.~\eqref{eqn:1and4} is modified
to include scale dependence, but the structure functions $f_1(x,\mu^2)$ and $f_4(x,\mu^2)$ do not mix under 
evolution.
}
This simplification is a peculiarity of the light-cone power counting\footnote{%
Notice the vector $p^\mu$ has only a minus component and hence projects out the plus component of a vector dotted into it, while 
the $n^\mu$ vector has only a plus component and projects out the minus component. In older terminology, $p^\mu$ gives one
the \emph{good} components, while $n^\mu$ gives one the \emph{bad} components. 
} 
because the overall
matrix element has one mass dimension but two irreducible structure functions.

To summarize, we have reduced the virtual forward Compton amplitude in the Bjorken limit to one unknown structure function $f_1(x)$. 
In terms of proton matrix elements, this structure function is
\begin{equation} \label{eqn:f1}
f_1(x) = \frac{1}{2} \int \frac{d \l}{2 \pi} e^{ i \l x} \langle P | \ol \psi(0) \gamma^+ \psi(\l n) | P \rangle
.\end{equation}
Conventionally $f_1(x)$ is referred to as a parton distribution function (PDF). This is technically only true
in light-cone quantization in light-cone gauge (see Appendix \ref{chap:LCQ}). 
Nevertheless, to adequately model and physically interpret the PDF as a distribution one needs to know properties of the
bound state (here the proton) on the light front. Such properties are encoded in the light-cone wavefunctions. 
We shall give a preliminary introduction to these light-cone wavefunctions in Section \ref{bound}.

\subsection{Power counting with a twist}

Above we have reviewed DIS in a manner intrinsically tied to the light cone. This will guide us throughout the first few Chapters
as we investigate properties of bound states and wavefunctions on the light cone.  
There exists an alternate, covariant way to view DIS structure functions that will be required in the later Chapters
when we take up the calculation of double distributions (DDs). 
Moreover this covariant representation will allow us some insight as to the peculiarities of light-cone power counting.
Instead of re-deriving DIS amplitudes in the new framework, we shall take our result above Eq.~\eqref{eqn:f1} and 
merely extract from it a new perspective.

Consider the $n^{\text{th}}$ moment of the PDF $f_1(x)$. We can write it in the form
\begin{equation}
\int dx \, x^n f_1(x) = \frac{1}{2} \int dx  \int \frac{d\l}{2 \pi} 
\left[
\left( -i \frac{d}{d \l} 
\right)^n 
e^{i \l x}
\right]
\langle P | \ol \psi(0) \gamma^+ \psi(\l n) | P \rangle
.\end{equation}
Performing $n$-integrations by parts enables evaluation of the $x$-integral, which finally appears as 
$\int \frac{dx}{2 \pi} e^{i \l x} = \delta(\l)$. 
Thus we have
\begin{equation} \label{eqn:nnn}
\int dx \, x^n f_1(x) = \frac{1}{2} n_\mu n_{\mu_1} \ldots n_{\mu_n} 
\langle P | \ol \psi(0) \gamma^\mu (i \partial^{\mu_1}) \cdots (i \partial^{\mu_n}) \psi(0) | P \rangle
.\end{equation}

This suggests we consider the tower of local operators $\cO^{(n)}$ defined by
\begin{equation} \label{eqn:twisttwo}
\cO^{\mu \mu_1 \ldots \mu_n} = \ol \psi(0) \gamma^{\{\mu} i \overset{\leftrightarrow}{D}{}^{\mu_1}  \cdots i \overset{\leftrightarrow}{D}{}^{\mu_n\}} \psi(0)
,\end{equation}
where the action of ${}^{\{ \ldots \}}$ on Lorentz indices produces the symmetric traceless part of the tensor.
The derivative $D^\mu = \partial^\mu + i g A^\mu$ is the QCD gauge covariant derivative. 
In extracting the form of $\cO^{(n)}$ in Eq.~\eqref{eqn:twisttwo} from Eq.~\eqref{eqn:nnn}, we made a number of trivial modifications for 
free. Firstly we made the partial derivative act symmetrically $\overset{\leftrightarrow}{\partial} = \overset{\rightarrow}{\partial} - \overset{\leftarrow}{\partial}$
since the $\ol \psi$ field has no $\l$ dependence in Eq.~\eqref{eqn:f1}. Next we upgraded the partial derivatives to gauge covariant derivatives to render the  
tower of operators $\cO^{(n)}$ gauge invariant. Symmetrization with respect to all Lorentz indices was also done at no cost since the
prefactor of $(n+1)$-$n_\mu$ vectors in Eq.~\eqref{eqn:nnn} filters out the symmetric part of the tensor operators. 
Notice that symmetrization allows us to identify the operator $\cO^{(1)}$ with the quark part of the QCD energy-momentum tensor.
Lastly tracelessness, which allows us to classify $\cO^{(n)}$ irreducibly as spin-$(n+1)$ operators, is a free choice since any 
trace terms are proportional to $g^{\mu_i \mu_j}$ and must vanish when contracted with $n_{\mu_i} n_{\mu_j}$ since $n^2 = 0$.

Now we merely observe that because of Lorentz invariance, proton matrix elements of the $\cO^{(n)}$ operators have the form
\begin{equation} \label{bunchofPs}
\langle P | \cO^{\mu \mu_1 \ldots \mu_n} | P \rangle = 2 a_n P^{\{\mu} P^{\mu_1} \cdots P^{\mu_n \} }
,\end{equation}
where $a_n$ is a Lorentz scalar.\footnote{%
In general $a_n = a_n(\mu^2)$. As an alternative to our above discussion of DIS, we can view the scale dependence
as arising from the renormalization of the composite operators in Eq.~\eqref{eqn:twisttwo}.
Notice $\frac{d}{d \mu^2} a_0  = 0$ since $\cO^{(0)}$ is an operator with vanishing anomalous dimension.
} Recall that $n \cdot P = n \cdot p  = 1$. Thus we have
\begin{equation}
\int dx \, x^n f_1(x) = a_n
.\end{equation}
Moments of PDFs are proton matrix elements of local operators $\cO^{(n)}$. A natural question to ask, however, 
is why there are infinitely many such operators. Power counting seems to suggest that more $P^{\mu}$'s
in Eq.~\eqref{bunchofPs} means more suppression in the Bjorken limit.

The resolution of this paradox (and also the peculiarities of light-cone power counting) involves
fully appreciating the nature of the Bjorken limit. Let us denote an arbitrary operator of mass 
dimension $d$ and spin $s$ as $\cO(d,s)$. Unpolarized proton matrix elements of this operator 
will pick up $s$-powers of $P$ due to Lorentz covariance. 
But the net effect on the Compton amplitude $T^{\mu \nu}(q,P)$ must come from contracting these
$P$'s with $q$'s. Hence contributions to the physical cross section from $\cO(d,s)$ are proportional to
\begin{equation}
\langle P | \cO(d,s) | P \rangle \to 
\left( 
\frac{2 q \cdot P}{M Q}
\right)^s
\left(
\frac{M}{Q}
\right)^{d} 
=
\frac{1}{x^s} \left( \frac{M}{Q} \right)^{d - s}
,\end{equation}
where the equality utilizes the definition of Bjorken $x$, Eq.~\eqref{eqn:Bjorken}. 
Thus the relevant power counting in DIS comes with a \emph{twist}
\begin{equation}
\tau = d - s
.\end{equation}
The operators with the lowest twist $\tau$ give the leading contribution to DIS.\footnote{%
The renormalization of these operators does not modify the twist expansion because operators of different
twist do not mix under evolution. One has the additional feature, however, that operators of the same twist 
(and quantum numbers) do mix under evolution, e.g.~to $\cO(\a_s)$ one must consider mixing of $\cO^{(n)}$
above with the twist-two gluon operators
\begin{equation} \notag
\cO_g^{\mu \mu_1 \ldots \mu_n \nu} = F^{\a \{ \mu} i \overset{\leftrightarrow}{D}{}^{\mu_1}  \cdots i \overset{\leftrightarrow}{D}{}^{\mu_n}
F^{\nu \}} {}_{\a}
.\end{equation}
} 
The operators $\cO^{(n)}$
are twist-$2$ contributions and are thus \emph{all} relevant for DIS. The function $f_4(x)$
encountered above is a twist-$4$ contribution. In summary, the calculation of model quark distributions
from their moments is an interesting study of the interplay between the light-cone formalism and covariance.
We shall take this much further in Chapters \ref{chap:DDs} and \ref{chap:proton}, where we 
obtain model double distribution functions.

\section{Bound states} \label{bound}

Above we have reviewed DIS from two perspectives: a covariant development formulated in terms of matrix elements of local twist-2
operators, and also from the perspective of dynamics along the light cone. For applications throughout the rest of this 
work, we need to introduce light-cone wavefunctions and investigate how the features of covariant field theory manifest 
themselves in the many-body light-cone wavefunctions. To this end, we first discuss in Section \ref{BSE} two-body bound states in covariant 
field theory. Next in Section \ref{toyboost} we provide an exactly soluble model wherein one can compare the covariant, light-cone 
and familiar instant form wavefunctions. This involves solving for the model's wavefunction in the rest frame and then boosting to 
an arbitrary frame.

\subsection{Bethe-Salpeter equation} \label{BSE}

\begin{figure}
\begin{center}
\epsfig{file=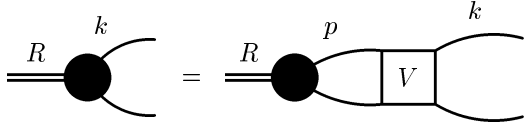}
\end{center}
\caption[Diagrammatic representation of the Bethe-Salpeter equation.]{Diagrammatic representation of the Bethe-Salpeter equation. 
The blob represents the vertex function $\Gamma$.}
\label{fBS}
\end{figure}

In terms of fully covariant operators, the Lippmann-Schwinger equation for the two-particle transition matrix $T$ appears as
\begin{equation} \label{LS}
T = V + V G T.
\end{equation}
Above, $V$ is the irreducible two-particle scattering kernel and $G$ is the completely disconnected two-particle propagator (which 
is merely the product of two single-particle propagators). 
A pole in the $T$-matrix (at some $R^2 = M^2$, say) corresponds to a two-particle bound state. Investigation of the pole's residue
gives an equation for the bound state vertex $\Gamma$
\begin{equation} \label{eqn:Gamma}
\Gamma = V G \Gamma.
\end{equation}
The amplitude $\Psi$ is defined as $G \Gamma$ and hence satisfies a similar equation, the Bethe-Salpeter equation \cite{Schwinger:1951ex,Gell-Mann:1951rw,Salpeter:1951sz}
\begin{equation} \label{eqn:BetheSalpeter}
\Psi = G V \Psi
.\end{equation}

Following \cite{Sales:1999ec}, it is convenient to denote quantities able to be rendered in position or momentum space with 
bras and kets. We will employ this notation only for quantities that have been stripped of their overall momentum-conserving delta functions, 
for example $\Gamma(k,R)$ is defined by $\Gamma(k,R) = \langle k | \Gamma_{R} \rangle $. Here we have used $R$ as a label for the bound state for which $R^2 = M^2$,
and $k$ labels the four momentum of the first particle. The same is analogously true for $\Psi(k,R)$.  
For the disconnected two-particle propagator, we define first $G(R)$ which is the disconnected propagator of total momentum $R$ defined through the relation
$\langle R^\prime | G | R \rangle = (2 \pi)^4 \delta^{(4)}(R^\prime - R) G(R)$. Additionally we remove the momentum conserving delta function 
between initial and final states of particle one in the disconnected Green's function to arrive at $G(k,R)$, i.e.~$ \langle k | G(R) | p \rangle = 
 (2\pi)^4 \delta^{(4)}(k - p) G(k,R)$. Using $\langle R^\prime | V | R \rangle = (2\pi)^4 \delta^{(4)}(R - R^\prime) V(R)$, we see the momentum-space 
kernel $V(k,p;R)$ is $V(k,p;R) = \langle k | V(R) | p \rangle$. 
Thus rendered in momentum space, the Bethe-Salpeter equation \eqref{eqn:BetheSalpeter} reads (see Figure \ref{fBS})
\begin{equation} \label{eqn:BSE}
\Psi(k,R) = \int \frac{d^4 p}{(2\pi)^4} G(k,R) V(k,p;R) \Psi(p,R)  
.\end{equation}

Armed with the Bethe-Salpeter amplitude $\Psi$, one can calculate field-theoretic bound-state matrix elements by taking the appropriate residues of 
four-point Green's functions. These matrix elements may ultimately require knowledge of higher-point functions which then must be solved for 
consistently in the same dynamics. The Bethe-Salpeter amplitude $\Psi$ is in some ways the covariant analogue of the Schr\"odinger wavefunction. 
While the features of relativistic field theory (in particular: particle creation and annihilation, retardation effects$\ldots$) 
make the exact analogy impossible, in the non-relativistic limit, one can show that the BSE reduces to the Schr\"odinger equation.

\subsection{Toy wavefunction and boost} \label{toyboost}

\begin{figure}
\begin{center}
\epsfig{file=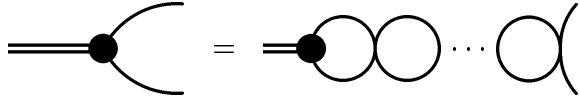,width=3in}
\end{center}
\caption[Bethe-Salpeter equation for a point interaction.]{Bethe-Salpeter equation for a point interaction. The state is bound by the infinite chain of bubbles.}
\label{f:toy}
\end{figure}

Above we have derived the covariant BSE for two-body bound states.  
In this Section, we consider a toy model for the BSE that is exactly soluble. We shall for simplicity
consider a scalar bound state composed of two scalar particles.  The solution will enable us 
to compare and contrast instant form dynamics and light-front dynamics all while maintaining exact covariance. 
The instant and front form of dynamics are reviewed in Appendix \ref{chap:forms}.

One can obtain the simplest soluble BSE equation by choosing a point-like interaction for $V(k,p;R)$ in Eq.~\eqref{eqn:BSE}, 
namely $V(k,p;R) = i g$, where $g$ is a coupling constant. 
The two scalar particles that make up the scalar bound state thus interact infinitely many times according to the BSE
to bind the state. For a point-like interaction, a bubble chain is generated by the BSE and is shown in Figure \ref{f:toy}. 
With this choice of interaction, the bound state equation simplifies tremendously. Since the kernel is independent of momentum, 
the only $p$ dependence that remains in Eq.~\eqref{eqn:BSE}
is in $\Psi(p,R)$ and this quantity is subsequently integrated over all $p$. The integration merely produces some number that can be 
absorbed into the overall normalization of the wavefunction (along with the coupling constant in the kernel). 
Thus we are left with the solution
\begin{equation} \label{eqn:soln}
\Psi(k,R) = i G(k,R)
,\end{equation}
where the overall constant is a matter of taste. Comparing with Eq.~\eqref{eqn:Gamma}, we can also trivially solve for the
bound state vertex function: $\Gamma(k,R) = i$. We shall see later that a point-like vertex is a convenient choice 
for phenomenological applications because the crossing properties are trivial. The Bethe-Salpeter equation 
for $\Gamma(k,R)$ determines the mass $M^2 = R^2$ of the bound state via the consistency equation
\begin{equation}
1 = i g \int \frac{d^4 p}{(2\pi)^4} G(k,R) 
.\end{equation}

Now for simplicity we shall neglect self-energy contributions in the propagators and choose the single 
particle propagator to have the basic Klein-Gordon form. The two-particle disconnected propagator 
is a product of these Klein-Gordon propagators and hence by virtue of Eq.~\eqref{eqn:soln} the covariant wavefunction is
\begin{equation}
\Psi(k,R) = -i [k^2 - m^2 + i \varepsilon]^{-1} [(k-R)^2 - m^2 + i \varepsilon]^{-1}
.\end{equation}
Here we have labeled the constituent mass by $m$. 
If we imagine taking the Fourier transform of this wavefunction, we have a four-dimensional 
analogue of the Schr\"odinger wave function. There is an important distinction, however. We also know the
time dependence of the wave function---the time evolution governed by the Hamiltonian operator
is automatically included because of the necessity of covariance. Moreover, we know from the Poincar\'e
algebra that there are other dynamical operators besides the energy\footnote{%
Dirac advocated the use of ``Hamiltonian''
for all operators that are dynamical, of which the energy operator is one.
}.
As to which operators are Hamiltonians and thus which are kinematical depends upon the form of dynamics chosen.
The instant and front forms are reviewed in Appendix \ref{chap:forms}.
We shall consider first the familiar instant form.

In the instant form of dynamics, the energy and the Lorentz boosts are the dynamical operators. The initial 
conditions are specified on the boundary $x^0 = 0$, as in classical and quantum mechanics. The time evolution of the system
is governed by the energy operator and involves non-trivial effects due to the system's dynamics. Moreover
the relativistic properties of the system also are complicated by the dynamics in the instant form. This is 
reflected in the fact that the generators of Lorentz boosts are Hamiltonians. If we write out the Fourier transform 
of the Bethe-Salpeter wavefunction
\begin{equation} \label{eqn:fourier}
\Psi(x,R) = \int \frac{d^4 k}{2 \pi}  e^{i x \cdot k} \, \Psi(k,R)
,\end{equation}
we see that the instant form wavefunction can be obtained by integration over the energy $k^0$.
Thus at time $x^0 =0$ and in the rest frame of the bound state, we have
\begin{equation} \label{eqn:rested}
\psi^0(\mathbf{k},\mathbf{0}) = \int \frac{d k^0}{2 \pi} \Psi(k,R) \Big|_{R^\mu = (M,\mathbf{0})} 
.\end{equation}

Given our solution to the BSE Eq.~\eqref{eqn:soln}, we can carry out this projection onto the initial surface.
The integration can be done using the residue theorem bearing in mind the four poles of the integrand
\begin{align}
k^0_{a,\ol a} & = \pm \sqrt{\mathbf{k}^2 + m^2} \mp i \varepsilon \\
k^0_{b,\ol b} & = M \pm \sqrt{\mathbf{k}^2 + m^2} \mp i \varepsilon
.\end{align} 
The instant form wavefunction is $2 \pi i [ \res(k^0_{\ol a}) + \res(k^0_{\ol b})]$. Explicitly
we have
\begin{equation} \label{eqn:rest}
\psi^0 (\mathbf{k},\mathbf{0}) = \frac{1}{\sqrt{\mathbf{k}^2 + m^2}} \, \, \frac{1}{M^2 - 4 (\mathbf{k}^2 + m^2)}
\end{equation}
and overall multiplicative constants have been neglected. 
Notice the wavefunction is manifestly rotationally invariant. 
This is indicative of the kinematic nature of the generators of rotations in the instant form.

From the point of view of relativistic quantum mechanics, one might wonder what the wavefunction is in an 
arbitrary frame. For example, in considering electromagnetic form factors, the final state is boosted relative to the initial one
and thus to model the form factors phenomenologically, one requires knowledge of the boost.
To deduce the boosted wavefunction, one must solve the complicated dynamical equation
\begin{equation} \label{eqn:boosted}
|\psi^0(\mathbf{k},\mathbf{R}) \rangle = e^{i \mathbf{K} \cdot \mathbf{R}} |\psi^0(\mathbf{k},\mathbf{0}) \rangle
,\end{equation}
where $\mathbf{K}$ are the generators of Lorentz boosts. For any three-vector $\mathbf{k}$, let us define
$E(\mathbf{k}) = \sqrt{\mathbf{k}^2 + m^2}$. 
The boosted wavefunction which solves Eq.~\eqref{eqn:boosted} can be compactly written as
\begin{multline} \label{eqn:boost}
\psi^0(\mathbf{k},\mathbf{R}) = \frac{1}{E(\mathbf{R}) + E(\mathbf{k}) - E(\mathbf{R} - \mathbf{k})} 
\left\{
\frac{1}{E(\mathbf{k})
\big[
E(\mathbf{R}) + E(\mathbf{k}) + E(\mathbf{R} - \mathbf{k})
\big]}
\right. \\
+
\left.
\frac{1}{ E(\mathbf{R} - \mathbf{k})
\big[
E(\mathbf{R}) - E(\mathbf{k}) - E(\mathbf{R} - \mathbf{k})
\big]}
\right\}
.\end{multline}
To obtain the boosted wavefunction, we chose not solve Eq.~\eqref{eqn:boosted}. We merely returned to the covariant Bethe-Salpeter
wavefunction and evaluated it at $x^0 = 0$ using Eq.~\eqref{eqn:rested} in an arbitrary frame. The non-transparent 
relation between the rest frame wavefunction Eq.~\eqref{eqn:rest} and the boosted wavefunction Eq.~\eqref{eqn:boost}
indicates the dynamical nature of the boost.

In the front form of dynamics, one is interested in the properties of physical states along the advance of a wavefront of light.
The objects of front form dynamics are the light-cone wave functions which are projections onto the initial surface $x^+=0$.
In analogy with the instant form, one refers to $x^+$ as light-cone time, and its Fourier conjugate $k^-$ as light-front
energy. In the front form, the energy is a Hamiltonian along with two rotation operators corresponding to two independent 
rotations of the wavefront of light.  In contrast with the instant form, light-front Lorentz boosts are kinematical.

By virtue of Eq.~\eqref{eqn:fourier}, the light-cone wavefunction is
\begin{equation}
\psi(x, \mathbf{k}^\perp_{\text{rel}}) = \int \frac{d k^-}{2 \pi} \Psi(k,R)
,\end{equation} 
where $x$ is the fraction of the plus momentum carried by the first particle, $x = k^+ / R^+$ and
the relative transverse momentum is $\mathbf{k}^\perp_{\text{rel}} = \kperp - x \mathbf{R}^\perp$. 
Evaluation of the integral is standard and actually simpler than in the instant form because, on the light-cone, 
there are only two energy poles. Taking the appropriate residue, the light-cone wave function for our toy model is
\begin{equation}
\psi(x, \mathbf{k}^\perp_{\text{rel}}) = \frac{\theta[x(1-x)]}{R^+ x(1-x)}
\,  \frac{1}{M^2 - \frac{\mathbf{k}^\perp_{\text{rel}}{}^2 + m^2}{x(1-x)}}
.\end{equation}
Notice the wave function maintains only a cylindrical symmetry and further that the relation of the 
rest frame wave function to the boosted wavefunction is all compactly written above in 
terms of $\mathbf{k}^\perp_{\text{rel}}$. These facts are
of course indicative of the kinematic subgroup of the Poincar\'e generators
(the subgroup includes rotation within the initial surface and light-front boosts). Accordingly 
the full rotational symmetry of the rest frame wavefunction is not manifest, \emph{cf} Eq.~\eqref{eqn:rest}
and Eq.~\eqref{eqn:soln}.

One should now inquire as to how the two wavefunctions $\psi^0$ and $\psi$ are related to each other.
We can try to turn the light-cone wavefunction into the rest frame wavefunction. 
In the literature, this is accomplished by introducing an auxiliary variable $k^3$. 
So that $k^3$ has a physical interpretation in terms of the $z$-component of momentum, 
one uses the two-body center of mass relation 
\begin{equation}  \label{eqn:bad}
x = \frac{k^+}{R^+} = \frac{1}{2} + \frac{k^3}{2\sqrt{\mathbf{k}_\perp^2 + (k^3)^2 + m^2}} 
.\end{equation}
Here $R^+$ is taken to be the sum of the free constituents' plus momentum. 
There is no binding effect in the total plus momentum because longitudinal 
translations are kinematic in the front form of dynamics.
Inverted this relation between $x$ and $k^3$ reads \cite{Tiburzi:2000je}
\begin{equation} \label{eqn:variable}
k^3 = \left( x - \frac{1}{2} \right) \sqrt{\frac{\mathbf{k}_\perp^2 + m^2}{x(1-x)}}
.\end{equation}

Simple algebra yields the relation
\begin{equation}
x(1-x) = \frac{\mathbf{k}_\perp^2 + m^2}{4 E(\mathbf{k}^2)}
\end{equation}
from which we deduce
\begin{equation}
\psi(k^3,\kperp) =  \frac{E(\mathbf{k})^2}{\mathbf{k}_\perp^2 + m^2} \,  \frac{1}{M^2 - 4 E(\mathbf{k})^2}
.\end{equation}
This bears a resemblance to the instant form wavefunction in the rest frame Eq.~\eqref{eqn:rest}. 
If we include the Jacobian 
\begin{equation}
\frac{\partial x}{\partial k^3} = \frac{\mathbf{k}_\perp^2 + m^2}{2 E(\mathbf{k})^3}
\end{equation}
as a multiplicative prefactor, then the wavefunctions agree. There is, however, no reason why this should work:
only a factor of $\sqrt{\frac{\partial x}{\partial k^3}}$ is justified to preserve the norm. 
When one considers not the wavefunctions themselves but the relevant amplitudes, enforcing the constraint of Poincar\'e
invariance on the physical amplitude can lead to the proper factors to include, e.g.~for form factors 
in the Drell-Yan frame the proper factor is $\sqrt{\frac{\partial x}{\partial k^3}}$.
The variable transformation Eq.~\eqref{eqn:variable} is additionally convenient for the evaluation of integrals.
Notice, the light-cone wavefunction itself can be evaluated in the rest frame, however, it is not connected 
to the rest frame wavefunction in the instant form by a non-singular boost. For amplitudes involving a fixed 
number of particles, the above correspondence Eq.~\eqref{eqn:variable} can be used along with constraints from 
Poincar\'e covariance to determine the relation to the instant form rest-frame wavefunction. 
In general, however, there is no relation between field theoretic amplitudes and 
Poincar\'e covariant representations of relativistic dynamics.

While there is no general way to unboost the light-cone wavefunctions,
the instant form wavefunction $\psi^0(\mathbf{k},\mathbf{R})$ can be boosted to the infinite momentum 
frame (IMF).  As a result of this procedure, one recovers the light-cone wavefunction $\psi(x,\kperp)$.  
Originally field theories were
re-formulated in the infinite momentum frame \cite{Chang:1973xt,Chang:1973qi,Yan:1973qf,Yan:1973qg}
as a way to simplify covariant perturbation theory. The connection to light-cone quantization was made later.

To boost the wavefunction to infinite momentum, we first simplify matters by choosing $\mathbf{R}^\perp =0$. 
In the IMF, $R^3 \to \infty$ while $R^2 = M^2$ is kept fixed. Thus one has
\begin{equation}
R^0 = R^3 
\left[
1 + \frac{M^2}{2 (R^3)^2} 
+ \ldots
\right]
\end{equation} 
Notice in this limit, we have $R^+ \approx \sqrt{2} R^3 \to \infty$ while
$R^- \approx M^2 / 2 \sqrt{2} R^3 \to 0$.

Working with the constituent's momentum, we use the parametrization
\begin{equation}
\mathbf{k} = \kperp + x R^3 \, \mathbf{\hat{z}}
.\end{equation}
The energy factors then simplify in the IMF as
\begin{eqnarray}
E(\mathbf{k}) &=& x R^3 + \frac{\mathbf{k}^2_\perp + m^2}{ 2 x R^3} + \ldots \label{Eone} \\
E(\mathbf{R -k}) &=& (1-x)R^3 + \frac{\mathbf{k}^2_\perp + m^2 }{2 (1-x) R^3} + \ldots \label{Etwo}
.\end{eqnarray}
Carrying out the boost to infinite momentum on the wavefunction in Eq.~\eqref{eqn:boost} yields
the IMF wavefunction which we denote $\psi^0(\mathbf{k},\infty)$. Using Eqs.~\eqref{Eone} and \eqref{Etwo}
we find
\begin{equation}
\psi^0(\mathbf{k},\infty) = \frac{1}{R^3 x(1-x)} \, \frac{1}{M^2 - \frac{\mathbf{k}^2_\perp + m^2}{x (1-x)}}
,\end{equation}
which since $R^+ \propto R^3$ is just the light-cone wavefunction evaluated for 
$\mathbf{R}^\perp = 0$. Unboosting the wavefunctions from the IMF is generally impossible but can be 
done in the context of a few specific amplitudes for which a correspondence between relativistic quantum mechanics and 
field theory can be made.

In summary, this simple covariant model for the BSE Eq.~\eqref{eqn:soln} has allowed us to explore both the instant and front form 
wavefunctions. We saw how the structure of these wavefunctions is related to the respective kinematic subgroups of the Poincar\'e algebra. 
Moreover, a fully covariant starting point allowed us a simple way to correctly formulate the three-dimensional dynamics. We shall employ this
philosophy throughout the work.

\chapter{Light-Front Bethe-Salpeter Equation} \label{chap:LFBS}

Above we have seen the utility of light-front dynamics in processes at large momentum transfer. Additionally we investigated the properties of 
light-cone wavefunctions in a simple exactly soluble model for the covariant BSE. 
This Chapter concerns bound states of two particles in the light-front formalism, specifically of interest are current matrix elements between bound states. 
We approach the topic, however, from covariant perturbation theory in order to dispel rampant misconceptions about bound states
on the light front. 
As demonstrated by the tremendous undertaking of \cite{Ligterink:1995tm,Ligterink:1995wk}, one can derive light-front perturbation theory for scattering states
by projecting covariant perturbation 
theory onto the light cone, thereby demonstrating their equivalence---including the delicate issue of renormalization. As to 
the issue of light-front bound states, a reduction scheme for the Bethe-Salpeter equation recently appeared \cite{Sales:1999ec,Sales:2001gk} 
that produces a kernel calculated in light-front perturbation theory. This reduction scheme makes formal earlier observations about the 
connection between the Bethe-Salpeter equation and the light-front bound state equation \cite{Brodsky:1985vp}. 
For the purpose of simplicity, in this Chapter we consider only bound states of two scalar particles interacting via the exchange of a massive scalar 
in the $(1+1)$-dimensional ladder model.

Our main consideration is to extend the reduction to current matrix elements to investigate carefully valence and non-valence contributions in the 
light-front Bethe-Salpeter formalism. To do this, we calculate our model's form factor. As calculations in the light-front reduction
are covariant only when summed to all orders, our results violate Lorentz symmetry and we choose to extract the form factor from the 
plus-component of the electromagnetic current in order to make contact with the Fock space representation. 
Moreover,  in calculating $(1+1)$-dimensional form factors we cannot choose a frame of reference where Z-graphs vanish. 
This enables us to investigate their contribution, 
which in $(3+1)$ dimensions has a variety of applications such as to generalized parton distributions, which we will detail in Chapter \ref{chap:GPDs}. 
Z-graph contributions haunt 
light-front dynamics since non-valence properties of the bound state are involved, so that 
valence wavefunction models cannot be utilized directly.  On one hand, the light-cone Fock representation provides expressions for the 
Z-graph contributions in terms of Fock component overlaps which are non-diagonal in particle number \cite{Diehl:2000xz,Brodsky:2000xy}. 
While on the other hand, vertices which cannot be related to the valence wavefunction 
(coined as \emph{non-wavefunction vertices} 
in \cite{Bakker:2000rd}) 
appear in the light-front Bethe-Salpeter formalism. A variety of ways have been proposed for dealing with these non-wavefunction vertices
\cite{Einhorn:1976uz,Tiburzi:2001je} including attempts to model the 
covariant vertex \cite{deMelo:1997cb,Jaus:1999zv}, or (when possible) estimating the contribution from higher Fock states \cite{Demchuk:1996zx}.
In a Poincar\'e covariant framework, Z-graph contributions involve a dynamical light-front rotation and represent a similar formidable problem.

Below we show that non-wavefunction vertices are supplanted by contributions from higher Fock states in light-front 
time-ordered perturbation theory (provided the interaction has light-cone time dependence). 
In essence contributions from non-wavefunction vertices are reducible and should 
only be used when the interaction is (or is approximately) instantaneous\footnote{We shall often refer to interactions and
vertices merely as instantaneous if they are independent of \emph{light-cone} time, or equivalently \emph{light-cone} energy.
Similarly we are not careful about referring to light-cone time and light-cone energy as time and energy, respectively.}. 
This constitutes a replacement theorem for non-wavefunction vertices which trivially extends to $(3+1)$ dimensions.
When one works from covariant perturbation theory, the coupled tower of Dyson-Schwinger equations gives rise to the light-cone Fock components
\cite{Frederico:2003zk,Frederico:2003tu}.
Thus when the Green's functions are treated properly and the light-front BSE is derived, instead of postulated, non-wavefunction contributions
are absent.

The organization of this Chapter is as follows. First in Section \ref{LFCQM} we present the issue of non-wavefunction vertices and 
energy poles of the Bethe-Salpeter vertex focusing on a typical light-front constituent quark model as an example. For such models, 
non-wavefunction vertices are required to express the form factor. We find the commonly used assumptions in quark models necessitate 
vertices not only without energy poles but without energy dependence. Hence covariance is generally lost.
Next in Section \ref{reduce},
we review the reduction of the Bethe-Salpeter equation presented in \cite{Sales:1999ec} focusing on the energy poles of the vertex.
We derive an interpretation of the reduction as a procedure for approximating the poles of the vertex.  
Additionally in Section \ref{normmy}, we show how the normalization of the covariant Bethe-Salpeter equation 
turns into a familiar many-body normalization in the light-front reduction. 
The normalization condition resembles a diagonal matrix element of a pseudo current
and has contributions from higher Fock states.  We calculate the explicit normalization condition for the ladder model at 
next-to-leading order in  perturbation theory. The connection to the familiar many-body Fock state normalization is made in 
Appendix \ref{oftopt}.
In the following Section (\ref{curry}), 
we construct the current to be used with the reduced formalism. 
In Section \ref{meat} the $(1+1)$-dimensional ladder model is presented and in Section \ref{formmy} we compare the calculation of the form factor for the model using two 
different paths to the reduction. The comparison allows us to see when non-wavefunction vertices can be efficiently used. 
Lastly we summarize our findings in Section \ref{sum}.

\section{Light-front reduction}

To derive a bound state equation on the light front, we must investigate the reduction of the covariant BSE Eq.~\eqref{eqn:BetheSalpeter}
to the hypersurface $x^+=0$. Following the toy wavefunction example presented in Section \ref{toyboost}, we must integrate over the light-cone
energy to obtain the proper three-dimensional boundary value. The clear first step to handling such a reduction approximately 
is to make simplifying assumptions about the dependence on the light-cone energy to enable the projection. Near independence from light-front energy corresponds 
to an instantaneous approximation with respect to light-front time. This is where we begin our discussion.

\subsection{Poles of the Bethe-Salpeter vertex} \label{LFCQM}

To introduce the reader to non-wavefunction vertices and instantaneous approximations, we focus on
light-front constituent quark models. We start by writing down the covariant equation for the meson\footnote{%
Here we use meson to denote a general two-particle bound state. Mesons just happen to be convenient examples.}
vertex function $\Gamma$. In $(1+1)$ dimensions, it satisfies a simple BSE (see Figure \ref{fBS}):
\begin{equation} \label{BS}
\Gamma(k,R) =  i \int \frac{d^2p}{(2\pi)^2} V(k,p) \Psi_{BS}(p,R),
\end{equation}
in which we have defined the Bethe-Salpeter wavefunction $\Psi_{BS}$ as
\begin{equation}\label{psiBS}
\Psi_{BS}(k,R) = G(k,R) \Gamma(k,R),
\end{equation}
with the two-particle disconnected propagator $G(k,R) = d(k) d(R-k)$. For scalars of mass $m$, the renormalized, single-particle propagator $d$ 
has a Klein-Gordon form
\begin{equation}
d(k) = \frac{i}{(k^2 - m^2)[1+ (k^2 - m^2)f(k^2)]+ i \epsilon},
\end{equation}
where the residue is $i$ at the physical mass pole and the function $f(k^2)$ characterizes the renormalized, one-particle irreducible self-interactions. 
To simplify the comparison carried out in Section \ref{meat}, we shall ignore $f(k^2)$. Above $V$ is the irreducible two-to-two scattering
kernel which we shall refer to inelegantly as the interaction potential.

Now we imagine the initial conditions of our system are specified on the hypersurface $x^+= 0$. 
As we saw for DIS, correlations between field operators 
evaluated at equal light-cone time turn up in hard processes for which this choice of initial surface is natural. In order to project out
the initial conditions (wavefunctions, \emph{etc}.) of our system, we must perform the integration over the Fourier conjugate to $x^+$, namely $k^-$.
For instance, our concern is with the light-front wavefunction $\psi(x)$ defined as the projection of the covariant Bethe-Salpeter wavefunction onto $x^+ = 0$,
\begin{equation} \label{psilf}
\psi(x) = 2 R^+ x(1-x) \int \frac{dk^-}{2\pi} \Psi_{BS}(k,R),
\end{equation}
with $ x = \kplus / R^+$. The overall multiplicative factor in the definition of $\psi(x)$ is an arbitrary choice.

Looking at Eq. \eqref{psiBS}, in order to project the wavefunction exactly, we must know the analytic structure of the bound-state vertex function. If the vertex function $\Gamma(k,R)$ had no poles in $\kminus$, then our task would be simple: the light-front projection of $\Psi_{BS}$ would pick up contributions only from the poles of the propagator $G(k,R)$. Next we observe from Eq.~\eqref{BS}, that the $\kminus$ dependence of the interaction $V(k,p)$ must give rise to the $\kminus$ poles of the vertex function $\Gamma(k,R)$. Hence an instantaneous interaction gives rise to an instantaneous vertex and a simple light-front 
projection (see, e.g., \cite{Einhorn:1976uz}).

On the other hand, constituent quark models often assume a less restrictive simplification of the analytic structure of the vertex 
(see, e.g., \cite{Jaus:1999zv}) in order to permit the light-cone projection. We shall show that this assumption
along with the presumed covariance of the quark model implies instantaneous vertices. For any momentum $p$, let us denote
the $(1+1)$-dimensional on-shell energy $p^-_{\text{on}} = m^2 / 2 p^+$. The propagator $G(k,R)$ has two poles
\begin{equation} \label{valpoles}
 \begin{cases}
 k^-_{a} = k^-_{\text{on}} - \frac{\ie}{x}\\
 k^-_{b} = R^{-} + (k - R)^{-}_{\text{on}} - \frac{\ie}{x - 1}.
 \end{cases} 
\end{equation} 
Notice that although we work in $(1+1)$ dimensions, the results are trivial to extend to $(3+1)$ dimensions because the
imaginary parts of poles have precisely the same dependence on (only) the plus-momenta. This remark applies not only to this
Section but to the rest of this Chapter.

In constituent quark models that attempt to find a field theoretic basis, $\Gamma$ is assumed to have no poles in the upper-half $k^-$-plane for $0<x<1$. 
Since the light-front wavefunction is 
proportional to $\theta[x(1-x)]$, in light of Eq.~\eqref{valpoles} we further require any possible poles of $\Gamma$ to lie in the upper-half plane for $x<0$ 
and in the lower-half plane for $x>1$. With these restrictions Eq.~\eqref{psilf} dictates the form of the constituent quark wavefunction
\begin{align} \label{g}
\psi(x) = \dw (x | M^2) \Gamma(k_{b},R) \theta[x(1-x)],
\end{align}
where we have defined the Weinberg propagator as
\begin{equation}
\dw (x|M^2) = \frac{1}{M^2 - \frac{m^2}{x(1-x)}}
\end{equation}
and used the abbreviation $\Gamma(k_{b},R)$ to denote evaluation at the energy pole $k^- = k^-_{b}$ appearing in Eq.~\eqref{valpoles}.

\begin{figure}
\begin{center}
	\epsfig{file=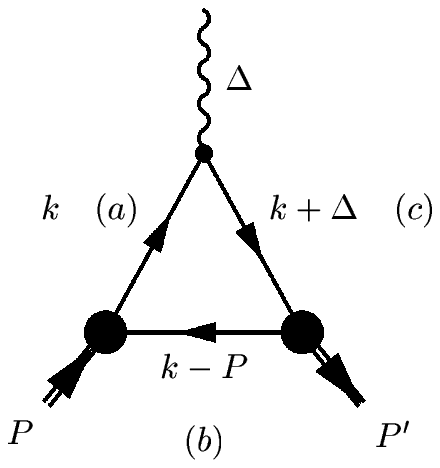,height=1.5in}
  	\caption[Covariant diagram for the electromagnetic form factor.]{\label{f:tri} Covariant diagram for the electromagnetic form factor. Labels $a,b$ and $c$ denote
subscripts used for on-shell energy poles.}
\end{center}
\end{figure}

When we calculate the (elastic) electromagnetic form factor for these constituent quark models (see Figure \ref{f:tri}), we are confronted with more poles. 
Let $\Delta^+ = - \zeta P^+ < 0$ be the plus-component of the momentum transfer between initial and final state mesons.  
The form factor is then
\begin{equation} \label{ff1}
F(t) \propto \int \frac{(2 x - \zeta) \Gamma(k,P) \Gamma^*(k+\Delta,P+ \Delta) \; d^2 k}{[k^2 -m^2][(P-k)^2 - m^2][(k+\Delta)^2 -m^2]}
\end{equation}
The $k^-$-poles from the propagators are $k^-_{a}, k^-_{b}$ defined in \eqref{valpoles} and 
$k^-_{c} = - \Delta^- + (k+\Delta)^-_{\text{on}} - \frac{\ie}{x - \zeta}$.

Given this pole structure, the contributions to $F(t)$ are proportional to $\theta[x(1-x)]$. In the region $x<\zeta$, closing the contour in the 
upper-(lower-) half plane will enclose possible poles of the final (initial) vertex. 
As one can see from considering the region $x>\zeta$, the form factor  can be determined solely from $\Res(k^-_{b})$. 
In $(3+1)$ dimensions, where we are free to choose frames in which $\zeta = 0$, this becomes the only contribution to the form factor. 
But if $\zeta \neq 0$, the additional poles from vertices in the region $x < \zeta$ are required by Lorentz invariance.

If one advocates no modification to the pole structure of Eq.~\eqref{ff1} due to the vertices,  then by closing the contour in the lower-half
plane, one picks up the residue at $k^-_{a}$ without any contribution from poles of the initial-state vertex. 
Such poles cannot lie in the upper-half plane
for $x<\zeta$ since the form of $\psi$ would be both frame dependent and contrary to that of Eq.~\eqref{g} 
[the $(3+1)$-dimensional version of which is employed by all light-front quark models]. 
Thus one is actually assuming there are no poles of the Bethe-Salpeter vertex, which we originally pointed out in \cite{Tiburzi:2001je}.

Returning to the definition of the wavefunction \eqref{psilf} under the premise of a vertex devoid of poles, we now find $\Gamma(k_b,P) = 
\Gamma(k_a,P)$, which we shall call pole symmetry. This symmetry is essential for making contact with the 
Drell-Yan formula \cite{Drell:1970km}, since for $x>\zeta$ both 
initial- and final- state vertices may be expressed in terms of wavefunctions. When $x<\zeta$, however, the final-state vertex
becomes $\Gamma^*(k_a + \Delta, P + \Delta)$ which cannot be expressed in terms of $\psi^*(x^\prime)$ (where $x^\prime = \frac{x - \zeta}{1 - \zeta}$)
even with pole symmetry. Such a vertex we refer to as a non-wavefunction vertex. Understanding and dealing with such objects from 
the perspective of time-ordered perturbation theory is the primary goal of this Chapter.

Naturally constituent quark models presume the integral in Eq.~\eqref{ff1} converges. This in turn enables us to relate an initial-state
non-wavefunction vertex to the final-state non-wavefunction vertex encountered above via integrating around a circle at infinity. 
Equating the sum of residues $\Res (k^-_a) + \Res (k^-_b) + \Res (k^-_c) = 0$, we find
\begin{equation} \label{loop}
\frac{\Gamma^*(k_a + \Delta, P^\prime)}{\Gamma^*(k_b+\Delta,P^\prime)} (k^-_b - k^-_c)
- \frac{\Gamma(k_c,P)}{\Gamma(k_b,P)} (k^-_b - k^-_a) = k^-_a - k^-_c.
\end{equation}
This relation holds for the class of models for which the pole structure of Eq.~\eqref{ff1} is not modified by the vertices. 
The ratio structure of the vertices does not allow
for a common factor $k^-_a - k^-_c$ in the two terms in Eq.~\eqref{loop}. The equality then depends on delicate cancellations
between initial- and final- state vertices, which in general are unrelated.\footnote{%
In particular, one could consider the final state to be an excited state. While the process is not elastic, the analysis
goes through unchanged and the initial and final state vertices would not permit the delicate cancellations required by Eq.~\eqref{loop}.
} 
The philosophy of constituent quark model 
phenomenology is to choose the form of $\psi$ and hence the form of $\Gamma$. 
This rules out any such delicate cancellation.
Treating $\Gamma$ as free to be chosen, the equality Eq.~\eqref{loop} can only 
hold if both ratios are one. This yields the restriction
\begin{equation}
\frac{\partial}{\partial \Delta^-} \Gamma(k_c,P) = 0.
\end{equation}
But $\Delta^-$ only enters $\Gamma(k_c,P)$ through $k^- = k^-_c$. Hence $\Gamma$ is independent of $k^-$, which 
is a hidden assumption in light-front quark models.\footnote{%
There is another way to reveal this hidden assumption. The vertex $\Gamma$ cannot be both
covariant and devoid of $k^-$ poles for the integral in Eq.~\eqref{ff1} to exist.
}
Clearly this assumption is problematic for the calculation of form 
factors in frames where $\zeta \neq 0$. Furthermore, covariance cannot be borne in unless the vertex $\Gamma(k,R)$ is completely 
independent of $k^\mu$.

In general, of course, 
the vertex $\Gamma$ not only has light-front energy dependence but poles as well. As we shall see below, the light-front reduction 
of the Bethe-Salpeter equation is a procedure for approximating the poles of the vertex function. Moreover when applied to current 
matrix elements, which require further contributions from higher Green's functions, these poles generate higher Fock state contributions.

\subsection{The reduction scheme} \label{reduce}

To reduce the Lippmann-Schwinger equation (or any such Dyson-Schwinger equation) to a light-front version, 
we must introduce an auxiliary Green's function
$\Gt$ in place of $G$ (as in \cite{Woloshyn:1973wm}). Thus we have
\begin{equation} \label{TW}
T = W + W \Gt T,
\end{equation}
provided that
\begin{equation} \label{W}
W = V + V (G - \Gt) W. 
\end{equation}
Taking residues of Eq.~\eqref{TW} gives us an alternate way to express the bound state vertex function 
\begin{equation} \label{regamma}
\Gamma = W \Gt \Gamma. 
\end{equation}

To choose a light front reduction, $\Gt$ must inherently be related to projection onto the initial surface $x^+ = 0$. For simplicity, we denote
the integration $\int \frac{d\kminus}{2\pi} \langle \kminus | \mathcal{O}(R) = \Big| \mathcal{O}(R)$. With this notation, 
we will always work in $(1+1)$-dimensional momentum space for which the only sensible matrix elements of $\Big| \mathcal{O}(R)$ are of the form 
$\langle \kplus | \;  \Big| \mathcal{O}(R) | \pminus, \pplus \rangle$. 
The operator $\mathcal{O}(R) \Big|$ is defined similarly. 
The generalization of both operators to $(3+1)$ dimensions is obvious.  
For a useful reduction scheme (one that preserves unitarity), 
we must have $\Big| G(R) \Big| = \Big| \Gt(R) \Big|$. 
The simplest choice of $\Gt$ that results in time-ordered perturbation theory requires
\begin{equation} \label{gt}
\Gt(R) = G(R) \Big| g^{-1}(R) \; \Big| G(R),
\end{equation} 
where the reduced disconnected propagator $g(R)$ is defined by the matrix elements
\begin{equation} \label{lilg}
\langle x R^+ | \; \Big| G(R) \Big| \; |  y R^+ \rangle =  
\langle x R^+ | g(R) | y R^+ \rangle.
\end{equation}
Explicitly this forces
\begin{equation}
\langle x R^+ | g(R) | y R^+ \rangle =  2\pi \delta(xR^+ - yR^+) 
\theta[x(1-x)] \frac{2 \pi i}{2 R^+ x(1-x)} \dw(x|R^2).
\end{equation}
The inverse propagator $g^{-1}(R)$ can then be constructed, bearing in mind $g^{-1}(R)$ only exists in the subspace
where $g(R)$ is non-zero. Explicitly
\begin{equation} \label{lilginv}
\langle x R^+ | g^{-1}(R) | y R^+ \rangle =  2\pi \delta(xR^+ - yR^+) 
\theta[x(1-x)] \frac{2 R^+ x(1-x)}{2 \pi i} \dw^{-1}(x|R^2).
\end{equation}
This forces 
\begin{equation}
\langle x R^+ | g^{-1}(R) g(R) | y R^+ \rangle =  2\pi \delta(xR^+ - yR^+) \theta[x(1-x)],
\end{equation}
which is unity restricted to the subspace where the operators $g(R)$ and $g^{-1}(R)$ are defined.
We are more careful about this point than the authors \cite{Sales:1999ec} 
since the consequences of Eq.~\eqref{lilginv} are essential
for dealing with instantaneous interactions. Notice $\Gt(R)$ defined in Eq.~\eqref{gt} 
is consequently non-zero only for plus-momentum fractions between zero and one.

The reduced transition matrix $t(R)$ is
\begin{equation} \label{tred}
t(R) = g^{-1}(R) \Big| \Gt(R) T(R) \Gt(R) \Big| g^{-1}(R) 
\end{equation}
Taking the residue of Eq.~\eqref{tred} at $R^2 = M^2$, gives a homogeneous equation for the reduced vertex function $\gamma$
\begin{equation} \label{gammaredu}
| \gamma_{R} \rangle = w(R) g(R) | \gamma_{R} \rangle,
\end{equation}
where the reduced auxiliary kernel is
\begin{equation} \label{wred}
w(R) = g^{-1}(R) \Big| G(R) W(R) G(R) \Big| g^{-1}(R).
\end{equation}
Given this structure, the reduced kernel $w(x,y|R^2) \equiv \langle x R^+| w(R) | y R^+ \rangle$
will always be $\propto \theta[x(1-x)] \theta[y(1-y)]$. Moreover the reduced vertex $\gamma(x|M^2) 
\equiv \langle x R^+ | \gamma_R \rangle \propto \theta[x(1-x)]$ as a result of Eq.~\eqref{gammaredu}.

From \eqref{gammaredu} we can define the light-front wavefunction $|\psi_{R}\rangle \equiv g(R) |\gamma_{R}\rangle$, notice this too 
restricts the momentum fraction $x$: $\psi(x) \propto \theta[x(1-x)]$. 
By iterating the Lippmann-Schwinger equation for $T$ twice, it is possible to relate $T$ to $t$ and thereby construct $T$ given $t$, which is 
clearly not possible from the definition \eqref{tred}. Taking the residue of this relation between $T$ and $t$ yields the  
reduced-to-covariant conversion between bound-state vertex functions, namely
\begin{equation} \label{convert}
| \Gamma_{R} \rangle = W(R) G(R) \Big| \;| \gamma_{R} \rangle.
\end{equation}
Finally, we can manipulate the covariant Bethe-Salpeter amplitude into the form
\begin{equation} \label{324}
| \Psi_{R} \rangle = \Bigg( 1 + \Big(G(R) - \Gt(R)\Big)W(R) \Bigg) G(R) \Big| \; |\gamma_{R} \rangle,
\end{equation}
which justifies the interpretation of $|\psi_{R}\rangle$ as the light-front wavefunction since
it is easy to demonstrate that
$\Big| \; |\Psi_{R}\rangle = 
|\psi_{R}\rangle$.

While all light-front reduction schemes when summed to all orders yield the 
$x^+ = 0$ projection of the Bethe-Salpeter equation,  the choice of $\Gt(R)$
in Eq.~\eqref{gt} generates a kernel calculated in light-front time-ordered perturbation theory.
The normalization of the covariant and reduced wavefunction will be discussed below.

\subsection{An interpretation for the reduction} \label{intuit}

The heart of our intuition about the light cone lies in integrating out the minus-momentum dependence of the 
covariant wavefunction.  So we merely cast the formal reduction in a way which highlights the contributions 
from poles of the covariant vertex function.

Utilizing Eqs. \eqref{regamma} and \eqref{convert},
we can show
\begin{equation} \label{key}
\Gt (R) | \Gamma_R \rangle = G(R) \Big| \; |\gamma_R \rangle.
\end{equation}  
Thus the appearance of $\Gt(R) | \Gamma_R \rangle$ has the form of an instantaneous approximation since Eq.~\eqref{key} shows
that it has no minus-momentum poles besides those of the propagator $G(R)$.

This instantaneous approximation appears in determining the light-front wavefunction. Using Eqs.~\eqref{W} and \eqref{324}, we
have
\begin{equation}
| \psi_R \rangle 
= \Big| G(R) V(R) 
\sum_{j = 0}^{\infty} \Big[ \Big( G(R) - \Gt(R) \Big) V(R) \Big]^j G(R) \Big| \;  |\gamma_R \rangle.
\end{equation}
From truncating the series in $G(R) - \Gt(R)$ at some $j = n - 1$ and using a consistent approximation to Eq.~\eqref{convert}, 
we are led to the $n^{\text{th}}$-order approximate solution $\psi^{(n)}$
\begin{equation} \label{interpretation}
|\psi^{(n)}\rangle  = \Big| (GV)^n \Gt | \Gamma \rangle,
\end{equation}
after having used Eq.~\eqref{key}.
Thus at any order $n$ in the formal reduction scheme, 
we have iterated the covariant Bethe-Salpeter equation $n$-times and subsequently made an instantaneous approximation via Eq.~\eqref{key}.
Retaining the minus-momentum dependence in $V$ to $n^{\text{th}}$ order allows for an $n^{\text{th}}$-order approximation to the vertex function's poles.

\section{Normalization} \label{normmy}

The relation Eq.~\eqref{324} 
between the four-dimensional Bethe-Salpeter wavefunction and the reduced light-front wavefunction was presented in \cite{Sales:1999ec}. 
The normalization of the reduced wavefunction, however, was not discussed in much detail and we address this below
following our work in \cite{Tiburzi:2002sx}.

The reducible four-point function defined by $G^{(4)} = G + G T G$ has the behavior
\begin{equation}
G^{(4)}(R) =  - i \frac{|\Psi_{R} \rangle \langle \Psi_{R} |}{R^2 - M^2 + \ie} + \; \text{finite},
\end{equation} 
near the bound state pole. 
Since the four-point function satisfies the equation
\begin{equation}
G^{(4)}(R) = G^{(4)}(R) \Big( G^{-1}(R) - V(R) \Big) G^{(4)}(R),
\end{equation}
the Bethe-Salpeter amplitude must satisfy
\begin{equation}
1 = \lim_{R^2 \to M^2} - i  \frac{\langle \Psi_{R}| \Big( G^{-1}(R) - V(R) \Big) |\Psi_{R}\rangle}{R^2 - M^2},
\end{equation}
which is necessarily finite since $|\Psi_{R}\rangle$ satisfies the bound state
equation: $|\Psi_{R}\rangle = G V |\Psi_{R}\rangle$. Application of l'H\^opital
yields the covariant normalization condition \cite{Itzykson:1980rh}
\begin{equation} \label{normcov}
2 i R^\mu =  \langle \Psi_{R} | \frac{\partial}{\partial R_{\mu}} \Big( G^{-1}(R) - V(R)  \Big) |\Psi_{R} \rangle \Bigg|_{R^2 - M^2}
.\end{equation}
The normalization \eqref{normcov} takes the form of a diagonal 
matrix element of a pseudo current.

In what follows we shall omit total four-momentum labels since they are all identically $R$. 
As we saw above, the light-front reduction is performed by integrating out the minus-momentum
dependence with the help of an auxiliary 
Green's function $\Gt(R)$. 
The normalization condition for the reduced wavefunction is then deduced by using the conversion Eq.~\eqref{324}
and the definition of the reduced wavefunction $|\psi_R\rangle$, namely
\begin{equation} \label{psi}
|\psi_R\rangle = \Big| \; | \Psi_R \rangle = g(R) |\gamma_R \rangle.
\end{equation}
Hence taking the plus component of Eq.~\eqref{normcov}, we arrive at
\begin{equation} \label{normlf}
2 i R^+ =  \langle \gamma_{R} | \; \Big| G \Bigg( 1 + W ( G - \Gt ) \Bigg) 
\Bigg( \frac{\partial}{\partial R^-} \Big[ G^{-1} - V  \Big] \Bigg) \Bigg( 1 + (G - \Gt ) W \Bigg) G \Big| \; |\gamma_{R}\rangle.
\end{equation}
The complicated normalization condition is indicative of the effects of higher Fock space components.

To see this explicitly, we must know the minus-momentum dependence of the interaction. We therefore adopt a weakly coupled, 
one-boson exchange model for $V$ (the so-called ladder approximation). Supposing the boson mass is $\mu$ and the coupling constant $g$, we have
\begin{equation} \label{ladder}
V(k,p) =  \frac{- g^2}{(p-k)^2 - \mu^2 + \ie}  
.\end{equation}
In considering the normalization condition, we will work in $(3+1)$ dimensions.
The bound state equation and wavefunctions which result from the leading-order
ladder exchanges are collected in Appendix \ref{fff}.
Notice $\partial V/\partial R^\mu = 0$. For our purposes, it suffices to assume we have a light-front
wavefunction $\psi(x,\kperp)$ for the lowest Fock state and then investigate the form of the constraint given 
in Eq.~\eqref{normlf}.

Let us start with the contribution at leading order in $G - \Gt$ to the reduced wavefunction's normalization. 
\begin{equation} \label{normLO}
\frac{-i}{2 R^+}\langle \gamma_{R}  | \; \Big| G 
\Big( \frac{\partial}{\partial R^-} G^{-1} \Big)  G \Big| \; | \gamma_{R} \rangle = 1.
\end{equation}
To perform the integration, we note
\begin{equation}
\frac{\partial}{\partial R^-} G^{-1}(k,R) = - 2 i R^+ d^{-1}(k) (1 - x), 
\end{equation}
where we have customarily chosen $x = \kplus/R^+$. Evaluation of the integral in equation \eqref{normLO} is standard and yields
\begin{equation} \label{2to2}
N^{\text{LO}} \equiv \int \frac{dx d\kperp}{2 (2\pi)^3 x (1-x)} \psi^*(x,\kperp) \psi(x,\kperp) = 1,
\end{equation}
a simple overlap of the two-body wavefunction.

To analyze the normalization to first order in $G - \Gt$, we expand equation \eqref{normlf} to first order  
\begin{equation} 
N^{\text{LO}} + \delta N  + \ldots = 1,
\end{equation}
where $N^{\text{LO}}$ is the integral appearing in \eqref{2to2} and the first-order correction arising from \eqref{normlf} is
\begin{equation} \label{deltaN}
\delta N = \frac{-i}{2 R^+} \langle \gamma_{R} | \; \Big| G \Big( \frac{\partial}{\partial R^-} G^{-1} \Big) \Big( G - \Gt \Big) V G  
+  G V \Big( G - \Gt \Big) \Big( \frac{\partial}{\partial R^-} G^{-1} \Big) G \Big| \; | \gamma_{R} \rangle.
\end{equation}
The presence of $\Gt$ merely subtracts the leading-order result $N^{\text{LO}}$. 
Considering for the moment just the first term in the above equation (and omitting the subtraction $\Gt$), we have
\begin{equation}
- i \int \frac{d^4 k}{(2\pi)^4} \; \frac{d^4p}{(2\pi)^4} (1-x) \gamma^* (x,\kperp|M^2) d(k) 
d(R-k)^2 V(k,p) d(p) d(R-p) \gamma (y,\pperp|M^2).
\end{equation}
The minus-momentum integrals above are similar to those considered in deriving the bound-state equation to leading order in Appendix \ref{fff}. 
The only difference is the double pole due to the extra propagator $d(R-k)$. With $x = \kplus/R^+$ and $y = \pplus/R^+$, for $x>y$
we avoid picking up the residue at the double pole and the result is the same as in Eq.~\eqref{normLO} after using the bound-state equation.
This term is then subtracted by the $\Gt$ term in equation \eqref{deltaN}. On the other hand, when $x<y$ we pick up the residue at the double
pole. Part of the residue is subtracted by the $\Gt$ term; the other half depends on $\partial V(k,p)/\partial k^-$. The second term in Eq.~\eqref{deltaN}
is evaluated identically up to $\{k \leftrightarrow p\}$.  Now combining the two terms and their relevant 
$\theta$ functions, we can rewrite the result using the explicit form of the light-front time-ordered 
one-boson-exchange potential Eq.~\ref{OBE}, namely
\begin{equation} \label{nonN}
\delta N = \int \frac{dx d\kperp}{2(2\pi)^3 x (1-x)} \; \frac{dy d\pperp}{2 (2\pi)^3 y (1-y)} \psi^*(x,\kperp) \Bigg( - \frac{\partial}{\partial M^2}      
V(x, \kperp; y, \pperp |M^2) \Bigg)  \psi(y, \pperp).  
\end{equation} 
Thus although the covariant derivative's action on the potential vanishes, we can manipulate the correction to the normalization into the form of a 
derivative's action on the light-front kernel. With this form, we can compare to the familiar nonvalence probability discussed in 
Appendix \ref{oftopt} [in the frame where $R^+ = R^- = M/\sqrt{2}$, with $M/\sqrt{2}$ as the eigenvalue of the light-front Hamiltonian, denoted 
$\pminus$ in Eq.~\eqref{quspace}]. This correction to the normalization has been discussed earlier \cite{Lepage:1980fj}.

Here we have seen that the normalization of the light-cone wavefunction includes effects from higher Fock states. Since this normalization
condition stemmed from a diagonal matrix element of a pseudo current, we should not be surprised that matrix elements of the electromagnetic 
current, when treated in this reduction scheme, pick up contributions from higher Fock states. 
These contributions to current matrix elements are the subject of the next Section.

\section{Current} \label{curry}

Here we extend the formalism presented so far to include current matrix elements between bound states. We do so in a gauge invariant 
fashion following \cite{Gross:1987bu}. Our notation, however, is more in line with the elegant method of gauging equations presented in 
\cite{Kvinikhidze:1998xn,Kvinikhidze:1999xp}. This latter general method extends to bound systems of more than two particles. 
Our goal here is to see how higher Fock contributions appear from the covariant starting point and rule out the use of non-wavefunction vertices. Once one
accepts the results, one may work directly with the three-dimensional light-front kernel and gauge it. The results are the same \cite{Kvinikhidze:2003de}.

\begin{figure}
\begin{center}
\epsfig{file=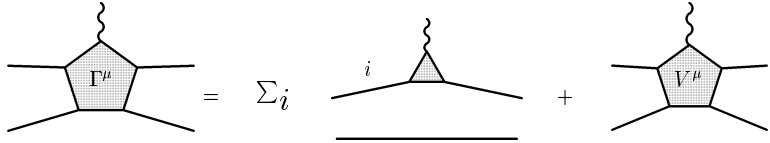,width=4in}
\end{center}
\caption[Graphical depiction of the irreducible five-point function.]{Graphical depiction of the irreducible five-point function $\Gamma^\mu$ as 
sum of impulse terms and a gauged interaction. By construction, 
$\Gamma^\mu$ is gauge invariant.}
\label{fgamma}
\end{figure}

Consider first the full four-point function $G^{(4)}$ defined by
\begin{equation}\label{g4}
G^{(4)} = G + G T G. 
\end{equation}
For later use, it is important to note that the residue of $G^{(4)}(R)$ at the bound state pole $R^2 = M^2$ is $- i |\Psi_{R}\rangle\langle \Psi_{R}|$. 
Using the Lippmann-Schwinger equation for $T$, we can show the four-point function satisfies 
\begin{equation}\label{LSg4}
G^{(4)} = G + G V G^{(4)}. 
\end{equation}

To discuss electromagnetic current matrix elements, we will need the three-point 
function for the $i^\text{th}$ constituent, which we write as $d_{i}^\mu$. 
We define an irreducible
three-point function for the $i^\text{th}$ constituent $\Gamma_{i}^\mu$ in the obvious way
\begin{equation}
d_{i}^\mu = d_{i} \Gamma_{i}^\mu d_{i}.
\end{equation}
Now we need to relate the one-particle electromagnetic vertex function to the $T$ matrix. Let $j^\mu$ denote the electromagnetic coupling to the constituent
particles (since our particles are scalars $j^\mu = \overset{\leftrightarrow}\partial{}^\mu$). 
Since the electromagnetic three-point function $\Gamma_{i}^\mu$ is irreducible, we have
\begin{equation}
\Gamma_{i}^\mu  = G^{-1} G^{(4)} j^\mu
\end{equation}
 and by using the definition of $G^{(4)}$ Eq.~\eqref{g4}, we have the desired relation
\begin{equation}
\Gamma_{i}^{\mu} = 
j^\mu + T G j^\mu
\end{equation}
Notice the right hand side lacks the particle label $i$. In the first term, the bare coupling acts on the $i$th particle while in the second term the bare coupling does not act on the $i$th particle.
For this reason we have dropped the label which will always be clear from context.

In considering two propagating particles' interaction with a photon, the above definitions lead us to the impulse approximation $\Gamma_{0}^\mu$
to the current
\begin{equation}
\Gamma_{0}^\mu = \Gamma_{1}^\mu d_{2}^{-1} + d_{1}^{-1} \Gamma_{2}^\mu.
\end{equation}
Additionally the photon could couple to interacting particles. Define a gauged interaction $V^\mu$ topologically
by attaching a photon to the kernel in all possible places. This leads us to the irreducible electromagnetic vertex $\Gamma^\mu$ defined as (see Figure \ref{fgamma})
\begin{equation} \label{emvertex}
\Gamma^\mu = \Gamma_{0}^\mu + V^\mu,
\end{equation}
which is gauge invariant by construction.

\begin{figure}
	\begin{center}
	\epsfig{file=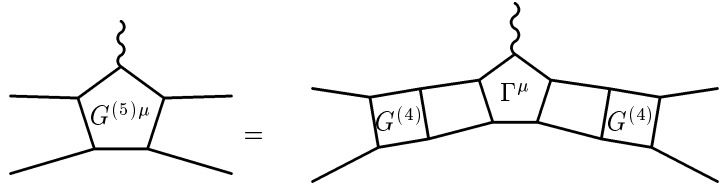,width=3.75in}
		\end{center}
	\caption[Graphical depiction of the reducible five-point function.]{Graphical depiction of the reducible five-point function $G^{(5) \; \mu}$. The 
	irreducible five-point function is the gauge invariant $\Gamma^\mu$.}
	\label{f5pt}
\end{figure}

Lastly to calculate matrix elements of the current between bound states it is useful to define a reducible five-point function
(see Figure \ref{f5pt})
\begin{equation} \label{5alive}
G^{(5) \; \mu} = G^{(4)} \Gamma^\mu G^{(4)}.
\end{equation}
Having laid down the necessary facts about electromagnetic vertex functions and gauge invariant currents, we can now specialize to their matrix elements 
between bound states by taking appropriate residues of Eq.~\eqref{5alive}. The form factor is then
\begin{equation} \label{me}
- i (P^{\prime\mu} + P^\mu) F(t) = \langle \Psi_{P^\prime} | \Gamma^\mu(-\Delta) | \Psi_{P} \rangle, 
\end{equation}
where $P^{\prime \mu} = P^\mu + \Delta^\mu$ and $t = \Delta^2$.

\section{Two models} \label{meat}

In this Section, we use the formalism thus far developed for two models. The first model has an instantaneous interaction
while the second is not instantaneous. Contributions to form factors are contrasted and the issue of non-wavefunction contributions is resolved.

\subsection{Wavefunctions}

\begin{figure}
\begin{center}
\epsfig{file=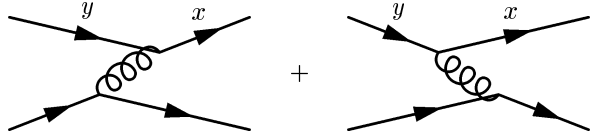}
\end{center}
\caption[Diagrammatic representation of the one-boson exchange potential.]{Diagrammatic 
representation of the one-boson exchange potential $w(x,y|M^2)$ appearing in Eq. \eqref{OBEP1}.}
\label{fOBEP1}
\end{figure}

To be specific, we again work in the ladder approximation for the kernel Eq.~\eqref{ladder}
for which the energy pole (with respect to $p$) of the interaction $V(p,k)$ is 
\begin{equation} \label{pv}
p^-_{v} = \kminus + \frac{\mu^2}{2 (\pplus - \kplus)} -\frac{\ie}{2(\pplus - \kplus)}.
\end{equation}
This interaction is completely non-local in space-time and hence does not have an instantaneous piece. The reduced kernel Eq.~\eqref{wred} 
is consequently made up of retarded terms (i.e.~dependent on the eigenvalue $M^2$) where higher order in $G - \Gt$ 
means more particles propagating at a given instant of light-cone time (see, e.g., \cite{Sales:1999ec}). 
The leading-order equation for $\psi$ from Eq.~\eqref{gammaredu} 
is the Weinberg equation \cite{Weinberg:1966jm}
\begin{equation} \label{OBE1}
\psi(x) = - \dw(x|M^2) \int_0^1 \frac{w(x,y|M^2) \psi(y)}{2 (2\pi) y(1-y)} dy.
\end{equation}
with the time-ordered one-boson exchange potential (see Figure \ref{fOBEP1}) calculated to leading order from 
Eq.~\eqref{wred}
\begin{equation} \label{OBEP1}
w(x,y|M^2) = \frac{- g^2}{x - y} \theta[x(1-x)] \theta[y(1-y)] 
\Big[ \theta(x-y) D(x,y|M^2) - \{x \leftrightarrow y\}  \Big],
\end{equation}
where
\begin{equation}
D^{-1}(x,y|M^2) = M^2 - \frac{m^2}{y} - \frac{\mu^2}{x - y} - \frac{m^2}{1-x}.
\end{equation}
We can obtain Eq.~\eqref{OBE1} most simply by iterating the Bethe-Salpeter equation once (see Section \ref{intuit}) 
and then projecting onto the light cone. See Appendix \ref{fff} for the $(3+1)$-dimensional derivation.

In the limit $\mu^2 \gg m^2, M^2$, the interaction becomes approximately 
instantaneous, which suggests we separate out an instantaneous piece $V_o$:
\begin{equation} \label{instladder}
V(k,p) = V_o + [V(k,p) - V_o],
\end{equation}
where
\begin{equation} \label{Vo}
V_o = V_o(x,y) = \frac{- g^2 \theta(x - y)}{E^2(x - y) - \mu^2} + \frac{-g^2 \theta(y-x) }{E^2(y - x) - \mu^2},
\end{equation}
with $E$ as a constant parameter to be chosen. Of course other choices of $V_o$ are possible. 
We choose the above form of $V_o$ for two reasons.

First is the form of the instantaneous approximation wavefunction, which we denote by $\phi(x)$. 
When we write the potential as Eq.~\eqref{instladder}, we expand Eq.~\eqref{W} to first order in $g^2$ as
\begin{equation} \label{instW}
W = V_o + (V - V_o ) + (V - V_o)\Big( G - \Gt \Big) (V - V_o). 
\end{equation}
To zeroth order in $G - \Gt$ and $V - V_o$, the integral equation for $\phi$ is
\begin{equation} \label{nOBE}
\phi(x) = - \dw(x|M^2) \int_0^1 \frac{ w_o(x,y) \phi(y)}{2(2\pi) y(1-y)} dy, 
\end{equation} 
where by Eq.~\eqref{wred}, the reduced instantaneous potential is merely $w_o(x,y) = \theta[x(1-x)] \theta[y(1-y)] V_o(x,y)$.
In the instantaneous limit, $\mu^2 \gg m^2, M^2$ solutions to Eqs.~\eqref{OBE1} and \eqref{nOBE} 
coincide---both wavefunctions approach $\sim \dw(x,4m^2)$.

Secondly, we 
preserve the contact interaction limit by excluding from $V_o$ the factor $\theta[x(1-x)]\theta[y(1-y)]$.
That is, when $\{\mu^2,g^2\} \to \infty$ with $g^2/\mu^2$ fixed (or equivalently $E \to 0$) we have the contact interaction\footnote{This 
exactly soluble $(1+1)$-dimensional light-cone model, in which $\psi(x) \propto \dw(x|M^2)$, 
has been considered earlier in \cite{Sawicki:1992qj}.}
$V_o \to g^2 /\mu^2$, which rightly knows nothing about the momentum fractions $x$ and $y$.    
We shall now investigate contributions to the 
form factors for each of the models Eqs.~\eqref{ladder} and \eqref{Vo}.

\subsection{Form factors} \label{formmy}

From Eq.~\eqref{me} we can calculate the form factors for each of the models \eqref{ladder} and \eqref{Vo} in the reduced formalism. 
Working in perturbation theory, we separate out contributions up to first order
by using Eq.~\eqref{324} to first order in $G - \Gt$ and $\Gamma^\mu$ \eqref{emvertex} in the first Born approximation. The 
matrix element $J^\mu = \langle \Psi_{P^\prime} | \Gamma^\mu(-\Delta) | \Psi_{P} \rangle$ then appears for a model with some kernel $V$
\begin{eqnarray}
J^\mu  &\approx&  \langle \gamma_{P^\prime} | \; \Big| G(P^\prime) \Big( 1 + V(P^\prime) ( G(P^\prime) - \Gt(P^\prime)) \Big) \notag
\\ && \times \Big( \overset{\leftrightarrow}\partial{}^\mu(-\Delta)  d_{2}^{-1} + V(-\Delta) G(-\Delta) 
\overset{\leftrightarrow}\partial{}^\mu(-\Delta) d_{2}^{-1} \Big) \notag
\\ && \times \Big( 1 + (G(P) - \Gt(P)) V(P)  \Big) G(P) \Big| \; | \gamma_{P} \rangle \notag  \\
	&=& \Big( J^\mu_{\text{LO}} + \delta J^\mu_{i} +  \delta J^\mu_{f} + \delta J^\mu_{\gamma} \Big) + \mathcal{O}[V^2],
\end{eqnarray}
with the leading-order result
\begin{equation} \label{LO1}
J^\mu_{\text{LO}} = \langle \gamma_{P^\prime} | \; \Big| G(P^\prime) \overset{\leftrightarrow}\partial{}^\mu (-\Delta) 
d_{2}^{-1} G(P) \Big| \; | \gamma_{P} \rangle.
\end{equation}
The first-order terms are
\begin{eqnarray}    
		\delta J^\mu_{i} &=&  \langle \gamma_{P^\prime} | \; \Big| G(P^\prime) \overset{\leftrightarrow}\partial{}^\mu (-\Delta) 
		d_{2}^{-1} \Big(G(P) - \Gt(P) \Big) V(P) G(P) \Big| \; |\gamma_{P} \rangle \notag \\
		\delta J^\mu_{f} &=& \langle \gamma_{P^\prime} | \; \Big| G(P^\prime) V(P^\prime) \Big(G(P^\prime) - \Gt(P^\prime) \Big) 
		\overset{\leftrightarrow}\partial{}^\mu (-\Delta) d_{2}^{-1} G(P) \Big| \; | \gamma_{P} \rangle \notag \\
		\delta J^\mu_{\gamma}  &=& \langle \gamma_{P^\prime} | \; \Big| G(P^\prime) \Big(V(-\Delta) G(-\Delta) 
		\overset{\leftrightarrow}\partial{}^\mu (-\Delta) \Big) 
		d_{2}^{-1} G(P) \Big| \; | \gamma_{P} \rangle. \label{NLO1}
\end{eqnarray} 
The labels indicate the intuition behind the reduction scheme (seen in Section \ref{intuit}):
the term $\delta J^\mu_{f}$ arises from one iteration of the covariant Bethe-Salpeter equation for the
final-state vertex followed by an instantaneous approximation \eqref{key}, $\delta J^\mu_{i}$ arises in 
the same way from the initial state and $\delta J^\mu_\gamma$ comes from one iteration of the Lippmann-Schwinger
equation \eqref{LS}.

At this point, we must specify which component of the current vector we are using to calculate the form factor. 
In Section \ref{curry}, we started with the manifestly Lorentz invariant decomposition of the current matrix element, 
\emph{cf}.~Eq.~\eqref{me}. Working to first-order in the reduction scheme, however, covariance has been lost, e.g., the kernel Eq.~\eqref{OBEP1}
breaks rotational invariance \cite{Cooke:1999yi}. Thus all components of the current are no longer equivalent. In order to make connection 
with the Drell-Yan formula and the Fock space decomposition in \cite{Diehl:2000xz,Brodsky:2000xy}, we must 
choose the plus-component of the current. Agreement with the form factor calculated from the minus-component
is only achieved when one works to all orders in the reduction scheme. Further discussion can be found in \cite{Tiburzi:2002sx} and also
below in Section \ref{gpds}.

Because the leading-order expression \eqref{LO1} is independent of the kernel $V$, the result will have the same form for both models. 
Using the effective resolution of unity, the above expression converts into
\begin{equation}
J^+_{\text{LO}} = \int \frac{d^2 p}{(2\pi)^2} \; \frac{d^2k}{(2\pi)^2} \langle \gamma_{P^\prime} | \pplus\rangle 
\langle p | G(P^\prime)\overset{\leftrightarrow}\partial{}^+ (-\Delta) d_{2}^{-1} G(P) | k \rangle \; \langle \kplus | \gamma_{P} \rangle.
\end{equation} 
Bearing in mind the delta function present in $G(R)$, we have the factor
\begin{equation}
\langle p | \overset{\leftrightarrow}\partial{}^+ (-\Delta)| k \rangle = -i (2 \kplus + \Delta^+) (2\pi)^2 \delta^{2}(p - k - \Delta) .  
\end{equation}
Now define $ x = \kplus/P^+$ and $\Delta^+ = - \zeta P^+$ as above in Section \ref{LFCQM} and denote the reduced vertex 
$\langle \kplus| \gamma_{R} \rangle = \gamma(x|M^2)$. The leading-order contribution is then
\begin{equation} \label{formLO}
J^+_{\text{LO}} = \int \frac{d^2 k}{(2 \pi)^2}  \gamma^*(x^\prime|M^2)  d(k + \Delta) 
\big( 2 x - \zeta \big) d(k) d(P-k) \gamma(x|M^2),
\end{equation}
where $x^\prime = \frac{x - \zeta}{1-\zeta}$. Notice Eq.~\eqref{formLO} is quite similar to Eq.~\eqref{ff1}. For $x>\zeta$ evaluation
is straightforward and leads to 
\begin{equation}\label{FFLO1}
F_{LO}(t) = \frac{\theta(x - \zeta)}{1 - \zeta/2} \int \frac{dx}{2 (2\pi)} \frac{2 x - \zeta}{x (1-x) x^\prime} \psi^*(x^\prime) \psi(x),
\end{equation}
for the non-instantaneous case. For the instantaneous case $F_{LO}(t)$, merely replace $\psi$ with $\phi$.

\begin{figure}
\begin{center}
\epsfig{file=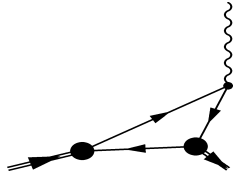,width=1.2in,height=0.9in}
\caption{The $Z$-graph confronting evaluation of the electromagnetic form factor.}
\label{fZ}
\end{center}
\end{figure}

On the other hand, we know from Section \ref{LFCQM} the region $x<\zeta$ contains a non-wavefunction vertex
(see Figure \ref{fZ}).
However, evaluation of Eq.~\eqref{formLO} with reduced vertices is quite different than in Section \ref{LFCQM}. 
Rather simply, the term $\gamma^*(x^\prime|M^2) = 0$ by virtue of Eq.~\eqref{gammaredu} because $x^\prime < 0$. Thus there
is no contribution at leading order for $x < \zeta$.

The first-order terms depend explicitly on the interaction and will hence be considerably different for each of the models. 
We consider each model separately.

\subsubsection{Non-instantaneous case}

\begin{figure}
\begin{center}
\epsfig{file=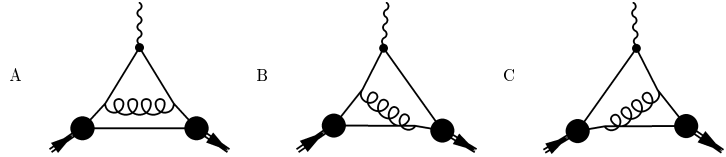,width=3.75in}
\caption[Diagrams which contribute to the form factor to first order, for $x >\zeta$.]{Diagrams 
which contribute to the form factor to first order in $G - \Gt$ for the non-instantaneous case (for $x >\zeta$).}
\label{ftri2}
\end{center}
\end{figure}

\begin{figure}
\begin{center}
\epsfig{file=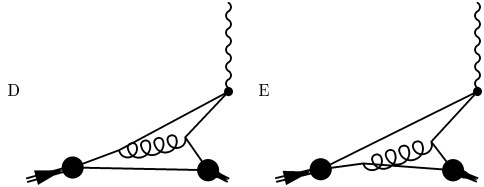,width=2.5in}
\caption[The remaining diagrams for the first-order form factor.]{The 
remaining diagrams (characterized by $x < \zeta$) for the electromagnetic form factor at first order in $G - \Gt$ for the 
non-instantaneous case.}
\label{fZZZ}
\end{center}
\end{figure}

Evaluating contributions at first order for the non-instantaneous interaction Eq.~\eqref{ladder} is complicated
by the presence of poles in the interaction [\emph{cf} Eq.~\eqref{pv}]. First we evaluate the first Born term
$\delta J^+_{\gamma}$ in Eq.~\eqref{NLO1}. After careful evaluation of the two minus-momentum integrals,
we have the contribution to $\delta J^+_\gamma$ for $x>\zeta$
\begin{equation}
\delta J^+_{A} = \int \frac{\theta(x - \zeta) \; dx dy \; (2 x - \zeta)}{16\pi^2 x x^\prime y(1-y) y^\prime} \psi^*(y^\prime)  
D(y^\prime,x^\prime|M^2) \frac{g^2 \theta(y - x)}{y - x} D(y,x|M^2) \psi(y),
\end{equation}
where $y^\prime = \frac{y - \zeta}{1 - \zeta}$. This contribution corresponds to diagram $A$ in Figure \ref{ftri2}. 
Additionally using $x^{\prime\prime} = x / \zeta$, we have for $x<\zeta$
\begin{equation}
\delta J^+_D = \int \frac{\theta(\zeta -x) \; dx dy \; (2 x - \zeta)/\zeta}{16\pi^2  y(1-y) y^\prime x^{\prime\prime} (1 - x^{\prime\prime})}
\psi^*(y^\prime) \dw(x^{\prime\prime}|t) \frac{g^2 \theta(y - x)}{y - x} D(y,x|M^2)  \psi(y),
\end{equation}
which corresponds to diagram $D$ in Figure \ref{fZZZ}.

The initial-state iteration term $\delta J^+_{i}$ in Eq.~\eqref{NLO1} is complicated by the subtraction of the two-particle
reducible contribution
\begin{eqnarray} \label{isreduce}
&-& \langle \gamma_{P^\prime} | \; \Big| G(P^\prime) \overset{\leftrightarrow}\partial{}^+ (-\Delta) d_{2}^{-1} \Gt(P) V(P) G(P) 
\Big| \; |\gamma_{P} \rangle \notag
\\  = &-& \langle \gamma_{P^\prime} | \; \Big| G(P^\prime) \overset{\leftrightarrow}\partial{}^+ (-\Delta) 
d_{2}^{-1} G(P) \Big| \; |\gamma_{P} \rangle. 
\end{eqnarray}
Thus this term merely removes contributions which can be reduced into the initial-state wavefunction. 
Evaluation of the two minus-momentum integrals 
yields a contribution for $x>\zeta$ to $\delta J^+_{i}$
\begin{equation}
\delta J^+_{B} =  \int \frac{\theta(x-\zeta) \; dx dy \; (2 x - \zeta)}{16\pi^2 x x^\prime (1-x^\prime) y (1-y)}
\psi^*(x^\prime) D(y^\prime,x^\prime|M^2) \frac{g^2 \theta(y - x)}{y - x} D(y,x|M^2) \psi(y),
\end{equation}
which corresponds to diagram $B$ in Figure \ref{ftri2}. On the other hand,  for 
$x<\zeta$, we have the non-wavefunction vertex $\gamma^*(x^\prime|M^2)$ for the final state, which of course vanishes.
Similarly the $\Gt(P)$ term vanishes. Thus there is no contribution to $\delta J^+_{i}$ for $x<\zeta$.

Finally there is the final-state iteration term $\delta J^+_f$ in Eq.~\eqref{NLO1}. There are only two types of contributions. For
$x>\zeta$, that which can be reduced in to the final-state wavefunction is subtracted by $\Gt$. 
The remaining term is:
\begin{equation}
\delta J^+_{C} =  \int \frac{\theta(x - \zeta) \;dx dy^\prime (2 x - \zeta)}{16\pi^2 x(1-x) x^\prime y^\prime (1- y^\prime)} \psi^*(y^\prime)
D(y^\prime,x^\prime|M^2) \frac{g^2 \theta(y - x)}{y - x} D(y,x| M^2) \psi(x),
\end{equation}
where $y = \zeta + (1-\zeta) y^\prime$. This corresponds to diagram $C$ in Figure \ref{ftri2}.  For $x<\zeta$, 
the subtraction term vanishes since $x^\prime < 0$ for which $\Gt(P^\prime) = 0$. 
The remaining term in $\delta J^+_{f}$ gives a contribution
\begin{equation}
\delta J^+_{E} =  \int \frac{\theta(\zeta - x) \;dx dy^\prime  (2 x - \zeta)/\zeta}{ 16\pi^2 (1-x) x^{\prime\prime} (1 - x^{\prime\prime})
y^\prime (1-y^\prime)} \psi^*(y^\prime) \dw(x^{\prime\prime}|t)  \frac{g^2 \theta(y - x)}{y - x} 
D(y,x| M^2) \psi(x),
\end{equation}
which corresponds to diagram $E$ in Figure \ref{fZZZ}. To summarize, the non-valence correction to the form factor in the non-instantaneous
case is
\begin{equation} \label{NVNI}
\delta F_{NI} =  \frac{1}{1-\zeta/2} \Big[ \delta J^+_A + \delta J^+_B + \delta J^+_C + \delta J^+_D + \delta J^+_E \Big],
\end{equation}
and there are no non-wavefunction terms.

\subsubsection{Instantaneous case}

\begin{figure}
\begin{center}
\epsfig{file=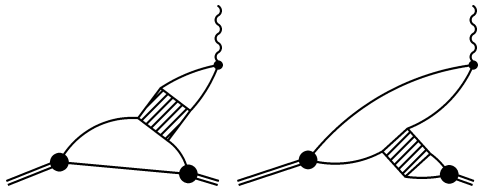,width=2.5in}
\caption[Diagrams with crossed interactions.]{Diagrams with crossed interactions necessary to calculate the form factor in the region $x < \zeta$ for the instantaneous case.}
\label{f:cross}
\end{center}
\end{figure}

The case of an instantaneous interaction is quite different due to the absence of light-front energy poles in $V_o$. 
Note we are working with Eq.~\eqref{instW} to zeroth order in $G - \Gt$ and $V - V_o$. 
The first term in Eq.~\eqref{NLO1} we consider is the Born term $\delta J^+_\gamma$. 
The pole structure leads only to a contribution for $x<\zeta$:
\begin{equation} \label{btinst}
\delta J^+_1 = - \int \frac{\theta(\zeta - x) \theta(y - x) \; dx dy \; (2 x - \zeta)/ \zeta}{16\pi^2 y (1-y) y^\prime x^{\prime\prime}
(1- x^{\prime\prime})} 
\phi^*(y^\prime) \dw(x^{\prime\prime}|t) V_o(x^{\prime \prime},y^{\prime \prime}) \phi(y),
\end{equation}
and is depicted by the first diagram in Figure \ref{f:cross}. The interaction along the way to the photon vertex is crossed and
represents pair production off the quark line ($x^{\prime \prime} > 1$).

The next term $\delta J^+_{i}$ simplifies considerably due to the absence of light-cone time dependence in $V_o$.
As above, the final state vertex restricts $\zeta < x < 1$. But then we are confronted with a factor
$ \langle k | \Big(G(P) - \Gt(P)\Big) \Big| = 0$ since $x > 0$. Thus $\delta J^+_i = 0$.

The last term we must consider is $\delta J^+_f$, in which we have an analogous factor for the final state
$\Big| \Big( G(P^\prime) - \Gt(P^\prime) \Big) | k + \Delta \rangle$.
This is zero for $x - \zeta > 0$, else $\Gt(P^\prime) = 0$ by 
virtue of Eq.~\eqref{gt} and Eq.~\eqref{lilginv}. Thus we only have a contribution for $x<\zeta$ which is from 
$\Big| G(P^\prime) | k+\Delta \rangle$. The expression for this contribution is
\begin{equation} \label{fsiinst}
\delta J^+_2 = - \int \frac{\theta(\zeta - x) \theta(y - x) dx d\yp (2 x - \zeta) / \zeta}{16 \pi^2 (1-x) \xpp (1-\xpp) \yp (1-\yp)}
\phi^*(\yp) V_o(y^\prime,x^\prime) \dw(\xpp|t) \phi(x),
\end{equation}
and is depicted on the right in Figure \ref{f:cross}. The interaction again is crossed ($x^\prime <0$) and represents pair production.
With Eqs.~\eqref{btinst} and \eqref{fsiinst}, we have both bare-coupling pieces of the 
full Born series for the photon vertex (further terms in the series, which result from higher-order terms in the expansion of $W$, 
add interaction blocks to each diagram on the quark-antiquark pair's 
path to annihilation). In this way we recover the full four-point Green's function from summing the Born series \cite{Einhorn:1976uz,Tiburzi:2001je}. 
Notice also the above form of $\delta J^+_2$ is what one would obtain from extending the definition of $\gamma$ as a non-wave
function vertex \cite{Einhorn:1976uz}. For the case of an instantaneous interaction, the light-front Bethe-Salpeter formalism 
automatically incorporates crossing.

To summarize: we have found the non-valence contribution to the instantaneous model's form factor, namely
\begin{equation} \label{NVI}
\delta F_I = (\delta J^+_1 + \delta J^+_2)/(1 - \zeta/2)
.\end{equation}
This expression involves crossed interactions or equivalently terms which could be called non-wavefunction contributions.

\subsubsection{Comparison}

Now we compare the form factors for the cases of instantaneous and non-instantaneous interactions. 
In the instantaneous limit $\mu^2 \gg m^2, M^2$, however, we understand the behavior of both wavefunctions.  
The wavefunctions Eq.~\eqref{OBE1} and Eq.~\eqref{nOBE} both become narrowly peaked about $x = 1/2$ in the large $\mu^2$ 
limit, \emph{cf} the behavior of $\dw(x,4m^2)$. 
Since we are investigating non-valence contributions to \emph{form factors}, we choose additionally 
to solve for the instantaneous wavefunction to first
order in $V - V_o$ (see Eq.~\ref{instW}) to put the wavefunctions on equal footing: i.e.~$\phi(x) \approx \psi(x)$, 
because the difference $V - V_o$ is presumed small. 
This has the efficacious consequence of producing identical leading-order terms, \emph{cf} Eq.~\eqref{FFLO1}, and eliminates
the issue of normalization. 
Furthermore, the optimal choice of the instantaneous interaction is not under investigation here. So we shall simplify the issue
by choosing $E = 0$ in Eq.~\eqref{Vo}.

\begin{table}
\caption[Numerical solution of the bound-state equation.]{Numerical solution of the bound-state equation Eq.~\eqref{OBE1} for various values of $\mu^2$.
The coupling constant $\alpha = 0.100$. }  
\begin{center}
	\begin{tabular}{|| l | r ||}
	\hline
	                $ \mu^2/m^2 $        &   $ 4 - M^2/m^2  $   \\ 
	\hline
			     $0.100$                  &   $6.57 \times 10^{-1}$     \\
	\hline
			     $0.316$                  &   $2.72 \times 10^{-1}$     \\	
	\hline
	                     $1.00$                   &   $6.34 \times 10^{-2}$     \\ 
        \hline
	                     $3.16$                   &   $8.79 \times 10^{-3}$     \\ 
	\hline
	                     $10.0$                   &   $9.57 \times 10^{-4}$     \\ 
	\hline
	                     $31.6$                   &   $4.81 \times 10^{-5}$     \\ 
	\hline
	\end{tabular}
\end{center}

\label{bsvals}
\end{table}

\begin{figure}
\begin{center}
\epsfig{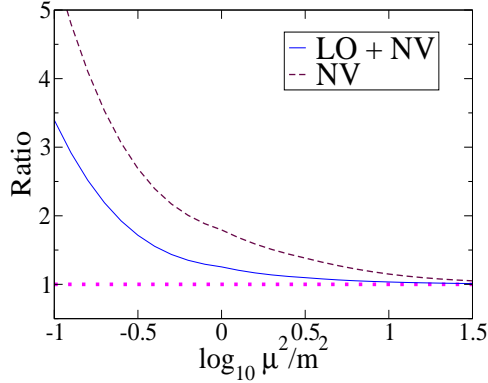}
\caption[Comparison of form factors in instantaneous and non-instantaneous cases]{Comparison of 
form factors in instantaneous and non-instantaneous cases. The ratio Eq.~\eqref{ratio} of form factors (LO + NV)
 is plotted versus $\log_{10}(\mu^2/m^2)$ at fixed $\zeta = 0.707$. 
Additionally the ratio Eq.~\eqref{ratioprime} of non-valence contributions to the form factor (NV) is plotted. 
For $\mu^2 \gg m^2, M^2$, the non-instantaneous contributions becomes approximately instantaneous.}
\label{it}
\end{center}
\end{figure}

For a few values of $\mu^2$, we solve for the wavefunction using Eq.~\eqref{OBE1}. In Table \ref{bsvals}, we list the values of
$\mu^2$ used as well as the corresponding eigenvalue $M^2$ (all for the coupling constant $\alpha = 0.100$, where 
$\alpha = g^2 / 4 \pi m^4$). We then calculate the form
factors in the instantaneous and non-instantaneous cases. We arbitrarily choose $\zeta = 0.707$. 
In $(1+1)$ dimensions, this fixes $\Delta^2 = - \zeta ^2 M^2 / (1-\zeta)$. 
The ratio of these form factors---defined using Eqs.~(\ref{FFLO1}), (\ref{NVNI}), and (\ref{NVI}), namely
\begin{equation} \label{ratio}
(F_{LO} + \delta F_I)/(F_{LO} + \delta F_{NI} )
\end{equation} 
---is plotted versus $\log_{10}(\mu^2/m^2)$ 
in Figure \ref{it}.  The Figure indicates the form factors are the same in the large $\mu^2$ limit. 
Since the leading-order contributions are identical, the Figure additionally shows 
the ratio of non-valence contributions to the form factor, namely
\begin{equation} \label{ratioprime}
\delta F_I / \delta F_{NI}.
\end{equation} 
This ratio too tends to one as $\mu^2$ becomes large, of course not as rapidly. 
Thus for finite $\mu^2$, we must add the correction $V - V_o$ in Eq.~\eqref{instW}
which results in higher Fock contributions present in the non-instantaneous case.

Lastly we remark that all of this is analytically clear: having eliminated different $\mu^2$ dependence hidden 
in $\psi$ and $\phi$, expressions for form factors in the instantaneous
and non-instantaneous cases are identical to leading order in $1/\mu^2$, i.e.~$\delta J^+_A, \delta J^+_B, \delta J^+_C
\sim 1/\mu^4$ while the remaining non-valence contributions $\delta J^+_D, \delta J^+_E$ match up 
with $\delta J^+_1$ and $\delta J^+_2$, respectively, to $1/\mu^2$.

Of course, we have demonstrated the replacement of non-wavefunction vertices only to first order in perturbation theory.  
Schematically, we have required
\begin{equation}
| V - V_o | \ll |V ( G - \Gt) V|.
\end{equation}
This is merely the condition that corrections to $V_o$ from finite $\mu^2$ are larger than second-order corrections
from the reduction scheme in Eq.~\eqref{instW}. 
Quantitatively this condition translates to $m/ 4 \pi \mu \ll \alpha$. 
If this condition is met, the leading corrections are all of the form $V - V_o$
and the instantaneous model becomes the ladder model (and non-wavefunction vertices disappear when calculating the form factor). 
When the condition is not met, but holds to $[V(G - \Gt)]^2 V$, say, we 
must work up to second order in the reduction scheme and use the second Born approximation. 
We do not pursue this lengthy endeavor here, since for the non-instantaneous case $\gamma(x|M^2) \propto \theta[x(1-x)]$ 
necessarily excludes non-wavefunction contributions in time-ordered perturbation theory.
In fact at any order in time-ordered perturbation theory we expect to find a complete expression for the form 
factor in terms of Fock component overlaps devoid of non-wavefunction vertices, crossed interactions, \emph{etc}. 
We shall make this correspondence explicit when considering the application of the light-front BSE to GPDs in 
Chapter \ref{chap:GPDs}.

\section{Summary} \label{sum}

Above we investigate current matrix elements in the light-front Bethe-Salpeter formalism.
First we present the issue of non-wavefunction vertices by taking up the common
assumptions of light-front constituent quark models. By calculating the form factor in frames where
$\zeta \neq 0$, Lorentz invariance mandates contributions from the vertex function's poles. 
Quark models which neglect these residues and postulate a form for the wavefunction are assuming not only a pole-free vertex,
but a vertex independent of light-front energy. These assumptions are very restrictive.

This leads us formally to investigate instantaneous and non-instantaneous contributions
to wavefunctions and form factors necessitating the reduction formalism Eq.~\eqref{324}.   
We provide an intuitive interpretation for the light-front reduction, namely it is a procedure 
for approximating the poles of the Bethe-Salpeter vertex function. This procedure consists of covariant 
iterations followed by an instantaneous approximation, \emph{cf} Eq.~\eqref{interpretation}, where the
auxiliary Green's function $\Gt$ enables the instantaneous approximation, see Eq.~\eqref{key}.  
In order to calculate form factors in the reduction formalism, we construct the gauge invariant
current Eq.~\eqref{emvertex}. As results in the reduction scheme are only covariant when summed to all
orders, our expressions violate Lorentz symmetry. To extract the form factor we use the plus-component of 
the current in order to make contact with the Drell-Yan formula and the Fock space representation.

Using the ladder model \eqref{ladder} and an instantaneous approximation \eqref{Vo} we compare 
the calculation of wavefunctions and form factors in the light-front reduction scheme. 
Calculation of form factors is dissimilar for the two cases. 
In the ladder model, which is not instantaneous,
non-wavefunction vertices are excluded
in time-ordered perturbation theory. 
For the instantaneous model, however, contribution from crossed interactions
is required. Moreover, 
these instantaneous contributions derived
are identical in form to non-wavefunction vertices used in constituent quark models. 
As a crucial check on our results, we take the limit 
$\mu^2 \gg m^2, M^2$ for which 
the ladder model becomes approximately instantaneous. In this limit, calculation of the form factors for the two models is identical.

The net result is an explicit proof, for $(1+1)$ dimensions, that non-wavefunction vertices are replaced by contributions form higher
Fock states if the interactions between particles are not instantaneous. 
We call this a replacement theorem \cite{Tiburzi:2002sw}. 
The analysis relies on general
features of the pole structure of the Bethe-Salpeter equation in light-front time-ordered perturbation theory. Therefore it is trivial to 
extend the theorem to $(3+1)$ dimensions. 
When one properly considers the entire coupled tower of Dyson-Schwinger equations, the light-front reduction produces
light-front versions in light-cone time-ordered perturbation theory \cite{Frederico:2003zk,Frederico:2003tu}. 
We will make this connection in our approximation of 
neglecting the self energy when we apply the light-front BSE  and its reduction scheme to GPDs in Chapter \ref{chap:GPDs}.
\chapter{Generalized Parton Distributions} \label{chap:GPDs}

The light-front BSE development and resolution of the non-wavefunction controversy 
discussed above have natural applications. 
In Chapter \ref{chap:LFBS}, we worked through matrix elements of the electromagnetic current in a frame where $\Delta^+ \neq 0$; 
we can now make the connection to generalized parton distributions (GPDs). These distributions, which in some sense 
are the natural interpolating functions between form factors and quark distribution functions, turn up in a variety of hard exclusive processes, 
e.g.,  deeply virtual Compton scattering (DVCS) and hard electro-production of mesons \cite{Muller:1994fv,Radyushkin:1996nd,Radyushkin:1996ru,Ji:1997ek,Ji:1997nm}. 
There are a number of insightful review articles on this subject from a variety of perspectives 
\cite{Ji:1998pc,Radyushkin:2000uy,Goeke:2001tz,Belitsky:2001ns,Diehl:2003ny}.
The scattering amplitude for these processes factorizes into a convolution
of a hard part (calculable from perturbative QCD) and a soft part which the GPDs encode. 
We will use DVCS to motivate the introduction of GPDs. 
Since light-cone correlations are probed in these 
hard processes, the soft physics has a simple interpretation and expression in terms of light-front wavefunctions of the initial and final
states \cite{Diehl:2000xz,Brodsky:2000xy}.
After analyzing DVCS, we will see this explicitly for the Wick-Cutkosky model.

Below, we take the light-front BSE model of Chapter \ref{chap:LFBS} into $(3+1)$ dimensions and make clear the connection to higher Fock 
components by explicitly constructing the three-body wavefunction from our expressions for 
form factors. 
The immediate application 
of our development for form factors (in a frame where the plus-component
of the momentum transfer is non-zero) is to compute GPDs. Connection 
is made to the Fock space representation and continuity of the distributions is put under scrutiny.
We show how continuity is ensured by light-cone time-ordered perturbation theory for the Wick-Cutkosky model.
The light-front Bethe-Salpeter formalism can also be employed to obtain time-like form factors.
This latter application is interesting since no Fock space expansion in terms of bound
states is possible. In the interest of organization, a summary of our investigation in~\cite{Tiburzi:2002sx}
for timelike processes.
appears in Appendix~\ref{chap:time}.

The Chapter is organized as follows.
First we review the basics of GPDs in Section \ref{gpdrev}, including the field theoretic properties they must obey.
The introduction of GPDs is framed in the context of DVCS. 
This discussion of DVCS is the natural extension of our review of deep-inelastic scattering in Section \ref{DIS}.
Next in Section \ref{gpds}, we use the light-front reduction scheme to calculate generalized parton distributions
in the Wick-Cutkosky model.
We do so by using the integrand of the electromagnetic form factor calculated in an arbitrary frame
[these expressions in $(3+1)$ dimensions are collected in Appendix \ref{fff}, where the leading-order bound-state 
equation for the wavefunction also appears].
We discuss the continuity of these distributions in terms of relations between Fock components at vanishing plus momentum. 
Connection is also made to the overlap representation of generalized parton distributions.
Construction of the higher Fock components directly from light-front perturbation theory is reviewed in Appendix \ref{oftopt}.
Further applications of the reduction formalism to timelike processes are contained in Appendix~\ref{chap:time}.
We conclude our study of the light-front BSE with a brief summary (Section \ref{summy1}).    
Finally we discuss limitations in modeling GPDs from a few different perspectives.

\section{Definitions and properties} \label{gpdrev}

In this Section we define GPDs in terms of a light-like correlation of hadronic matrix elements. In order 
to motivate this definition, we consider DVCS which is the theoretically cleanest process into which the 
GPDs enter. The discussion of DVCS below builds on that of DIS presented in Section \ref{DIS}. 
In contrast to DIS, DVCS is an exclusive process where the final states are detected. Despite this difference, 
DVCS and DIS are quite similar theoretically.

\subsection{DVCS}

\begin{figure}
\begin{center}
\epsfig{file=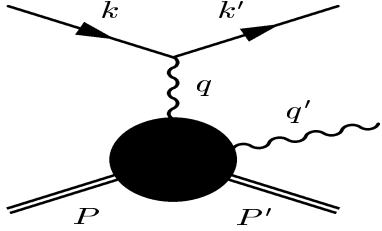,width=2in,height=1.2in}
\end{center}
\caption[Diagrammatic depiction of the DVCS process.]{Diagrammatic depiction of the DVCS process. 
An electron of momentum $k$ is incident on a proton of momentum $P$. The leading-order
electromagnetic process is virtual photon exchange between the electron and proton. As a result, the electron's momentum is altered to $k'$. 
The reaction produces an intact proton of momentum $P'$ and a real photon, $q'^2 =0$. }
\label{f:DVCS}
\end{figure}

In DVCS off the proton (or other targets), one is interested lepton scattering at high momentum transfer. 
The basic reaction is (see Figure \ref{f:DVCS})
\begin{equation}
e(k) + p(P) \longrightarrow e(k') + p(P') + \gamma(q')
.\end{equation}
Thus the situation of the reaction is similar to DIS in setup, however, afterward one is interested in 
measuring not only the final electron, but also recovering an intact proton (altered to some momentum $P'$)
in addition to a real photon, $q'^2 = 0$. Experimentally it is easier to deduce the existence of the final state photon 
by missing mass techniques rather than to detect it directly. Furthermore the existence of a plethora of other channels
requires high luminosity beams and long run times in order to filter out enough DVCS events exclusively.

As in DIS, we label the momentum transfered to the proton as $q = k - k'$. 
We will again work to leading order in QED, for which the momentum is transfered via
a single virtual photon exchange. In the deeply virtual limit, the photon's virtuality becomes arbitrarily large, $q^2 \to - \infty$.
The amplitude for this process can be written in terms of the off-diagonal Compton amplitude
$T^{\mu \nu}(q,P,\D)$, where $\D = P' - P$ is the momentum transfered to the final state proton. 
The off-diagonal Compton amplitude is
\begin{equation}
T^{\mu \nu}(q,P,\D) = \int d^4 y \, e^{i q \cdot y} \langle P' | T \Big\{ J^\mu (0) J^\nu (y) \Big\} | P \rangle
,\end{equation} 
and differs from the DIS case only by the final state proton being off the momentum-space diagonal. 
The overall amplitude $\mathcal{A}$ for the DVCS process has the form
\begin{equation} \label{eqn:DVCSamp}
\mathcal{A}_{\text{DVCS}} =  \epsilon^*_\mu (q') L_{\nu}(k,q) T^{\mu \nu}(q,P,\D)
,\end{equation}
where $\epsilon^*_\mu(q')$ is the polarization vector for the final state photon,
and $L_\nu(k,q)$ describes the QED virtual photon exchange off the electron line.
Notice $q' = q - \D$ by momentum conservation.
We include the form of the amplitude Eq.~\eqref{eqn:DVCSamp} in order to contrast 
with the case of DIS, where the amplitude squared was under investigation. Moreover
we are interested in the amplitude as a whole, not just the imaginary part.

Our setup so far has paralleled the case of DIS. We shall thus choose our variables in a similar 
fashion. This choice corresponds to the asymmetric frame employed by Radyushkin. Nearing the end of this
section we will also provide expressions in the symmetric frame of Ji. In the rest of this work we 
will make use of both frames and their correspondingly different sets of variables. 
In the asymmetric frame, we have $\mathbf{P}^\perp = 0$, as in Section \ref{DIS}; and the decomposition 
of the initial state's momentum $P^\mu$ is identical to Eq.~\eqref{eqn:P}. 
Let us define the Bjorken variable for DVCS as 
\begin{equation}
\zeta = - \frac{q^2}{2 P \cdot q}
.\end{equation}
Further let us choose $q \cdot n = 0$, so that $- q^2 = \mathbf{q}_\perp^2$ gives the photon's virtuality.
This fixes the photon momentum to be
\begin{equation}
q^\mu = q_\perp^\mu  - \frac{q^2}{2 \zeta} n^\mu
.\end{equation}
Now for the $t$-channel momentum transfer $\D^\mu$, we have $\D^2 = t$. Furthermore $\D$ is restricted
from the constraint $q'^2 =0$ from the final photon.  
Thus we have the decomposition
\begin{equation} 
\label{eqn:D}
\D^\mu = - \zeta \, p^\mu + \D_\perp^\mu  + \frac{\D_\perp^2 - t}{\zeta} n^\mu
,\end{equation}
which holds only as $q^2 \to - \infty$. Notice the plus component of the final state photon
momentum is $q'^+ = \zeta P^+$.

\begin{figure}
\begin{center}
\epsfig{file=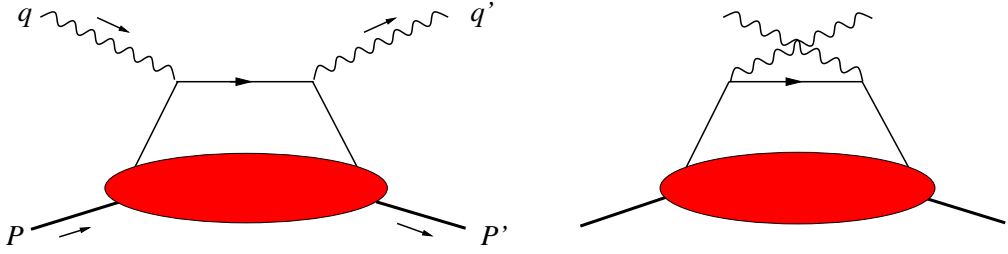,width=5.25in}
\end{center}
\caption[Graphical depiction of the handbag mechanism in DVCS.]{Graphical 
depiction of the handbag mechanism in DVCS. 
There are two contributions to the off-diagonal Compton amplitude: 
the handbag shown on the left, and the crossed
bag shown on the right. Figure adapted from \cite{Diehl:2003ny}}
\label{f:hand2}
\end{figure}

As we know from DIS, in the deeply virtual limit the Compton amplitude is 
dominated by the leading light-cone singularity in the product of currents. 
Taking this contribution generates the so-called handbag dominance due to asymptotic freedom,
see Figure \ref{f:hand2}. The only differences compared to Section \ref{DIS} are that 
the final states have altered momenta, the final photon is real, and we are not just looking at 
the cuts of the diagrams. There is an additional requirement so that the amplitude 
maintains a clear separation of scales, 
namely the stipulation that $|t / q^2| \ll 1$.

Using the handbag dominance, we arrive at the leading contribution to the off-diagonal Compton 
amplitude in the deeply virtual limit
\begin{equation} \label{eqn:Toff}
T^{\mu \nu}(q,P,\D) = - i \int \frac{d^4 k}{(2\pi)^4} 
\tr 
\Bigg\{ \Bigg[ 
\gamma^\mu \frac{i (\rlap \slash k + \rlap \slash q)}{(k + q)^2 + i \varepsilon }
\gamma^\nu 
+ 
\gamma^\nu \frac{i (\rlap \slash k - \rlap \slash q + \rlap \slash \Delta)}{(k - q + \D)^2 + i \varepsilon}
\gamma^\mu
\Bigg]
M(k, P, \D)
\Bigg\}
,\end{equation}
where we have written down contributions from both handbags. The off-diagonal
free quark-proton Green's function appearing in Eq.~\eqref{eqn:Toff} is 
\begin{equation}
M_{\a \b}(k,P,\D) = \int d^4 y  \, e^{i k \cdot y} \langle P' | T \Big\{ \ol \psi_\b (0) \psi_\a (y) \Big\} | P \rangle
,\end{equation}
which again differs from the DIS case only by the momentum $\D$ transfered to the final state.

We now use the collinear approximation to obtain the leading contribution from 
this Green's function in the deeply virtual limit.  The leading contribution in Eq.~\eqref{eqn:Toff}
can be isolated by taking $k^\mu \approx (k \cdot n) \, p^\mu$. Further we shall 
define $X  = k \cdot n$, i.e., $X = k^+ / P^+$. In the collinear approximation, the 
off-forward Compton amplitude now appears as\footnote{%
As in DIS, the are perturbative QCD corrections to the formula Eq.~\eqref{eqn:Toff2} 
that result in the factorization scale dependence of both the hard scattering kernel 
and the soft matrix element, here $M(x,\zeta, t, \mu^2)$, such that $T^{\mu \nu}(q,P,\D)$ 
is scale independent. Additionally similar to DIS, we can interpret $\mu$ as the resolution
scale for the partonic process. For a complete discussion of $\cO(\a_s)$ corrections to DVCS
and beyond, see, e.g.~\cite{Belitsky:2001ns,Diehl:2003ny}.
}
\begin{equation}
\label{eqn:Toff2}
T^{\mu \nu}(q,P,\D) = \frac{1}{2} \left( 
g^{\mu \nu} - p^\mu n^\nu - p^\nu n^\mu
\right)
\int dX 
\left( 
\frac{1}{X - \zeta + i \varepsilon} 
+ 
\frac{1}{X + i \epsilon}
\right)
M(X,\zeta,t)
,\end{equation}
where we have introduced the function $M(X,\zeta,t)$ defined by
\begin{equation}
M_{\a \b}(X,\zeta,t) = \int \frac{d^4 k}{(2\pi)^4} \delta( X - k \cdot n) M_{\a \b}(k,P,\D) 
.\end{equation}
As in the DIS, the integral over $y$ that is buried in  $M_{\a \b}(k,P,\D)$ can be evaluated trivially 
and we thus arrive at
\begin{equation}
M_{\a \b}(X,\zeta,t) = \int \frac{d \l}{2\pi} \, e^{i \l X} \, \langle P' | \ol \psi_\b (0) \psi_\a ( \l n ) | P \rangle
.\end{equation}

Following Section \ref{DIS}, we now separate $M_{\a \b}(X,\zeta,t)$ into irreducible structure functions. We will 
do this only schematically. First we write out the general dependence on Dirac structures
\begin{equation} 
M_{\a \b}(X,\zeta,t) 
= 
\frac{1}{2} \rlap \slash p_{\a \b} \,  F(X,\zeta,t) 
+ 
\frac{1}{2} \rlap \slash \Delta_{\a \b}^\perp \, F_3(X,\zeta,t) 
+ 
\frac{1}{2} \rlap \slash n_{\a \b} \, M^2 F_4(X,\zeta,t)
.\end{equation}
Next we observe that by taking various traces we can write
\begin{equation} \label{eqn:GPDdecompo}
\int \frac{d \l}{2\pi} \, e^{i \l X} \, \langle P' | \ol \psi (0) \gamma^\mu \psi ( \l n ) | P \rangle
= 
2 \left[
F(X,\zeta,t) \, p^\mu 
+ 
F_3(X,\zeta,t) \, \D_\perp^\mu
+
M^2 F_4(X,\zeta,t) \, n^\mu
\right]
.\end{equation}
Finally we identify $F(X,\zeta,t)$ as the leading structure function for DVCS.\footnote{%
As in DIS, beyond leading order in QCD perturbation theory, the bilocal operator in Eq.~\eqref{eqn:GPDdecompo}
must be renormalized. This renormalization leads to $\mu^2$ dependence of the structure functions
but does not affect the twist decomposition.
} 
The contributions from 
the other structure functions $F_3(X,\zeta,t)$ and $F_4(X,\zeta,t)$ are of higher twist (suppressed 
by $|\D_\perp|/Q$ and $M^2 / Q^2$, respectively). The GPDs are contained in the function $F(X,\zeta,t)$.

\subsection{GPDs}

In the last Section, we reduced the amplitude for DVCS down to one leading contribution in the deeply virtual limit. 
This leading contribution is a convolution of a hard scattering part [seen in Eq.~\eqref{eqn:Toff2}], and a 
soft hadronic matrix element $F(X,\zeta,t)$.  The latter contains the GPDs. In this Section, we will inspect the GPDs
more closely and we shall also switch frames from Radyushkin's asymmetric frame to Ji's symmetric one. 
The switch to the symmetric frame illuminates further properties of the GPDs.

Utilizing  Eq.~\eqref{eqn:GPDdecompo}, we can write the leading-twist structure function as
\begin{equation} \label{eqn:Fdef}
F(X,\zeta,t) = \frac{1}{2} \int  \frac{d \l}{2\pi} \, e^{i \l X} \, \langle P' | \ol \psi (0) \, \rlap \slash n  \, \psi ( \l n ) | P \rangle
.\end{equation}
From this relation we see immediately that the PDFs are contained in the limit $\D^\mu \to 0$, namely $F(X,0,0) = f_1(X)$. 
For the case of the proton (or any other spin-$\frac{1}{2}$ target), we can decompose the matrix element in terms of 
two independent GPDs $H(X,\zeta,t)$ and $E(X,\zeta,t)$, \emph{viz}.
\begin{equation} \label{eqn:RadGPDs}
F(X,\zeta,t) = \ol u(P') \left[
\rlap \slash n \, H(X,\zeta,t) + \frac{i \sigma^{\mu \nu} n_\mu \D_\nu}{2 M} E(X,\zeta,t)
\right] u(P)
.\end{equation}
Notice by virtue of the Gordon identity (see Appendix \ref{sec:gamma}), 
a possible third structure proportional to $n \cdot \D$ is not independent.
Furthermore, the reduction relation to PDFs involves only $H(X,\zeta,t)$, i.e.,
$f_1(X) = H(X,0,0)$
The independent function $E(X,\zeta,t)$ is simply not visible in DIS because
its presence in any amplitude requires momentum transfer.

The decomposition of the function $F(X,\zeta,t)$ into GPDs $H(X,\zeta,t)$ and
$E(X,\zeta,t)$ in Eq.~\eqref{eqn:RadGPDs} is reminiscent of the Dirac and Pauli electromagnetic form factors.
Moreover, in DVCS the proton remains intact which suggests an even deeper connection with the proton's elastic form factors.  
This connection is afforded by the ``sum rule'' that produces the local limit of the bilocal field operators in Eq.~\eqref{eqn:Fdef}. 
Notice the effect of integration over $X$ in Eq.~\eqref{eqn:Fdef} produces the local limit
\begin{equation}
\int \frac{dX}{2 \pi} \, e^{i \l X} \, \ol \psi(0) \, \rlap \slash n \, \psi (\l n)
= 
\delta (\l) \, \ol \psi(0) \, \rlap \slash n \, \psi (0)
.\end{equation}
Thus we find the sum rules
\begin{align} 
\int dX \, H(X,\zeta,t) &= F_1(t) \label{one} \\
\int dX \, E(X,\zeta,t) &= F_2(t) \label{two}
,\end{align}
where $F_1(t)$ is the Dirac form factor and $F_2(t)$ is the Pauli form factor. 
The $\zeta$ dependence drops out\footnote{%
The local limit also produces an operator of zero anomalous dimension and 
hence the sum rules Eq.~\eqref{one} and \eqref{two} are independent
of the renormalization scale $\mu$. Physically the form factors do not depend
upon the resolution scale chosen because it is the (transverse) distribution of 
total charge that is probed. 
} 
upon integration over all $X$. This is
a manifestation of Lorentz invariance because in calculating form factors, 
one can choose a frame in which the momentum transfer $\D^+ = - \zeta P^+ = 0$. 
For DVCS, however, 
there is no remaining freedom to use such a frame since the 
virtual photon's momentum was chosen so that $q^+ = 0$. 
The requirement $\zeta \neq 0$ implies that DVCS is sensitive to physics
otherwise absent from electromagnetic form factors due to Lorentz invariance.

In fact, there are more stringent constraints imposed by Lorentz invariance. 
The $n^\text{th}$ moment of each GPD with respect to $X$ must be
a polynomial in  $\zeta$ of order $n$. This constraint is referred to as the 
``polynomiality'' property of GPDs. We will devote ample space to the issue of polynomiality
and the special case of  sum rules in Chapters \ref{chap:DDs} and \ref{chap:proton}.

There is a further property that the GPDs must satisfy: they are bounded by the positivity of the 
norm on Hilbert space. This condition implies a bound, referred to generally as a positivity bound or very imprecisely 
as just the positivity property of GPDs.  Looking at the field operators in Eq.~\eqref{eqn:Fdef}, we see
that the definition of $F(X,\zeta,t)$ is similar to that of a mixed density. As such it is constrained
by the PDFs. 
Using the Cauchy-Schwartz inequality, one finds the bound
\begin{equation} \label{posbo}
\theta ( X - \zeta ) \Big| F(X,\zeta,t) \Big| 
\leq 
\sqrt{f_1 \Big( X \Big) f_1 \Big( \frac{X - \zeta}{1 - \zeta} \Big) }
\end{equation}
that we shall refer to as positivity.\footnote{%
The positivity bound is stable under QCD evolution, i.e.~with Eq.~\eqref{posbo} modified to 
include $\mu^2$ dependence in all three functions. This can be demonstrated by studying the evolution 
kernels. Alternately one can use the light-cone wavefunction representation where the Fock 
components themselves acquire scale dependence (because gluon radiation mixes Fock components). 
Since the positivity bound is manifest in the wavefunction representation, writing down $\mu^2$ dependence
in the light-cone wavefunctions in $F(X, \zeta, t, \mu^2)$, one immediately sees
\begin{equation} \notag
\theta ( X - \zeta ) \Big| F(X,\zeta,t,\mu^2) \Big| 
\leq 
\sqrt{f_1 \Big( X ,\mu^2 \Big) f_1 \Big( \frac{X - \zeta}{1 - \zeta}, \mu^2  \Big) }
.\end{equation}  
} 
The positivity bound is a property that shall haunt us
throughout Chapters \ref{chap:DDs} and \ref{chap:proton} and thus we only briefly mention it here.

As a final point in our introductory discussion of GPDs,
we switch reference frames and variables to define GPDs that are more symmetric with 
respect to the initial and final states. To do so, we define the average momentum 
between the states as $\ol P {}^\mu = \frac{1}{2} ( P + P')^\mu$ so that the initial momentum $P = \ol P - \D / 2$
and the final momentum $P' = \ol P + \D / 2$.  Now we choose a frame of 
reference in which $\mathbf{ \ol P} {}^\perp = 0$.  This frame is related to the asymmetric frame by a boost
and so strictly speaking $\D_\perp$ is different from above.
It is then more convenient to use a modified Bjorken variable. Recall $\zeta = - \D^+ / P^+$ is a measure
of the longitudinal momentum transfer relative to the initial proton's momentum. In the symmetric frame, 
a useful quantity to consider is the amount of longitudinal momentum transfer relative to the average proton momentum.
Thus we define the variable $\xi$ by
\begin{equation}
\xi = - \frac{\D \cdot n}{2 \ol P \cdot n} = - \frac{\D^+}{ 2 \ol P {}^+}
.\end{equation}
The conversion between $\zeta$ and $\xi$ is
\begin{equation}
\zeta = \frac{2 \xi}{1 + \xi}
\end{equation}

The properly symmetric definition for the GPDs appears as
\begin{equation} \label{eqn:Fsymm}
F(x,\xi,t) = \frac{1}{2} \int  \frac{d \l}{2\pi} \, e^{i \l x} \, \langle P' | \ol \psi(-\l n  / 2) \, \rlap \slash n  \, \psi ( \l n /2 ) | P \rangle
,\end{equation}
which features the variable $x$ instead of $X$. The two are related by
\begin{equation}
X = \frac{x + \xi}{1 + \xi}
,\end{equation}
because $x$ is measured relative to the average plus momentum.
The operator combination in Eq.~\eqref{eqn:Fsymm} is Hermitian and so $F(x,\x,t)$ is real. 
Under the transformation $\D^\mu \to - \D^\mu$, we thus have $F(x,-\x,t) = F(x,\x,t)$
which is commonly referred to as $\x$-symmetry.  This symmetry naturally extends to the 
GPD moments which are hence required to be even polynomials in $\x$.

As was the case for PDFs, a natural physical picture for the GPDs emerges when we use 
light-cone quantized fields. This is because the bilocal operator entering into the GPD 
definition is an equal light-cone time operator and not an equal time operator. 
Despite this similarity to PDFs, 
there is not 
a general probabilistic interpretation for the GPDs because the requisite matrix element 
has more the form of a mixed density operator. Decomposition of the GPD
in terms of light-cone quark and anti-quark modes is presented in Appendix \ref{chap:LCQ}.
In the special case $\x = 0$, one can manipulate the GPDs into true probability distributions of 
quarks with momentum fraction $x$ in impact parameter space \cite{Burkardt:2000za,Burkardt:2002hr}. The case for $\x \neq 0$
does not have a probabilistic interpretation, however, some physical insight can be gained \cite{Diehl:2002he}.

Having motivated the definitions of GPDs from analyzing DVCS and having presented a brief description of their 
properties, it makes sense to consider a toy model example of GPDs. We do so in the following Section.

\section{GPDs and the light-front BSE} \label{gpds}

In this Section, we cast our results for form factors in the Wick-Cutkosky model \cite{Tiburzi:2002sx} in the language of GPDs
and the light-cone Fock space expansion. The $(3+1)$-dimensional expressions for form factors are presented in Appendix \ref{fff}. 
These results are the higher dimensional generalizations of the $(1+1)$-dimensional expressions we derived in Section \ref{meat}. 
Additionally one can obtain these results directly from time-ordered perturbation theory using two-body projection operators 
as explicated in Appendix \ref{oftopt}.

In contrast to the proton case (to which we return in Chapter \ref{chap:proton}), the GPD\footnote{%
Notice we have readily abused notation. Here we use the variable $x$ to denote the active quark's plus momentum fraction in the asymmetric 
frame. Above in Section \ref{gpdrev}, we used $X$ but now return to the Wick-Cutkosky model of Chapter \ref{chap:LFBS} where we 
employed $x$ throughout.
}
$F(x,\zeta,t)$ for our meson model is defined by a non-diagonal matrix 
element of bilocal field operators
\begin{equation} \label{bilocal}
F(x, \zeta, t) = \int \frac{dy^-   }{4\pi} e^{i x P^+ y^-}
\langle \Psi_{P^\prime} | \; \phi(0)   i \overset{\leftrightarrow}\partial{}^+ \phi(y^-) \; | \Psi_{P} \rangle,
\end{equation}
where $\phi(x)$ denotes the quark field operator and $\overset{\leftrightarrow}\partial{}^\mu = \overset{\rightarrow}\partial{}^\mu - 
\overset{\leftarrow}\partial{}^\mu$ is the electromagnetic coupling to scalar particles. 
Here we are considering only the twist-two contribution\footnote{%
Additionally by ignoring self-energy corrections, there is no $\mu^2$ dependence in the model. 
} 
and have taken accordingly the plus-component
of the vector current in Eq.~\eqref{bilocal}.
Comparing to the current matrix element $J^\mu$ in Appendix \ref{fff}, the definition of the GPD leads immediately to the sum rule
\begin{equation} \label{sumrule}
\int \frac{dx}{1 - \zeta/2} F(x, \zeta, t) = F(t)
,\end{equation}
where $F(t)$ is the electromagnetic form factor for our model.
Hence one can calculate these distributions from the integrand of the form factor because the non-local light-cone operator
merely fixes the plus momentum of the active quark. In this way, the light-cone correlation defined in 
Eq.~\eqref{bilocal} has a natural description in terms of light-front time-ordered perturbation theory, e.g., for $x>\zeta$ the 
relevant graphs contributing to the GPD are in Figure \ref{ftri2}, and those for $x<\zeta$ are in Figure \ref{fZZZ}. 
Conversion of the contributions to the form factor into GPDs is straightforward using Eq.~\eqref{sumrule}; we merely 
remove $- 2 i P^+$ and $\int dx$ from Eqs.~(\ref{FFLO}-\ref{fs2}). These cumbersome expressions are not worth generating here.

\subsection{Continuity}

In order for the deeply virtual Compton scattering 
amplitude to factorize into hard and soft pieces (at leading twist), 
the GPDs $F(x,\zeta,t)$ must be continuous at $x = \zeta$, see Eq.~\eqref{eqn:Toff2}. 
There is no proof of this from QCD, but the power-law scaling seen experimentally allows us to infer the GPDs continuity. 
Maintaining continuity at the crossover is more pressing because  
experiments which measure the beam-spin asymmetry are limited to the crossover \cite{Diehl:1997bu}. 
Let us scrutinize continuity in our results for the Wick-Cutkosky model.

The leading-order expressions we have derived for the Wick-Cutkosky model are continuous. This 
is easy to see since the contribution for $x < \zeta$ is identically zero. 
The valence wavefunction contribution for $x>\zeta$ is a convolution of wavefunctions
one of which is $\psi^*(x^\prime,\ldots)$ which is probed at the end point since $x^\prime \equiv \frac{x - \zeta}{1-\zeta} \to 0$. 
From the bound-state equation Eq.~\eqref{wavefunction}, 
we see the two-body wavefunction vanishes quadratically at the end points. Taking into account the overall weight $x^{\prime -1}$, the 
valence piece vanishes linearly at the crossover. At leading order then, continuity is maintained at the crossover, while the derivative
is discontinuous. Working only in the valence sector, valence quark models will never be of any use to beam-spin asymmetry measurements 
since the value at the crossover requires one wavefunction to be at an end point.\footnote{In the three-body bound state problem (e.g. the nucleon), 
the valence GPD will vanish only if the three-body interaction 
is non-singular at the end points (which is physically reasonable and perturbatively true)
\cite{Tiburzi:2002fq}.}

Let us now check the next-to-leading order contributions to the GPD for continuity. First we shall deal with the term stemming from iterating 
the Bethe-Salpeter equation for the initial state \eqref{is1} (see diagram $B$ of Figure \ref{ftri2}).
 Since there is no corresponding Z-graph generated from iterating the initial state, we expect this contribution to vanish. 
Looking at the expression, we see again the factor 
$\psi^*(x^\prime,\ldots)/x^\prime$ which vanishes linearly as $x \to \zeta$. Moreover there are the interaction terms: 
first $D(y,\pperp;x,\kperp|M^2) $ which is finite as $x \to \zeta$, and then 
\begin{equation}
D(\yp,\ppp;\xp,\kpp|M^2) \overset{x\to \zeta}{=} \frac{-\xp}{(\kperp + \mathbf{\Delta}^\perp)^2 + m^2}, 
\end{equation}
which vanishes at the crossover. Thus not only does the initial-state iteration term vanish at the crossover, its derivative does so as well.

Now we investigate the Born terms Eq.~\eqref{bt1} (see diagram $A$ of Figure \ref{ftri2})  
and Eq.~\eqref{bt2} (see diagram $D$ of Figure \ref{fZZZ}) at the crossover. Approaching $\zeta$ from above,
we use Eq.~\eqref{bt1}, and we 
have the finite contribution at the crossover
\begin{multline}\label{bcross}
F(\zeta, \zeta, t)^{\text{Born}} =  \int \frac{d\kperp dy \, d\pperp}{(16\pi^3)^2  y (1-y)\yp }  
\psi^*(y^\prime, \ppp)  \\ \times \frac{g^2 \theta(y - \zeta)/(y - \zeta) }{(\kperp + \mathbf{\Delta}^\perp)^2 + m^2} 
D(y,\pperp;\zeta,\kperp|M^2) \psi(y, \pperp). 
\end{multline} 
On the other hand, approaching the crossover from below in Eq.~\eqref{bt2} we have to deal with singularities as $x^{\prime\prime} = x/ \zeta \to 1$. Writing out
the propagator for the quark-antiquark pair heading off to annihilation, we see
\begin{equation} \label{qqbarprop}
\dw(x^{\prime\prime}, \kppp | t ) \to -  \frac{1 - x^{\prime\prime}}{(\kperp + \mathbf{\Delta}^\perp)^2 + m^2}.
\end{equation}
This linear vanishing cancels the weight $(1 - x^{\prime\prime})^{-1}$. Taking the limit $x \to \zeta$ then produces
equation \eqref{bcross} and thus the Born terms are continuous.

Lastly we must see how the final-state iteration terms match up at the crossover. 
Using Eq.~\eqref{fs1} to approach $\zeta$ from above (see diagram $C$ of Figure \ref{ftri2}), we have the contribution
\begin{multline}\label{fcross}
F(\zeta, \zeta, t)^{\text{final}} = \int \frac{d\kperp d\yp d\ppp}{(16\pi^3)^2 (1-\zeta)\yp (1-\yp) } 
\psi^*(y^\prime, \ppp) 
\\ \times \frac{g^2/y }{(\kperp + \mathbf{\Delta}^\perp)^2 + m^2} 
D(y,\pperp;\zeta,\kperp|M^2) \psi(\zeta, \kperp).
\end{multline}
Approaching $\zeta$ from below (see diagram $E$ of Figure \ref{fZZZ}), we utilize equation \eqref{qqbarprop} in taking the limit of \eqref{fs2}. The result is \eqref{fcross}
and hence we have demonstrated continuity to first order, i.e.
\begin{equation} \label{contequ}
F(\zeta,\zeta,t) = F(\zeta,\zeta,t)^{\text{Born}} +   F(\zeta,\zeta,t)^{\text{final}}
\end{equation}
no matter how we approach $x = \zeta$.

\subsection{Fock space representation}

We now write the derived GPDs in terms of Fock component overlaps. In the diagonal overlap region $x > \zeta$ this will be a mere rewriting 
of our results, while there is a subtlety for the non-diagonal overlaps. To handle the zeroth-order term in Eq.~\eqref{FFLO}, we 
define the two-body Fock component as
\begin{equation} \label{twofock}
\psi_{2}(x_{1},\mathbf{k}^\perp_1, x_{2}, \mathbf{k}^\perp_2) = \frac{1}{\sqrt{ x_{1} x_{2} }} \psi(x_{1}, \mathbf{k}^\perp_{\text{rel}}), 
\end{equation}
noting that the relative transverse momentum can be defined as $\mathbf{k}^\perp_{\text{rel}} = x_{2} \mathbf{k}^\perp_{1} - x_{1} \mathbf{k}^\perp_{2}$. 
In terms of Eq. \eqref{twofock}, the GPD appears as 
\begin{equation} \label{222}
F(x, \zeta, t)^{\text{LO}} = \frac{\theta(x - \zeta)}{\sqrt{1-\zeta}} \int [dx]_{2} [d\kperp]_{2} \sum_{j = 1,2} \delta(x - x_{j})
  \psi_{2}^*(x^{\prime}_{i}, \mathbf{k}^\prime_{i}{}^\perp) \frac{2 x_j - \zeta}{\sqrt{ x^{\prime}_{j} x_j }}  \psi_{2}(x_{i},\mathbf{k}_{i}^\perp),  
\end{equation}
where the primed variables are given by
\begin{equation} \label{primed}
\begin{cases}
x^\prime_{i} = \frac{x_{i}}{1-\zeta}\\
\mathbf{k}^\prime_{i}{}^\perp  = \mathbf{k}_{i}^\perp - x^\prime_{i} \mathbf{\Delta}^\perp, 
\end{cases}
\; \text{for} \; i \neq j 
\qquad
\text{and}
\begin{cases}
x^\prime_{j} = \frac{x_{j} - \zeta}{1 - \zeta}\\
\mathbf{k}^\prime_{j}{}^\perp = \mathbf{k}_{j}^\perp + (1 - x^\prime_{j}) \mathbf{\Delta}^\perp
\end{cases}
\end{equation}
and the integration measure is given by
\begin{align}
[dx]_{N} & = \prod_{i = 1}^{N} dx_{i} \; \delta \Big( 1 - \sum_{i = 1}^{N} x_{i} \Big)\\
[d\kperp]_{N} & = \frac{1}{[2(2\pi)^3]^{N-1}} \prod_{i=1}^N d\mathbf{k}^\perp_{i} \; \delta \Big( \sum_{i=1}^N \mathbf{k}^\perp_{i} \Big).
\end{align}
Notice the sum over transverse momenta in the delta function is zero since our initial meson has $\mathbf{P}^\perp = 0$.
The sum over $j$ in Eq. \eqref{222} produces the overall factor of two for our case of equally massive (equally charged) constituents.

To cast the next-to-leading order expressions for $x > \zeta$ in terms of diagonal Fock space overlaps, we must write out the three-body Fock component.
Looking at the diagrams in Figure \ref{ftri2}, it is constructed from the two-body wavefunction in the following manner
\begin{multline} \label{threefock}
\psi_{3} (x_{i},\mathbf{k}_{i}^\perp) = g \frac{2 (2\pi)^3 }{\sqrt{x_{1} x_{2} x_{3}}} 
\int [dy]_{2} [d\pperp]_{2} \frac{\psi_{2}(y_{j},\mathbf{p}_{j}^\perp)}{\sqrt{y_{1} y_{2}}} \\
\Bigg[ \theta(y_{1} - x_{1}) x_{3} \delta(y_{2} - x_{3}) 
\delta(\mathbf{p}^\perp_{2} - \mathbf{k}^\perp_{3}) D(y_1,\mathbf{p}^\perp_1;x_1,\mathbf{k}^\perp_1|M^2)
\\ + \theta(y_2 - x_3) x_1 \delta(y_{1} - x_{1}) \delta(\mathbf{p}^\perp_{1} - \mathbf{k}^\perp_{1}) 
D(y_2,\mathbf{p}^\perp_2;x_3,\mathbf{k}^\perp_3|M^2)
\Bigg] ,
\end{multline}
where $i$ runs from one to three and the label $j$, which stems from the integration measure, runs from one to two. 
We discuss how to obtain this three-body wavefunction directly from time-ordered perturbation theory in 
Appendix \ref{oftopt}. Using $\psi_3$ in Eq. \eqref{threefock}, the terms in the GPD at first 
order in the weak coupling can then be written compactly as
\begin{equation} \label{323}
F(x,\zeta,t)^{\text{NLO}} = \frac{\theta(x - \zeta)}{1 - \zeta} \int [dx]_3 [d\kperp]_{3} \sum_{j = 1,3} 
\delta( x - x_{j}) \psi_{3}^*(x_{i}^\prime,\mathbf{k}_{i}^\prime{}^\perp) \frac{2 x_j - \zeta}{\sqrt{ x^{\prime}_{j} x_j }}
\psi_{3}(x_{i},\mathbf{k}^\perp_i).
\end{equation}
One can verify that the diagrams in Figure \ref{ftri2} are generated by Eq.~\eqref{323}. 
Additionally there is a fourth diagram generated by Eq. \eqref{323} which does not appear in the figure. This missing diagram is characterized
by the spectator quark's one-loop self interaction and is absent since we have ignored $f(k^2)$ and the scale dependence of light-cone wave 
functions.\footnote{%
Recently the scale dependence of light-cone Fock components in QCD has been investigated \cite{Burkardt:2002uc}.
In the Wick-Cutkosky model, the scale dependence is of course different and leads to different evolution
compared to perturbative QCD. 
}
The absence of this diagram does not affect continuity at the crossover. The missing diagram vanishes at $x = \zeta$ since the 
final-state wavefunction is $\psi^*(x^\prime,\ldots)$, which is probed at the end point. Lastly we note the above Fock component overlaps satisfy the positivity
constraint for a composite scalar composed of scalar constituents \cite{Tiburzi:2002kr}. Moreover, with the inclusion of the self-energy loop, 
the form factor is gauge invariant \cite{Kvinikhidze:2003de} and hence the GPDs are properly gauged, in the sense that the bilocal
light-cone operator is inserted in all possible places.

Now we must come to terms with the non-diagonal overlap region, $x < \zeta$. At first order, the diagrams of Figure \ref{fZZZ}
correspond to four-to-two Fock component overlaps. We have been cavalier about time ordering, however. The expressions  
Eqs.~\eqref{bt2} and \eqref{fs2} do not correspond to time-ordered graphs. Both terms contain a product of time-ordered propagators:
one for the two quarks leading to the final-state vertex and another for the quark-antiquark pair heading off to annihilation. But for an interpretation
in terms of a four-body wavefunction, all four particles must propagate at the same time. This is a subtle issue as a graph containing the product 
of two independently time-ordered pieces (where one leads to a bound-state vertex) corresponds to a sum of infinitely many time-ordered graphs. 
It is easiest to write out the relevant pieces in terms of the propagators' poles. The quark, anti-quark heading to annihilation
have propagators $d(k)$ and $d(k+\Delta)$ and poles we label $k^-_a$ and $k^-_c$, respectively. The remaining propagators of interest
$d(p^\prime)$ and $d(P^\prime - p^\prime)$ have poles $p^-_a$ and $p^-_b$, respectively.
We can then manipulate as follows
\begin{equation} 
\frac{1}{k^-_{c} - k^-_{a}} \; \frac{1}{p^-_{b} - p^-_{a}} 
	=  \frac{1}{p^-_{b} - p^-_{a} + k^-_{c} - k^-_{a}} \; \Bigg( 
\frac{1}{p^-_{b} - p^-_{a}} + \frac{1}{k^-_{c} - k^-_{a}} \Bigg).   \label{trickery}
\end{equation}
In this form, we have produced the correct energy denominator for the instant of light-front time where four particles are propagating. 
Multiplying this denominator by the three-body wavefunction yields the four-body wavefunction (up to constants).
This is the part of the four-body wavefunction relevant for GPDs (there are additional pieces for two-quark, two-boson states, see Appendix 
\ref{oftopt}).  
In the resulting sum Eq.~\eqref{trickery}, the first term will produce the two-body wavefunction for the final state and we have a
genuine four-to-two overlap. We do not write this out explicitly.

The second term in Eq.~\eqref{trickery}, however, contains again the propagator for the pair heading to annihilation. Using the light-front 
Bethe-Salpeter equation for the vertex (which contains infinitely many times) we can introduce a factor of the time-ordered interaction. 
The resulting product of independent time orderings can again be manipulated as in Eq. \eqref{trickery}. The result produces another overall four-body
denominator which contributes to the four-body Fock component of the initial state. Since we iterated the interaction, however, this new contribution
is no longer at leading order and can be neglected. Thus the second term in \eqref{trickery} does not contribute at this order.

Having manipulated the GPDs into non-diagonal overlaps for $x < \zeta$, we must wonder if continuity at the crossover is still maintained. In the limit 
$x \to \zeta$ the light-front energy of the struck quark goes to infinity. 
Consequently $k^-_{c}$, 
which contains this on shell energy, is infinite and dominates the four-body energy denominator. This is identical to the reasoning
in Eqs.~\eqref{bt2} and \eqref{fs2} where instead of the four-body denominator, we have the propagator
$\dw(x^{\prime\prime}, \kperp + x^{\prime\prime} \mathbf{\Delta}^\perp |t)$,
which is dominated by $k^-_{c}$ at the crossover. Either way, 
we arrive at the expressions found above for the crossover Eqs.~\eqref{bcross} and \eqref{fcross}.

Having cast our expressions for generalized parton distributions in terms of the Fock components, 
we can enlarge our understanding of the sum rule and continuity at the crossover. Both must deal with the relation between higher Fock components. The way 
$\zeta$-dependence disappears from Eq.~\eqref{sumrule} mandates a relation between the diagonal and non-diagonal Fock component overlaps that make up the 
GPD. Such a relation between Fock components must follow from the field-theoretic equations of motion. Continuity itself is a special
case of the relation between Fock components, specifically at the end points. Above we have seen our expressions are continuous (and non-vanishing)
at the crossover and explicitly that the three- and four-body components match at the end point (where $x - \zeta = 0$). This weak binding model for behavior 
at the crossover is a simple example of the relations between Fock components at the end points\footnote{%
The known
Fock component end-point relations in gauge theory derived in \cite{Antonuccio:1997tw} do not shed light on the continuity of GPDs.
}. More general 
relations must be permitted from the equations of motion to guarantee Lorentz covariance (e.g. in the structure of the Mellin moments of the GPDs, of which 
the sum rule is a special case). Here, of course, Lorentz symmetry is violated. Infinitely many light-cone time-ordered graphs are needed
in the reduced kernel to reproduce the covariant one-boson exchange \eqref{ladder}. Thus exactly satisfying polynomiality requires not only infinitely many 
exchanges in the kernel but contributions from infinitely many Fock components. It should be possible, however, to show how
the sum rule and polynomiality are improved order by order.

\subsection{Summary} \label{summy1}

Above we have investigated the application of the light-front reduction of current matrix elements to GPDs. 
In Appendix \ref{fff}, we reviewed the derivation of the form factor at next-to-leading order
in the $(3+1)$-dimensional ladder model. These expressions were then converted into the GPD
for the model. Continuity of these distributions at the crossover (where the plus momentum of 
the struck quark is equal to the plus component of the momentum transfer) was explicitly demonstrated,
\emph{cf} Eq.~\eqref{contequ}. Connection was made to the overlap representation of GPDs
by constructing the three-body wavefunction to leading order in perturbation theory. 
As a check on our results, we also reviewed the construction of higher Fock states from the valence sector in
old-fashioned time-ordered perturbation theory (Appendix \ref{oftopt}). The derived
overlaps Eqs.~\eqref{222} and \eqref{323} satisfy the relevant positivity constraint. 
The non-vanishing of the GPDs at the crossover could then be tied to higher Fock components,
specifically at vanishing plus momentum, and are hence essential for any phenomenological modeling of these distributions.  
This rewriting allowed us to understand how continuity arises perturbatively from the small-$x$ behavior of Fock state wavefunctions. 
In perturbation theory, the diagonal valence overlap vanishes at the crossover, 
while the higher Fock component overlaps do not. In general 
the $n$-to-$n$ overlap matches up with the $(n+1)$-to-$(n-1)$ overlap at the crossover due to the
dominance of the rebounding quark's infinite energy. 
In essence, no matter how complicated the higher Fock component is, the events in the last instant of light-cone time
must simplify. Either the active quark does not interact in this time, for which the contribution vanishes, or else 
there is an interaction which then reduces to a contribution of a Fock component with two fewer particles. 
This is true to all orders in perturbation theory and holds for 
bound states of any number of particles due to the nature of the time-ordered kernel and light-cone perturbation theory.

Unfortunately issues involving Lorentz invariance (such as the sum rule for the electromagnetic form
factor and the polynomiality constraints) are left untouched. To maintain covariance 
one would need infinitely many time-ordered exchanges in the kernel as well as 
infinitely many Fock components. It should be possible, however, to 
understand perturbatively how the $\zeta$-dependence disappears from Eq.~\eqref{sumrule}. 
This requires further relations between Fock components and these should be afforded
by the field-theoretic equations of motion. 
An additional application of the reduction formalism for currents
is to calculate the time-like form factor.   
As seen in Appendix \ref{chap:time} and in \cite{Tiburzi:2002sx}, we calculate the ladder model's generalized distribution amplitude, 
which is related to the time-like form factor via a sum rule similar to Eq.~\eqref{sumrule}.
This is in contrast to the non-existent Fock space expansion for these types of processes. 
Thus the reduction formalism succeeds in generating contributions where the light-cone Fock component wavefunctions cannot be employed.

With the formalism explored here, one could use phenomenological Lagrangian based models to explore both generalized
parton distributions and generalized distributions amplitudes within the light-front framework. Such an investigation
is interesting not only for testing phenomenological models, but also for anticipating problems for approximate non-perturbative
solutions for the light-cone Fock states. Nonetheless more model studies are warranted before truly realistic calculations
can be pursued. In particular, in light of our discussion continuity of GPDs may be transparent in light-cone time-ordered perturbation theory, but one 
must go beyond the toy model considered above and include spin degrees of freedom.

\section{Limitations}

In order to assess future data on GPDs one needs phenomenological parameterizations for the new distributions. 
The goal of the rest of this work is construct one. For ordinary quark distributions, one has a variety of forms invented by hand, as it were. For GPDs
however, the constraints do not permit such cavalier parameterizations. Additionally the parametrization should be simple
enough to be readily accessible to experimentalists. Let us first review a few types of models that could be utilized for GPDs. 
This will lead us to double distributions (DDs) which are taken up in subsequent Chapters.

Above we have seen the difficulties encountered when applying light-cone wavefunctions to calculate GPDs. 
The kinematical region $- \x < x < \x$ necessarily involves higher Fock components of the bound state.
In light-front phenomenology these are rarely considered. Although we derived these components perturbatively in the Wick-Cutkosky 
model, one ultimately must sacrifice Lorentz covariance to make the problem numerically tractable, even at weak coupling. 
Violation of Lorentz symmetry is undesirable for GPD models. The essence of the distributions is the dependence on the skewness 
variable $\x$. In terms of the form factor, we can understand GPDs via the calculation of the form factor in an arbitrary frame
labeled by $\x$ (we will appreciate this better in Chapter \ref{chap:DDs}). Viewed from this perspective, 
the sum rule's lack of dependence on $\x$ is due to the freedom to choose arbitrary frames in which to calculate the Lorentz 
invariant form factor.  A virtue of GPDs is the ability to measure correlations which otherwise vanish due to Lorentz invariance 
in single current matrix elements. The polynomiality restrictions of GPD moments are further manifestations of Lorentz invariance
and are additionally spoiled by light-front calculations.

These observations about light-cone wavefunction calculations of GPDs are generally true of quark models as well. 
The reason quark models fail to be useful for GPDs stems from their non-relativistic nature coupled with the restriction to a fixed number of particles. 
If one were to utilize a Poincar\'e covariant framework to calculate GPDs in terms of a fixed number of constituents, 
the kinematic region $- \x < x < \x$ will be a complication. In this framework, one must ultimately deal with constructing dynamical operators
consistent with Poincar\'e covariance in order to express the GPDs. In order to ensure Lorentz invariance, one must return to the 
general two photon amplitude and construct according to the Poincar\'e symmetries. 
This procedure is likely to be inconsistent with asymptotic freedom as the dynamics required for Poincar\'e invariance of the two 
photon amplitude have nothing to do with perturbative QCD in the asymptotic limit.

In non-relativistic scenarios, one also has the problem of support. 
Ordinary PDFs calculated in non-relativistic quark models extend beyond $x \in [0,1]$ 
and this trouble naturally carries over to GPDs. 
Moreover, one is lead non-relativistically to little or no contribution in the region $- \x < x < \x$, and 
the distributions will likely vanish at the crossover $x = \x$. 
Removing these weaknesses of model 
GPDs by using evolution is unsatisfactory because one then relies solely upon perturbative QCD for 
the non-perturbative physics absent from the original model. 
Besides, this is considerable work when alternatively the gross qualitative
features of PDFs can be obtained more simply from suitable evolution of the distribution $f_1(x) = \delta ( x - 1/3)$.
By analogy one can forgo the entire the issue of $\x$ dependence,
use the \emph{Ansatz}: $H(x,\x,t) = f_1(x) F_1(t)$, and hope for the best. 
Ultimately this parametrization will not be quantitative enough for interpreting data.

Beyond quark models, an approach that we considered in some detail was the calculation of GPDs from 
Dyson-Schwinger equation (DSE) models \cite{Roberts:1994dr,Maris:2003vk}. The Dyson-Schwinger approach
features Lorentz covariance and attempts to model self-consistently the solutions to field theories. 
Both of these qualities are desirable from the point of view of GPDs because relativity is borne in and 
there is no fixed particle restriction in field theory. 
Unfortunately, current Dyson-Schwinger model approaches, which fall
into two classes, are incompatible with the properties of GPDs as we now explain.

The first class of DSE models involves solving the covariant DSEs in a truncation scheme, such 
as the rainbow-ladder truncation. The solution of these equations requires treatment in Euclidean space
and the connection to light-cone dominated amplitudes is hence generally lost. The singularity 
structure of propagators and vertices must be known in the whole complex plane
to enable calculation of PDFs or GPDs. Clearly this cannot be done with a numerical solution in a limited domain. 
Imagining that GPDs could be obtained in the rainbow-ladder truncation scheme, one has the additional
complication of the positivity bounds. It is unclear whether non-perturbative truncation schemes 
are consistent with positivity.  Part of this uncertainty stems from a lack of knowledge of the analytic continuation, in particular
the way to assess contributions from the infinite light-cone singularities of the bound-state vertex. 
On the other hand, an interesting application of these DSE models would be to calculate the moments of PDFs and GPDs since 
the calculation of matrix elements of twist-two operators 
can be carried out in Euclidean space. Surprisingly this territory is largely unexplored
.

\begin{figure}
\begin{center}
\epsfig{file=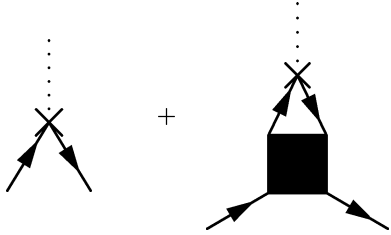}
\end{center}
\caption[Proper gauging of the three-point function with the non-local operator.]{Proper 
gauging of the DSE for the three-point function with the non-local light-cone operator (denoted by a cross). 
The Ball-Chiu \emph{Ansatz} cannot be used to resolve the active quark .  
}
\label{f:BCmess}
\end{figure}

The second class of DSE models relies upon using \emph{Ans\"atze} for the various vertex functions and propagators. 
Additionally field theoretic constraints arising from Ward-Takahashi identities are imposed, although the solutions
to these constraints are not uniquely specified by the model elements. This class of models has the same 
drawbacks for calculating GPDs as the first class. One again has difficulty connecting the Euclidean space formulation of these models 
with light-cone correlation functions. Moreover, entire functions are often used for model elements and these in turn 
defy a description of PDFs and GPDs because the integral over the relative light-cone energy does not converge in the complex plane. 
It is clear for this class of models that the positivity bounds must be generally violated. This is because an \emph{ad hoc}
bound state vertex cannot be gauged with the non-local light-cone operator. Correctly inserting the non-local operator at the bounds state vertex, 
which ensures the positivity bound is respected, requires knowledge of the underlying field theoretic dynamics.

Lastly there is an additional 
problem present in this second class of DSE models due to the incorrect treatment of higher point functions.   
The well-known Ball-Chiu form for the electromagnetic vertex \cite{Ball:1980ay}
is used in these models to describe an electromagnetic three-point function that is consistent with the Ward-Takahashi identity. 
The Ball-Chiu vertex, however, cannot be used for PDFs or GPDs because one needs to gauge the three-point function with the non-local light-cone operator 
and this requires knowledge of the four-point function \cite{Einhorn:1976uz,Burkardt:2000uu,Tiburzi:2001je}, see Figure \ref{f:BCmess}. 
As is, the Ball-Chiu \emph{Ansatz} cannot fix the plus momentum of the active quark because it models the sum of the ungauged diagrams in the Figure. 
The individual contributions cannot be resolved 
and hence cannot be properly gauged.

Clearly the consistent modeling of GPDs is a difficult problem.  Most standard hadronic models cannot be used in their 
present forms to calculate GPDs. Furthermore investigating how to rectify such models with the properties required of GPDs
is lengthy and not particularly illuminating, when one considers that the models have no transparent relation to QCD.
Thus it makes sense to consider only simple models whereby one can guarantee the properties of GPDs will be satisfied from the onset,
and the resulting parametrization will be efficaciously tractable for comparison with experiment. To devise such models, 
we turn our attention to double distributions.

\chapter{Double Distributions} \label{chap:DDs}

Modeling GPDs directly from their definition leads to light-cone quantization and light-cone wavefunctions.
Another approach is to use the formalism of double distributions (DDs) \cite{Radyushkin:1997ki,Radyushkin:1998es}
and this is the discourse taken up in the rest of this work.
The formalism of DDs elegantly explains the polynomiality conditions required of GPDs and thus gives one the ability to construct 
models consistent with known properties---although insight into model construction has often been limited to factorization \emph{Ans\"atze}.
Very few DDs have actually been calculated, although \emph{ad hoc} DDs are almost exclusively used for phenomenology.
Recently two-body light-front wavefunction models of the 
pion have been used to obtain GPDs \cite{Mukherjee:2002gb} based on DDs. Without modifying the quark distribution, 
the resulting GPDs violate the positivity constraints \cite{Tiburzi:2002kr}. This inconsistency is attributed to missing
contributions from non-wavefunction vertex diagrams the contribution of which are unknown when one uses non-covariant vertices. 
In general these diagrams are a substitute for higher Fock components, see \cite{Tiburzi:2002mn}. 
Thus the method used by~\cite{Mukherjee:2002gb} must at least use covariant wavefunctions to be sensible.
The covariant models we used~\cite{Tiburzi:2002tq}, however, 
allowed us to test the uniqueness of this ostensible construction. In this Chapter, 
we present the two models which were used to test the construction of DD functions. The two models are both simple models for
scalar bound states, one with scalar constituents and the other with spin-$\frac{1}{2}$ constituents.

Indeed we found that appealing to Lorentz invariance is enough to determine only one component of the
double distribution in the two-component formalism (even for $C$-odd distributions, where the Polyakov-Weiss $D$-term~\cite{Polyakov:1999gs} 
is absent). Moreover, the component determined from the reduction relations in the scheme presented 
by~\cite{Mukherjee:2002gb} is completely ambiguous.
Below we show that in the scalar constituent model, missing the second component leads to incorrect GPDs. 
The same is true for a spin-$\frac{1}{2}$ constituent model of the pion, where additionally the positivity bound is violated. 
The correct DDs unique to each model can be calculated from non-diagonal matrix elements of twist-two
operators and is demonstrated. 
On the other hand, exploiting the ambiguity inherent in 
defining one component DDs (which is akin to gauge freedom \cite{Teryaev:2001qm}) 
one can generate infinitely many different GPD models which share 
the same form factor and quark distribution as well as satisfy polynomiality (and likely positivity).
We also calculate DDs for a model with a propagator having complex conjugate energy poles. 
Such a parametrization is an efficacious way to model confinement. The calculation 
produces no trouble in Euclidean space, while the difficulties in Minkowski space are dealt with in Appendix \ref{minkowski}.

The organization of the Chapter is as follows. After defining DDs in Section \ref{def4}, we explicitly derive the GPD for the scalar triangle
diagram with point-like vertices in Section \ref{scalar4}. This is the GPD for the toy BSE model in Section \ref{toyboost}. 
Next we show the DD for this model extracted from the form factor in the Drell-Yan frame via~\cite{Mukherjee:2002gb}
leads to incorrect GPDs. Having encountered this problem, we calculate the missing component of the DD for the scalar triangle diagram
in Section \ref{scalarrev}. 
Then in Section \ref{ccpoles}, we present a calculation of the DD for a similar model where the propagator is extended
to have complex conjugate poles. The calculation is done in Euclidean space here and in Minkowski space in Appendix~\ref{minkowski}.
In  Section \ref{spinor}, we move on to the model 
for the spin-$\frac{1}{2}$ case.
Next in Section \ref{form}, we regularize the current and then extract this model's GPD.
Although not manifest, this model satisfies polynomiality, which is demonstrated in Section \ref{prop}.
Using \cite{Mukherjee:2002gb} as a guide, we construct a DD for this model in Section \ref{double}. 
Similar to Section \ref{scalar4}, this one-component DD too gives rise to a different GPD than the light-front projection.  
Additionally positivity is not satisfied by this one-component DD (Section \ref{pc}). 
We calculate the complete two-component DD from matrix elements of twist-two operators in Section \ref{real}. 
Lastly we conclude with a brief summary (Section \ref{summy}).

\section{Definitions} \label{def4}

In this Section we define DDs for scalar bound states via their moments by following the two-component formalism of \cite{Polyakov:1999gs,Teryaev:2001qm}. 
Focusing on the ambiguities inherent in defining one-component DDs from two component objects, we will understand why the DD 
constructed according to \cite{Mukherjee:2002gb} leads to an incorrect GPD. Moreover, we shall calculate the correct DD for the scalar triangle diagram from 
matrix elements of twist-two operators. We remark in passing that the two components of the DD (below $F$ and $G$, or $F$ and $D$-term 
in the standard formalism) can be viewed as projections of a single function of two variables \cite{Belitsky:2000vk}.

In defining GDPs and DDs there are two convenient choices of variables to use.
Above we have worked in Radyushkin's asymmetrical frame and asymmetrical variables which are ideal for perturbation theory and
are a natural generalization of ordinary parton distributions. 
The non-diagonal matrix elements of twist-two operators are, however, more conveniently expressed in 
variables symmetric with respect to initial and final states.\footnote{Good discussion of the conversion from symmetrical and asymmetrical 
variables and distributions can be found in \cite{Golec-Biernat:1998ja}. Additionally advantages and disadvantages of both are presented.}
To this end we define the average momentum $\ol P {}^\mu  = (P + P^\prime)^\mu / 2$.
Let $\overset{\leftrightarrow}{D^\mu} = \frac{1}{2} (\overset{\rightarrow}{\partial^\mu} - \overset{\leftarrow}{\partial^\mu} )$. Then for 
a pion of spin-$\frac{1}{2}$ constituents we have
\begin{multline} \label{decomp1}
\langle \; \ol P  + \Delta/2 | \bar{\psi}(0) \gamma^{\big\{\mu} i\overset{\leftrightarrow}{D}{}^{\mu_1}
\cdots i \overset{\leftrightarrow}{D}{}^{\mu_n}{}^{\big\}} \psi(0) | \ol P - \Delta/2 \; \rangle 
= \\ 2 \ol P {}^{\big\{ \mu} \sum_{k = 0}^n \frac{n!}{k!(n-k)!} A_{nk}(t) \; \ol P {}^{\mu_1} \cdots \ol P {}^{\mu_{n-k}} 
\Big(-\frac{\Delta}{2}\Big)^{\mu_{n - k + 1}} \cdots \Big(-\frac{\Delta}{2}\Big)^{\mu_n \big\}}  \\
- \Delta^{\big\{ \mu} \sum_{k = 0}^n \frac{n!}{k!(n-k)!} B_{nk}(t) \; \ol P {}^{\mu_1} \cdots \ol P {}^{\mu_{n-k}} 
\Big(-\frac{\Delta}{2}\Big)^{\mu_{n - k + 1}}  \cdots \Big(-\frac{\Delta}{2}\Big)^{\mu_n \big\}},
\end{multline}
where the action of ${}^{\{ \cdots \} }$ on Lorentz indices produces only the symmetric traceless part and $T$-invariance restricts 
$k$ in the first sum to be even and odd in the second. 
For a pion of scalar constituents, replace $\gamma^\mu$ with $2 i\overset{\leftrightarrow}{D^\mu}$. As it 
stands there is considerable freedom in the above decomposition,  e.g., one could rewrite the above with 
\hbox{$k B_{n,k-1}(t)/ (n - k + 1)$} as  a contribution to $A_{nk}(t)$. 
Carrying this out for all $k$, puts the bulk in the first term and 
renders the second term proportional to the symmetric traceless part of only ($n+1$) $\Delta$'s--- moments of the Polyakov-Weiss 
$D$-term \cite{Polyakov:1999gs}. This is the usually encountered form of the DD with $D$-term. Calculationally, however, we find
our results directly in the form of Eq.~\eqref{decomp1}.

The DDs are defined as generating functions for the moments\footnote{%
In general the twist-two form factors and hence the double distributions have renormalization scale dependence
that arises from perturbative QCD corrections.
}
\begin{align}
A_{nk}(t) & = \int_{-1}^{1} d\beta \int_{-1 + |\beta|}^{1 - |\beta|} d\alpha \; \beta^{n - k} \alpha^k F(\beta,\alpha;t) \label{Fdef1}\\
B_{nk}(t) & = \int_{-1}^{1} d\beta \int_{-1 + |\beta|}^{1 - |\beta|} d\alpha \; \beta^{n-k}   \alpha^k G(\beta,\alpha;t) \label{Gdef1}.
\end{align}
As a consequence of the restriction on $k$ in the sums, the function $F(\beta,\alpha;t)$ is even in $\alpha$ 
while $G(\beta,\alpha;t)$ is odd. Also for $n$-even, there is no contribution from the $D$-term to the function $G(\beta,\alpha;t)$.

These DD functions then appear in the decomposition of  matrix elements of light-like 
separated operators. Summing the moments to produce these operators, we have
\begin{multline} \label{DDdecomp1}
\langle \; \ol P + \Delta/2 \; | \; \bar{\psi}(-z^-/2) \; \rlap\slash z  \; \psi(z^-/2) \; | \; \ol P  - \Delta/2 \; \rangle 
\\ = 2 \ol P \cdot z \int_{-1}^{1} d \beta \int_{-1 + |\beta|}^{1 - |\beta|} d \alpha \; 
e^{-i \beta \ol P \cdot z + i \alpha \Delta \cdot z /2} \, F(\beta,\alpha;t) \\
- \Delta \cdot z \int_{-1}^{1} d\beta \int_{-1 + |\beta|}^{1 - |\beta|} d\alpha \; e^{-i \beta \ol P \cdot z + i \alpha \Delta \cdot z /2}
\, G(\beta,\alpha;t),
\end{multline}
where $z^2 = 0$.

Denoting $\xi = - \Delta^+/ 2 \ol P {}^+$, the GPD in symmetric variables reads
\begin{equation}
H(\tilde{x},\xi,t) = \frac{1}{4\pi} \int dz^- e^{i \xt \Pbar {}^+ z^-} \langle \; \ol P  + \Delta/2  |\bar{\psi}(-z^-/2) \gamma^+  
\psi(z^-/2) |  \ol P  - \Delta/2 \; \rangle. 
\end{equation}
Inserting Eq.~\eqref{DDdecomp1} into this definition yields
\begin{equation} \label{JiGPD1}
H(\xt,\xi,t) = \int_{-1}^{1} d\beta \int_{-1 + |\beta|}^{1 - |\beta|} d\alpha \; \delta(\xt - \beta - \xi \alpha) 
\Big[ F(\beta,\alpha;t) + \xi G(\beta, \alpha;t) \Big].  
\end{equation}
By integrating over $\xt$, we uncover \emph{two} sum rules: the sum rule for the form factor
\begin{equation} \label{Fsum1}
\int d\beta \int d\alpha \; F(\beta,\alpha;t) = F(t)
\end{equation} 
and what we call the $G$-sum rule
\begin{equation} \label{Gsum1}
\int d\beta \int d\alpha \; G(\beta,\alpha;t) = 0,
\end{equation}
which is trivial since $G$ is an odd function of $\alpha$. Eq.~\eqref{Gsum1} has non-trivial consequences however, e.g., it 
shows the method employed by \cite{Mukherjee:2002gb} leads only to the $F$ DD in Eq.~\eqref{DDdecomp1}. This function 
integrates to the form factor via Eq.~\eqref{Fsum1} and in the forward limit $\{\xi,t \to 0\}$ reduces to the quark distribution
(when integrated over $\alpha$). Thus $F(\beta,\alpha;t)$ should be properly termed the forward-visible DD, which encompasses 
\emph{more} than just neglecting the $D$-term. From Eq.~\eqref{decomp1}, we see that $G(\beta,\alpha;t)$ does affect
higher moments of the GPD. This will be the source of the discrepancy between the method employed by~\cite{Mukherjee:2002gb}
and the correct calculation of the DDs from Eq.~\eqref{decomp1}.
We now calculate model DDs and expose this discrepancy.

\section{Simple models}

\subsection{Scalar model} \label{scalar4}

For the meson model with scalar constituents, we choose a point-like Bethe-Salpeter vertex $\Gamma(k,P) = 1$, where
the coupling constant is assumed to be absorbed into the overall normalization. 
This is the model considered in Section \ref{toyboost}.
Furthermore, we choose derivative coupling of the photon to charged scalar particles.

\begin{figure}
\begin{center}
\epsfig{file=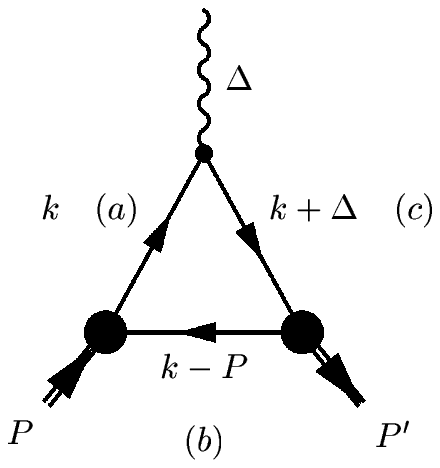}
\caption{\label{ftri4} Covariant triangle diagram for the meson electromagnetic form factor.}
\end{center}
\end{figure}

The meson electromagnetic form factor for this model can be calculated from the Feynman triangle diagram. In order to derive
the GPD, however, we need to choose the kinematics specified in Figure \ref{ftri4} with $k$ as the momentum of the struck 
quark. Using the stated meson vertex and 
taking the plus-component of the current, we have
\begin{equation} \label{sff}
F(t) = \int d^4k \frac{-i |N|^2(2 k^+ + \Delta^+)}{(1 - \zeta/2)P^+}  \Big\{ \big[ k^2 - m^2 + \ie\big] 
\big[ (k+\Delta)^2 - m^2 + \ie\big]  \big[ (P-k)^2 - m^2 + \ie\big] \Big\}^{-1},
\end{equation}
where the momentum transfer is $t = \Delta^2$ and the skewness is defined relative to the initial state $\Delta^+ = - \zeta \Pplus < 0$.
Physically $\zeta$ plays the role of Bjorken variable for DVCS. Additionally we work in the asymmetric frame where $\mathbf{P}^\perp = 0$.

To turn Eq.~\eqref{sff} into an expression for the GPD $H(x,\zeta,t)$, 
we insert $\delta(k^+/ P^+ - x)$ to fix the momentum of the struck quark and keep $\zeta \neq 0$. This forces
\begin{equation} \label{defn}
F(t) = \int  H(x,\zeta,t) dx.
\end{equation}
Lastly we integrate over $\kminus$ to project onto the light cone. 
Doing the contour integration to extract $H(x,\zeta,t)$ in Eq.~\eqref{sff}, we are confronted with the poles
\begin{equation}
\begin{cases}
	k^-_{a} = k^-_{\text{on}}  -  \frac{\ie}{x}\\
	k^-_{b} = P^- + (k-P)^-_{\text{on}}  -  \frac{\ie}{x-1}\\	
	k^-_{c} = - \Delta^- + (k + \Delta)^-_{\text{on}}  -  \frac{\ie}{x^\prime}\\
\end{cases},
\end{equation}  
where the on-shell energies are $p^-_\text{on} = \frac{\pperp^2 + m^2}{2 p^+}$ and the abbreviation 
$x^\prime = \frac{x - \zeta}{1-\zeta}$ is used. Thus the non-vanishing contribution to the integral is
\begin{equation}
2 \pi i \; \theta[x(1-x)] \left[ \theta(x - \zeta) \res(k_b^-)  - \theta(\zeta - x) \res(k^-_a) \right],
\end{equation}
which leads to
\begin{equation} \label{sgpd}
(1 - \zeta/2) H(x,\zeta,t) = \theta(x - \zeta) H_1(x,\zeta,t) + \theta(\zeta - x) H_2(x,\zeta,t). 
\end{equation}
Using $\mathbf{k}^{\prime \perp} = \kperp + (1-x^\prime) \dperp$ for the relative transverse momentum of the final state, 
the functional forms are
\begin{align}
H_1(x,\zeta,t)  & = (2 x - \zeta) |N|^2 \int d\kperp \dw(x,\kperp|M^2_\pi) \dw(\xp,\kpperp|M^2_\pi)/ x(1-x) \xp 
\label{H1} \\
H_2(x,\zeta,t)  & = (2 x - \zeta) |N|^2 \int d\kperp \dw(x,\kperp|M^2_\pi) 
\dw(\xpp,\mathbf{k}^{\prime\prime\perp}|t)/\zeta \xpp (1 - \xpp)(1-x) \label{H2},
\end{align}
where $\xpp \equiv x / \zeta$ and $\mathbf{k}^{\prime\prime\perp} \equiv  \kperp + \xpp \dperp$ are the relative momenta
of the photon. Additionally, we use the replacement 
\begin{equation}
\dw (x, \kperp | M^2)^{-1}  = M^2 - \frac{\kperp^2 + m^2}{x(1-x)},
\end{equation}
which is the propagator of the Weinberg equation \cite{Weinberg:1966jm}.

Comments about the GPD $H(x,\zeta,t)$ in Eq.~\eqref{sgpd} are in order. The model is covariant and thus the 
sum rule and polynomiality conditions are met (see Section \ref{prop} below for clarification). We have checked this
explicitly and suitable discussion can be found in \cite{Pobylitsa:2002vw,Theussl:2002xp}. 
Consideration of this model without derivative coupling 
was first done from the perspective of DDs, see e.g.~the toy model of \cite{Radyushkin:1997ki}. 
This DD model was revisited recently with derivative coupling at the photon vertex in the Appendix of \cite{Mukherjee:2002gb}
and the same DD also appears in \cite{Pobylitsa:2002vw}. 
To derive the DD for this simple model, we first appeal to Lorentz invariance as suggested in \cite{Mukherjee:2002gb}, 
recalling along the way the relevant properties of DDs in asymmetric variables.

First consider the form factor. In the $\zeta = 0$ frame, Eqs.~(\ref{H1}--\ref{H2}) reduce to the Drell-Yan formula \cite{Drell:1970km,West:1970av} 
via the definition in Eq.~\eqref{defn}
\begin{equation} \label{sFF}
F(t) =  |N|^2 \int \frac{dx d\kperp}{x(1-x)} \dw(x,\kperp|M^2_\pi) \dw(x,\kperp + (1-x) \dperp|M^2_\pi),
\end{equation}
where $t = - \D_\perp^2$. 
The form factor is Lorentz invariant and has a decomposition in terms of the Lorentz invariant
DD $F(x,y;t)$, namely
\begin{equation} \label{DDform}
F(t) = \int_0^1  \int_0^1  F(x,y;t) \, dy \, dx.
\end{equation}
The DD satisfies the following relations \cite{Radyushkin:1998es}: support property
\begin{equation} \label{A}
F(x,y;t) \propto \theta(1-x-y),
\end{equation}
reduction to the quark distribution at zero momentum transfer
\begin{equation} \label{B}
q(x) = \int_0^{1-x} F(x,y;0) \, dy
\end{equation}
and is \emph{M\"unchen} symmetric \cite{Mankiewicz:1998uy}
\begin{equation}
F(x,y;t) = F(x,1-x-y;t).        \label{C}
\end{equation}
Using Eq.~\eqref{sFF}, we can write $F(t)$ in the form \eqref{DDform} with
\begin{equation} \label{sDD}
F(x,y;t) = \frac{x |N|^2 \theta( 1 - x - y)}{m^2 - x(1-x) M^2_\pi - y(1-x-y) t},
\end{equation}
which satisfies Eqs.~(\ref{A}--\ref{C}). The ingenuity of DDs comes about when we construct the GPD via
\begin{equation} \label{prescrip}
H(x,\zeta,t) = \int_0^1  \int_0^1  F(z,y;t) \delta(x - z - \zeta y ) \, dy \, dz.
\end{equation}
In this form the sum rule and polynomiality conditions follow trivially.

\begin{figure}
\begin{center}
\epsfig{file=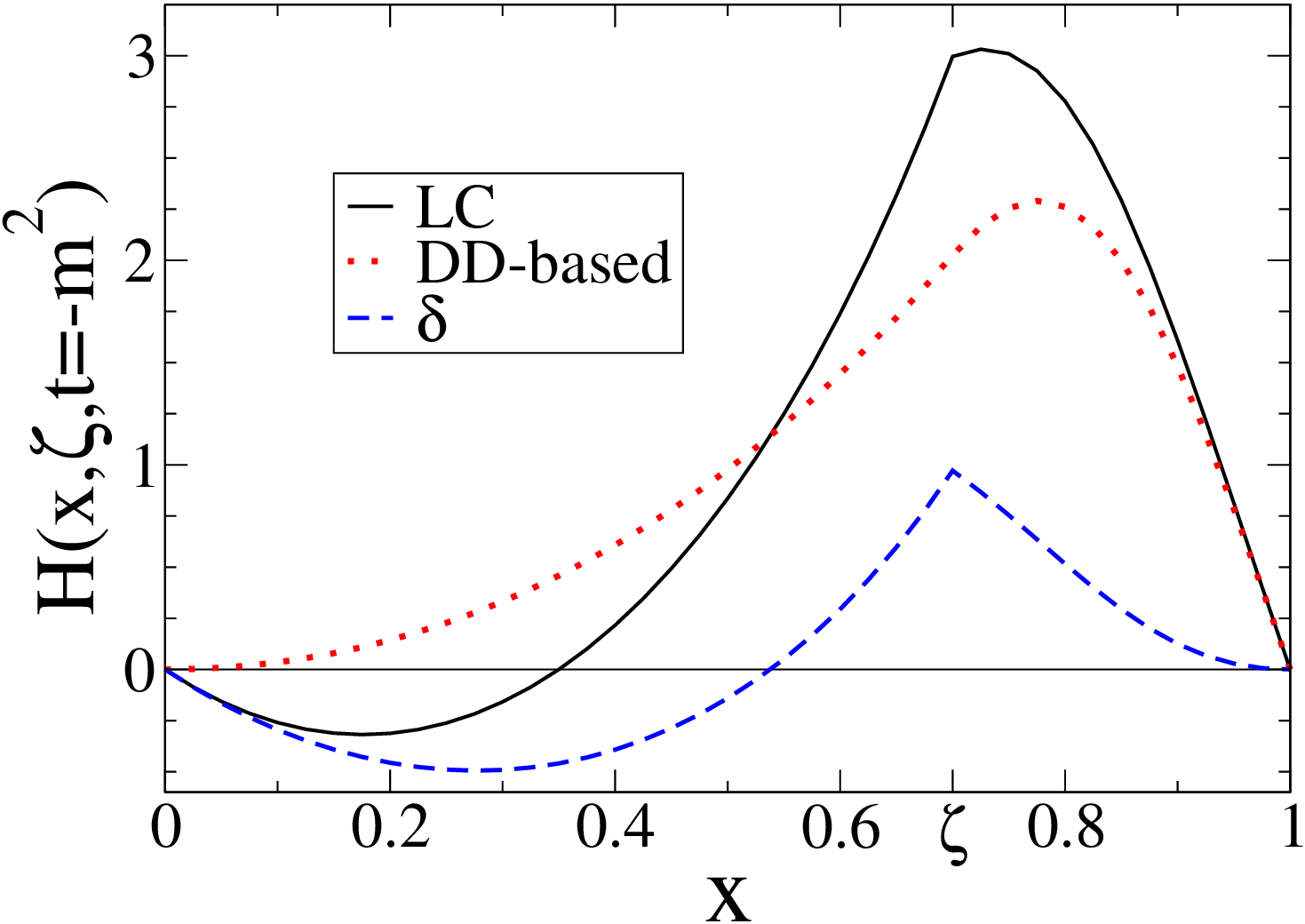,width=2.5in,height=1.5in}
\caption[Comparison of covariant GPDs for the scalar triangle diagram.]{Comparison of covariant 
GPDs for the scalar triangle diagram. The GPDs Eq.~\eqref{sgpd} (denoted LC) and 
Eq.~\eqref{prescrip} (DD-based) are plotted as a function of $x$ for fixed $\zeta = 0.7$ and $t = - m^2$ for the mass $M_\pi = m < 2m$.
We also plot the difference between the two curves ($\delta$). The area under the curves is identically $F(-m^2)$ for LC and DD-based GPDs, 
and hence zero for their difference $\delta$.}
\label{surprise}
\end{center}
\end{figure}

In Figure \ref{surprise}, we plot the GPD Eq.~\eqref{sgpd} as well as the GPD derived from DD via Eq.~\eqref{prescrip}.
Surprisingly the two are different despite the fact both models are covariant and posses the same form factor and quark distribution.
Additionally we plot their difference ($\delta$) as a function of $x$.

\subsection{Scalar model, revisited} \label{scalarrev}

Based on the definitions in Eq.~\eqref{decomp1}, we need more than the form factor to determine the GPDs. The procedure
above thus missed one component of the DDs. 
To correctly derive both $F$ and $G$ DDs for the scalar triangle diagram of Section \ref{scalar4}, we must now consider the action
of the operator $2 i \overset{\leftrightarrow}{D}{}^{\{\mu} i\overset{\leftrightarrow}{D}{}^{\mu_1}
\cdots i \overset{\leftrightarrow}{D}{}^{\mu_n}{}^{\}}$. This produces a factor
\begin{equation}
\frac{1}{2^n} (2 k + \Delta)^{\{ \mu}(2 k + \Delta)^{\mu_1} \cdots (2 k + \Delta)^{\mu_n \}}
\end{equation}
in the integrand of Eq.~\eqref{sFF}, which we now take in the symmetric frame. After the integration over $k$ is performed, 
we are left only with 
\begin{equation}
\int_{0}^{1} d\beta \int_{-1 + \beta}^{1 - \beta} d\alpha \;  D(\beta,\alpha;t) \, 
(2 \beta \ol P - \alpha \Delta)^{\{ \mu}(\beta \ol P - \alpha \Delta/2)^{\mu_1}
\cdots(\beta \ol P - \alpha \Delta/2)^{\mu_n \}},
\end{equation}
where we have used the replacement
\begin{equation} \label{dba}
D(\beta,\alpha;t) = \frac{1}{m^2 - \beta (1-\beta) M^2_\pi - t [(1-\beta)^2 - \alpha^2]/4}.
\end{equation}

Using the binomial expansion, we can identify $F(\beta,\alpha;t)$ and $G(\beta,\alpha;t)$ via Eqs.~\eqref{Fdef1} and \eqref{Gdef1}, 
namely\footnote{%
Notice in this case we can see that $F$ and $G$ are projections of a single function $D(\beta,\alpha;t) = f(\beta,\alpha)$ in the notation
of \cite{Belitsky:2000vk}. Determination of $f(\beta,\alpha)$ for the spin-$\frac{1}{2}$ model of Section \ref{spinor}, however, 
cannot be done by inspection.
}
\begin{align}
F(\beta,\alpha;t) & = \beta \; \theta(1 - \beta - \alpha) D(\beta,\alpha;t) |N|^2 \label{F1}\\
G(\beta,\alpha;t) & = \alpha \; \theta(1 - \beta - \alpha) D(\beta,\alpha;t) |N|^2  \label{G1}.
\end{align}
To compare with the results of Section \ref{scalar4}, we must revert to asymmetric variables which is accomplished
by the transformation $\b \to x$, $y \to ( \a - \b + 1)/2$. The denominator common to both terms becomes
\begin{equation} \label{dxy}
D(x,y;t) = \frac{1}{m^2 - x(1-x) M^2_\pi - y (1-x-y) t},
\end{equation}
and we have
\begin{align}
F(x,y;t) & = x \; \theta(1 - x - y) D(x,y;t) |N|^2 \label{F2}\\
G(x,y;t) & = (2 y + x - 1) \; \theta(1- x- y) D(x,y;t)  |N|^2 \label{G2}.
\end{align}
Notice the function $G(x,y;t)$ is \emph{M\"unchen} antisymmetric, i.e.~$G(x,y;t) = - G(x, 1 - x - y;t)$, 
which is required because $G(\beta,\alpha;t)$ is odd with respect to $\alpha$.

To construct the GPD $H(x,\zeta,t)$ we must also convert Eq.~\eqref{JiGPD1} to asymmetric variables.
\begin{equation}  \label{correct}
H(x,\zeta,t) = \int_0^1 dz \int_0^{1-z} dy \; \delta(x - z - \zeta y) \Big[ F(x,y;t) + \frac{\zeta}{2 - \zeta} G(x,y;t) \Big]
\end{equation} 
We can now plot the GPD calculated from Eq.~\eqref{correct} using the two DDs in Eqs.~\eqref{F2} and \eqref{G2}. 
The result agrees with Eq.~\eqref{sgpd} depicted in Figure \ref{surprise}. Moreover, the contribution 
from $\frac{\zeta}{2 - \zeta} G$ is identically the difference $\delta$ plotted in the figure.
In the DD formalism, the reduction relations do not determine the GPD. We will illustrate this further
with a spin-$\frac{1}{2}$ model for  the pion.  Before doing so, however, we will take a detour to pursue
an interesting extension of the above scalar model.

\subsection{Complex conjugate poles} \label{ccpoles}

In this Section, we extend the calculation of the model DD above to include complex conjugate singularities in the 
constituents' propagators.
Complex conjugate singularities present in solutions to Dyson-Schwinger
equations have been studied in the connection with the violation of Osterwalder-Schrader reflection positivity 
and confinement \cite{Atkinson:1979tk,Habel:1990aq,Habel:1990tw,Burden:1992gd,Krein:1992sf,Maris:1995ns,Stingl:1996nk,Gribov:1999ui}. 
Recent work \cite{Bhagwat:2003wu} in solving the Bethe-Salpeter equation with a quark propagator consisting of pairs of complex
conjugate singularities shows that the width for meson decay (into free quarks) generated from one pole is exactly canceled by 
the contribution from its complex conjugate. Additionally recent studies have modeled Euclidean space lattice data 
with propagators that have time-like complex conjugate singularities \cite{Bhagwat:2003vw,Alkofer:2003jj,Alkofer:2003jk}. 
In \cite{Tiburzi:2003ja} we pursued 
the calculation of space-like amplitudes in Minkowski space using a simple model propagator consisting of a pair of complex 
conjugate poles. Specifically, we were interested in the calculation of light-cone dominated amplitudes for this type of model which 
necessitates a treatment in Minkowski space. This treatment is contained in Appendix \ref{minkowski} because there is considerable
subtlety. Nonetheless, covariant calculation in Euclidean space is straightforward and provides another chance to 
obtain DDs properly.

In Euclidean space, the model propagator is chosen to be
\begin{equation}
S_{E}(k) = \sum_{\varepsilon = \pm } \frac{1/2}{k^2 + m^2 - i \eps}.
\end{equation}
Here and below we use the shorthand $\eps = \pm$ to denote the pair of complex conjugate poles.
Notice $\eps$ is finite and can be chosen to be positive.
Unlike in Minkowski space where the measure is imaginary, contributions to Euclidean space amplitudes are real 
and one has no difficulty in calculating form factors 
and distribution functions using $S_{E}(k)$ in the relevant diagrams. The simplicity of the model at hand will allow us to 
calculate its double distribution analytically
and thereby determine the quark distribution and electromagnetic form factor, since these functions are related to the 
double distribution by the so-called reduction relations. The difficult task of 
calculating of these quantities in Minkowski space by projecting onto the light-cone is contained in Appendix \ref{minkowski}.

GPDs are not Lorentz 
invariant objects, however, they stem from a projection of a Lorentz invariant double distribution  function
\cite{Radyushkin:1997ki}. These functions are particularly attractive from the perspective of model building \cite{Mukherjee:2002gb}, 
though one must be careful that the starting point is indeed covariant \cite{Tiburzi:2002kr}, otherwise desirable distribution 
properties and straightforward physical interpretation may be sacrificed. The model under consideration is fully covariant, and thus the 
DD representation is an ideal testing ground for 
our model propagator.  Hence we proceed to calculate the model's Euclidean space DD, recalling along the way the 
relevant properties of DDs.
In this Section, by contrast to Eq.~\eqref{decomp1}, we decompose the matrix elements of twist-two operators in the asymmetric 
frame, with asymmetric variables.

Let $\overset{\leftrightarrow}{D^\mu} = \overset{\rightarrow}{\partial^\mu} - \overset{\leftarrow}{\partial^\mu} $.
For this scalar model, we define the twist-two operator of spin-$(n+1)$ as above
\begin{equation}
\mathcal{O}^{\mu \mu_1 \ldots \mu_n} = \phi(0) i \overset{\leftrightarrow}{D} { }^{\{ \mu}  i\overset{\leftrightarrow}{D}{}^{\mu_1}
\cdots i \overset{\leftrightarrow}{D}{}^{\mu_n}{}^{\}}    \phi(0),
\end{equation}
where the action of ${}^{\{ \cdots \}}$ on Lorentz indices produces only the symmetric traceless part.

We work in Radyushkin's asymmetric frame
with $P$ as the momentum of the initial state, $P+\Delta$ that of the final and $t = \Delta^2$. The initial and final 
states are on-shell: $P^2 = (P + \Delta)^2 = M^2$. 
The non-diagonal matrix element of $\mathcal{O}^{\mu \mu_1 \ldots \mu_n}$ can be decomposed into Lorentz invariant moment functions
$A_{n k}(t)$ and $B_{n k}(t)$. Asymmetrically this decomposition reads
\begin{multline} \label{decomp2}
\langle \; P + \Delta | \mathcal{O}^{\mu \mu_1 \ldots \mu_n} | P\; \rangle = \\ 
(2 P + \Delta)^{\{ \mu} \sum_{k = 0}^n \frac{n!}{k!(n-k)!} A_{nk}(t) \; (2P + \Delta)^{\mu_1} \cdots (2P + \Delta)^{\mu_{n-k}} 
(-\Delta)^{\mu_{n - k + 1}} \cdots (-\Delta)^{\mu_n \}}  \\
- \Delta^{\{ \mu} \sum_{k = 0}^n \frac{n!}{k!(n-k)!} B_{nk}(t) \; (2 P + \Delta)^{\mu_1} \cdots (2 P + \Delta)^{\mu_{n-k}} 
(-\Delta)^{\mu_{n - k + 1}}  \cdots (-\Delta)^{\mu_n \}},
\end{multline}
Hermiticity forces the matrix elements of $\mathcal{O}^{\mu \mu_1 \ldots \mu_n}$ to be invariant under the transformation 
\begin{equation} \notag
\begin{cases} 
	P \to P + \Delta \\ 
	\Delta \to - \Delta\\
\end{cases}.  
\end{equation} 
Consequently the values of $k$ are restricted to be even in the first sum  and odd in the second. 
As it stands there is again considerable freedom in writing Eq.~\eqref{decomp2},   e.g., one could rewrite the above with 
\hbox{$k B_{n,k-1}(t)/ (n - k + 1)$} as  a contribution to $A_{nk}(t)$. 
Carrying this out for all $k$, puts the bulk in the first term and 
renders the second term proportional only to the symmetric traceless part of ($n+1$) $\Delta$'s--- moments of the Polyakov-Weiss 
$D$-term \cite{Polyakov:1999gs}.  This is the usually encountered form of the DD with $D$-term. Alternatively one can also 
express the moments as projections of a single Lorentz invariant function \cite{Belitsky:2000vk}. 
Calculationally, however, Eq.~\eqref{decomp2} becomes the most practical to work with \cite{Tiburzi:2002tq}.

The $F$ and $G$ DDs can be defined as generators of the coefficient functions 
\begin{align}
A_{nk}(t) & = \int_{0}^{1} dx \int_{0}^{1 - x} dy \; x^{n - k} (x + 2 y -1 )^k F(x,y;t) \label{Fdef2}\\
B_{nk}(t) & = \int_{0}^{1} dx \int_{0}^{1 - x} dy \; x^{n-k}   (x + 2 y - 1)^k G(x,y;t) \label{Gdef2}.
\end{align}
As a consequence of the restriction on $k$ in the sums, the function $F(x,y;t)$ is \emph{M\"unchen} 
symmetric \cite{Mankiewicz:1998uy}, i.e.~$F(x,y;t) = F(x,1-x-y;t)$, while $G(x,y;t)$ is \emph{M\"unchen} antisymmetric. 
Also for $n$-even, there is no contribution from the $D$-term to the function $G(x,y;t)$.

\begin{figure}
\begin{center}
\epsfig{file=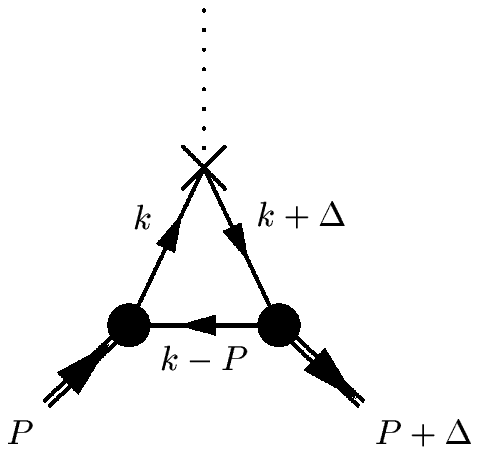,height=1.6in,width=1.6in}
\caption[Diagram used to calculate non-diagonal matrix elements of twist-two operators.]{Diagram 
used to calculate non-diagonal matrix elements of twist-two operators (denoted by a cross).}
\label{ftwist2}
\end{center}
\end{figure}

These functions then appear in the decomposition of  matrix elements of light-like 
separated operators
\begin{multline} \label{DDdecomp2}
\langle \; P +\Delta \; | \; \phi(0) \;  i z \cdot \overset{\leftrightarrow}{D}  \; \phi(z^-) \; | \; P \; \rangle = \\ 
(2 P \cdot z + \Delta \cdot z) \int_{0}^{1} d x \int_{0}^{1 - x} d y \; 
e^{-i x P \cdot z + i y \Delta \cdot z } F(x,y;t)  \\
- \Delta \cdot z \int_{0}^{1} d x \int_{0}^{1 - x} d y \; e^{-i x P \cdot z + i y \Delta \cdot z}
G(x,y;t),
\end{multline}
where $z^2 = 0$.

Denoting $\zeta = - \Delta^+/ P^+ > 0$, the GPD in asymmetric variables reads
\begin{equation} \label{GPDdef2}
H(x,\zeta,t) =  \int \frac{dz^- e^{i x P^+ z^-}}{2 \pi( 2 - \zeta)} \langle \; P + \Delta  \; | \; \phi(0)  
i \overset{\leftrightarrow}{D^+}  \phi(z^-) \; | \;  P  \; \rangle. 
\end{equation}
Physically $\zeta$ plays the role of Bjorken variable for deeply virtual Compton scattering.
Inserting Eq.~\eqref{DDdecomp2} into this definition yields
\begin{equation} \label{JiGPD2}
H(x,\zeta,t) = \int_{0}^{1} dz \int_{0}^{1 - z} dy \; \delta(x - z - \zeta y) 
\Big[ F(z,y;t) + \frac{\zeta}{2 - \zeta} G(z,y;t) \Big].  
\end{equation}
which is what we found circuitously above.
By integrating Eq.~\eqref{JiGPD2} over $x$, we uncover two familiar sum rules for the DDs: the sum rule for the form factor
\begin{equation} \label{Fsum2}
\int_0^1 d x \int_0^{1-x} d y \; F(x,y;t) = F(t)
\end{equation} 
and the $G$-sum rule
\begin{equation} \label{Gsum2}
\int_0^1 d x \int_0^{1-x} d y \; G(x,y;t) = 0,
\end{equation}
which follows after a change of variables since $G$ is \emph{M\"unchen} antisymmetric. 
Eq.~\eqref{Gsum2} is important and mandated by current conservation. 
As we saw above, the $G$ DD function is all too frequently overlooked and treated as identically zero.  
Lastly, the model quark distribution function $q(x)$ can be found from the DD at 
zero momentum transfer, see Eq.~\eqref{GPDdef2}, 
\begin{equation} \label{qred}
q(x) = \int_0^{1-x} dy \, F(x,y;0).
\end{equation}

We can use the decomposition in Eq.~\eqref{decomp2} 
to calculate our simple model's DD. Parameterizing the momenta as in Figure \ref{ftwist2}, the non-diagonal
matrix element of $\mathcal{O}^{(n)}$ reads
\begin{multline} \label{bob}
\langle \; P + \Delta | \mathcal{O}^{\mu \mu_1 \ldots \mu_n} | P\; \rangle  = \\
\frac{2 N}{\pi^2}
\sum_{\eps,\epsp,\epspp = \pm} \int d^4 k \frac{(2 k + \Delta)^{\{ \mu}(2 k + \Delta)^{\mu_1} \cdots (2 k + \Delta)^{\mu_n\}}}
{[k^2 + m^2 - i \eps] [ (k+\Delta)^2 + m^2 - i \epsp] [(k-P)^2 + m^2 - i \epspp]}
\end{multline}
The normalization constant $N$ is chosen by the condition $F(0) = 1$.
Let us denote the propagators
simply by $\mathfrak{A} = (k - P)^2 + m^2 - i \epspp$, $\mathfrak{B} = (k+\Delta)^2 + m^2 - i \epsp$ and  
$\mathfrak{C} = k^2 + m^2 - i \eps$. We introduce two Feynman parameters $\{x,y\}$ 
to render the denominator specifically in the form $[x \mathfrak{A} + y \mathfrak{B} + (1-x-y) \mathfrak{C}]^{-3}$.
One then translates $k^\mu$ to render the integral (hyper-) spherically symmetric via the definition 
$k^\mu = l^\mu + x P^\mu -  y \Delta^\mu$. The resulting integral over $l$ can be evaluated directly (remember we are in Euclidean space).

Binomially expanding the result of the integral, we can make contact with Eq.~\eqref{decomp2} and subsequently determine the
$F$ and $G$ double distributions by inspection from Eqs.~\eqref{Fdef2} and \eqref{Gdef2}. Defining the auxiliary functions
\begin{equation} \label{dzero}
D_o(x,y;t) = m^2 - x ( 1- x) M^2 - y (1- x - y) t
\end{equation}
and
\begin{equation}
D(x,y;t) = N \sum_{z = 0, x, y, x+y} \frac{D_o(x,y;t)}{D_o(x,y;t)^2 + \eps^2 (1 - 2 z)^2},
\end{equation}
the DDs can be written simply as
\begin{align}
F(x,y;t) & = x D(x,y;t) \label{FDD} \\
G(x,y;t) & = (x + 2 y - 1) D(x,y;t) \label{GDD}.
\end{align}
We treat factors of $\theta( 1 - x - y)$ as implicit above.
Accordingly $F$ is \emph{M\"unchen} symmetric and $G$ is antisymmetric.
Notice although $\eps$ is finite, corresponding results using the standard perturbative propagator 
can always be recovered in the limit $\eps \to 0$. For example, 
the correct $F$ and $G$ DDs are recovered in the limit $\eps \to 0$ \cite{Tiburzi:2002tq}.

The model GPD can be derived by utilizing Eq.~\eqref{JiGPD2}, although the integral must be performed numerically. The quark distribution
can be found via the reduction relation Eq.~\eqref{qred}, namely
\begin{equation} \label{qofx}
q(x) = N \sum_{z = 0, x}\Bigg[ \frac{x(1-x) D_o(x,0;0)}{D_o(x,0;0)^2 + \eps^2 (1 - 2 z)^2}  + \frac{x}{\eps} \tan^{-1} 
 \frac{\eps ( 1 - 2 z) }{D_o(x,0;0)}  \Bigg]
\end{equation}
Lastly the form factor can be found from the sum rule Eq.~\eqref{Fsum2}

\begin{figure}
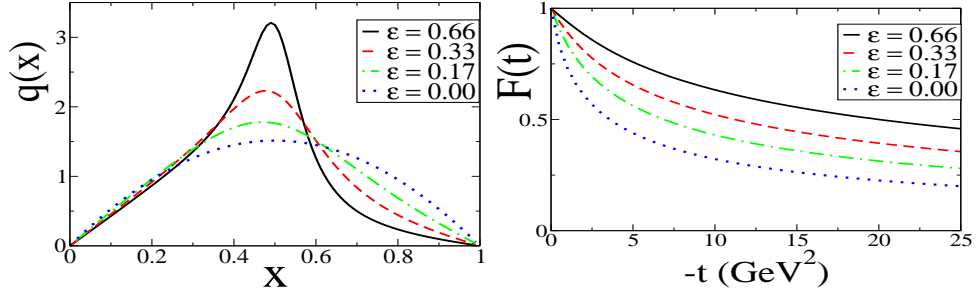

\begin{center}
\epsfig{file=q.eps,height=1.5in,width=2.5in}
\epsfig{file=form.eps,height=1.5in,width=2.5in}
\caption[Quark distribution and form factor for model with complex conjugate poles.]{\label{figqandF} On the left, the quark distribution Eq.~\eqref{qofx} is plotted as a 
function of $x$ for a few values of $\eps$ in GeV${}^2$.  On the right, the form factor 
calculated from Eqs.~\eqref{Fsum2} and \eqref{FDD} is plotted as a function 
of $-t$ for a few values of $\eps$ in GeV${}^2$. 
The model parameters are arbitrarily chosen as: $M = 0.14$ GeV 
and $m = 0.33$ GeV.}
\end{center}
\end{figure}

\begin{figure}
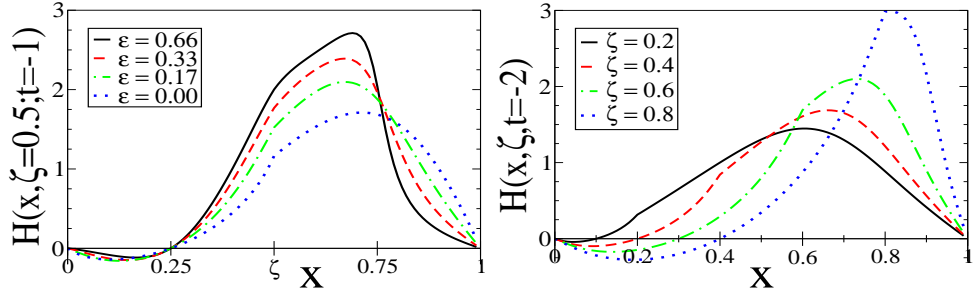

\begin{center}
\epsfig{file=gpd.eps,height=1.5in,width=2.5in}
\epsfig{file=gpd2.eps,height=1.5in,width=2.5in}
\caption[GPDs for model with complex conjugate poles.]{\label{figGPD} Plots of the GPD calculated from Eqs.~\eqref{FDD},\eqref{GDD} and \eqref{JiGPD2}. 
On the left, the GPD is plotted as a 
function of $x$ for a few values of $\eps$ (in Gev${}^2$) at fixed $\zeta = 0.5$ and $t = - 1.0$ GeV${}^2$. On the right, 
the GPD appears at fixed $\eps = 0.17$ GeV${}^2$ and $t = - 2.0$ GeV${}^2$ and is plotted
as a function of $x$ for a few values of $\zeta$.
The model parameters are arbitrarily chosen as: $M = 0.14$ GeV and 
$m = 0.33$ GeV.}
\end{center}
\end{figure}

In Figure \ref{figqandF}, we plot the quark distribution and electromagnetic form factor for various values of $\eps$ in GeV${}^2$. 
We have arbitrarily chosen the other model parameters as $M = 0.14$ GeV and $m = 0.33$ GeV. Additionally in Figure \ref{figGPD}, the
GPD is plotted: first at fixed $\zeta$ and $t$ for various values of $\eps$ and then at fixed $t$ and $\eps$ for various values 
of $\zeta$. Curves corresponding to $\eps = 0$ are the standard results for a propagator with one real pole.

Above we considered calculation of amplitudes for space-like processes using a scalar propagator with one pair of complex conjugate poles. 
Moreover the analysis can be extended easily to the case where vertex functions have complex conjugate singularities. 
Such models cannot be directly employed in Minkowski space, they must be analytically continued from Euclidean space. 

In Appendix \ref{minkowski}, the problems of using a propagator with complex conjugate poles in Minkowski space 
are discussed at the level of the quark distribution function. If the model is defined in Minkowski space, one 
will generally violate the support and positivity properties of the quark distribution. Above we
showed the model is perfectly well defined in Euclidean space by calculating non-diagonal matrix elements of twist-two operators. 
This leads us to the model's double distribution which we used to calculate parton and generalized parton distributions
as well as the electromagnetic form factor.

\section{Double Distributions for the pion} \label{spinor}

Having calculated model DDs for scalar constituent models, we shall now add spin degrees of freedom to the constituents. 
In this Section, we explore a model for the pion both on the light cone and from the covariant description in terms of
DDs. To calculate the DD functions, we again highlight the ambiguity and limitations inherent in the method employed by
\cite{Mukherjee:2002gb} and then properly determine the DDs.

\subsection{Bethe-Salpeter amplitude and pion wavefunction} \label{pion}

For the spin-$\frac{1}{2}$ model, we choose the trivially $q\bar{q}$ symmetric Bethe-Salpeter vertex
\begin{equation}
\Gamma(k,P) = - i g \g5.
\end{equation}
Here we have assumed only $\g5$ coupling at the quark-pion vertex with coupling constant $g$, 
whereas four Dirac structures exist \cite{Llewellyn-Smith:1969az}. 
This simple coupling is suggested by an effective interaction Lagrangian of the form 
(see, e.g., \cite{Frederico:1992ye,Frederico:1994dx})
\begin{equation} \label{lint}
\mathcal{L}_{\text{I}} = - i g \; \bm{\pi} \cdot \bar{q} \g5 \bm{\tau} q, 
\end{equation}
where the coupling constant $g = m / f_{\pi}$, with $m$ the constituent mass and $f_{\pi}$ the pion decay constant. 
Notice the (ladder approximation) kernel is independent of light-cone time. Thus this model
(as well as the scalar triangle model in Section \ref{scalar4}) are special cases of the instantaneous formalism described by 
\cite{Tiburzi:2001je} in the impulse approximation. The relation of the vertex to the Bethe-Salpeter wavefunction is given by
\begin{equation} \label{bswfn}
\Psi(k,P) = \frac{i}{\rlap\slash k - m + \ie} \; (-ig) \gamma^5 \frac{i}{\rlap\slash k  - \rlap\slash P - m + \ie}.
\end{equation}

The valence wavefunction can be found by projecting the Bethe-Salpeter wavefunction onto the 
light-cone $x^+ = 0$, see e.g.~\cite{Liu:1993dg}. Using the normalization convention of \cite{Lepage:1980fj}, we have
\begin{equation} \label{val}
\psi(x,\kpr;\lambda,\lamp) = \frac{1}{2 P^+} \int \frac{d \kminus}{2 \pi} \frac{ \bar{u}_{\lambda} ( xP^+, \kperp ) }
{ \sqrt{x} } \gp \Psi(k,P) \gp \frac{ v_{\lamp} \big( (1-x)P^+, \Pperp - \kperp \big) }{ \sqrt{1-x}},
\end{equation}
where $x$ is the fraction of the pion's plus momentum carried by the quark ($x = \kplus/\Pplus$), and the relative transverse momentum is $\kpr = \kperp - x \Pperp$. The valence wavefunction is found from Eq.~\eqref{val} to be
\begin{equation} \label{wfn}
\psi(x,\kperp;\lambda,\lamp) = \frac{g \sqrt{N_c/2} \; \mathcal{C}}{x(1-x)} \dw(x,\kperp|M^2_\pi)
\Big[  k_{-\lambda} \; \delta_{\lambda,\lamp} - \lambda m \; \delta_{\lambda,-\lamp}  \Big] , 
\end{equation}
where we have employed the notation $k_{\lambda} = k^1 + i \lambda k^2$. 
As a result of the contour integration, we have a factor of $\theta[x(1-x)]$ implicitly in Eq.~\eqref{wfn}.
Additionally the wavefunction is symmetric under interchange of $x$ and $1-x$. As is known, introduction of orbital angular momentum 
into this wavefunction leads to divergent quark distributions and form factors which will be handled below. 
Since this model is non-renormalizable, the choice of regularization scheme influences the dynamics.

\subsection{Form factor and generalized parton distribution} \label{form}

The pion electromagnetic form factor for this model can be calculated from the Feynman triangle diagram. In order to extract the GPD, however, 
we need to choose the kinematics specified in Figure \ref{ftri4} with $k$ as the momentum of the struck quark. As it stands, using the wavefunction
Eq.~\eqref{wfn}, the triangle diagram diverges.  Following the approach of \cite{Bakker:2000pk}, we covariantly smear the point-like 
photon vertex in Figure \ref{ftri4} in a way reminiscent of Pauli-Villars regularization
\begin{equation} \label{ansatz}
\gamma^\mu \to \Gamma^\mu_\Lambda = \frac{\Lambda^2}{k^2 - \Lambda^2 + \ie} \gamma^\mu \frac{\Lambda^2}{(k+\Delta)^2 -\Lambda^2 + \ie}.
\end{equation}
Eq.~\eqref{ansatz} is a simple way to model non-$q\bar{q}$ components of the wavefunction.\footnote{%
For all its positivity preserving virtues, the vertex smearing has unphysical drawbacks. Firstly the Ward-Takahashi identity
is messed up. Secondly the large momentum behavior of the induced constituent quark form factor is inconsistent with 
asymptotic freedom. We will come back to these issues in Chapter \ref{chap:proton} for the proton.
} 
Alternatively one could smear the $q\bar{q}-\pi$ vertex in a covariant manner \cite{deMelo:1997cb,Jaus:1999zv,deMelo:2002yq}. This smearing should 
additionally respect the $q\bar{q}$ symmetry of the vertex. 
We do not pursue this option here since positivity constraints (see Section \ref{pc}) are generally violated. 
On the other hand, one could use Pauli-Villars subtractions to regulate the theory, however, positivity would also be put into question. 
Because our concern is with model comparisons not phenomenology, we shall choose $\Lambda = m$ merely for simplicity. Although not obvious
from inspection, results for $\Lambda \neq m$ exhibit the same features investigated below. Most noteworthy, positivity remains 
satisfied when $\Lambda \neq m$.

Considering matrix elements of the current operator $J^\mu$ between pion states, the model Eq.~\eqref{ansatz} conserves current. 
This can be demonstrated most easily by calculating $\Delta \cdot J$ in the Breit frame. Additionally since the model is fully 
covariant, we can extract the electromagnetic form factor 
from any component of the current. In particular, potential end-point singularities present in matrix elements of $J^-$ have been removed 
by the photon vertex smearing  Eq.~\eqref{ansatz} \cite{Bakker:2000pk}. Using the plus-component of the current, we have the expression
\begin{equation} \label{ff}
F(t) =  \frac{i g^2 N_c |\mathcal{C}|^2 m^4}{1 - \frac{\zeta}{2}} \int \frac{d^4 k}{(2\pi)^4} \frac{\tr\Big[(\rlap\slash k + m )\g5  
(\rlap\slash k - \rlap\slash P  + m) 
\g5 (\rlap\slash k + \rlap\slash \Delta + m) \gp \Big]}
{\big[ k^2 - m^2 + \ie\big]^2 \big[ (k+\Delta)^2 - m^2 + \ie\big]^2 \big[ (P-k)^2 - m^2 + \ie\big]},
\end{equation}
where the momentum transfer is $t = \Delta^2$ and the skewness $\zeta$ is defined relative to the initial state: 
$\Delta^+ = - \zeta P^+ < 0$.

To calculate the GPD, we follow the procedure described in Section \ref{scalar4}. The result can be written as
\begin{equation} \label{gpd}
(1 - \zeta/2) H(x,\zeta,t) = \theta(x - \zeta) H_{\text{eff}}(x,\zeta,t) 
+ \theta(\zeta - x) \left[( H_{\text{inst}}(x,\zeta,t) + H_{\text{nval}}(x,\zeta,t) \right].
\end{equation}
where $H_{\text{eff}}$ is the piece determined by the effective two-body wavefunction, 
$H_{\text{inst}}$ is the contribution from instantaneous propagation of the spectator quark,
and the remaining contributions we term non-valence (although strictly speaking the instantaneous piece is 
also of the non-valence variety). It is a peculiarity of this model that explicit instantaneous terms
are not present for $x > \zeta$.

The functional forms are
\begin{align} \label{Feff}
 H_{\text{eff}}(x,\zeta,t)  & = \int \frac{d\kperp}{(2\pi)^3} \sum_{\lambda,\lamp} 
\psi^*_{\text{eff}}(x^\prime,\kpperp ;\lambda,\lamp) \psi_{\text{eff}}(x,\kperp;\lambda,\lamp),\\
 H_{\text{inst}}(x,\zeta,t) & =  - \mathcal{A} \int d\kperp \frac{4 \zeta \dw^3(\xpp,\kperp|t)}{(1-x)\xpp(1-\xpp)} \label{Finst}\\
 H_{\text{nval}}(x,\zeta,t) & =  - \mathcal{A} \int d\kperp \frac{2(\kperp \cdot \kpperp + m^2) \dw(x,\kperp|M_\pi^2) 
\dw^2(\xpp,\mathbf{k}^{\prime\prime\perp}|t)}{x(1-x)\xpp(1-\xpp)x^\prime(1-x^\prime)(1-\zeta)} \notag \\
& \times \Big[2\zeta \dw(\xpp,\mathbf{k}^{\prime\prime\perp}|t) + \dw(x,\kperp|M^2_\pi) \Big], 
\label{Fnval}
\end{align}
where we have defined the effective wavefunction
\begin{equation} \label{psieff}
\psi_{\text{eff}}(x,\kperp; \lambda,\lamp) = \frac{g \sqrt{N_c/2} \; \mathcal{C} m^2 }{x^2(1-x)}  
\Big[  k_{-\lambda} \; \delta_{\lambda,\lamp} - \lambda m \; \delta_{\lambda,-\lamp}  \Big] \dw^2(x,\kperp|M^2_\pi). 
\end{equation}
and made the abbreviation $\mathcal{A} = g^2 |\mathcal{C}|^2 N_c m^4 / 2 (2\pi)^3$.
It is sensible to think of Eq.~\eqref{psieff} as an effective wavefunction since $x \to 1-x$ symmetry has been lost. 
Moreover the wavefunction vanishes at $x = 1$ and is non-vanishing at $x = 0$. This is a desirable addition to the dynamics
stemming from the regularization. Notice the ladder kernel \eqref{lint} is momentum independent and hence does not vanish at $x = 0,1$. 
This is the dynamical reason why the un-regularized wavefunction Eq.~\eqref{wfn} does not vanish at the end points.  
Continuity of the GPD at $ x = \zeta$ follows directly from Eqs.~(\ref{Feff}--\ref{Fnval}).

\begin{figure}
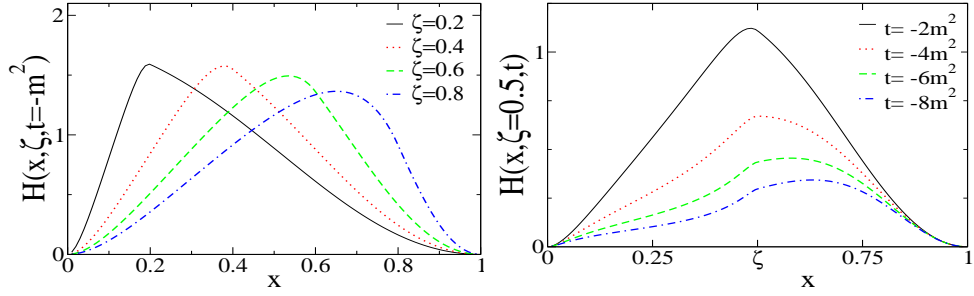

\begin{center}
\epsfig{file=lct.eps,height=1.5in,width=2.5in}
\epsfig{file=lcz.eps,height=1.5in,width=2.5in}
\caption[GPDs for toy model of the pion.]{\label{figLC} On the left, the GPD Eq.~\eqref{gpd} is plotted at fixed $t = - m^2$ for 
a few values of $\zeta$. On the right, the same GPD is plotted for fixed $\zeta = 0.5$ for a few values of $t$.
The model parameters are arbitrarily chosen as: $M = 0.14$ MeV and $m = 0.33$ MeV.}
\end{center}
\end{figure}

In Figure \ref{figLC}, we plot the GPD for the parameters: $M = 0.14$ MeV and $m = 0.33$ MeV. On the left, the graph
shows the GPD for a few values of $\zeta$ as a function of $x$ for fixed $t$, while on the right we have fixed $\zeta$ and $t$ varying.

\subsection{Sum rule and polynomiality} \label{prop}

\begin{figure}
\begin{center}
\epsfig{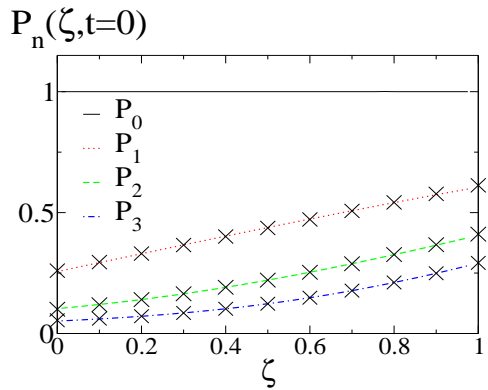}
\caption[Polynomiality conditions checked for the pion model's GPD.]{\label{figpoly} 
Polynomiality conditions checked for the GPD Eq.~\eqref{gpd}. The moments $P_{n}(\zeta,t=0)$
from Eq.~\eqref{poly} are plotted as a function of $\zeta$ for $n=0,1,2,3$. Additionally X's denote the simple $(n+1)$-point
polynomial fit to the moment $P_n$.}
\end{center}
\end{figure}

Since not manifest, one should check the covariance of the model Eq.~\eqref{gpd}. With the covariant starting point Eq.~\eqref{ff}, we 
anticipate polynomiality will be satisfied which provides a useful check on our expressions Eqs.~(\ref{Feff}--\ref{Fnval}). 
First we define the moments of the GPD with respect to asymmetric variables
\begin{equation} \label{poly}
P_n(\zeta,t) = \int x^n H(x,\zeta,t) dx.
\end{equation}
Polynomiality requires the moments $P_n$ to be of the form
\begin{equation} \label{polycon}
P_n(\zeta,t) = \sum_{j = 0}^{n} a_{n j}(t) \; \zeta^j.
\end{equation}
The zeroth moment is merely the sum rule for the form factor, hence $a_{0 0}(t) = F(t)$. 
For simplicity, we check a few of the lowest moments for the polynomiality condition
Eq.~\eqref{polycon} at $t = 0$. In Figure \ref{figpoly}, we plot the moments $P_n(\zeta,0)$
for $n = 0,1,2,3$ which appear as smooth functions. Additionally we plot simple $(n+1)$-point polynomial 
fits to the moments, which line up nicely with the integrals \eqref{poly}.

\subsection{Double distribution} \label{double}

To construct the DD, we shall first proceed incorrectly by appealing to the Lorentz invariance of the form factor as in Section 
\ref{scalar4}. This will at least lead to one component of the DD satisfying the reduction relations and can be compared to Eq.~\eqref{gpd}.

Using Eq.~\eqref{Feff} in the $\Delta^+ = 0$ (Drell-Yan) frame, we can write $F(t)$ in the form \eqref{DDform} with
\begin{equation} \label{DD}
F(x,y;t) =  \Big( 3 m^2 - M^2_\pi x(1-x) + y(1-x-y) t \Big)  
\frac{2 \pi \mathcal{A} \; \theta(1-x-y)\;  y(1-x-y)}{(1-x) \Big[m^2 - M^2_\pi x(1-x) - y (1-x-y) t  \Big]^3}.
\end{equation}
Aside from factors arising from spin, this DD is basically the same as that considered \cite{Mukherjee:2002gb} 
which one can realize by utilizing $\lambda^2 = - M^2_\pi /4 + m^2$. Not surprisingly, then, 
this DD satisfies Eqs.~(\ref{A}--\ref{C}). For reference we give the quark distribution function 
\begin{equation} \label{q}
q(x) = \frac{2\pi \mathcal{A}}{6} \frac{(1-x)^2 \Big[ 3 m^2 - M^2_\pi x(1-x) \Big]}{[m^2 - M^2_\pi x(1-x)]^3},
\end{equation}
which could be calculated directly from $\psi_{\text{eff}}$ in Eq.~\eqref{psieff}.

\begin{figure}
\begin{center}
\epsfig{file=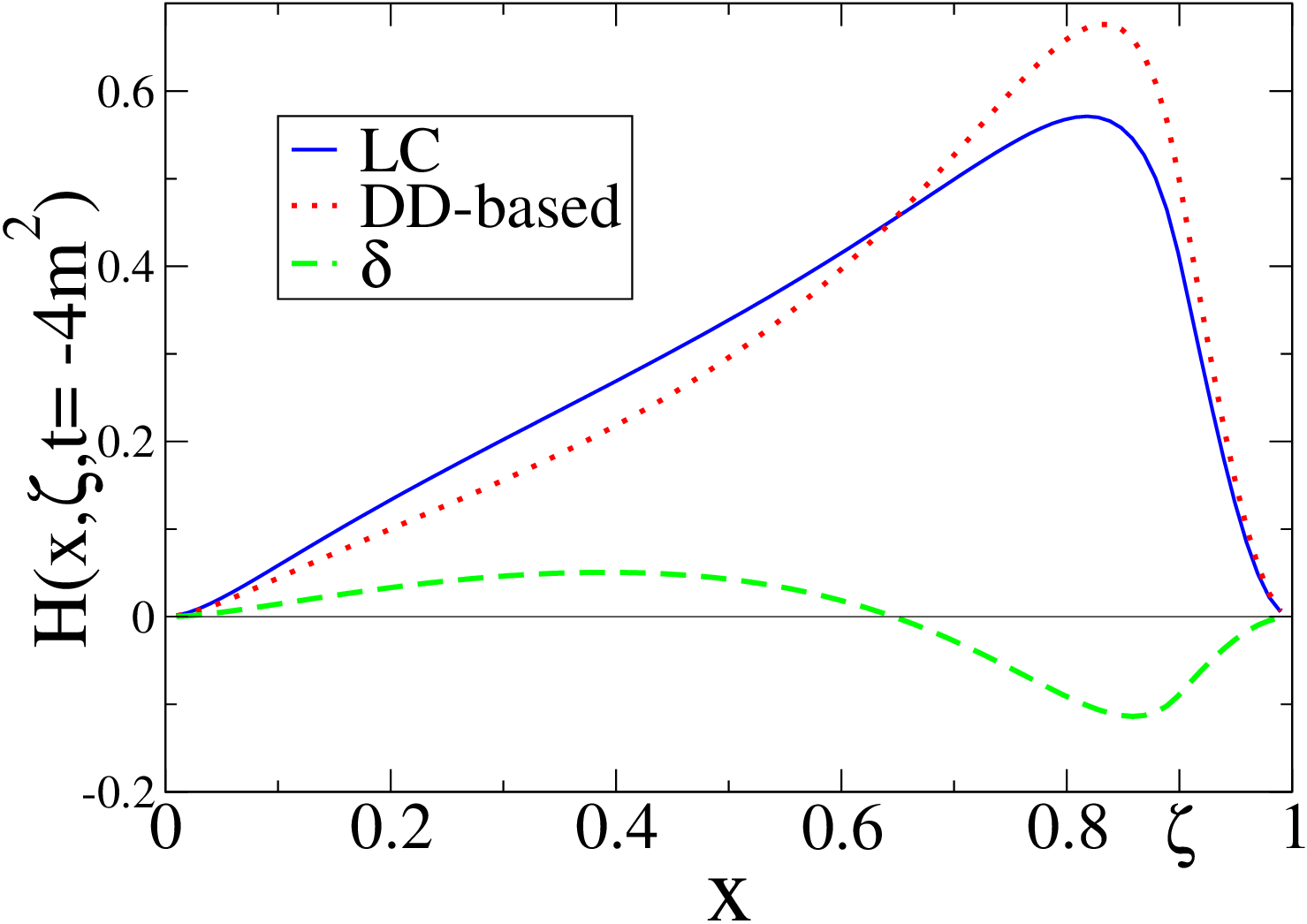,width=2.5in,height=1.5in}
\caption[Comparison of covariant GPDs for the pion model.]{Comparison of covariant GPDs 
for the spinor triangle diagram. The GPDs Eq.~\eqref{gpd} (denoted LC) and 
Eq.~\eqref{DD} (DD-based) are plotted as a function of $x$ for fixed $\zeta = 0.9$ and $t = - 4 m^2$ for the mass $M_\pi = 0.15 m$.
We also plot the difference between the two curves ($\delta$). The area under the curves is identically $F(-4m^2)$ for LC and DD-based 
GPDs, and hence zero for their difference $\delta$.}
\label{surprise2}
\end{center}
\end{figure}

In Figure \ref{surprise2}, we plot the GPD Eq.~\eqref{gpd} as well as the GPD derived from DD Eq.~\eqref{DD} via Eq.~\eqref{prescrip}.
As in Section \ref{scalar4}, the two are different despite the fact both models are covariant and posses the same form factor 
and quark distribution. Additionally we plot their difference as a function of $x$.

\subsection{Positivity constraints} \label{pc}

Here we demonstrate another difference between the GPD in \eqref{gpd} and the one stemming from   
the one-component DD Eq.~\eqref{DD}. To do so, we look at the positivity constraints. Originally these constraints appeared in 
\cite{Radyushkin:1998es,Pire:1998nw} and were derived from the positivity of the density matrix by restricting 
the final-state parton to have positive plus-momentum 
[and ignoring the contribution from $E(x,\zeta,t)$ for the spin-$\frac{1}{2}$ case].  Although the 
matrix elements involved for GPDs are off diagonal, they are still restricted by positivity and their diagonal elements. Correcting the constraints
for the presence of the $E$-distribution was first done in \cite{Diehl:2000xz}. By considering the positivity of the norm on Hilbert space, 
stricter constraints for the spin-$\frac{1}{2}$ distributions $H$ and $E$ have recently appeared as well as constraints for 
the full set of twist-two GPDs \cite{Pobylitsa:2001nt}.

\begin{figure}
\begin{center}
\epsfig{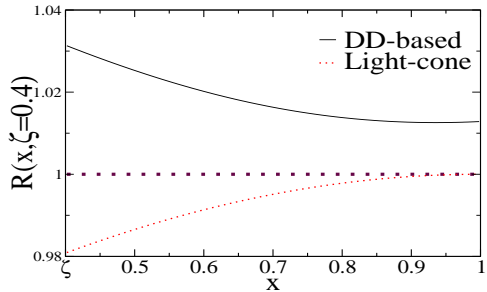}
\caption[Comparison of the positivity of GPDs.]{\label{figpos}
Comparison of GPDs: GPD calculated from the DD Eq.~\eqref{DD} via Eq.~\eqref{prescrip} (DD-based) compared with
the light-cone projection of the form factor Eq.\eqref{gpd} (Light-cone) for fixed $\zeta =0.4$ at $t = 0$. Here we plot
the ratio $R(x,\zeta)$ appearing in Eq.~\eqref{posi} as a function of $x>\zeta$. Positivity constrains this ratio 
to be less than one.}
\end{center}
\end{figure}

For the scalar pion case there is of course no contribution from the non-existent $E$-distribution and hence the original bounds are actually correct 
(modulo factors due to the difference of a spin-$\frac{1}{2}$ proton versus a spin-$0$ pion). Given the matrix element definition 
of the GPD consistent with equation \eqref{defn}, namely
\begin{equation}
H(x,\zeta,t) = \frac{1}{1-\zeta/2}  \int \frac{dz^-}{4\pi} e^{i x P^+ z^-}  
\langle \pi(P^\prime) | \bar{\psi}(0) \gp \psi(z^-) | \pi(P) \rangle,
\end{equation} 
the spin-$0$ positivity constraint (for $x>\zeta$) reads
\begin{equation} \label{posi}
R(x,\zeta) \equiv \big( 1 - \zeta/2 \big) \frac{\big|H(x,\zeta,0)\big|}{\sqrt{q (x) q(x^\prime)}}   \leq 1,
\end{equation}
where $q(x)$ is the model distribution function in Eq.~\eqref{q}. Of course the above result holds for finite $-t$, 
however since the function $\mathcal{F}$ decreases with $-t$, Eq.~\eqref{posi} is the tightest constraint. Notice
for $M^2_\pi \neq 0$, the limit $t = 0$ is in an unphysical region. If we treat this limit as formal, however, and analytically
continue our expressions, we can use Eq.~\eqref{posi}. Such continuation is consistent with the light-cone Fock space
representation of GPDs \cite{Brodsky:2000xy,Diehl:2000xz}.

Given the constraint Eq.~\eqref{posi}, we can test whether GPDs calculated from the light-cone projection
\eqref{Feff} and DD \eqref{DD} satisfy positivity. In Figure \ref{figpos}, we plot $R(x,\zeta)$ for each 
GPD as a function of $x$ for the fixed value of $\zeta = 0.4$. There is noticeably different behavior in the figure: 
positivity is violated by the DD-based model. As above (Section \ref{scalarrev}), we must carefully derive contributions 
from the other component $G(x,y;t)$.

\subsection{Derivation of the correct DDs}\label{real}

To derive both $F$ and $G$ DDs for the spin-$\frac{1}{2}$ pion model, we must consider the action of the
operator $\gamma^{\{\mu} i\overset{\leftrightarrow}{D}{}^{\mu_1}
\cdots i \overset{\leftrightarrow}{D}{}^{\mu_n}{}^{\}}$ between non-diagonal pion states. 
Inserted into Eq.~\eqref{ff} which is now taken in the symmetrical frame,  this operator produces
\begin{equation} \label{nspinor}
\int d^4 k \frac{\tr
\left[
(\rlap\slash k + m )\g5  
(\rlap\slash k - \rlap\slash \ol P  + \rlap\slash \Delta / 2 + m) 
\g5 (\rlap\slash k + \rlap\slash \Delta + m) 
\gamma^{\{ \mu} 
\right]
(k + \Delta/2)^{\mu_1} \cdots (k + \Delta/2)^{\mu_n \}}}
{\big[ k^2 - m^2 + \ie\big]^2 \big[ (k+\Delta)^2 - m^2 + \ie\big]^2 \big[ (k - \Pbar + \Delta/2)^2 - m^2 + \ie\big]}.
\end{equation}
The presence of the trace
\begin{equation}
\tr [ \ldots ] {}^\mu = 4 \Big[ \ol P {}^\mu (m^2 - k^2 - k \cdot \Delta) + \frac{\Delta^\mu}{2} (m^2 - k^2 + 2 k \cdot \ol P ) 
+ k^\mu (m^2 - k^2 - \frac{t}{2} + 2 k \cdot \ol P  - k \cdot \Delta)  \Big]
\end{equation}
complicates evaluating Eq.~\eqref{nspinor} by requiring contributions from diagrams reduced by one propagator. Since we
have smeared the photon via \eqref{ansatz}, the reduced diagrams are finite. Let us denote the propagators
simply by $\mathfrak{A} = (k - \ol P  + \Delta/2)^2 - m^2 + \ie$, $\mathfrak{B} = (k+\Delta)^2 - m^2 + \ie$ and  
$\mathfrak{C} = k^2 - m^2 + \ie$. To correctly evaluate Eq.~\eqref{nspinor}, we must write the trace as
\begin{equation} \label{redtrace}
\tr\{ \ldots \}{}^\mu = 4 \Big[ \ol P {}^\mu \big(\frac{t}{2} - \frac{1}{2} (\mathfrak{B} + \mathfrak{C}) \big) + 
\frac{\Delta^\mu}{2} \big( M^2 - \frac{t}{4} - \mathfrak{A} + \frac{1}{2} (\mathfrak{B} - \mathfrak{C}) \big) 
+ k^\mu ( M^2 - \frac{t}{2} - \mathfrak{A})  \Big]
\end{equation}
and evaluate each term separately canceling propagators in the denominator of Eq.~\eqref{nspinor}. 
These cancellations enable the DD to be read off by inspection.

These integrals can easily be evaluated using Feynman parameters. For example, let us consider the non-reduced 
contribution from Eq.~\eqref{redtrace}. The denominator of Eq.~\ref{nspinor} appears as
$\mathfrak{A} \; \mathfrak{B}^2 \; \mathfrak{C}^2$ and so we introduce two Feynman parameters $\{x,y\}$ 
to render the denominator specifically in the form $[x \mathfrak{A} + y \mathfrak{B} + (1-x-y) \mathfrak{C}]^{-5}$.
One then translates $k^\mu$ to render the integral (hyper-) spherically symmetric via the definition 
$k^\mu = l^\mu + \beta \ol P {}^\mu - (\alpha + 1) \Delta^\mu / 2$. Here $\beta = x$ and $\alpha = x + 2y - 1$. Using a Wick rotation
to evaluate the resulting integral over $l$, we can cast the contribution to Eq.~\eqref{nspinor} from non-reduced terms
in the form
\begin{multline} \label{nonred}
\frac{\pi \mathcal{A}}{4} \int_{0}^{1} d\beta \int_{-1 + \beta}^{1 - \beta} d\alpha 
[(1-\beta)^2 - \alpha^2] D(\beta, \alpha;t)^3
\Big[ 2 \ol P (\beta M_\pi^2 + (1-\beta)t/2) - \Delta \alpha (M_\pi^2 - t/2)  \Big]^{\{ \mu}
\\ \times \sum_{k = 0}^n \frac{n!}{k! (n-k)!} \beta^{n-k} \alpha^k \ol P {}^{\mu_1} \cdots \ol P {}^{\mu_{n-k}} 
\Big(-\frac{\Delta}{2}\Big)^{\mu_{n - k + 1}} \cdots \Big(-\frac{\Delta}{2}\Big)^{\mu_n \} }
\end{multline}
with $D(\beta,\alpha;t)$ given by Eq.~\eqref{dba}. Given the form of Eq.~\eqref{nonred}, we can identify $\{\beta,\alpha\}$
as DD variables and hence read off contributions to $F$ and $G$ DDs. 
\begin{align}
\delta F(\beta,\alpha;t)  & = \pi \mathcal{A} \; [(1-\beta)^2 - \alpha^2] \; D(\beta,\alpha;t)^3 \big(\beta M_\pi^2 + ( 1- \beta) t/2\big)
\\ 
\delta G(\beta,\alpha;t) & = \pi \mathcal{A} \; \alpha \; [(1-\beta)^2 - \alpha^2] \; D(\beta,\alpha;t)^3 \big( M^2_\pi - t/2\big). 
\end{align}
Notice that these contributions respect the properties of DDs, namely $\delta F$ is even in $\alpha$ while
$\delta G$ is odd. This need not be the case, however, for each intermediate step of the calculation, e.g.~contributions
from $\mathfrak{B}$-reduced terms and $\mathfrak{C}$-reduced terms are individually neither even nor odd in $\alpha$ while 
their sum is even and difference is odd.

Ignoring for the moment contributions from $\mathfrak{A}$-reduced terms, we arrive at the DDs
\begin{align}
\label{FDD2}
F(\beta,\alpha;t) & = \pi \; \mathcal{A}  \; D(\beta,\alpha;t)^3  \Big[ (1-  \beta) m^2 
 - \beta \alpha^2 M_\pi^2 + (1-\beta) [(1-\beta)^2 - \alpha^2] t/ 4 \Big]
\\ 
G(\beta,\alpha;t) & = - \pi \; \mathcal{A} \; \alpha  \; D(\beta,\alpha;t)^3 \Big[ m^2  
- M^2_\pi  (1 - \beta - \alpha^2)  + [(1-\beta)^2 - \alpha^2] t / 4  \Big] \label{GDD2},
\end{align} 
The contribution from $\mathfrak{A}$-reduced terms has the form of a $D$-term in that it is proportional to $\delta(\b)$. 
Using Feynman parameters for the denominator
$\mathfrak{B}^2 \; \mathfrak{C}^2$ and suitable changes of variables, we arrive at the contribution to Eq.~\eqref{nspinor}
\begin{equation}
- \Delta^{\{\mu} \int_{-1}^{1} d\alpha  \frac{\pi \mathcal{A} \alpha^{n+1} (1 - \alpha^2)}{[m^2 - (1-\alpha^2) t/4]^2} 
 \Big(-\frac{\Delta}{2}\Big)^{\mu_{1}} \cdots \Big(-\frac{\Delta}{2}\Big)^{\mu_n \}}
\end{equation} 
from which we can identify the $D$-term
\begin{equation}
D(\alpha;t) = \pi \mathcal{A} \frac{\alpha (1-\alpha^2)}{[m^2 - (1-\alpha^2) t/4]^2}. 
\end{equation}
Although strictly speaking, the $D$-term is a contribution to the $G$-DD, we shall treat it separately for ease.
Furthermore, the complete $D$-term arises from the $\b$-integral of $G(\b,\a;t)$, see Appendix \ref{sec:amb} and \cite{Teryaev:2001qm}.

Switching now to asymmetric variables, we have
\begin{eqnarray}
F(x,y;t)  & = &  \pi \; \mathcal{A} \;  D(x,y;t)^3  \notag \\
& & \times \Big[ (1  -x) m^2 - x (x  +  2 y - 1)^2 M_\pi^2 + (1-x) y(1-x-y)t \Big] \label{Fdd} \\
G(x,y;t)  & = & - \pi  \; \mathcal{A} \; (x + 2 y - 1) \; D(x,y;t)^3  \notag  \\ 
& & \times \Big[ m^2 + y(1 -x-y) t  - M^2_\pi (1 - x - (x + 2y - 1)^2)  \Big] 
\label{Gdd} \\
D(y;t) & = & \pi \mathcal{A} y(1-y) (2 y - 1) \; [m^2 - y (1-y) t]^{-2} \label{Dy}  
\end{eqnarray}
where the function $D(x,y;t)$ is given by Eq.~\eqref{dxy}. Accordingly $F(x,y;t)$ is \emph{M\"unchen} symmetric and $G(x,y;t)$ is 
antisymmetric, while $D(y;t)$ is antisymmetric about $y = 1/2$. Notice $F(x,y;t)$ in Eq.~\eqref{Fdd} is not that of 
Eq.~\eqref{DD}. Given the ambiguity inherent in defining $F$ versus $G$ DDs [cf. Eq.~\eqref{Gsum1}], there is no reason to believe 
the $F$s would be the same. In principle, we could construct a quasi-gauge transformation \cite{Teryaev:2001qm} to render the $F$s 
the same. This would enable identification of the missing $G$ function unique to Section \ref{double}. We shall not pursue this 
tangential point,\footnote{%
Notice the contribution to the GPD from our 
$D$-term resembles that of $H_\text{inst}$ in Eq.~\eqref{Finst} but is not identical. Both terms originate from a reduction of the 
spectator's propagator. In the case of the $D$-term, the spectator's propagator is completely removed by the $\mathfrak{A}$-reduction. 
For $H_{\text{inst}}$, there is residual $x$-dependence stemming from the light-cone instantaneous propagator $\gp / 2 P^+ (1 - x)$.
}
but simply point out that the procedure in \cite{Mukherjee:2002gb} leads only to one component of the DD and precisely which component [in the 
decomposition of Eq.~\eqref{decomp1}] is unknown and hence useless.

The function $F(x,y;t)$ satisfies the reduction relations: it reduces to the quark distribution via Eq.~\eqref{B} and
integrates to the form factor Eq.~\eqref{C}---the latter can only be checked numerically. Lastly then it remains to see whether the 
DD-based GPD lines up with true GPD calculated in Section \ref{form}. To construct the GPD we use the form 
of Eq.~\eqref{correct} modified to handle the extra $D$-term in Eq.~\eqref{Dy} separately 
\begin{equation} \label{recorrect}
H(x,\zeta,t) = \int_0^1 dz \int_0^{1-z} dy \; \delta(x - z - \zeta y) 
\Big[ F(x,y;t) 
+ \frac{\zeta}{2 - \zeta} G(x,y;t) \Big] 
+ \frac{\theta[\xpp(1 - \xpp)]} {2(2 - \zeta)} D(\xpp;t)
\end{equation}

\begin{figure}
\begin{center}
\epsfig{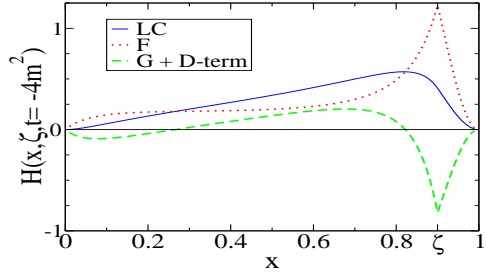}
\caption[Contributions to the covariant GPD from DDs.]{Contributions to the covariant GPD from DDs. The light-cone GPD Eq.~\eqref{gpd} 
and the DD-based Eq.~\eqref{recorrect} are identical, denoted (LC) and plotted as a function of $x$ for fixed $\zeta = 0.9$ 
and $t = - 4 m^2$ for the mass $M_\pi = 0.15 m$. We also plot the individual contributions from $F$ in Eq.~\eqref{Fdd} and  
from $\frac{\zeta}{2 - \zeta} G + \frac{1}{2(2 - \zeta)} D$ in Eqs.~\ref{Gdd} and \ref{Dy}, denoted (G + D-term).}
\label{surprise3}
\end{center}
\end{figure}

In Figure \ref{surprise3}, we plot the GPD Eq.~\eqref{gpd} and the DD-based Eq.~\eqref{recorrect}. They are identical. 
We also plot the individual $F$ and $G + D$ contributions to the GPD. 
Even in the absence of the $D$-term, the contribution from $G$ cannot be neglected in ascertaining the DD. Again calculation of the
GPD from the full two-component DD agrees with the light-cone GPD definition. The argument of Lorentz invariance used to 
calculate an $F$-type DD that satisfies the reduction relations does not determine the GPD.

For a given model, there exists a unique two-component DD in the decomposition of Eq.~\eqref{decomp1}. This DD produces
a GPD which is then consistent with the light-cone projection and satisfies all the necessary reduction relations. 
On the other hand, if one is interested in 
constructing \emph{different} GPD models satisfying the reduction relations, polynomiality, \emph{etc.}, 
one could exploit the ambiguity of one-component DDs.  Consider the following.
The contribution to the GPD from $F$ in Figure \ref{surprise3} is markedly different 
from Eq.~\eqref{gpd} and the analogous contribution from Eq.~\eqref{DD}.  These three GPDs, however, share the 
same form factor and quark distribution, since their respective $G$ functions do not contribute to the reduction relations. 
Thus one can always use DD-gauge freedom \cite{Teryaev:2001qm} to transform a given $F$ into a new function 
and throw away contributions from $G$. The result is an infinite set of different GPDs with identical form factors 
and quark distributions. As pointed out in Section \ref{pc}, one must be careful to maintain positivity although
it is likely that there still is an infinite set of GPDs which would.

\section{Summary} \label{summy}

Above we consider two covariant models for pions: one with scalar constituents and the other with spin-$\frac{1}{2}$.
The spin-$\frac{1}{2}$ model requires regularization and we choose the method in \cite{Bakker:2000pk} in order to maintain positivity. 
For each case we derive the GPD from its matrix element definition which forces us to consider the triangle diagram for the
form factor with the plus momentum of the struck quark kept fixed in a general $\Delta^+ \neq  0$ frame.

We also construct the DDs for each model. The approach of \cite{Mukherjee:2002gb} leads only to one component 
of the DD (the ``forward-visible'' piece) which is itself ambiguous, thus resulting GPDs are incorrect and need not satisfy positivity.
This fact remains true even for $C$-odd distributions. This means the reduction relations alone are not enough from which to calculate
a model's DD (and hence GPD). To obtain both components of the DD unambiguously, 
we calculate the matrix elements of twist-two operators. The resulting  DD-based GPDs then agree with those calculated 
on the light cone. The ``gauge freedom'' inherent in defining $F$ vs. $G$ DDs
could be exploited, however, for phenomenological studies where one is interested in a mathematical fit to data, rather than
a calculation of a given model's DD.  Having investigated the calculation of DDs for pions, we can now turn to the case of the 
proton.

\chapter{Double Distributions for the Proton} \label{chap:proton}

In this Chapter, we extend the simple scenarios so far pursued to build a DD model for the proton.
We treat the proton as a bound state of a residual quark and two quarks strongly coupled in both
the scalar and axial-vector diquark channels. The resulting light-cone wavefunction of the 
proton has appropriate spin structure: containing correlations where the residual quark carries the
spin of the proton as well as correlations where the quark and diquark are in a relative $p$-wave.
We use this model to derive DDs for the proton. Inclusion of the spin structure into double distributions 
is crucial if one wishes to make contact with the spin sum rule for hadrons. While this (in essence two-body) 
model for the proton is crude, model parameters can be tuned so that the electromagnetic form factors are fit
at small momentum transfer. Quark distributions can be matched and resulting GPDs are suitable for phenomenology.

This Chapter has the following organization. First we review our conventions for DDs for spin-$\frac{1}{2}$ particles
and the relation of these DDs to GPDs in Section~\ref{sec:def}. 
In order to deal with $D$-terms efficiently, we opt to calculate three DD functions. The ambiguity inherent in defining DDs for 
spin-$\frac{1}{2}$ systems is detailed in Appendix \ref{sec:amb}. Next in Section~\ref{sec:DDcalc}, we present the quark-diquark model 
under consideration.  DDs are calculated in this model in both the scalar and axial-vector diquark channels. Relevant 
identities are gathered in Appendix~\ref{sec:gamma}, while the details of the derivation appear in Appendix~\ref{sec:DDDD}.
Intuition about the model and its construction is provided in Appendix~\ref{sec:LC}, where the effective light-cone wavefunction
is extracted from projecting onto the light-cone.  Section~\ref{sec:fit} presents potential phenomenological uses for the model. 
The model is tuned to reproduce the Dirac and Pauli form factors of the proton for small momentum transfer. Resulting 
quark distributions and GPDs are sensible at a low scale but do not contain the relevant non-perturbative small-$x$ behavior 
relevant for GPD phenomenology. One could then follow \cite{Mukherjee:2002gb} and augment the DDs with empirically parametrized
quark distributions and re-tune the model's parameters or use perturbative evolution. 
Consequently these simple model GPDs then satisfy all known constraints, including positivity.

\section{Definitions} \label{sec:def}

To begin, we set forth our conventions for DDs and their relation to GPDs. 
Moments of DDs appear naturally in the decomposition of twist-two operators' matrix elements that are non-diagonal
in momentum space; moreover, 
they provide an elegant explanation of the polynomiality property of GPDs.  
To make any progress in calculating DDs, we must use the parton model simplification for the gauge covariant derivative: 
$D^\mu \to \partial^\mu$.

The non-diagonal proton matrix elements of twist-two operators can be decomposed in a fully Lorentz 
covariant fashion in terms of various twist-two form factors $A_{nk}(t)$, $B_{nk}(t)$ and $C_{nk}(t)$, namely\footnote{%
Beyond the parton model, this decomposition still holds. The twist-two form factors, however, acquire scale dependence due
to the renormalization of the twist-two operators. This scale dependence must ultimately be dealt with for phenomenological 
applications.
}
\begin{multline} 
\langle P^\prime,\lp | 
\ol \psi (0) \gamma^{\{\mu}i \overset{\leftrightarrow}{D} {}^{\mu_1} \cdots i \overset{\leftrightarrow}{D} {}^{\mu_n\}} \psi(0)
| P,\l \rangle  
 \\ 
= \ol u_{\lp}(P') \gamma^{\{ \mu } u_{\l}(P) 
\sum_{k=0}^{n} \frac{n!}{ k! (n-k)!} A_{nk}(t) 
\ol P {}^{\mu_1} \cdots \ol P {}^{\mu_{n-k}} 
\left( - \frac{\D}{2}\right)^{\mu_{n-k+1}} \cdots \left( - \frac{\D}{2}\right)^{\mu_{n}\}}
\\
+ \ol u_{\lp}(P') \frac{i \sigma^{ \{ \mu \nu } \D_\nu}{2 M}  u_{\l}(P) 
\sum_{k=0}^{n} \frac{n!}{ k! (n-k)!} B_{nk}(t) 
\ol P {}^{\mu_1} \cdots \ol P {}^{\mu_{n-k}} 
\left( - \frac{\D}{2}\right)^{\mu_{n-k+1}} \cdots \left( - \frac{\D}{2}\right)^{\mu_{n}\}}
\\
- \ol u_{\lp}(P') \frac{\D^{ \{ \mu} }{4 M}  u_{\l}(P) 
\sum_{k=0}^{n} \frac{n!}{ k! (n-k)!} C_{nk}(t) 
\ol P {}^{\mu_1} \cdots \ol P {}^{\mu_{n-k}} 
\left( - \frac{\D}{2}\right)^{\mu_{n-k+1}} \cdots \left( - \frac{\D}{2}\right)^{\mu_{n}\}} \label{eqn:moments}
,\end{multline}
where the action of ${}^{\{}\cdots{}^{\}}$ on Lorentz indices produces the symmetric traceless part of the tensor
and $\ol P$ is defined to be the average momentum between the initial and final states $\ol P {}^\mu = \frac{1}{2} ( P' + P)^\mu$
and $\D$ is the momentum transfer, $\D^\mu = ( P' - P)^\mu$.  
$T$-invariance restricts $k$ in the first two sums to be even and odd in the last sum. 
There are three Dirac structures in the above decomposition since in general the 
twist-two currents are not conserved thus allowing for a structure proportional to $\D^\mu$. 
The ambiguity of DDs for spin-$\frac{1}{2}$ systems is addressed in Appendix \ref{sec:amb}.
Such difficulties in constructing DDs have been addressed in the literature \cite{Polyakov:1999gs,Belitsky:2000vk,Teryaev:2001qm,Tiburzi:2002tq},
and we find the construction in Eq.~\eqref{eqn:moments} the easiest to work with in actual calculations.

The above decomposition can 
be used to define three double distributions as generating functions for the twist-two form factors
\begin{align} \label{eqn:generate}
A_{nk}(t) &= \int_{-1}^{1} d\b \int_{-1 + |\b|}^{1 - |\b|} d\a \,
\b^{n - k} \a^k F(\b,\a;t) \\
B_{nk}(t) &= \int_{-1}^{1} d\b \int_{-1 + |\b|}^{1 - |\b|} d\a \,
\b^{n - k} \a^k K(\b,\a;t) \\
C_{nk}(t) &= \int_{-1}^{1} d\b \int_{-1 + |\b|}^{1 - |\b|} d\a \,
\b^{n - k} \a^k G(\b,\a;t) \label{eqn:generated}
.\end{align} 
As a consequence of the restriction on $k$ in the sums, the functions $F(\b,\a;t)$ and $K(\b,\a;t)$ are even in $\a$, while
$G(\b,\a;t)$ is odd. The $F(\b,\a;t)$ and $K(\b,\a;t)$ DDs are similar in form to the functions originally employed in \cite{Radyushkin:1997ki}. 
The difference is due to the third DD, $G(\b,\a;t)$, which incorporates the $D$-term \cite{Polyakov:1999gs} among other things.

Summing up the moments in Eq.~\eqref{eqn:moments}, 
these DD functions then appear in the decomposition of the light-like separated quark bilinear operator
\begin{multline} \label{eqn:bilocal}
\langle P^\prime,\lp | 
\ol \psi_q \left( - z/2 \right) \rlap \slash z
\psi_q \left( z/2 \right) 
| P,\l \rangle 
=
\int_{-1}^{1} d\b \int_{-1 + |\b|}^{1 - |\b|} d\a \;
e^{ - i \b \ol P \cdot z + i \a \D \cdot z / 2}
\\
\times 
\ol u_{\lp}(P') 
\Bigg[ \rlap \slash z  F_q(\b,\a;t)
+
\frac{i \sigma^{\mu \nu} z_\mu \D_{\nu}}{2 M}
K_q(\b,\a;t) 
-
\frac{\D \cdot z}{4 M} G_q(\b,\a;t)
\Bigg] u_{\l}(P) 
,\end{multline}
where we have appended a flavor subscript $q$ in the relevant places and $z^\mu$ is a lightlike vector, $z^2 = 0$.

Now we define the light-cone correlation function by Fourier transforming with respect to the light-cone separation $z^-$
\begin{equation} \label{eqn:lcc}
\mathcal{M}_q^{\lp,\l}(x,\x,t) = \frac{1}{4 \pi} \int dz^- 
e^{i x \ol P {}^+ z^-}
\langle P^\prime, \lp | 
\ol \psi_q \left( - z^-/2 \right) \gamma^+ 
\psi_q \left( z^-/2 \right) 
| P, \l \rangle
.\end{equation}
Above the variable $\x$, or skewness parameter, is defined by $\x = - \D^+ / 2 \ol P {}^+$. As is customarily done, 
we assume without loss of generality that $\x > 0$. 
The correlation function in Eq.~\eqref{eqn:lcc} can be written in terms of the two independent GPDs $H_q(x,\x,t)$ and $E_q(x,\x,t)$ as
\begin{equation} \label{eqn:lccor}
\mathcal{M}_q^{\lp,\l}(x,\x,t) = 
\frac{1}{2 \ol P {}^+} \ol u_{\lp}(P') \left[ 
\gamma^+ H_q(x,\x,t)
+ 
\frac{i \sigma^{+\nu} \D_\nu }{2 M}
E_q(x,\x,t)     
\right] u_{\l}(P)   
.\end{equation}
Unlike the DDs, these GPDs are quantities that enter directly into the amplitude for deeply virtual Compton scattering (DVCS), for example.
Inserting the DD decomposition Eq.~\eqref{eqn:bilocal} into the correlator in Eq.~\eqref{eqn:lccor}, we 
can express the GPDs as projections of the DDs
\begin{align} \label{eqn:Hq}
H_q (x,\x,t) = \int_{-1}^{1} d\b \int_{-1 + |\b|}^{1 - |\b|} d\a \;
\delta(x - \b - \x \a) 
\left[ F_q(\b,\a;t) + \x G_q(\b,\a;t)
\right]
\\ \label{eqn:Eq}
E_q (x,\x,t) = \int_{-1}^{1} d\b \int_{-1 + |\b|}^{1 - |\b|} d\a \;
\delta(x - \b - \x \a) 
\left[ K_q(\b,\a;t) + \x G_q(\b,\a;t)
\right]
\end{align}
from which we can view the $\x$-dependence of GPDs as arising from different slices of the underlying Lorentz invariant DDs. 
Due to the symmetry of the DDs with respect to $\a$, the GPDs $H_q(x,\x,t)$ and $E_q(x,\x,t)$ 
are both even functions of the skewness parameter $\x$.

The GPD $H_q(x,\x,t)$ has an important reduction property. Taking the diagonal limit of the light-cone correlator
Eq.~\eqref{eqn:lccor}, we have
\begin{equation} \label{eqn:quark}
f_q(x) = H_q(x,0,0) = \int_{-1 + |x|}^{1 - |x|} d\a \,  F_q(x,\a;t)
\end{equation}
In DVCS, the relevant current operators produce the charge and flavor structure $\sum_q e_q^2$ since there are two photons
and thus the charge squared weighted GPDs enter in relevant physical amplitudes.
To discover the relation of GPDs to electromagnetic form factors it is advantageous to consider instead the 
single photon structure $\sum_q e_q$ and define
\begin{align}
H(x,\x,t) & = \sum_q e_q H_q (x,\x,t) \\  
E(x,\x,t) & = \sum_q e_q E_q (x,\x,t)
.\end{align}
Since $G_q(\b,\a;t)$ is an odd function of $\a$, we have $\int_{-1}^{1} d\b \int_{-1 + |\b|}^{1 - |\b|} d\a  \, G_q(\b,\a;t)  = 0$ and consequently the sum rules
\begin{align}
\int_{-1}^{1} dx \, H (x,\x,t) =  \sum_q e_q \int_{-1}^{1} d\b \int_{-1 + |\b|}^{1 - |\b|} d\a  \, F_q(\b,\a;t)  &= F_1 (t) \\
\int_{-1}^{1} dx \, E (x,\x,t) =  \sum_q e_q \int_{-1}^{1} d\b \int_{-1 + |\b|}^{1 - |\b|} d\a  \, K_q(\b,\a;t)  &= F_2 (t)
\end{align}
which relate the zeroth moments to Dirac and Pauli form factors.  Higher $x$-moments of $H_q(x,\x,t)$ and $E_q(x,\x,t)$
are even polynomials in $\x$ as can be seen directly from Eqs.~\eqref{eqn:Hq} and \eqref{eqn:Eq}.

The GPDs satisfy further constraints: the positivity bounds 
\cite{Radyushkin:1998es,Pire:1998nw,Diehl:2000xz,Pobylitsa:2001nt,Pobylitsa:2002gw,Pobylitsa:2002iu,Pobylitsa:2002ru} . 
These bounds are particularly important for comparing with experiment. Model GPDs which reduce to the experimental 
quark distribution in Eq.~\eqref{eqn:quark} but violate the positivity bounds are not worth considering because one 
knows from the start that rate estimates predicted by the model GPDs are automatically wrong. Usually violation of 
the positivity bounds is a signal that the model is inconsistent with the underlying field theory. Perhaps surprisingly
violation occurs frequently in more-or-less all standard hadronic models.
Of interest to us are the basic bounds for both the spin-flip and 
non-flip amplitudes
\begin{equation} \label{eqn:positivity}
\theta ( x - \x) \Big| \mathcal{M}_q^{\l,\pm \l}(x,\x,t) \Big| \leq 
\sqrt{ f_q\left( \frac{x- \x}{1 - \x}\right) f_q\left( \frac{x + \x}{1 + \x}\right)}
,\end{equation}
which we use as a stipulation in constructing our model.

Lastly we need to address the negative range of the DD variable $\b$. Experimentally and diagrammatically $\b$ is strictly positive and crossing
symmetry can be used to extend the range to positive and negative $\b$, or in our case reduce the range to strictly positive. 
To this end, we define two functions for each DD
\begin{eqnarray} \label{eqn:Fplus}
F_q^{\pm}(\b,\a;t) & = &  F_q(\b,\a;t) \pm  F_{\ol q} (\b,\a;t) \\
K_q^{\pm}(\b,\a;t) & = &  K_q(\b,\a;t) \pm K_{\ol q} (\b,\a;t)  \\
G_q^{\pm}(\b,\a;t) & = &  G_q(\b,\a;t) \pm  G_{\ol q} (\b,\a;t) \label{eqn:Gplus}  
,\end{eqnarray}
where the antiquark contributions are defined by crossing 
\begin{eqnarray}
F_{\ol q}(\b,\a;t)  &=& - F_{q}(-\b,\a;t)  \notag \\
K_{\ol q}(\b,\a;t) &=& - K_{q}(-\b,\a;t)  \notag \\
 G_{\ol q}(\b,\a;t) &=& - G_{q}(-\b,\a;t) \notag
.\end{eqnarray}
Thus the plus DDs [$F_q^+(\b,\a;t)$, $K_q^+(\b,\a;t)$, and $G_q^+(\b,\a;t)$] are odd functions of $\b$ 
and the minus DDs are even functions. In partonic language, the minus DDs correspond to 
a difference in quark and antiquark DDs (the valence configuration) while the plus DDs 
are a sum of quark and anitquark DDs.

By virtue of the above definitions Eqs.~\eqref{eqn:Fplus}--\eqref{eqn:Gplus}, we can remove the 
explicit negative-$\b$ parts from DDs and consequently the GPDs. We do so by defining
\begin{eqnarray}
H^\pm_q(x,\x,t) &=& H_q(x,\x,t) \pm H_{\ol q} (x,\x,t)  \\
E^\pm_q(x,\x,t) &=& E_q(x,\x,t) \pm E_{\ol q} (x,\x,t)
,\end{eqnarray}
where the antiquark contributions are defined by the crossing analogous to the DDs above. 
In this form, we can rewrite the reduction relations in a more familiar way
\begin{equation} \label{eqn:qval}
f^\pm_q(x) = H^\pm_q(x,0,0) = \int_{-1 + x}^{1 - x} d \a \, F^\pm_q(x,\a;t)
,\end{equation}
where $f^-_q(x)$ is the valence quark distribution, namely $f^-_q(x) = f_q(x ) - f_{\ol q} (x)$.
The sum rules are thus
\begin{align}
\int_{0}^{1} dx \, H^- (x,\x,t) =  \sum_q e_q \int_{0}^{1} d\b \int_{-1 + \b}^{1 - \b} d\a  \, F^-_q(\b,\a;t)  &= F_1 (t), \\
\int_{0}^{1} dx \, E^- (x,\x,t) =  \sum_q e_q \int_{0}^{1} d\b \int_{-1 + \b}^{1 - \b} d\a  \, K^-_q(\b,\a;t)  &= F_2 (t)
.\end{align}
In our simple valence model for the proton, we only address quark configurations. The double distribution variable 
$\b$ as well as the momentum fraction $x$ are positive below.
The above positivity bounds in Eq.~\eqref{eqn:positivity} also hold for valence and plus amplitudes.

\section{Model double distribution}\label{sec:DDcalc}

To calculate DDs for the proton, as a first step we use only a simple model consisting of two quarks strongly coupled 
in the scalar and axial-vector diquark channels along with a residual quark. 
This model can be considered as loosely based on relativistic quark models~\cite{Chung:1991st}
or on the Nambu-Jona Lasinio model of the proton in the static approximation, see, e.g., \cite{Buck:1992wz,Mineo:1999eq}.
We keep full Lorentz covariance in order to preserve the polynomiality of the moments Eq.~\eqref{eqn:moments}. 
Without covariance, we would not even be able to deduce the DDs.

In order to further simplify this calculation, we mandate only tractable one loop contributions. 
To this end, we treat the scalar and axial-vector diquark $T$-matrices as free particle propagators, namely of the forms
\begin{align} \label{eqn:scalar}
D(k) & = \frac{i}{k^2 - m_{SD}^2 + i \varepsilon} \\
D^{\mu \nu}(k) & =  \frac{- i g^{\mu \nu}}{k^2 - m_{AD}^2 + i \varepsilon}
\label{eqn:vector}
\end{align}
respectively. For simplicity we have neglected the term proportional to $k^\mu k^\nu$ in the massive vector propagator.
Evaluation of the contributions generated by this term are considerably complicated and will be the subject of a future publication.
The proton Bethe-Salpeter vertex for our model is thus
\begin{equation} \label{eqn:vertex}
\Gamma (k,P) = \frac{1}{\sqrt{2}} \, \chi^{(s)}  D(P-k)\otimes   \Gamma^{(s)}(k,P) 
+  \frac{1}{\sqrt{2}} \, \chi^{(a)}_{\mu,i} \,  D^{\mu \nu}(P - k) \otimes  \Gamma^{(a)}_{\nu,i}(k,P), 
\end{equation}
where the diquark vertices are direct products of spin and isospin factors
\begin{align}
\chi^{(s)} &= \frac{1}{\sqrt{2}} 
(i \gamma_5 C ) \otimes \frac{1}{\sqrt{2}}( i \tau_2 ) \\
\chi^{(a)}_{\mu, i} &= \frac{1}{\sqrt{6}}
( i \gamma_\mu C ) \otimes \frac{1}{\sqrt{6}} (i \tau_i \tau_2)  
,\end{align} 
and we do not append propagators for the first and second quarks. 
For simplicity we choose the quark-diquark vertex functions to be point-like, namely
\begin{align}
\Gamma^{(s)} &= \openone \otimes \openone \\
\Gamma^{(a)}_{\nu, i} &= \gamma_5 \gamma_\nu  \otimes \tau_i
.\end{align}
This choice corresponds to modeling only a subset of the possible structures for the proton wavefunction, 
see~\cite{Oettel:1998bk,Oettel:2000ig} for a complete discussion. The vertex function also contains an overall 
color anti-symmetrization which we suppress throughout. The conjugate vertex $\ol \Gamma(k,P) = C \, \Gamma(-k,-P)^{\text{T}} C^\dag$.
One could modify the point-like vertex with a form factor as is commonly done, however, this generally violates the positivity bounds.

In this model, the axial diquark channel does not contribute to the proton's electromagnetic form factors. Thus we can only determine the 
parameters $m$ and $m_{SD}$ by fits to the Dirac and Pauli form factors. The parameter $m_{AD}$ could be tuned by fitting the quark 
distributions at some scale, however, we shall pursue a simpler course and set $m_{SD} = m_{AD} = m_D$. 
Alternately $m_{AD}$ could be tuned from neutron form factor data.

\begin{figure}
\begin{center}
\epsfig{file=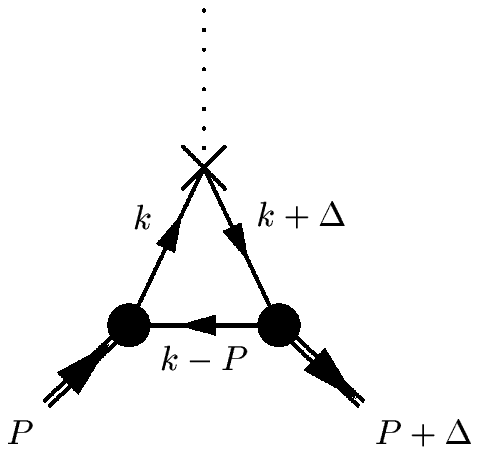}
\caption[Impulse approximation to the twist-two matrix elements.]{Impulse approximation 
to the twist-two matrix elements. Here the twist-two operators with momentum insertion are denoted
by a cross. The diquark spectator is depicted by a double line and the initial- (final-) state proton has momentum $P$ ($P'$).}
\label{ftwist}
\end{center}
\end{figure}

To derive the $F(\b,\a;t)$, $K(\b,\a;t)$ and $G(\b,\a;t)$ DDs, we consider the action of the twist-two operator $\mathcal{O}^{\mu \mu_1 \ldots \mu_n}$
between non-diagonal proton states. Working in the impulse approximation (see Figure \ref{ftwist}), 
we have the contribution in the scalar diquark channel, for example,
\begin{equation} \label{eqn:impulse}
\int  d^4 k 
\frac{\ol u_{\l'} (P') (\rlap\slash k + \rlap\slash \D + m) \Gamma^{\mu \mu_1 \ldots \mu_n} (\rlap\slash k + m ) u_\l(P)}
{[k^2 - m^2 + i \varepsilon]{}^a [(k+\D)^2 - m^2 + i \varepsilon]{}^a [(k - \ol P + \D/2)^2 - m_D^2 + i \varepsilon]}
,\end{equation}
where we have chosen an analytic regularization (so that $a >1$) respecting the positivity bounds.\footnote{%
Alternate schemes using Pauli-Villars subtractions often regulate such models
and are also attractive from the perspective of DDs since Lorentz covariance is maintained. These subtractions, however, generally
violate the bounds in Eq.~\eqref{eqn:positivity}. For example, in the NJL model of the pion with two subtractions \cite{Theussl:2002xp}, 
the positivity bounds, which were ignored by the authors, are violated for small values of $-t$. For the case of a quark-diquark model 
regularized via Pauli-Villars subtractions, the violations are more 
severe and persist for all values of $-t$ due to the mismatch of end-point and crossover behavior. This commonly encountered
problem is discussed in \cite{Tiburzi:2002kr}.
For these reasons, we have used the analytic regularization above, which we also employed previously for the pion \cite{Tiburzi:2002tq}
in Chapter \ref{chap:DDs}.
}
Since the NJL model is non-renormalizable, the choice of scheme is incorporated into the dynamics and hence
the choice of regularization should maintain desired properties. For phenomenological estimates of GPDs, the positivity
bounds are of crucial importance, and our regularization choice respects these bounds, see Appendix~\ref{sec:LC} for details.
One can view the regularization $a>1$ as mimicking the non-local, non-perturbative
structure of the twist-two, quark-antiquark vertex. 
This choice of smearing maintains current conservation but does not respect the Ward-Takahashi identity. Thus the normalization of 
amplitudes in Eq.~\eqref{eqn:vertex} is only approximately preserved. To remedy this feature, we add an $a$-dependent prefactor 
to the axial diquark contribution. This factor is then adjusted so that $N_d = 1$ and consequently the u-quark distribution is correctly 
normalized, $N_u = 2$. 
An additional drawback of the regularization scheme is that the induced quark form factors do not become point-like
for large momenta. While this is inconsistent with asymptotic freedom, the model is meant only for use in the low momentum
transfer region because the diquark substructure cannot be resolved.

The symmetric, traceless tensor $\Gamma$ in Eq.~\eqref{eqn:impulse} is
\begin{equation} \label{eqn:GAMMA}
\Gamma^{\mu \mu_1 \ldots \mu_n} = \gamma^{\{ \mu} (k + \D/2)^{\mu_1} \cdots (k + \D/2)^{\mu_n \}}
\end{equation}
The involved numerator structure complicates calculation of the symmetric traceless part of the tensor, however, 
this can be calculated directly without recourse to explicitly writing out such tensors of rank $n$. 
See Appendix~\ref{sec:DDDD}.
In order to compactly write out the DDs, we 
define the auxiliary functions
\begin{equation}
D_o(\b,\a;t) = \b m_D^2 + (1-\b) m^2 - \b (1-\b) M^2 + [(1-\b)^2 - \a^2] t / 4 
,\end{equation}
which is the typical energy denominator in both channels and
\begin{equation}
D(\b,\a;t) = 3 N \frac{ \Gamma(2 a - 1) }{ 2 (4)^{a - 1} \Gamma(a)^2 } 
\left[(1-\b)^2 - \a^2 \right]^{a -1} D_o(\b,\a;t)^{1-2a}
,\end{equation}
which is the typical prefactor for all DDs in this model.

Calculation of DDs for the scalar diquark yields
\begin{equation} \label{eqn:Fs}
\begin{pmatrix}
F^{(s)}_q(\b,\a;t) \\
K^{(s)}_q(\b,\a;t) \\
G^{(s)}_q(\b,\a;t)
\end{pmatrix}
= 
\delta_{q,u} D(\b,\a;t)
\begin{pmatrix}
(m + \b M )^2 +  \left[ (1-\b)^2 - \a^2 \right] \frac{t}{4} + \frac{D_o(\b,\a;t)}{2 a - 2} \\
2 M (1 - \b) ( m + \b M) \\
4 M  \a  ( m + \b M) 
\end{pmatrix}
.\end{equation}
In the axial-vector diquark channel, we find
\begin{multline}
\begin{pmatrix}
F^{(a)}_q(\b,\a;t) \\
K^{(a)}_q(\b,\a;t) \\
G^{(a)}_q(\b,\a;t)
\end{pmatrix}
= 
\frac{2}{9} Z(a) (2 \delta_{q,d} + \delta_{q,u})  D(\b,\a;t)
\\
\times 
\begin{pmatrix}
(2 m + \b M )^2 -  3 m^2 -  \left[ (1 + \b)^2 - \a^2 \right] \frac{t}{4} + \frac{D_o(\b,\a;t)}{2 a - 2}
\\
2 M  \b [ 2 m - (1 -  \b) M]
\\
- 4 M  \a  ( 2 m + \b M)
\end{pmatrix} 
,\end{multline}
where $Z(a)$ is the regularization dependent factor akin to wavefunction renormalization. As commented above, the value of $Z(a)$ is
chosen to preserve $N_d = 1$.

\section{Phenomenological applications} \label{sec:fit}

Our philosophy is to 
tune the parameters $m$ and $m_D$ so that proton electromagnetic 
form factor data at low momentum transfer are reproduced. This is particularly simple, 
since the axial diquark contribution cancels out of the
the proton form factor. We choose to fit to the charge radius and magnetic moment
and the experimental values we use are
$<r_E^2> = (0.870 \; \texttt{fm})^2$ and $\mu_p = 2.79 \, \mu_N$.
For the electromagnetic form factors of the proton, there is high precision data from JLAB~\cite{Jones:1999rz,Gayou:2001qd} 
and a recent global analysis and parametrization of~\cite{Arrington:2003qk}. The Sachs form factors are known 
experimentally to about $2\%$ accuracy in the small momentum transfer regime and are given by
\begin{align}
G_E(t) & = F_1(t) + \frac{t}{4 M^2} F_2(t) \\
G_M(t) & = F_1(t) + F_2(t)
.\end{align}

\begin{figure}
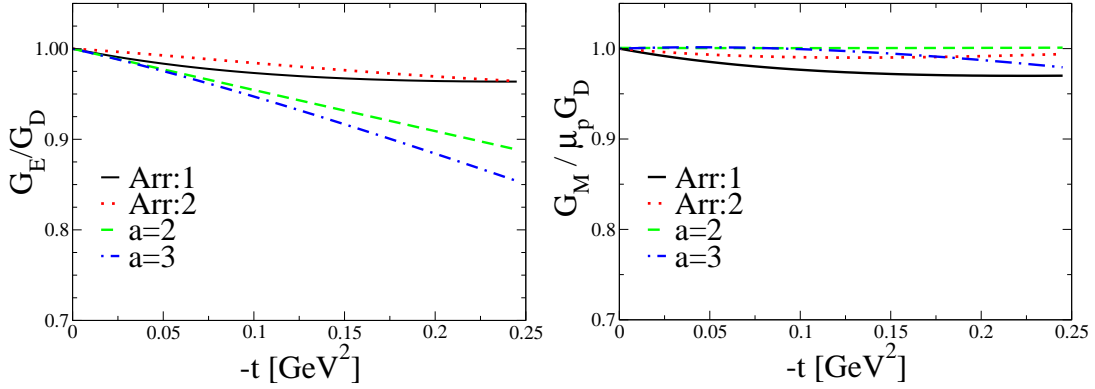

\begin{center}
\epsfig{file=GE.eps,height=2in}
\epsfig{file=GM.eps,height=2in}
\caption[Comparison of fits for $G_E(t)$ and $G_M(t)$ with empirical parameterizations.]{Comparison 
of fits for $G_E(t)$ and $G_M(t)$ with empirical parameterizations. 
The ratios $G_E(t)/G_D(t)$ and $G_M(t)/ \mu_p G_D(t)$ are plotted
vs.~$-t$ in $\texttt{GeV}^2$. The curves ``Arr:1'' and ``Arr:2'' correspond to the 
parameterizations of $G_E(t)$ given in Tables I and II of \cite{Arrington:2003qk}, respectively. The curves $a=2$ and $a=3$
are our model fits for different regularizations.}
\label{fGE}
\end{center}
\smallskip
\end{figure}

We treat the regulator parameter $a$ rather unconventionally; we use it as a means to explore
different covariant forms for the wavefunction.
For $a = 2$, magnetic moment and charge radius are reproduced $\lesssim 0.1 \%$ for $m = 0.437 \texttt{ GeV}$ 
and $m_D = 0.726 \texttt{ GeV}$. The Sachs electric and magnetic form factors match up $\lesssim 5 \%$
for $-t \lesssim 0.2 \texttt{ GeV}^2$.  
For $a = 3$, magnetic moment and charge radius are reproduced $\lesssim 0.1 \%$ for $m = 0.565 \texttt{ GeV}$ 
and $m_D = 0.825 \texttt{ GeV}$. For these parameters, the Sachs electric and magnetic form factors also match up $\lesssim 5 \%$
for $-t \lesssim 0.2 \texttt{ GeV}^2$. In the figures, we compare the phenomenological form factor fits to 
the two parameterizations of~\cite{Arrington:2003qk}. The fits for $G_E(t)$ and $G_M(t)$ are plotted in Fig.~\ref{fGE}.
As is standard, 
we plot ratios of electric and magnetic form factors to the empirical dipole form factor, namely
\begin{equation}
G_D(t) = \left(1 - \frac{t}{ M_D^2}\right)^{-2}
,\end{equation}
where the dipole mass squared is $M_D^2 = 0.71 \texttt{ GeV}^2$.

\begin{figure}
\begin{center}
\epsfig{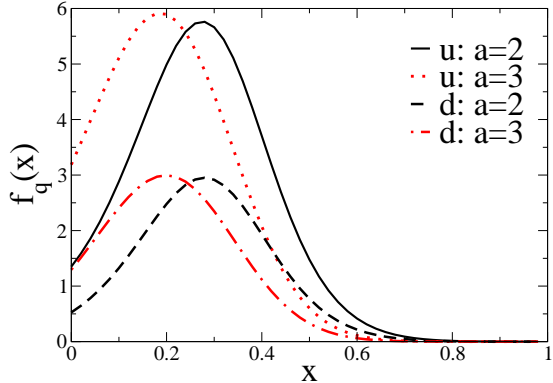}
\caption[Quark distributions for the proton model.]{Quark distributions for the proton model. The $u$- and $d$-quark distributions are plotted as a function 
of $x$ for the $a=2$ and $a=3$ fits to the form factors.}
\label{fquark}
\end{center}
\end{figure}

We can also determine the $u$ and $d$ quark distributions in our model. Since we do not have antiquarks,
the plus and minus distributions are identical $f^\pm_q(x) = f_q(x)$. In Fig.~\ref{fquark}, we
plot the $u$ and $d$ quark distributions as a function of $x$ for the $a=2$ and $a=3$ fits. 
The distributions are properly normalized 
so that $N_u = 2$ and $N_d = 1$. As commented above this normalization requires a relative $a$-dependent factor
for the axial-diquark contributions. Again this is required because the regularization scheme we choose 
does not preserve the Ward-Takahashi identity. Without the extra factor, the violation is $\sim 5$--$10 \%$.  
Notice the distributions do not vanish at the end-point $x = 0$. This is typical of NJL type calculations. The kernel is
independent of momentum and hence the wavefunction should be non-zero at both end-points. The fact that the distributions vanish at $x=1$
is due to our choice of regularization. Physically it is thus reasonable to think of the regularization as mimicking contributions 
from higher Fock states. Moreover, the model's scale is $a$-dependent, \emph{cf} Figure \ref{fquark}. 
\begin{figure}
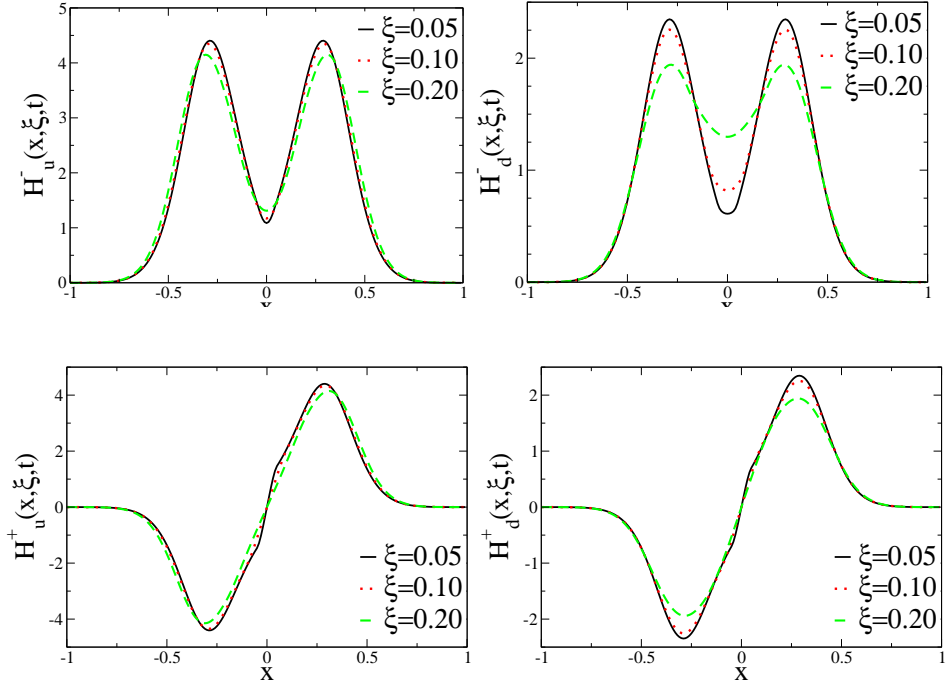

\begin{center}
\epsfig{file=hvalu.eps,height=1.65in}
\epsfig{file=hvald.eps,height=1.65in} \\
\bigskip 
\smallskip
\smallskip
\epsfig{file=hsingu.eps,height=1.65in}
\epsfig{file=hsingd.eps,height=1.65in}
\caption[$H$-GPDs for the proton model. ]{GPDs for the proton model. The $u$- and $d$-quark GPDs $H^\pm(x,\x,t)$ are plotted as a function 
of $x$ for the $a=2$ fit for a few values of $\x$ at $t = -0.1 \texttt{ GeV}^2$.}
\label{fHbare}
\end{center}
\end{figure}
\begin{figure}
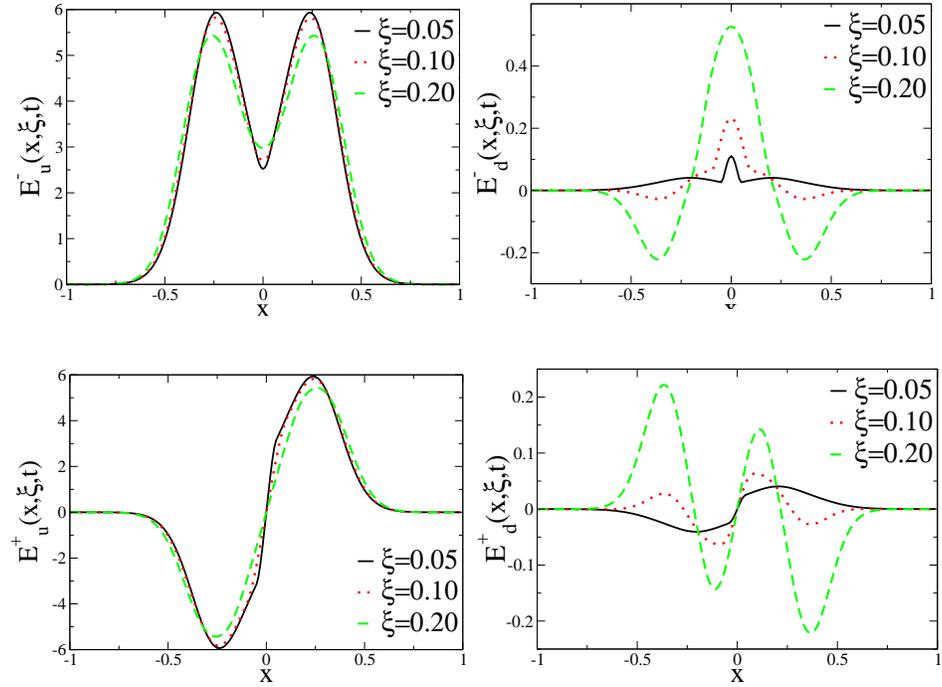

\begin{center}
\epsfig{file=evalu.eps,height=1.65in}
\epsfig{file=evald.eps,height=1.65in} \\
\bigskip 
\smallskip
\smallskip
\epsfig{file=esingu.eps,height=1.65in}
\epsfig{file=esingd.eps,height=1.65in}
\caption[$E$-GPDs for the proton model.]{GPDs for the proton model. 
The $u$- and $d$-quark GPDs $E^\pm(x,\x,t)$ are plotted as a function 
of $x$ for the $a=2$ fit for a few values of $\x$ at $t = -0.1 \texttt{ GeV}^2$.}
\label{fEbare}
\end{center}
\end{figure}
Additionally in Figs.~\ref{fHbare} and \ref{fEbare}, we have plotted the $H^\pm(x,\x,t)$ and $E^\pm(x,\x,t)$ GPDs. 
The figures show the GPDs at fixed $-t = 0.1 \texttt{ GeV}^2$ for a few values of $\x$. The GPDs plotted are for the $a=2$
fit; the distributions are qualitatively similar for the $a=3$ model parameters.

The distributions plotted in Figs.~\ref{fquark}--\ref{fEbare} are presumably at some low hadronic scale intrinsic to the model. 
One way to determine the scales of the $a=2$ and $a=3$ models is to use the evolution equations to evolve empirical parameterizations
down to a scale where the first few moments of our model distributions agree. This procedure is not unique: many models can reproduce
the empirical quark distributions at higher scales. Also the use of perturbative evolution is questionable at low scales. 
While the evolution kernels for GPDs are known at next-to-leading order \cite{Belitsky:1998gc,Belitsky:1998uk,Belitsky:1999gu,Belitsky:1999fu,Belitsky:1999hf}, 
perturbative evolution cannot generate the non-perturbative small-$x$ physics which our model lacks
and the small-$x$ physics is crucial for relating to DVCS data. The the leading-twist DVCS amplitude is a convolution of GPDs with a hard scattering kernel 
that emphasizes regions where the final-state wavefunction is evaluated near the end-point. In fact, the imaginary part of the amplitude
is directly proportional to an overlap of light-cone wavefunctions, where the final state is at zero $x$.

An attractive alternative to using perturbative evolution to define the scale of the model is to 
implant a realistic quark distribution at some scale $\mu$ via factorization of the DD \cite{Mukherjee:2002gb}. 
With a proper choice of scale, perturbative evolution can then be used from $\mu$ up to the experimental scale. 
This seems necessary for the parametrization to be most useful for experimental comparison. Ultimately the experiment
must guide phenomenology in constructing parameterizations that are general enough to account for the observed correlations 
among GPD variables. While the DD certainly does not factorize, the \emph{Ansatz} in \cite{Mukherjee:2002gb} avoids
factorizing the variable dependence of the GPDs. Further phenomenological consideration of this quark-diquark model 
is currently underway and will be presented elsewhere.

\chapter{Conclusions}

Above we have presented a comprehensive but more-or-less pedagogical treatment of modeling GPDs.  
In the course of attempting to model these distributions, we came across subtle issues 
about the light-front BSE and its reduction scheme. Bent on applications to GPDs, we 
reviewed our work concerning current in the reduction formalism in Chapter \ref{chap:LFBS}.  
There we proved a replacement theorem for apparent non-wavefunction type contributions. 
The resolution of this issue was crucial to enable proper calculation of GPDs in light-front perturbation 
theory.

In Chapter \ref{chap:GPDs} we detailed our calculation of GPDs in the light-front
Wick-Cutkosky model. This calculation provided a non-trivial example of the light-cone 
Fock space representation of DVCS, in which form the GPDs' positivity bounds are manifest. 
The non-vanishing of GPDs at the crossover between kinematic 
regions was tied to higher Fock components. Furthermore we saw how the features of 
light-cone time-ordered perturbation theory entail the continuity of GPDs. 
Continuity of GPDs ensures the QCD factorization theorem for DVCS. The proton case is likely 
to proceed along the lines outlined for the Wick-Cutkosky model, however, 
incorporation of spin degrees of freedom is difficult due to divergences.

Next in Chapter \ref{chap:DDs}, we considered an alternate approach to GPDs based on covariant DDs. 
This approach allowed us considerable insight as to the Lorentz covariance properties required 
of GPDs. Unfortunately in this formalism the positivity bounds are not transparent but can be used 
as a guide in model building or calculation of DDs. Using simple
models, we not only compared covariant and light-front calculations of GPDs, we also 
corrected the inconsistencies of the method proposed in \cite{Mukherjee:2002gb} for the 
calculation of DDs. Understanding the DD formalism is pertinent because \emph{ad hoc} 
forms for DDs are almost exclusively used to model GPDs. 
We have provided complete calculations of DDs for several models of scalar bound states.  
Such calculations fill a void in the literature.

Finally in Chapter \ref{chap:proton}, we used a simple two-body model to calculate DDs for the proton. 
This is the first such study. The model consists of two quarks
strongly coupled in the scalar and axial diquark channels along with a residual quark. The simplicity 
of such a model allows for analytic computation of the DD functions which contain appropriate spin 
structure. The simplicity also allows for us to make contact with the light-cone wavefunctions and the 
equivalent overlap representation of GPDs. This toy model study thus enables a comparison between 
the physical intuition of the light-cone Fock space representation of GPDs and the manifestly Lorentz invariant
decomposition of DD functions.

We were careful to choose a regularization scheme that satisfies both Lorentz invariance as well as
the positivity bounds required of GPDs. Our model, although toy-like in nature, can be used to match 
the electromagnetic form factors of the proton and can either be augmented with realistic valence quark distributions
or be evolved to higher scales.

The scope of our investigation is two fold. One direction is to calculate as many DDs in simple scenarios
as possible. This will give modelers a better sense of the form and behavior of DDs and may assist with empirical 
parameterizations of GPDs. In another nearly orthogonal direction, one can improve upon the proton model used
here in order to see how various features of a realistic proton wavefunction manifest themselves in processes like DVCS. 
Further extension of the model presented in Chapter \ref{chap:proton} and its application to 
future data are under investigation.

%
\nocite{*}   
\bibliographystyle{jhep}
\bibliography{uwthesis}
%
%
\appendix
\raggedbottom\sloppy
 
 
\chapter[Conventional Notations and Notational Conventions]{Conventional Notations and\\ Notational Conventions} \label{chap:notations}

For a general Lorentz vector $a^\mu = (a^0, a^1, a^2, a^3)$, we define the light-cone
variables $a^\pm$ as the linear combinations
\begin{equation}
a^\pm = \frac{1}{\sqrt{2}} (a^0 \pm a^3)
.\end{equation}
Now consider the scalar product of two Lorentz $4$-vectors, $a^\mu$ and $b^\mu$. 
\begin{align}
a \cdot b & = a^0 \, b^0 - a^1 \, b^1 - a^2 \,b^2 - a^3 \, b^3 \notag \\
          & = a^+ b^- - a^1 \, b^1 - a^2 \, b^2 + a^- b^+ 
.\end{align}
This implicitly defines the metric tensor $g_{\mu \nu}$, namely 
$g_{+-} = g_{-+} = 1$, $g_{11} = g_{22} = -1$ and all unspecified entries vanish.
Notice in our conventions $g_{\mu \nu} = g^{\mu \nu}$. For convenience we 
shall also refer to transverse two-vectors, namely $\mathbf{a}^\perp = (a^1, a^2)$.
For an on-shell particle of mass $m$, the dispersion relation $k^2 = m^2$
takes on a form devoid of the square root
\begin{equation}
k^- = \frac{\kperp^2 + m^2}{2 k^+}
.\end{equation}

For Dirac gamma matrices, when explicit forms are needed we use the Bjorken 
and Drell convention:
\begin{equation}
\gamma^0 = 
\begin{pmatrix}
1 & 0 & 0 & 0 \\
0 & 1 & 0 & 0 \\
0 & 0 & -1 & 0 \\
0 & 0 & 0 & -1 
\end{pmatrix}
\end{equation}
and
\begin{equation}
\gamma^i = 
\begin{pmatrix}
0 & \sigma^i \\
- \sigma^i & 0 
\end{pmatrix}
,\end{equation}
where $i = 1,2,3$ and the Pauli matrices $\sigma^i$ are given by
\begin{equation}
\sigma^1  = 
\begin{pmatrix}
0 & 1 \\ 
1 & 0 
\end{pmatrix}
,\end{equation}
\begin{equation}
\sigma^2  = 
\begin{pmatrix}
0 & -i \\ 
i & 0 
\end{pmatrix}
,\end{equation}
and
\begin{equation}
\sigma^3  = 
\begin{pmatrix}
1 & 0 \\ 
0 & -1 
\end{pmatrix}
.\end{equation}

One can similarly form plus and minus gamma matrices in the analogous way
\begin{equation}
\gamma^\pm = \frac{1}{\sqrt{2}} ( \gamma^0 \pm \gamma^3 )
.\end{equation}
Also useful when considering fermions are the projection operators $\Lambda_\pm$
that are defined by
\begin{equation}
\Lambda_\pm = \frac{1}{2} \gamma^\mp \gamma^\pm
.\end{equation}
Notice $(\gamma^+)^2 = (\gamma^-)^2 = 0$ and the projector is simple to express 
in terms of $\gamma^0$, namely $\Lambda_\pm = \frac{1}{\sqrt{2}} \gamma^0 \gamma^\pm$.

\chapter{Instant and Front Form Dynamics} \label{chap:forms}

In this Appendix we review the features of relativistic dynamics and the Poincar\'e algebra. 
Above in Section \ref{toyboost}, we used a toy model to illustrate the characteristics of 
both the instant and front form of dynamics. Here we give a more theoretical description
of the differences between the two forms and thereby define what it means to specify a form of dynamics.

A relativistic system is characterized by the ten operators of the Poincar\'e group. Full relativistic invariance
stems from the algebra formed from the Poincar\'e generators. The generators of the Poincar\'e group 
consist of the four momenta $P^\mu$, and the six operators embedded in the antisymmetric tensor $M^{\mu \nu}$. 
The latter are the angular momenta, specified by
\begin{equation}
L^i = \frac{1}{2} \epsilon^{ijk} M^{jk}
;\end{equation}
and the Lorentz boosts, specified by
\begin{equation}
K^i = M^{0i}
,\end{equation}
where the indices $i$, $j$, and $k$ run from $1$ to $3$.

These ten generators satisfy the Poincar\'e algebra, namely
\begin{eqnarray}
i [P^\mu,P^\nu] &=& 0, \\
i [M^{\mu \nu},P^\rho] &=& g^{\nu \rho} P^\mu - g^{\mu \rho} P^\nu, \\
i [M^{\mu \nu}, M^{\rho, \sigma}] &=& g^{\mu \sigma} M^{\nu \rho} - g^{\mu \rho} M^{\nu \sigma} - g^{\nu \sigma} M^{\mu \rho} + g^{\nu \rho} M^{\mu \sigma}
.\end{eqnarray}
This algebra enforces that the $P^\mu$ are independent and transform as a four-vector under the Lorentz transformation, 
while $M^{\mu \nu}$ must transform as an antisymmetric tensor. 
To find a representation of the Poincar\'e algebra, one constructs the generators from dynamical variables: positions, momenta, 
spins, \emph{etc}.

To make our discussion concrete, we restrict our attention to the case of a single free scalar particle.
The dynamical variables at our disposal are the position and momentum four-vectors $x^\mu$ and $p^\mu$
that obey the quantum mechanical commutator
\begin{equation}
i [x^\mu, p^\mu] = g^{\mu \nu}
.\end{equation} 
This commutation relation will furnish a representation of the Poincar\'e algebra provided that
we construct the generators as
\begin{align} \label{eqn:representation}
P^\mu &= p^\mu \\
M^{\mu \nu} & = x^\mu p^\nu - x^\nu p^\mu
.\end{align}
We also must impose the relativistic constraint $p^2 = m^2$ to have Lorentz covariance.

In order to specify the form of dynamics of the system, one needs to create a foliation of space-time into space and 
time.  This is done by choosing the initial surface $\Sigma$ on which the general time variable $\tau$ is zero. 
Choices for $\Sigma$ are restricted by the condition that all possible world lines may intersect $\Sigma$ exactly once.
Clearly this is true also in classical mechanics where Galilean time $t$ is completely uncoupled from three-dimensional 
Euclidean space and thus the instant $t = 0$ specifies a unique point on the space-time trajectory of all particles. 
As we know from relativity, the world lines of particles are restricted to lie on or within the fore and aft light cones. 
Because relativity restricts the class of possible world lines, it enlarges the possible initial surfaces $\Sigma$ and hence
broadens our ability to define ``time'' in relativistic dynamics.

Once we have specified the initial surface $\Sigma$, we have also to construct the normal $N^\mu$.  
The significance of the normal is that $N \cdot P$ serves as the Hamiltonian in the sense that
it evolves the system away from the initial surface on parallel slices in $\tau$. Thus in close analogy with 
non-relativistic quantum mechanics, the Hamiltonian is dynamical.

Now having specified the dynamics of the system, we imagine the initial conditions (wavefunctions, \emph{etc})
are defined on the initial surface $\Sigma$. The Poincar\'e generators now fall into two categories: kinematical
operators and dynamical operators.  The kinematical operators have the property of mapping $\Sigma \to \Sigma$. 
They are called kinematical because the behavior of states under these operators does not require dynamical input
since $\tau$ remains zero. On the other hand, operators that do not act transitively on $\Sigma$ are dynamical, i.e., 
the effects of these operators on states require knowledge of the system away from $\Sigma$.

To give a familiar example of a form of dynamics, we consider first the instant form. The initial surface
is specified by the conventional instant of time $\Sigma: x^0 =0$. The conjugate to time is the energy
\begin{equation}
p^0 = N \cdot p = \sqrt{\mathbf{p}^2 + m^2}
.\end{equation}
To find out which operators are kinematical and dynamical under this foliation, we use the representation
in Eq.~\eqref{eqn:representation} evaluated on the surface $\Sigma$.  
The Hamiltonian, which for the instant form is $P^0 = p^0$, is automatically dynamical in any system. 
Its appearance in the other Poincar\'e generators
indicates that they too are dynamical. Thus we find six kinematic operators
\begin{align}
P^i &= p^i \\
M^{ij} &= x^i P^j - x^j P^i
\end{align}
and additionally three dynamical ones
\begin{equation}
M^{i0} = x^i P^0
.\end{equation}
This situation of instant form dynamics could be anticipated directly from special relativity. We know that spatial translations
and rotations leave the surface $x^0 = 0$ unchanged. Hence the generators of these transformations must be kinematical. 
On the other hand, boosts in any direction mix the coordinate of that direction with $x^0$, hence dynamics are required in order 
to get the system off the surface $\Sigma$.

\begin{figure} 
   \begin{center} 
   \epsfig{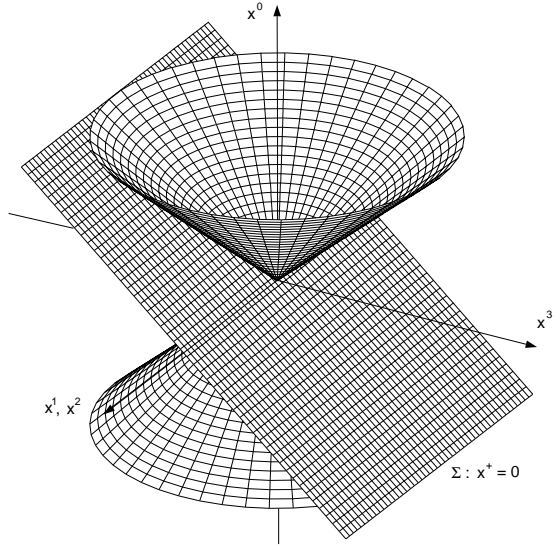}
   \end{center}    
\caption[The hypersurface that defines the front form of dynamics.]{\label{f:front} The hypersurface 
	$\Sigma : x^+ =0$ that defines the front form of dynamics. 
	The surface $\Sigma$ is the plane tangential to the
        light-cone, $x^2 =0$. Figure adapted from \cite{Heinzl:1998kz}}  
\end{figure}

Another example of a form of dynamics is the front form. In this form, the initial surface is specified by a 
plane tangential to the light-cone $\Sigma: x^+ = 0$, see Figure \ref{f:front}. The normal to this surface 
is given by the lightlike vector
\begin{equation}
N^\mu = \frac{1}{\sqrt{2}} (1, 0, 0, -1)
\end{equation}
and hence the Hamiltonian is given by
\begin{equation}
N \cdot P = P^- = p^-
.\end{equation}
The kinematic operators in the front form are 
the transverse and longitudinal momenta
\begin{equation}
P^\perp =p^\perp \;, \quad P^+ = p^+
,\end{equation}
an angular momentum
\begin{equation}
M^{12} = x^1 P^2 - x^2 P^1
,\end{equation}
and the light-front boosts
\begin{equation}
M^{+\perp} = - x^\perp P^+, \; \quad M^{+-} = x^- P^+
.\end{equation}
There are thus only two additional dynamical operators given by
\begin{equation}
M^{-\perp} = x^- P^\perp - x^i P^-
.\end{equation}

The physics of the partition into kinematic and dynamic classes of generators is clear. 
The two dynamical operators involve rotations about $x^1$ and $x^2$ axes. These rotations
change $\Sigma$ and hence require dynamics. 
There are three independent translations that leave the surface $x^+ = 0$ alone
and hence the three kinematical operators $P^\perp$ and $P^+$. Rotations within
the surface $\Sigma$ that leave $x^+=0$ unchanged are generated by the angular 
momentum $M^{12}$. There are two special combinations of rotations and boosts 
(from the instant form mindset) that make up the kinematical light-front boosts $M^{+\perp}$.
Finally there is the third kinematical boost $M^{-+} = -K^3$ that is identical (up to sign) to the instant form boost 
in the $x^3$ direction.  Under such a boost of rapidity $\eta$ the coordinates get mapped as follows
\begin{equation}
\begin{pmatrix}
x^0 \\
x^1 \\
x^2 \\
x^3
\end{pmatrix}
\longrightarrow
\begin{pmatrix}
x^0 \cosh \eta + x^3 \sinh \eta \\
x^1 \\
x^2 \\
x^3 \sinh \eta + x^0 \cosh \eta
\end{pmatrix}
.\end{equation}
Thus the light-cone time $x^+ \to e^\eta x^+$ is merely rescaled under this boost. The dynamics are diagonal with 
respect to $K^3$ boosts. Notice, consequently, the initial surface $\Sigma$ is indeed invariant under
the action of $M^{-+}$ since $x^+ = 0$ on $\Sigma$.

Of course our discussion of the forms of dynamics above was limited to the case of a free scalar particle. 
Inclusion of dynamics leads to interaction 
terms in the Poincar\'e generators. Finding such terms that are still consistent with the Poincar\'e algebra 
is the real struggle of constructing a relativistic dynamical system \cite{Dirac:1949cp}. Indeed 
Poincar\'e invariance alone is not enough to guarantee relativistic
covariance. Despite this, a few special cases that do maintain full relativistic covariance 
have been arduously worked out, see e.g. \cite{Leutwyler:1978vy}. 
This issue at play here is how to impose relativistic covariance on quantum mechanical, fixed particle number 
systems with interaction potentials.
At heart these two are incompatible: action at a distance and particle conservation have stepped aside in favor 
relativistic field theory. Nonetheless, the discussion in this Appendix motivates the use of front form dynamics
and the light-cone quantization of field theories.

\chapter{Light-Cone Quantization} \label{chap:LCQ}

In Section \ref{DIS}, we derived an expression for the DIS structure function $f_1(x)$. Here we elucidate the 
physics of this distribution function using light-cone quantization in the light-cone gauge. 
For reference the leading-twist function $f_1(x)$ encountered in DIS is defined as
\begin{equation} 
f_1(x) = \frac{1}{2} \int \frac{d \l}{2 \pi} e^{ i \l x} \langle P | \ol \psi(0) \gamma^+ \psi(\l n) | P \rangle
.\end{equation}
Notice that the correlator is not an equal-time Green's function since the quark field is annihilated 
at $x^0  =  \l / \sqrt{2} \Lambda$ and created at $x^0 = 0$. During this time, fluctuations of the
residual proton state occur that deny the matrix element a probabilistic interpretation. 
On the other hand, both quark fields are evaluated at $x^+ = 0$ which tells us the function $f_1(x)$ is an
equal light-cone time correlator. To see the density interpretation of $f_1(x)$ and hence justify the name PDF, 
we must accordingly use light-cone quantization.

On the light cone, spinor fields have components that are dynamically constrained. To separate out 
the dependent fields we write
\begin{equation}
\psi = \psi_+ + \psi_-
,\end{equation}
where $\psi_+$ is the good component of the spinor field, and $\psi_-$ is the bad component that 
is dynamically determined from the QCD equations of motion. Explicitly the good and bad
components are defined from the projectors (given in Appendix \ref{chap:notations}), namely
\begin{equation}
\psi_\pm = \Lambda_\pm \psi
.\end{equation}

The dynamically independent good fields $\psi_+$ are then quantized at equal light-cone time
\begin{equation}
\Big\{ 
\psi_{+\a}^\dagger (x),
\psi_{+\b}(y)
\Big\} \Bigg|_{x^+ = y^+}
=
\frac{1}{\sqrt{2}} (\Lambda_+)_{\a \b} \delta(x^- - y^-) \delta(\mathbf{x}^\perp - \mathbf{y}^\perp)
,\end{equation}
with all other possible anticommutators (among the good fields) vanishing. This anticommutation
relation is naturally solved by a Fourier decomposition in terms of plane-wave modes
\begin{multline} \label{eqn:mode}
\psi_+(x^-,\mathbf{x}^\perp) = \int \frac{dk^+ d\mathbf{k}^\perp}{2 \kplus (2\pi)^3} \theta(\kplus)
\sum_{\l = \pm} 
\left[ 
b_\l(\kplus,\mathbf{k}^\perp) u_{+\l}(k) e^{-i(x^- k^+ - \mathbf{x}^\perp \cdot \mathbf{k}^\perp)} \right.
\\ + \left.
d_\l^\dagger(\kplus,\mathbf{k}^\perp) v_{+\l}(k) e^{i(x^- k^+ - \mathbf{x}^\perp \cdot \mathbf{k}^\perp)}
\right]
.\end{multline}
The above plane-wave decomposition is merely a projection of the four-dimensional solutions to the Dirac equation 
onto the hypersurface $x^+ = 0$. The spinors in the above decomposition are good projections of the light-cone spinors in Appendix 
\ref{sec:LC}, namely $u_{+\l}(k) = \Lambda_+ u_\l(k)$ and similarly for $v_{+\l}(k)$.

The quark mode operators $b$ and $b^\dagger$ must satisfy the anticommutation relation
\begin{equation}
\Big\{
b_\l^\dagger(k^+,\kperp),
b_{\l'}(k'^+, \mathbf{k'}^\perp) 
\Big\}
=
2 k^+ \delta_{\l,\l'} \delta(k^+ - k'^+) \delta(\kperp - \mathbf{k'}^\perp)
,\end{equation}
and analogous relations hold for the antiquark creation and annihilation operators
$d^\dagger$ and $d$. Anticommutators between all remaining pairs of quark and antiquark 
mode operators vanish.

Using the relations in Appendix \ref{chap:notations}, we can convert the  
bilocal operator in Eq.~\eqref{eqn:f1} to a more useful form in terms of 
good fields
\begin{equation}
\ol \psi (0) \gamma^+ \psi(\l n) = \sqrt{2} \psi_+^\dagger(0) \psi_+(\l n)
.\end{equation}
Thus inserting the mode expansion Eq.~\eqref{eqn:mode} for the fields
above we arrive at
\begin{equation}
f_1(x) =  \frac{1}{4 \pi x} \sum_\l \int \frac{d\mathbf{k}^\perp}{(2\pi)^2}
\begin{cases}
\langle P | b_\l^\dagger( x P^+,\kperp)
b_{\l}(x P^+, \mathbf{k}^\perp) | P \rangle , \quad 0 < x < 1 \\
 \langle P | d_\l^\dagger( x P^+,\kperp)
d_{\l}(x P^+, \mathbf{k}^\perp) | P \rangle , \quad -1 < x < 0
\end{cases}
.\end{equation}
The above expressions show that $f_1(x)$ is determined from quark and antiquark density 
operators and hence has a probabilistic interpretation as the name PDF suggests. 
One can go further and expand the proton state in the light-cone Fock basis
to generate contributions to the PDF in the form of an infinite sum of squares of light-cone Fock
state wavefunctions.

We can attempt the same procedure for the higher-twist structure function $f_4(x)$
defined implicitly through Eq.~\eqref{eqn:1and4}
\begin{equation}
f_4(x) = \frac{1}{2} \int \frac{d \l}{2 \pi} e^{ i \l x} \langle P | \ol \psi(0) \gamma^- \psi(\l n) | P \rangle
.\end{equation}
The quark bilinear operator, however, only has bad components of the spinor field
\begin{equation}
\ol \psi(0) \gamma^- \psi(\l n) = \sqrt{2} \psi_-^\dagger(0) \psi_-(\l n)
.\end{equation}
These components are dynamically determined from the equations of motion for the $\psi_+$ field, namely
\begin{equation}
i \frac{\partial}{\partial \l} \psi_-(\l n) = \frac{1}{2} \gamma^+ ( - i \rlap \slash \mathbf{D}^\perp ) \psi_+(\l n)
.\end{equation}
Clearly there is no partonic density interpretation for the $f_4(x)$ structure function.
The partial derivative term is kinematically of higher twist.
The gluonic part of the covariant derivative leads to dynamical effects. Intermediate state gluons generally lead to more propagators
compared to the kinematic terms and usually more suppression.

A final point with respect to intermediate gluons must be made. One can imagine including longitudinal gluon scattering 
into the handbag diagram depicted in Figure \ref{f:hand}. Inclusion of these extra propagators does not necessarily lead 
to power-law suppression since large momentum flow can still be accommodated exclusively in the handle of the bag. 
These multiple gluon exchanges can thus be of the same order as the leading twist contribution. 
Summed to all orders these longitudinal gluon scatterings produce the gauge link
\begin{equation}
\ol \psi(0) \psi(\l n) \to \ol \psi(0) \mathcal{P}\left( e^{-i g \int_\l^0 n \cdot A d \l' } \right) \psi(\l n)
\end{equation}
which is necessary for the combination of bilocal fields to be gauge invariant in non-Abelian field theory. 
Above we have tacitly assumed light-cone gauge, $n \cdot A = 0$, in which longitudinal gluon scattering is 
a gauge artifact. There is more subtlety in this choice than we have presented here \cite{Ji:2002aa,Belitsky:2002sm}.

\begin{figure}
\begin{center}
\epsfig{file=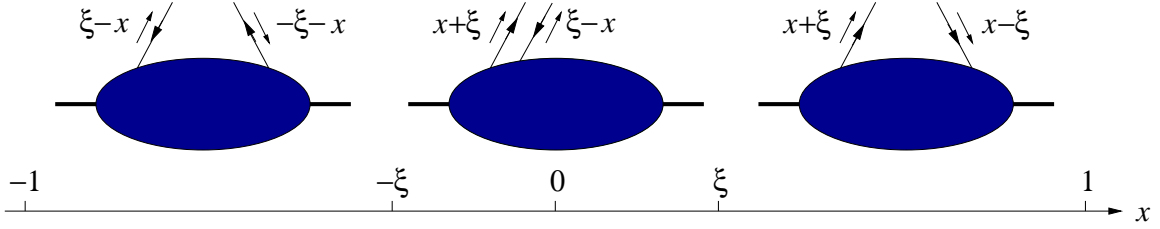,width=6in}
\end{center}
\caption[Diagrammatic depiction of the kinematic regions contributing to the GPDs.]{Diagrammatic depiction of the kinematic regions contributing to the GPDs. 
Figure adapted from \cite{Diehl:2003ny}.}
\label{f:regions}
\end{figure}

In summary we have seen the leading-twist PDF $f_1(x)$ describes the density of partons in 
the infinite momentum frame and in light-cone gauge. One can go further and ask for the decomposition
of GPDs in terms of the infinite momentum frame partons in light-cone gauge. Starting from the 
matrix element definition
\begin{equation}
F(x,\x,t) = \frac{1}{2} \int \frac{d \l}{2 \pi} e^{i \l x} 
\langle P' |
\ol \psi(-\l n /2) \gamma^+ \psi( \l n / 2)
| P \rangle
,\end{equation}
we again insert the mode expansion Eq.~\eqref{eqn:mode} for the good fields and 
arrive at
\begin{multline}
F(x,\x,t) = \frac{1}{4 \pi \sqrt{x^2 - \x^2}} \sum_\l \int \frac{d \kperp}{(2\pi)^2}
\\ \times
\begin{cases}
\langle P' | b_\l^\dagger [ (x-\x) \ol P^+,\kperp + \dperp]
b_{\l}[(x+\x) \ol P^+, \mathbf{k}^\perp] | P \rangle , \quad \quad \quad \quad \x < x < 1 \\
\langle P' | d_\l [ (-x+\x) \ol P^+,-\kperp - \dperp]
b_{-\l}[(x+\x) \ol P^+, \mathbf{k}^\perp] | P \rangle , \quad -\x < x < \x \\
\langle P' | d_\l^\dagger [ (-x-\x) \ol P^+,\kperp + \dperp]
d_{\l}[(-x+\x) \ol P^+, \mathbf{k}^\perp] | P \rangle , \quad -1 < x < -x
\end{cases}
.\end{multline}
Contributions to $F(x,\x,t)$ are shown as partonic cartoons in Figure \ref{f:regions}
For $x > \x$, one has the amplitude for removing a quark of a certain momentum
fraction from the initial state and then inserting it with altered momentum into 
the final state. The same interpretation is true for anitquarks in the region $x < -\x$. 
When $-\x < x < \x$, one has the amplitude for removing a quark-antiquark pair from 
the initial state. 
Notice as $\x \to 0$, one recovers a density of partons with momentum fraction $x$.  
Similar to the case of PDFs, one can expand the proton states in the light-cone Fock 
basis. Doing so, one derives the overlap representation of GPDs \cite{Diehl:2000xz,Brodsky:2000xy}.

\chapter{Wavefunctions and Form Factors in $(3+1)$ Dimensions} \label{fff}

In this Appendix we collect results relevant for Chapter \ref{chap:GPDs} 
for wavefunctions and form factors in the $(3+1)$-dimensional ladder model Eq.~\eqref{ladder}.

\section{Wavefunctions}
Using the Bethe-Salpeter equation \eqref{BS} and the definition of the light-cone wavefunction $|\psi_R\rangle$ Eq.~\eqref{psi}
we have the light-cone bound-state equation 
\begin{equation} \label{bse}
|\psi_{R}\rangle = g(R) w(R) |\psi_R \rangle,
\end{equation}
where $w(R)$ is the reduced auxiliary kernel
\begin{equation}
w(R) = g^{-1}(R) \Big| G(R) W(R) G(R) \Big| g^{-1}(R),
\end{equation}
with $W(R)$ defined in Eq.~\eqref{W}.

\begin{figure}
\begin{center}
\epsfig{file=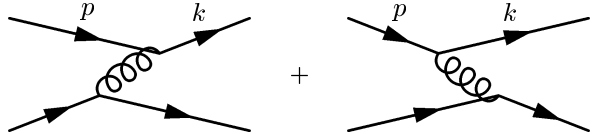}
\end{center}
\caption[Diagrammatic representation of the one-boson exchange potential.]{Diagrammatic representation of the one-boson exchange potential $V$ appearing in Eq. \eqref{OBE}.}
\label{fOBEP}
\end{figure}

To leading order in $G - \Gt$, one calculates $w(R)$ for the ladder model to be:
\begin{multline} \label{OBE}
V(x,\kperp;y,\pperp|M^2) \equiv - \langle \; xR^+,\kperp | \; w(R) \; | \; yR^+,\pperp \rangle \\
			 = \frac{g^2}{x-y}  \Big[ \theta(x-y) D(x,\kperp;y,\pperp|M^2) \\
- \big\{(x,\kperp) \longleftrightarrow (y, \pperp) \big\}    \Big] \theta[x(1-x)] \theta[y(1-y)],
\end{multline}
where we have defined 
\begin{equation}
D^{-1}(x,\kperp;y,\pperp|M^2) = 
M^2 - \frac{\pperp^2 + m^2}{y} 
- \frac{(\kperp - \pperp)^2 + \mu^2}{x - y} 
- \frac{\kperp^2 + m^2}{1-x},
\end{equation}
and taken $\mathbf{R}^\perp = 0$.  Graphically this one-boson exchange potential Eq.~\eqref{OBE} is depicted in Figure \ref{fOBEP}.

For reference, the bound-state equation \eqref{bse} appears as
\begin{equation} \label{wavefunction}
\psi(x,\kperp) = \dw(x,\kperp|M^2) \int \frac{dy d\pperp}{2(2\pi)^3 y (1-y)} 
V(x,\kperp;y,\pperp|M^2) \psi(y,\pperp)
\end{equation}
A few comments about these results are in order. Truncation of the series in $G - \Gt$ violates Lorentz symmetry
because $\Gt$ is intrinsically related to projection onto the light cone. Although the covariant one boson exchange
generates at leading order the two light-cone time-ordered one boson exchanges depicted in Figure \ref{fOBEP}, there are
infinitely many terms in the covariant ladder which are missing. To make this clear, we write out the second-order kernel
in the reduction scheme.

\begin{figure}
\begin{center}
\epsfig{file=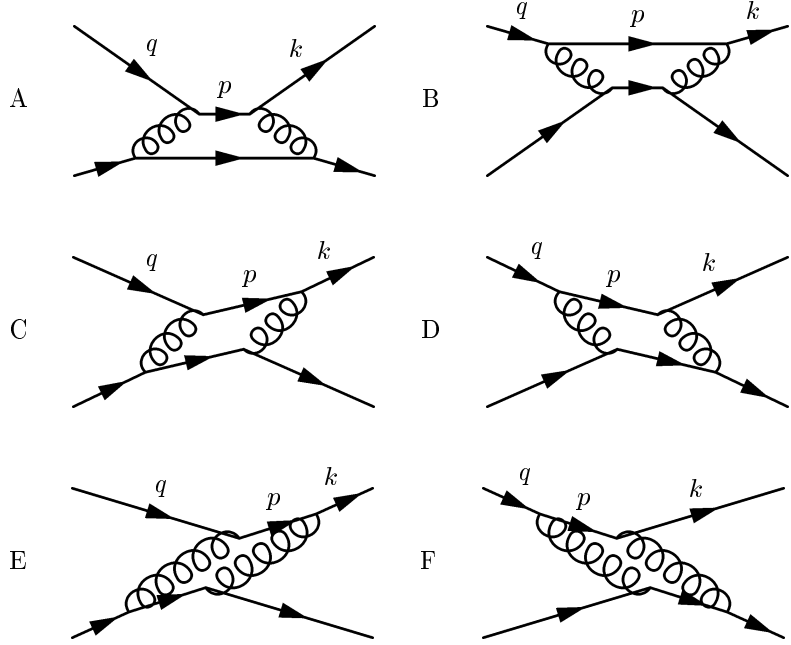}
\end{center}
\caption[Light-front time-ordered diagrams contributing to the second-order kernel.]{Light-front time-ordered 
diagrams contributing to the second-order kernel. Notice that diagrams A-D can be summed into the square of the 
first-order interaction and are consequently removed by the $\Gt$ subtraction term. The remaining diagrams E and F are the first of infinitely 
many required to maintain covariance.}
\label{fv2}
\end{figure}

At second order, the reduced auxiliary kernel reads
\begin{equation} \label{eqn:w2}
w^{(2)}(R) =  g^{-1}(R) \Big| G(R) V(R) \left[G(R) - \Gt(R)\right] V(R)  G(R) \Big| g^{-1}(R)
.\end{equation}
For the moment, we ignore the subtraction term, insert the effective resolution of unity
and evaluate the expression using Cauchy's residue theorem. The horrendous details are 
contained in \cite{Tiburzi:2002mn}. Contributions to $w^{(2)}$ are depicted as light-cone
time-ordered diagrams in Figure \ref{fv2}. 
Notice that the diagrams A-D can be summed into the square of the
first-order interaction. In this sense they are reducible contributions and are subtracted off by the 
$\Gt(R)$ term in Eq.~\eqref{eqn:w2}.  The remaining two diagrams (E and F) are characterized by four quanta propagating at a given 
instant of time. These contributions are irreducible and represent the
first of a string of infinitely many interactions omitted from the covariant ladder. 
To second order the bound state equation reads
\begin{equation}
\psi(x,\kperp) = \dw(x,\kperp|M^2) \int \frac{dy \; d\pperp}{2 (2\pi)^3 y(1-y)} \frac{dz \; d\qperp}{2 (2\pi)^3 z(1-z)} V(x,\kperp;y,\pperp;z,\qperp) 
\psi(z,\qperp),
\end{equation}
here the interaction $V(x,\kperp;y,\pperp;z,\qperp)$ is given by
\begin{multline}
 V(x,\kperp;y,\pperp;z,\qperp) = 2 (2\pi)^2 y (y-1) \delta(y-z) \delta(\pperp - \qperp) V(x,\kperp; z,\qperp|M^2) 
\\ - \langle \; xR^+,\kperp | \; w^{(2)}(R) \; | \; zR^+,\qperp \rangle 
,\end{multline}
and the explicit functional form of the matrix element above is given in \cite{Tiburzi:2002mn}
and is represented by diagrams E and F in the Figure. For brevity we have omitted writing the $M^2$ dependence.

\section{Form factors}

To calculate form factors, we use the electromagnetic vertex $\Gamma^\mu$ constructed in \cite{Tiburzi:2002sx} up to the first Born approximation (notice that 
the ladder model's gauged interaction $V^\mu = 0$)
\begin{equation}
\Gamma^\mu =  \Big( \overset{\leftrightarrow}\partial{}^\mu + V G \overset{\leftrightarrow}\partial{}^\mu  \Big) d^{-1}_2,
\end{equation}
where $\overset{\leftrightarrow}\partial{}^\mu$ denotes the electromagnetic coupling to scalars.

Now using Eq.~\eqref{324} to first order in $G - \Gt$, the matrix element $J^\mu = \langle \Psi_{P^\prime}| \Gamma^\mu(-\Delta) 
| \Psi_P \rangle$  then appears
\begin{eqnarray}
J^\mu  &\approx&  \; \langle \gamma_{P^\prime} | \; \Big| G(P^\prime) \Big( 1 + V(P^\prime) ( G(P^\prime) - \Gt(P^\prime)) \Big) \notag
\\ &&\times \Big( \overset{\leftrightarrow}\partial{}^\mu (-\Delta)  d_{2}^{-1} + V(-\Delta) G(-\Delta) \overset{\leftrightarrow}\partial{}^\mu (-\Delta) d_{2}^{-1} \Big) 
\notag
\\ &&\times \Big( 1 + (G(P) - \Gt(P)) V(P)  \Big) G(P) \Big| \; | \gamma_{P} \rangle  \notag
\\
	&=&  \Big( J^\mu_{\text{LO}} + \delta J^\mu_{i} +  \delta J^\mu_{f} + \delta J^\mu_{\gamma} \Big) + \mathcal{O}[V^2],
\end{eqnarray}
with the leading-order result
\begin{equation} \label{LO}
J^\mu_{\text{LO}} = \langle \gamma_{P^\prime} | \; \Big| G(P^\prime) \overset{\leftrightarrow}\partial{}^\mu (-\Delta)  d_{2}^{-1} G(P) \Big| \; | \gamma_{P} \rangle.
\end{equation}
The first-order terms are
\begin{eqnarray}    
		\delta J^\mu_{i} &=&  \langle \gamma_{P^\prime} | \; \Big| G(P^\prime) \overset{\leftrightarrow}\partial{}^\mu(-\Delta)  d_{2}^{-1} 
		\Big(G(P) - \Gt(P) \Big) V(P) G(P) \Big| \; |\gamma_{P} \rangle \notag \\
		\delta J^\mu_{f} &=& \langle \gamma_{P^\prime} | \; \Big| G(P^\prime) V(P^\prime) 
		\Big(G(P^\prime) - \Gt(P^\prime) \Big) 
		\overset{\leftrightarrow}\partial{}^\mu (-\Delta) d_{2}^{-1} G(P) \Big| \; | \gamma_{P} \rangle  \notag \\
		\delta J^\mu_{\gamma}  &=& \langle \gamma_{P^\prime} | \; \Big| G(P^\prime) \Big(V(-\Delta) G(-\Delta)  
		\overset{\leftrightarrow}\partial{}^\mu (-\Delta) \Big) d_{2}^{-1} G(P) \Big| \; | \gamma_{P} \rangle. \label{NLO}
\end{eqnarray}

\begin{figure}
\begin{center}
\epsfig{file=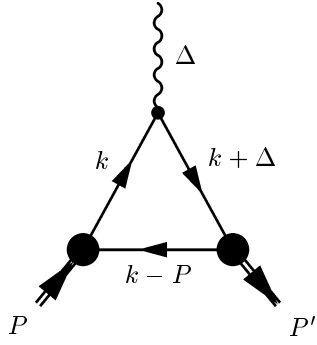}
\caption{The leading-order diagram for the electromagnetic form factor.}
\label{ftri}
\end{center}
\end{figure}

As outlined in \cite{Tiburzi:2002sx}, Eqs.~\eqref{LO} and \eqref{NLO} can be evaluated by residues being careful to remove two-particle reducible
contributions by utilizing \eqref{wavefunction}. 
Here we state the results of these calculations
in $(3+1)$ dimensions. We denote $\Delta^\mu$ as the momentum transfer and define $\Delta^+ = - \zeta P^+$ (see Figure \ref{ftri}).
The leading-order result appears
\begin{equation} \label{FFLO}
J^+_{LO} = - 2 i P^+ \int \frac{\theta(x - \zeta) \; dx \; d\kperp}{2(2\pi)^3 x(1-x) x^\prime} (2 x - \zeta)
\psi^*(x^\prime,\mathbf{k}^{\prime\perp}) \psi(x,\kperp),
\end{equation}
where $x^\prime = \frac{x - \zeta}{1-\zeta}$ and $\mathbf{k}^{\prime\perp} = \kperp + (1-x^\prime) \dperp$ denotes the momentum of the final state.
Using $J^\mu = - i (P + P^\prime)^\mu F(t)$, Eq.~\eqref{FFLO} reduces to the Drell-Yan formula \cite{Drell:1970km,West:1970av} for $\zeta = 0$.

The first of the leading order corrections is the Born term $\delta J^+_\gamma$. For $x>\zeta$ we have
\begin{multline} \label{bt1}
\delta J^+_{\gamma \; \; (x>\zeta)} = \frac{+ 2iP^+}{(16\pi^3)^2} \int \frac{\theta(x - \zeta) dx d\kperp dy d\pperp (2x - \zeta)}{x x^\prime y (1-y) y^\prime} 
\\ \times \psi^*(y^\prime,\mathbf{p}^{\prime\perp}) D(y^\prime,\mathbf{p}^{\prime\perp};x^\prime,\mathbf{k}^{\prime\perp} |M^2) 
\frac{g^2 \theta(y-x)}{y - x} D(y,\pperp;x,\kperp |M^2) \psi(y,\pperp),
\end{multline}
where $y^\prime = \frac{y - \zeta}{1-\zeta}$ and $\ppp = \pperp + (1- y^\prime) \dperp$. This contribution corresponds to 
diagram $A$ in Figure \ref{ftri2}.  On the other hand, for $x<\zeta$ we have
\begin{multline} \label{bt2}
\delta J^+_{\gamma \; \; (x<\zeta)} = \frac{+2 i P^+}{(16\pi^3)^2} \int \frac{\theta(\zeta - x) dx d\kperp dy d\pperp (2 x - \zeta)/\zeta}{y(1-y) \yp \xpp (1-\xpp)} \\ \times
\psi^*(\yp,\ppp) 
\dw(\xpp,\kppp|t) \frac{g^2 \theta(y - x)}{y - x} D(y,\pperp;x,\kperp|M^2) \psi(y,\pperp),
\end{multline}
where $\xpp = x / \zeta$ and $\kppp = \kperp + \xpp \dperp$ denotes the photon's relative momentum. This expression
is diagram $D$ in Figure \ref{fZZZ}.

The next leading-order term is the initial-state iteration $\delta J^+_i$. The only contribution is for $x>\zeta$, namely
\begin{multline} \label{is1}
\delta J^+_{i} = \frac{+2iP^+}{(16\pi^3)^2} \int \frac{\theta(x - \zeta) dx d\kperp dy d\pperp (2x - \zeta)}{x x^\prime (1- x^\prime) y (1-y)}  \\
\times
\psi^*(x^\prime,\mathbf{k}^{\prime\perp}) 
D(y^\prime,\mathbf{p}^{\prime\perp};x^\prime,\mathbf{k}^{\prime\perp} |M^2) 
\frac{g^2 \theta(y-x)}{y - x} D(y,\pperp;x,\kperp |M^2) \psi(y,\pperp),
\end{multline}
which corresponds to diagram $B$ of Figure \ref{ftri2}.

Lastly there is the final-state iteration term $\delta J^+_f$. For $x>\zeta$ we have
\begin{multline}  \label{fs1}
\delta J^+_{f \; \; (x>\zeta)} = \frac{+2iP^+}{(16\pi^3)^2} \int \frac{\theta(x - \zeta) dx d\kperp d\yp d\ppp (2 x - \zeta)}{x(1-x)\xp \yp (1-\yp)}
\\ \times \psi^*(\yp,\ppp) D(\yp,\ppp;\xp,\kpp|M^2) \frac{g^2 \theta(y - x)}{y - x} D(y,\pperp ;x,\kperp |M^2) \psi(x,\kperp), 
\end{multline}
where implicitly $y = \zeta + (1-\zeta) \yp$ and $\pperp = \ppp - (1- \yp) \dperp$. 
This corresponds to diagram $C$ in Figure \ref{ftri2}. While for $x<\zeta$, the expression
\begin{multline} \label{fs2}
\delta J^+_{f \; \; (x< \zeta)} = \frac{+2iP^+}{(16\pi^3)^2} \int \frac{\theta(\zeta - x) dx d\kperp d\yp d\ppp (2 x - \zeta)/\zeta}{(1-x) \xpp (1-\xpp) \yp (1-\yp)}
\\ \times \psi^*(\yp,\ppp) \dw(\xpp,\kppp|t) \frac{g^2 \theta(y - x)}{y - x} D(y,\pperp;x,\kperp|M^2) \psi(x,\kperp),
\end{multline}
corresponds to diagram $E$ of Figure \ref{fZZZ}. Eqs.~(\ref{FFLO}--\ref{fs2}) are then the complete 
expressions for the form factor up to first order ignoring the self-energy loop.

\chapter{Light-Front Time-Ordered Perturbation Theory} \label{oftopt}

The results of Chapters \ref{chap:LFBS} and \ref{chap:GPDs} 
can similarly be achieved directly from ``old-fashioned'' time-ordered perturbation theory in a form which utilizes
projecting onto the two-body subspace of the full Fock space. 
For a nice, complete discussion of this formalism for the light-cone ladder model, see \cite{Cooke:2000ef}. 
In this Appendix, we show how to derive higher Fock space components in this formalism, 
thereby demonstrating the generation of higher components from the lowest sector we found indirectly for GPDs 
and form factors in section \ref{gpds}.

We write the light-cone Hamiltonian as a sum of a free piece and an interacting piece which carries an explicit power of the weak coupling $g$. 
In an obvious notation this is
\begin{equation}
P^- = P^-_o + g P_{I}^-.
\end{equation}
The free term $P^-_o$ is diagonal in the Fock state basis, while the interaction generally mixes components of different particle number (in the scalar model we consider above, the interaction is completely off-diagonal since there are no instantaneous terms). Let us suppose that in the full Fock basis, we have an eigenstate of the Hamiltonian, i.e.
\begin{equation}
\Big( P^-_o + g P_{I}^- \Big) \psiket = \pminus \psiket, 
\end{equation}
where the eigenvalue is labeled by $\pminus$. Since the coupling is presumed small, the mixing of Fock components with a large number of 
particles will be small. Thus one imagines our bound state will be dominated by the two-body Fock component.

To make this observation formal, we define projection operators on the Fock space $\pe$ and $\qu$ in the usual sense. 
The operator $\pe$ projects out only the two-particle subspace of the full Fock space and hence $\qu$ projects out the compliment. Let us define the action
of these operators on our eigenstate
\begin{align}
\pe \psiket & = \psitwoket \label{2ket}\\
\qu \psiket & = \psiquket  \label{qket}.
\end{align}

As is well known, combination of Eqs.~\eqref{2ket} and \eqref{qket} leads to the following equation for the two-body Fock component
\begin{equation}
P^-_{\text{eff}} \psitwoket = \pminus \psitwoket,
\end{equation}
where the effective two-body Hamiltonian is
\begin{align} \label{veff}
P^-_{\text{eff}} & \equiv P^-_{\pe \pe} + V_{\text{eff}} \\
	& = P^-_{\pe \pe} + P^-_{\pe \qu} \frac{1}{\pminus - P^-_{\qu \qu}} P^-_{\qu \pe},
\end{align}
and we have defined the following notation for any operators $A$ and $B$, $P^-_{A B} \equiv A P^- B$.
The effective two-body interaction $V_{\text{eff}}$ defined in equation \eqref{veff} is dependent upon the energy eigenvalue $\pminus$ since we have 
suppressed the degrees of freedom of the $\qu$ subspace. The relation between the $\qu$-space probability (i.e. the non-valence contribution)
and the effective interaction appears
\begin{equation} \label{quspace}
\langle \psi_{\qu} \psiquket = - \frac{\partial}{\partial \pminus} \psitwobra V_{\text{eff}} \psitwoket.
\end{equation}

In a weak-binding limit, we can series expand the effective interaction in powers of the coupling and thereby re-derive the light-front potential. 
Given that every boson emitted must be absorbed in the two-quark sector, we can have only an even number of interactions and hence
\begin{equation}
V_{\text{eff}} = \pe g P^-_{I} \qu \frac{1}{\pminus - P_{o}^-} \sum_{n=0}^{\infty} \Bigg( \frac{g P^-_{I}}{\pminus - P^-_{o}} \Bigg)^{2 n} \qu g P^-_{I} \pe.
\end{equation}
So, for example, at leading order we have all possible ways to propagate from the two-body sector and back with only two interactions in between. The diagrams in Figure 
\ref{fOBEP} correspond to the two possibilities distinguished by the action of $\frac{1}{\pminus - P^-_{o}}$ between interactions. At the next order, we have all possible 
ways to propagate from two bodies to two bodies with four interactions in between, \emph{etc}.

 To generate higher Fock components from the two-body sector, we necessarily must look at the $\qu$-space state
arrived at from Eqs.~\eqref{2ket} and \eqref{qket}
 \begin{equation}
 \psiquket = \frac{1}{\pminus - P^-_{\qu \qu}} P^-_{\qu \pe} \psitwoket. 
 \end{equation}   
 To generate an $n$-body Fock component from this state, we merely act with an $n$-body projection operator which we shall denote $\qu_{n}$. 
Similar to the above, we expand in powers of the coupling to find
\begin{equation}
| \psi_{n} \rangle = \qu_{n} \frac{1}{\pminus - P^-_{o}} \sum_{n = 0}^{\infty} \Bigg(  \frac{g P^-_{I}}{\pminus - P^-_{o}} \Bigg)^n \qu g P^-_{I} \psitwoket.
\end{equation}
For example, the leading-order three-body state is obtained by attaching a boson to a quark line in the only two possible ways 
(and adding the light-front energy denominator at the end). With these three-body states, we can consider all possible three-to-three overlaps that 
would contribute to the form factor. These are depicted in Figure \ref{ftri2} (with the exception of a quark 
self-interaction).  The four-body sector is richer since there are two-boson, two-quark states as well as four-quark states. 
The two-to-four overlaps required for GPDs must have four quarks. At leading order, we generate the diagrams encountered above in 
Figure \ref{fZZZ}. Not surprisingly directly applying time-ordered perturbation theory from a light-front Hamiltonian agrees with our 
derivation above from covariant perturbation theory in the Bethe-Salpeter formalism projected onto the light cone.

\chapter{Application to Timelike Processes} \label{chap:time}

In this Appendix, we pursue timelike processes in light-front dynamics. These processes do not have a Fock space wavefunction representation.
Below we study time-like form factors in particular to demonstrate the versatility of the light-front reduction of the BSE. This will allow us to 
make the connection to the generalized distribution amplitudes (GDAs), which we explicitly calculate for the ladder model. 
Analogous to GPDs, GDAs encode the soft physics of two-meson production and can thus be thought of 
 as  crossed versions of the GPDs. The GDAs enter in convolutions for various two-meson production amplitudes, see e.g.~%
\cite{Diehl:1998dk,Diehl:2000uv,Polyakov:1998ze,Lehmann-Dronke:1999aq}. 
These distribution functions as well as time-like form factors are a theoretical challenge for light-front dynamics, 
since there is no direct decomposition in terms of meson Fock components alone. 
Furthermore, we shall see the leading-order expressions are non-valence contributions 
(which necessarily excludes a description in terms of constituent quark models).

\begin{figure}
\begin{center}
\epsfig{file=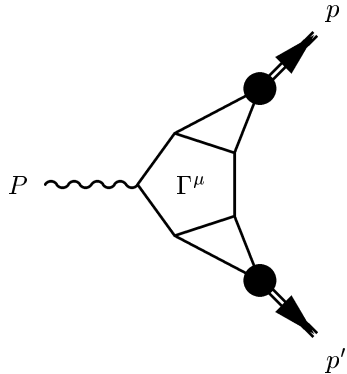}
\caption{The triangle diagram for the time-like, pion form factor.}
\label{ftimetri}
\end{center}
\end{figure}

The time-like form factor $F(s)$ for our model meson is defined by (see Figure \ref{ftimetri})
\begin{equation}
\langle \Psi_{p} \; \Psi_{p^\prime} | \; \Gamma^\mu \; | 0 \rangle = - i (p - p^\prime)^\mu F(s),
\end{equation}
where $s = (p + \pp)^2$ is the center of mass energy squared. Now define $P^\mu = p^\mu + \pp{}^\mu$ and $\zeta = \pplus / \Pplus$. We can work out the kinematics of this reaction in a frame where $\mathbf{P}^\perp = 0$ and see that 
\begin{align}
P^- & = \frac{s}{2 P^+}, \notag \\
p^- & = \frac{(1 - \zeta) s}{2 \Pplus}, \notag \\
\mathbf{p}_\perp^2 & = s (1 - \zeta) \zeta - M^2,
\end{align}
where $M$ is the meson mass.

Similar to GPDs, the GDA for our model has a definition in terms of a non-diagonal matrix element of bilocal field operators\footnote{%
And similar to our discussion of GPDs, perturbative QCD corrections lead to renormalization scale dependence
of the GDAs. For a complete discussion of $\cO(\a_s)$ corrections in the crossed channel, see, e.g.~\cite{Diehl:2003ny}.
}
\begin{equation} \label{bilocalgda}
\Phi(z,\zeta,s) = \int \frac{dx^-}{2 \pi}  e^{i z \Pplus x^-} \langle \Psi_{p} \; \Psi_{\pp} | \; q(0) i \
\overset{\leftrightarrow}\partial{}^+ q(x^-) \; | 0 \rangle.
\end{equation}
As in Section \ref{gpds}, our concern is only with the twist-two contribution and thus we have taken the plus-component of the 
vector current in Eq.~\eqref{bilocalgda}. The definition of the GDA leads directly to a sum rule for the time-like form factor
\begin{equation} \label{sumtime}
\int \frac{dz}{2 \zeta - 1} \Phi(z,\zeta,s) = F(s),
\end{equation}
and hence a means to calculate $\Phi$ from the integrand of the time-like form factor. Again this is because the effect of the 
non-local light-cone operator is to fix the plus momentum of the active quark.

Taking the appropriate residues of the five-point function, we arrive at Fig. \ref{ftimetri} for the time-like form factor. 
Keeping only the leading-order piece of the electromagnetic vertex $\Gamma^\mu$, we have
\begin{multline}
\Phi(z,\zeta,s) = i \Pplus (2z -1) \int \frac{d\kminus d\kperp}{(2\pi)^4} 
\gamma^*( \zpp, \kperp - \zpp \pperp | M^2 ) G(k,p) d^{-1}(k-p)
\\ \times  G(P-k, \pp)  \gamma^*\big( \zp, \kperp  - (1- \zp) \pperp \big| M^2 \big), 
\end{multline}
where we have made use of some auxiliary definitions: $z = \kplus/ \Pplus$, $\zp = \frac{z - \zeta}{1 - \zeta}$ and $\zpp = z / \zeta$. 
Recall $\gamma(x|M^2) \propto \theta[ x(1-x)]$. 
This translates to: $0< \zpp < 1$ and $0< \zp < 1$, and hence we do not pick up a contribution at zeroth order in the coupling.

To work at first order, we pick up three terms analogous to those in Appendix \ref{fff} for the spacelike form factor. 
We denote these as $\delta J^\mu_{\gamma}$, $\delta J^\mu_{p}$ and
$\delta J^\mu_{\pp}$. The Born term for the three-point electromagnetic vertex $\delta J^\mu_{\gamma}$ is quite simple. 
For the same reason as the zeroth-order result, the restriction of $\gamma(x|M^2) \propto \theta[x(1-x)]$ and momentum conservation
force the contribution $\delta J^\mu_{\gamma}$ to vanish. This leaves us to consider only diagrams that arise from iteration of the 
Bethe-Salpeter equation for either of the final-state mesons.

Considering first the term $\delta J^\mu_{p}$, we have the contribution to the GDA
\begin{multline} \label{blah}
\Phi_{p}(z,\zeta,s)  =  i \Pplus (2 z -1) \int \frac{d\kminus d\kperp}{(2\pi)^4} \; \frac{d^4q}{(2\pi)^4} 
\gamma^*(\zp, \kperp - (1-\zp) \pperp) d^{-1}(k-p)
\\ \times \gamma^*(\ypp, \qperp - \ypp \pperp) G(q,p) V(q,k) G(k,p) G(P - k, \pp) 
, \end{multline}
where we have chosen to abbreviate $y = \qplus/ \Pplus$ and hence the label $\ypp = y / \zeta$. We have customarily omitted the subtracted term containing $\Gt$, 
which is zero because there is no leading-order term to subtract.
Requiring wavefunction vertices mandates $0<\ypp<1$ and $\zeta < z < 1$.

\begin{figure}
\begin{center}
\epsfig{file=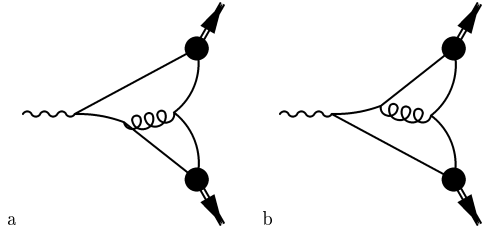,width=2.5in}
\caption{Leading-order diagrams for the time-like pion form factor.}
\label{fgda}
\end{center}
\end{figure}

Thus after integration over the loop energy, $\Phi_{p}$ produces one contribution to the GDA 
\begin{multline} \label{jp}
\Phi_{p}(z,\zeta,s)  = \frac{\theta(z - \zeta)}{(16\pi^3)^2 \zeta} \int \frac{d\kperp dy d\qperp (2 z - 1)}{z (1-z) \zp \ypp(1-\ypp)} 
 \dw(z, \kperp |s)  \frac{g^2 \theta(z - y)}{z - y} 
\\ \times  D(z,\kperp;y,\qperp|s) \psi^*(\zp, \kperp - (1-\zp) \pperp) \psi^*(\ypp, \qperp - \ypp \pperp).  
\end{multline}
As a contribution to the time-like form factor, we can interpret Eq. \eqref{jp} as the time-ordered diagram $b$ of Figure \ref{fgda}.

At first order in the weak coupling, we have one term remaining to consider $\delta J^\mu_{\pp}$. Again omitting the 
superfluous subtraction of $\Gt$, we have
\begin{multline}
\Phi_{p^\prime}(z,\zeta,s) = i \Pplus (2 z -1) \int \frac{d\kminus d\kperp}{(2\pi)^4} \; \frac{d^4 q}{(2 \pi)^4} \gamma^*(\zpp, \kperp - \zpp \pperp | M^2) G(k,p) 
d^{-1}(p-k) \\ \times G(P - k , \pp) V(P-k, P-q) G(P-q, \pp) \gamma^*(\yp, \qperp - (1- \yp ) \pperp|M^2), 
\end{multline}
with $\yp = \frac{y - \zeta}{1-\zeta}$.
Both $\yp$ and $z$ are restricted: $0 < \yp < 1$ and $0 < z < \zeta$, and  
the final contribution to the GDA is thus
\begin{multline} \label{jpp}
\Phi_{p^\prime}(z,\zeta,s) = \frac{\theta(\zeta - z)}{(16\pi^3)^2 \zeta} \int \frac{d\kperp d\yp d\qperp (2 z - 1)}{(1-z) \zpp (1-\zpp) \yp (1-\yp)} 
\dw(z, \kperp | s)  \frac{g^2 \theta(y - z)}{y-z} 
\\ \times  D(y,\qperp;z,\kperp|s) \psi^*(\zpp, \kperp - \zpp \pperp) \psi^*(\yp, \qperp + \yp \pperp), 
\end{multline} 
In this form we recognize this contribution as diagram $a$ of Figure \ref{fgda}.

Having found the leading non-vanishing contribution to the GDA namely $\Phi = \Phi_{p} + \Phi_{p^\prime}$, we observe that the higher Fock components derived in 
Section \ref{gpds} (as well as in Appendix \ref{oftopt}) do not fit naturally into Eqs.~\eqref{jp} or \eqref{jpp}. One needs a Fock space expansion for the photon 
wavefunction in order to have an expression for the GDA in terms of various Fock component overlaps. With the expressions derived for the GDA, we can use 
Eq. \eqref{sumtime} to obtain the time-like form factor. Notice the expressions derived for the GDA and hence also for the timelike form factor
depend on pieces of the light-front interaction kernel. Thus there is no way to directly utilize quark model wavefunctions for these observables.

\chapter{Complex Conjugate Poles in Minkowski Space} \label{minkowski}

In this Appendix 
we show that amplitudes cannot be directly calculated in Minkowski space when complex conjugate
poles are present in propagators and vertices.
We then investigate how to calculate amplitudes properly in Minkowski space. Firstly amplitudes 
dependent on the imaginary part of some set of diagrams can be calculated by using a straightforward generalization of the
cutting rules. We apply this to calculate the quark distribution function from the handbag diagram 
in the Bjorken limit in Section \ref{cutting}. Lastly we consider the analytic continuation of space-like amplitudes from 
Euclidean space. This is complicated by the presence of Wick poles and requires their residues to be appropriately added
when amplitudes are calculated. The details of the GPD calculation appear at the end of this Appendix in Section \ref{appapp}. 
Resulting functions calculated in Minkowski space after the Wick rotation agree with those
obtained from the model defined in Euclidean space.

This work lays the foundation for calculating light-cone dominated amplitudes using meromorphic model propagators and vertices
constrained by both lattice data\footnote{%
One must proceed with caution: the lattice calculations employ Landau gauge, while we
have tacitly used light-cone gauge above.
} 
and Ward-Takahashi identities. 
Distribution functions for such models could be calculated rigorously since the pole structure of the propagator 
is known and relevant integrals converge in the complex light-cone energy plane. 
As far as light-cone phenomenology is concerned,  resulting expressions would be truly Poincar\'e covariant (as opposed to 
diagonal with respect to the non-interacting operators) and would satisfy 
field-theoretic identities. Filling these two gaps is essential for adequate 
hadronic phenomenology for processes at large momentum transfer.  Moreover, such models would help light-cone methods 
and Dyson-Schwinger studies reach complementary standing.

\section{Model defined}

The model we take is $\phi^3$ theory with electromagnetic interactions. Equivalently we can view this model as a bound state of two scalar
particles with a trivial Bethe-Salpeter vertex $\Gamma(k,P) = 1$, where the coupling constant is assumed to be 
absorbed into the overall normalization, see Section \ref{toyboost}. 
We make a simple \emph{Ansatz} for the non-perturbative propagator consisting of a
pair of complex conjugate poles
\begin{equation} \label{Mprop}
S(k)  =  \frac{i (k^2 - a^2 + b^2)}{(k^2 - a^2 + b^2)^2 + 4 a^2 b^2},
\end{equation}
where $a^2 - b^2 > 0$. Defining for ease $m^2 = a^2 - b^2$ and $\eps = 2 a b$ (which is taken to be positive without any loss 
of generality), we can write the propagator as
\begin{equation}
S(k) = \frac{i/2}{k^2 - m^2 + i \eps} + \frac{i/2}{k^2 - m^2 - i \eps}.
\end{equation}
 The light-cone energy poles of the propagator are thus
\begin{equation} \label{apole}
k^-_{a} = \frac{\kperp^2 + m^2}{2 k^+} - \frac{i \eps}{2k^+}
\end{equation}
and $k^-_{a^*} = (k^-_{a})^*$, where $*$ denotes the complex conjugate. Although we use a scalar model, results straightforwardly extend
to spin-$\frac{1}{2}$ particles, e.g., since only the pole structure of Eq.~\eqref{Mprop} is relevant.

\begin{figure}
\begin{center}
\epsfig{file=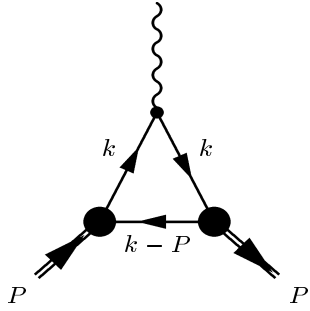,height=1.6in,width=1.6in}
\caption[Triangle diagram used to calculated the quark distribution.]{Triangle 
diagram at zero momentum transfer used to calculate the quark distribution function by projecting onto the light cone.}
\label{fqofx}
\end{center}
\end{figure}

Now let us consider calculating this model's quark distribution by projecting onto the light cone. The quark distribution
can be derived by fixing the plus-momentum of the active quark $x = k^+/P^+$, see Figure \ref{fqofx}, and taking the plus-component
of the current. Thus up to overall normalization, we have the expression
\begin{equation} \label{minq}
q(x) \propto \int d^4 k \; \delta(k^+ - x P^+) \; x S(k) S(k - P) S(k).
\end{equation}
The $k^-$ integral is then performed by residues. Choosing a frame in which $\Pperp = 0$, the spectator propagator
has light-cone energy poles
\begin{equation} \label{bpole}
k_b^- = P^- + \frac{\kperp^2 + m^2}{2 (k^+ - P^+)} - \frac{i \eps}{2(k^+ - P^+)},
\end{equation}
and $k_{b^*}^- = (k^-_b)^*$.

Performing the $k^-$ integral in Eq.~\eqref{minq}, we arrive at the quark distribution
\begin{multline}
q(x) \propto 2 \pi  i \Bigg\{ -\theta(-x) \big[\res(k^-_{a^*}) + \res(k^-_{b^*})\big] \\ 
+ \theta[x(1-x)] \big[\res(k^-_{a^*}) + \res(k^-_{b}) \big] 
+ \theta(x - 1) \big[ \res(k^-_{a^*}) + \res(k^-_{b^*}) \big]  \Bigg\}
\end{multline}
This distribution does not have proper support, i.e.~it is non-vanishing outside the interval $x \in [0,1]$. 
Moreover, the distribution is not real valued, whereas it should be positive definite. Thus the model based 
on the propagator in Eq.~\eqref{Mprop} cannot be suitably formulated in Minkowski space. We will find below that 
Eq.~\eqref{Mprop} makes sense as a Minkowski space propagator only after analytic continuation
from Euclidean space for the amplitude in question. In Section \ref{ccpoles}, we demonstrated that the model
is perfectly well defined in Euclidean space and thus the challenge of analytic continuation remains.

\section{Cutting rules} \label{cutting}

In this Section, we show there is some hope in working with the model propagator Eq.~\eqref{Mprop} in Minkowski space. 
We demonstrate that the quark distribution can be derived by a straightforward generalization of the cutting rules.

\begin{figure}
\begin{center}
\epsfig{file=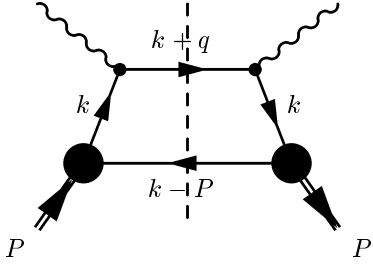}
\caption[Handbag diagram for the forward Compton amplitude. ]{Handbag diagram for the forward Compton amplitude. 
Dashed line denotes the cut which yields the quark distribution in the Bjorken limit.}
\label{fhand}
\end{center}
\end{figure}

Consider the forward Compton amplitude $T^{\mu \nu}(q,P)$ depicted in Figure \ref{fhand}. In the Bjorken limit, 
the imaginary part of this diagram is related to the quark distribution, see Section \ref{DIS}. 
For simplicity, we can choose a frame in which $\mathbf{q}^\perp = 0$. 
The minus-plus component of the forward Compton amplitude 
in such frames reads
\begin{equation} \label{fca}
i T^{-+}(q,P) = - \frac{16 N}{\pi^3} \int d^4 k \; (2 k^- + q^-)  S(k) \; S(k-P) \; S(k) \; (2 k^+ + q^+) S(k+q). 
\end{equation}

In the scalar particle case, the minus-plus component of the forward Compton amplitude can be used to define the quark distribution 
in a way analogous 
to the spin-$\frac{1}{2}$ case. The relation is simply $\Im \text{m} (T^{-+}) \propto q(x)$, see \cite{Tiburzi:2001je}. 
In Eq.~\eqref{fca}, we have adjusted the overall normalization so that equality between $\Im \text{m} (T^{-+})$ and $q(x)$ holds. 
With standard perturbative propagators, we could calculate the imaginary part of $T^{\mu \nu}(q,P)$
by using the cutting rules, whereby one replaces the cut propagators in Figure \ref{fhand} by an on-shell prescription, namely
\begin{equation} \label{cutrule}
S(k)_{\eps = 0}  \longrightarrow  - 2 \pi i \; \delta (k^2 - m^2).
\end{equation}
In Eq.~\eqref{cutrule}, we have specified $\eps = 0$ for the case of a free particle propagator. Since large momentum flows through
the handle of the handbag, we may neglect the mass of the struck quark and use the standard cutting rule for $S(k+q)$. In the Bjorken
limit, we define $x = - q^2 / 2 P \cdot q$  which remains finite as $q^2, P \cdot q \to \infty$ and is kinematically bounded
between zero and one. Further we orient our frame of reference so that  $q$ has a 
large minus component in this limit, and consequently  $q^+ = q^2 / 2 q^-$ is finite. Hence we have the familiar replacement
\begin{equation} \label{xcut}
S(k+q) \longrightarrow - \frac{\pi i}{2 q^-} \; \delta( k^+ - x P^+). 
\end{equation}

To complete the cut, we must deal with the spectator particle's complex mass shells. We must worry about the propagator Eq.~\eqref{Mprop}
only where the denominator is zero. 
Thus we are lead to the cutting rule for the propagator Eq.~\eqref{Mprop}
\begin{equation} \label{gencut}
S(k) \longrightarrow - \pi i \Big[  \delta(k^2 - m^2 + i \eps) + \delta(k^2 - m^2 -  i \eps )  \Big],
\end{equation}
which puts the intermediate state on its complex mass shells. Furthermore, the limit $\eps \to 0$ produces
the regular cutting rule Eq.~\eqref{cutrule}.

Using this cutting rule for the spectator particle along with Eq.~\eqref{xcut}, we can deduce the quark distribution 
from $\Im  \text{m} (T^{-+})$ in the Bjorken limit\footnote{%
This may seem obvious. But there are two inconsistencies which will ultimately be resolved in Section \ref{analytic}. 
First Eq.~\eqref{fca} is not the Minkowski space amplitude. As already mentioned it requires analytic continuation.
Second the rule Eq.~\eqref{gencut}, although plausible, does not produce the imaginary part of Eq.~\eqref{fca}.
The result Eq.~\eqref{LCqx}, however, leads us to believe these two wrongs make proverbial right.
}
\begin{equation} \label{qcut}
q(x) = \frac{4 N}{\pi} \int d^4 k  \; \delta(k^+ - x P^+) 
\Big[\delta(k^- - k_b^-) + \delta( k^-  - k^-_{b^*}) \Big] \frac{x S(k)^2}{1-x },
\end{equation}
where the light-cone energy poles are given in Eq.~\eqref{bpole}. Notice the resulting distribution is real and has proper
support due to the kinematic constraint $x \in [0,1]$ imposed by the Bjorken limit. Evaluation of the two trivial integrals
leaves only the transverse momentum integration
\begin{equation} \label{LCqx}
q(x) =  \frac{N}{\pi} \sum_{\eps,\epsp,\epspp = \pm} 
\int \frac{d\kperp}{x(1-x)} \dw(x,\kperp, \eps, \epspp |M^2) \dw(x,\kperp,\epsp,\epspp|M^2)
\end{equation}
where we have defined the Weinberg propagator generalized for complex masses as
\begin{equation} \label{weinprop}
\dw(x,\kperp,\eps,\epsp|M^2)^{-1} = M^2 - \frac{\kperp^2 + m^2}{x(1-x)} + \frac{i \eps}{x} + \frac{i \epsp}{1 - x}.
\end{equation}

Evaluation of the $\kperp$ integral yields an analytic
expression for $q(x)$, which is identical to that obtained from the DD Eq.~\eqref{qofx}. Notice Eq.~\eqref{qcut} is equivalent
to evaluating Eq.~\eqref{minq} at $\res(k^-_b) + \res(k^-_{b^*})$ and hence the quark distribution is derived as if the
complex conjugate spectator poles both lie in the upper-half complex plane! We will understand this better once we analytically 
continue from Euclidean space.

\section{Analytic continuation} \label{analytic}

Above we have seen that modified cutting rules can be used to derive the correct quark distribution function in Minkowski space.
In essence the result stems from putting the spectator particle on its complex mass shells. This must be justified by 
careful analytic continuation of Euclidean space amplitudes. Below we consider the Minkowski space calculation of the
generalized parton distribution which cannot be derived by cuts. This will force us to deal with the underlying Wick 
rotation necessary to define the model in Minkowski space.

\begin{figure}
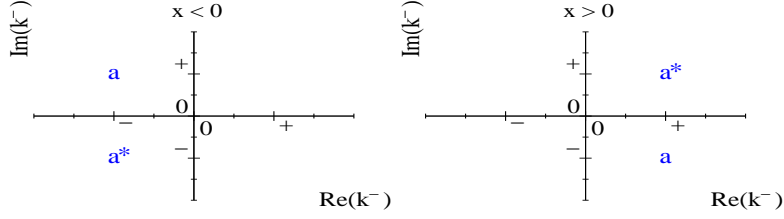

\begin{center}
\epsfig{file=k1.eps,width=2in,height=1.1in}
\epsfig{file=k2.eps,width=2in,height=1.1in}
\caption{Complex light-cone energy plane for the propagator.}
\label{fka}
\end{center}
\end{figure}

In Section \ref{ccpoles},  the model double distributions were calculated in Euclidean space. Thus to calculate related amplitudes
in Minkowski space, we must Wick rotate in the complex energy plane: $k^4 \to i k^0$. The analytic continuation can be 
viewed alternately in the complex light-cone energy plane. For $k^3 = 0$, the rotation is from the $\Im \text{m}(k^-)$ axis to the $\Re \text{e}(k^-)$ axis. 
In the general case, such a correspondence
can only be made precise by considering the Wick rotation in terms of $k^0$ and then boosting to the infinite momentum frame. 
This requires tedious algebraic manipulations and a proliferation of energy poles and time-ordered diagrams. Indeed it is 
easier just to imagine the rotation simply and deal with the light-cone singularities. This is the approach we present.

Before tackling the model's generalized parton distributions in Minkowski space, let us imagine a simpler fictitious example. Consider 
some well-defined Euclidean space amplitude having only poles at $k^-_a$ and $k^-_{a^*}$, see Eq.~\eqref{apole}. To calculate the
amplitude in Minkowski space, we naively integrate along the $\Re \text{e} (k^-)$ axis. In general the correct path on which to integrate is 
one which nears the $\Re \text{e} (k^-)$ axis except for detours around energy poles in the first and third quadrants. Such a path is 
correct since it can be continuously deformed into the Euclidean path. The difference between the naive integration and 
the correct path is a sum of residues of the Wick poles. The energy poles of our fictitious amplitude 
are depicted in Figure \ref{fka}. Their location depends upon the sign of $x = k^+ / P^+$. Thus for this amplitude 
the correct continuation from Euclidean space is
\begin{equation} \label{cont}
\int d \, \Im  \text{m} (k^-) \longrightarrow \int d \, \Re \text{e}  (k^-)  +  2 \pi i \big[ \theta(-x) - \theta(x) \big]  \res(k^-_{a^*}).  
\end{equation}
Notice only $k^-_{a^*}$ is a Wick pole; this is expected because we know the limit $\eps \to 0$ can be analytically continued in the naive 
fashion. Closing the contour in the upper-half plane to perform the Minkowski space energy integral\footnote{One must be careful of zero 
modes \cite{Yan:1973qg} for which $k^+ = 0$. In such cases, the pole lies on the contour at infinity and the integration cannot be performed by
residues. Since our fictitious example is only schematic, we are neglecting the issue of zero modes and are hence excluding amplitudes 
of the form
\begin{equation} \notag
\sum_{\eps = \pm} \int \frac{d^4 k}{(k^2 - m^2 + i \eps)^n}, \quad n>2,
\end{equation} 
which must be handled separately. The example
\begin{equation} \notag
\int \frac{d^4 k}{(k^2 - m^2 + i \eps)^2 (k^2 - m^2 - i \eps)^2}
\end{equation}
is devoid of zero-mode complications and more closely parallels the expressions encountered for GPDs.}
and evaluate our fictitious 
amplitude, we pick up $2 \pi i [\theta(-x) \res (k^-_a) + \theta(x) \res(k^-_{a^*})]$. The net result according to Eq.~\eqref{cont}
is thus
\begin{equation} \notag
2 \pi i \theta(-x) \big[ \res(k_a^-) + \res(k^-_{a^*}) \big].
\end{equation}
Looking back at Figure \ref{fka}, the net result after Wick rotation amounts to both poles lying in the same half-plane; 
or equivalently, we have effectively integrated in either the right- or left-half plane. The result that both poles 
lie in the same half-plane reflects quark confinement; the quark wavefunction must vanish as time becomes large~\cite{Gribov:1999ui}.

The space-like amplitude\footnote{There are additional complications for time-like amplitudes and for amplitudes involving
unstable bound states. In these cases Wick poles are present even when standard perturbative propagators are used. 
The analysis above must be more carefully considered in these cases where threshold effects are already inherent 
in the analytic continuation to, or from, Euclidean space.} for the generalized parton distribution can now
be continued to Minkowski space and hence be evaluated by projecting onto the light cone. To do so, we refer 
to Figure \ref{ftwist2} and insert the non-local light-cone operator $\phi(0) i \overset{\leftrightarrow}{D}{}^+ \phi(z^-)$ in 
place of the local twist-two operators denoted by a cross in the Figure. Here the plus-component picks out the leading-twist contribution 
according to light-cone power counting. Thus in momentum space, we arrive at
\begin{equation} \label{minkGPD}
H(x, \zeta,t) = \frac{2 N / \pi^2}{1 - \zeta / 2} \int d^4 k \; \delta(k^+ - x P^+) \;(2 k^+ + \Delta^+)\; S(k) S(k-P) S(k+\Delta).
\end{equation}
Above we have included the $\zeta$-dependent pre-factor to normalize the action of $\overset{\leftrightarrow}{D}{}^+$ between non-diagonal
states. The overall normalization is then the same as in Eq.~\eqref{bob}. By writing Eq.~\eqref{minkGPD} in Minkowski space, we must also
keep in mind the Wick residues implicitly necessary
so that Eq.~\eqref{minkGPD} is meaningful. In addition to the poles $k^-_a$, $k^-_b$ [given in Eqs.~\eqref{apole} and \eqref{bpole}, 
respectively] and their complex conjugates, the integrand of Eq.~\eqref{minkGPD} also has the poles
\begin{equation} \label{cpole}
k_c^- = - \Delta^- + \frac{(\kperp + \dperp)^2 + m^2}{2 (k^+ + \Delta^+)} - \frac{i \eps}{2 (k^+ + \Delta^+)} 
\end{equation}
and $k^-_{c^*} = (k^-_c)^*$. Carrying out the light-cone energy integration in Eq.~\eqref{minkGPD} as well as adding relevant
residues resulting from the Wick rotation produces (the subtle details of this calculation appear at the end of this Appendix
in Section \ref{appapp})
\begin{multline} \label{GPDres}
H(x, \zeta,t) = - 2 \pi i  \; \theta(x) \theta(\zeta - x) \Big[ \res(k^-_a) + \res(k^-_{a^*}) \Big] \\
+ 2 \pi i \; \theta(x - \zeta) \theta(1 - x) \Big[ \res(k^-_b) + \res(k^-_{b^*}) \Big],
\end{multline}
where the residue is of the integrand in Eq.~\eqref{minkGPD}. As a result of the effective relocation of poles
to the same half-plane as their complex conjugates, the resulting GPD Eq.~\eqref{GPDres} is real and vanishes outside $x$ from 
zero to one.

Using the Weinberg propagator Eq.~\eqref{weinprop}, the residues can be compactly written in terms of relative momenta.
Defining the relative momentum of the final state as
\begin{equation}
\xp = \frac{x - \zeta}{1 - \zeta},  \quad \quad \kpperp = \kperp + (1 - \xp) \dperp,
\end{equation}
and the relative momentum of the photon as
\begin{equation}
\xpp = \frac{x}{\zeta}, \quad \quad \kppperp = \kperp + \xpp \dperp ,
\end{equation}
the light-cone GPD can be expressed in the form
\begin{equation} \label{GPDresult}
( 1 - \zeta / 2) \; H(x,\zeta,t) =  \theta(x) \theta(\zeta - x) \; H_1(x,\zeta,t) + \theta( x - \zeta) \theta(1-x) \; H_2(x,\zeta,t),
\end{equation}
where we have made the abbreviations
\begin{align}
H_1(x,\zeta,t) & = ( 2 \xpp - 1) \frac{N}{2 \pi} \sum_{\eps,\epsp,\epspp = \pm} \int d\kperp 
\frac{\dw(x,\kperp,\eps,\epspp|M^2) \dw(\xpp,\kppperp, \eps, \epsp |\;t\;)}{\xpp (1 - \xpp) (1 - x)} 
 \\
H_2(x,\zeta,t) & = ( 2 x - \zeta) \frac{N}{2 \pi} \sum_{\eps,\epsp,\epspp = \pm} \int d\kperp 
\frac{\dw(x,\kperp, \eps,\epspp |M^2) \dw(\xp,\kpperp,\epsp, \epspp |M^2)}{x (1-x) \xp} 
\end{align}

Firstly one can see analytically that the correct quark distribution results at zero momentum transfer, namely
$H_2(x,0,0) = q(x)$, where $q(x)$ is given by Eq.~\eqref{LCqx}. As remarked in Section \ref{cutting}, this function is
identical to that obtained from the double distribution Eq.~\eqref{qofx}. Secondly, the resulting light-cone GPD Eq.~\eqref{GPDresult}
agrees numerically with that found from the double distribution, via Eq.~\eqref{JiGPD2}, which is plotted in Figure \ref{figGPD}. Finally 
the electromagnetic form factor found from the sum rule
\begin{equation} \label{GPDsum}
F(t) = \int_0^1 dx \; H(x, \zeta, t)
\end{equation}   
agrees numerically with the result of Eq.~\eqref{Fsum2} (which is plotted in Figure \ref{figqandF}). The $\zeta$-independence of Eq.~\eqref{GPDsum} 
stems from Lorentz invariance which is present, however, not manifest in Eq.~\eqref{GPDresult}. Thus with the calculation of 
Eq.~\eqref{GPDresult} from analytically continuing to Minkowski space, space-like amplitudes now agree with those calculated 
from the Euclidean space double distribution.

\section{Calculation of the GPD} \label{appapp}

Below we derive Eq.~\eqref{GPDres} for the GPD in Minkowski space. The details have been relegated here since there is some subtlety.
In order to evaluate Eq.~\eqref{minkGPD}, which implicitly needs analytic continuation, we must shift the energy integration variable
and define a prescription for dealing with vanishing real parts.  Let us see how these difficulties arise.

In considering the Wick rotation, one is usually only concerned with the single denominator that results from combining 
propagators via Feynman parameters. For the moment, let us ignore the complication of complex conjugate pairs of poles. 
In this case, combining the denominators of Eq.~\eqref{bob} using Feynman parameters results in $[l^2 - D_o(x,y;t) + i \eps]^{-3}$, 
where $D_o$ is given by Eq.~\eqref{dzero}. Since the bound state is stable and $t$ is space-like, $D_o(x,y;t)$ is always positive
and hence there are no Wick poles. Moreover, we have shifted the variable $k^\mu$ to arrive at $l^\mu$.

In analytically continuing the expression with uncombined propagators (and again no complex conjugate poles), we are confronted 
with a problem. The pole $k_b^-$, for example, is shifted by the energy $P^-$. Thus the location of the singularity in the complex plane
will be shifted parallel to the real axis depending upon the relative magnitude of the spectator's kinetic energy and $P^-$. 
In the schematic
example shown in Section \ref{analytic}, the pole $k_a^-$ does not have such a shift. Thus for the $b$ and $c$ poles, the location
of the singularities depicted in Figure \ref{fka} will shift along the real axis (depending on $P^-$ and $\Delta^-$) and there will 
be threshold Wick poles [the threshold is defined when $\Re \text{e} (k^-_{b,c} = 0)$]. This is unphysical: 
we just demonstrated the Wick rotation can be done for the combined denominators without
crossing any poles. To perform the same Wick rotation at the level of separate propagators, we must use the freedom to shift the energy 
variable as well as the stability of the bound state.

\begin{figure}
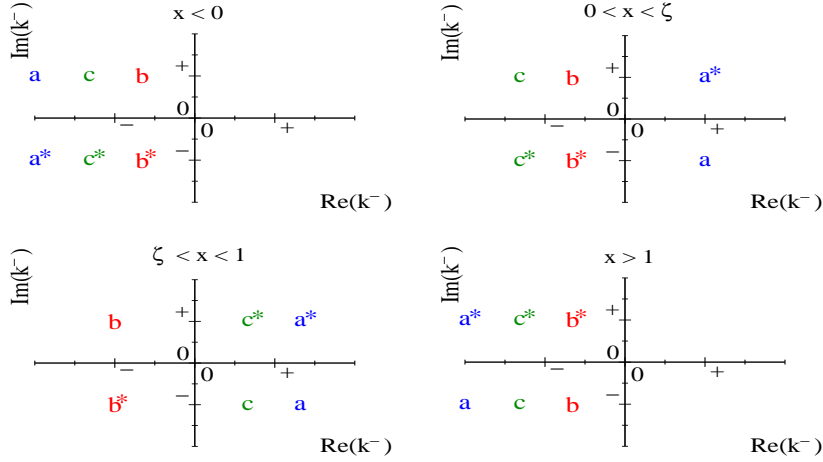

\begin{center}
\epsfig{file=poles1.eps,width=2in,height=1.1in} $\quad$
\epsfig{file=poles2.eps,width=2in,height=1.1in}

\smallskip\smallskip\smallskip\smallskip
\epsfig{file=poles3.eps,width=2in,height=1.1in} $\quad$
\epsfig{file=poles4.eps,width=2in,height=1.1in}
\caption[Complex light-cone energy plane for the shifted poles.]{Complex 
light-cone energy plane for the shifted poles Eq.~\eqref{newpoles} for the generalized parton distribution.}
\label{fpoles}
\end{center}
\end{figure}

On the light-cone, the bound state stability condition can be expressed as
\begin{equation} \label{bssc}
P^- < \frac{\kperp^2_{\text{rel}} + m^2}{2 P^+ x (1 -x)},
\end{equation}
where $\kperp_{\text{rel}} = \kperp -x \mathbf{P}^\perp$ is the relative transverse momentum of the two constituents. 
Because we consider the 
elastic electromagnetic form factor, there is an analogous relation for the final state $P^\prime$. 
Since each propagator contains the kinetic energy of a single particle, the bound state stability condition can never
be utilized without shifting $k^-$. Yet in order to perform such a shift, the real part of one pole must be zero and hence we must
invent a prescription for moving this pole off the Euclidean contour.

We choose the translation: $k^- \to k^- + \delta k^- + \eta$, where $\eta$ is a positive infinitesimal and
\begin{equation} \label{dk}
\delta k^- = P^- + \frac{\kperp^2 + m^2}{2 P^+ (x - 1)}. 
\end{equation}
The resulting poles of the integrand in Eq.~\eqref{minkGPD} we denote
\begin{equation} \label{newpoles}
\begin{cases}
\tilde{k}^-_a = k^-_a - \delta k^-  = - P^- + \frac{\kperp^2 + m^2}{2 P^+ x (1-x)} - \frac{i \eps}{2 P^+ x}\\
\tilde{k}^-_b = k^-_b - \delta k^- - \eta = - \eta - \frac{i \eps}{2 P^+ (x - 1)} \\
\tilde{k}^-_c = k^-_c - \delta k^-  = - P^{\prime-} + \frac{\kpperp^2 + m^2}{2 P^{\prime+} \xp (1- \xp) } - 
\frac{i \eps}{2 P^+ (x - \zeta)}
\end{cases},
\end{equation}
and similarly for their complex conjugate partners. Notice imaginary parts of the poles are unaffected by the energy translation. 
The $\eta$ prescription has displaced the resulting spectator pole away from the Euclidean path independent of $x$. Moreover the
$a$ and $c$ poles have non-zero real parts; so $\eta$ has been set to zero for these poles above.

Using the expressions for the new poles Eq.~\eqref{newpoles} and the bound state stability condition Eq.~\eqref{bssc}, we can 
determine the quadrant location of the singularities independent of $P^-$ and $\Delta^-$ (see footnote $2$). These quadrant locations 
are depicted for the full range of $x$ in Figure \ref{fpoles}. Accordingly poles in the first and third quadrants are Wick poles. As we saw in Section
\ref{analytic}, the net result
of the analytic continuation is to evaluate the integral by effectively closing the contour in the right- or left-half plane.  
Hence the GPD Eq.~\eqref{minkGPD} is
\begin{multline} \label{GPDrestilde}
H(x, \zeta,t) = - 2 \pi i  \; \theta(x) \theta(\zeta - x) \Big[ \res(\tilde{k}^-_a) + \res(\tilde{k}^-_{a^*}) \Big]  \\
+ 2 \pi i \; \theta(x - \zeta) \theta(1 - x) \Big[ \res(\tilde{k}^-_b) + \res(\tilde{k}^-_{b^*}) \Big].
\end{multline}
Evaluating the residues in Eq.~\eqref{GPDrestilde} yields the result of Section \ref{analytic}, namely  Eq.~\eqref{GPDresult} which 
is algebraically equivalent to Eq.~\eqref{GPDres}.

Notice from Figure \ref{fpoles}, other $\eta$ prescriptions for the shift, such as $+\frac{\eta}{2 P^+ (x - 1)}$, lead to an incorrect
result for the case when there are no complex conjugate pairs. The figure shows that the infinitesimal prescription must be positive 
and independent of the sign of $x$, $x - \zeta$, \emph{etc}, in order to reproduce the familiar result. It is interesting to note
that for $x > 1$ in the case without conjugate pairs, all poles of the integrand are Wick poles. Though since the integral is convergent,
the sum of these Wick residues vanishes.

The interested reader can verify that the alternate shifts which use the same pole prescription
\begin{equation}
\begin{cases}
	k^- \to k^- + \frac{\kperp^2 + m^2}{2 P^+ x} + \eta \\
	k^- \to k^- - \Delta^- + \frac{(\kperp+ \dperp)^2 + m^2}{2 P^+ ( x - \zeta)} + \eta
\end{cases}
\end{equation}
also yield the correct results provided $t$ is space-like.

\chapter{Ambiguity of the Double Distributions} \label{sec:amb}

The decomposition into $F(\b,\a;t)$, $G(\b,\a;t)$ and $K(\b,\a;t)$ DDs is ambiguous. Experimentally 
one cannot access the DDs directly, only the $H(x,\x,t)$ and $E(x,\x,t)$ GPDs. In Ref. \cite{Teryaev:2001qm}, the ambiguity inherent 
in defining DDs for the pion was likened to the gauge ambiguity of the vector potential of a two-dimensional magnetic field.
In this Appendix, we extend this analysis to the proton case. A full treatment has been given in \cite{Tiburzi:2004qr}. 

Utilizing the Gordon identities (which appear in Appendix \ref{sec:gamma}),  
we can rewrite the bilocal matrix element in Eq.~\eqref{eqn:bilocal} as 
\begin{multline} \label{eqn:curl}
\langle P^\prime,\lp | 
\ol \psi_q \left( - z/2 \right) \rlap\slash z
\psi_q \left( z/2 \right) 
| P,\l \rangle 
= 
\\
\frac{ 1 }{1 - \frac{t}{4 M^2} }
\ol u_{\lp}(\Pp) 
\int_{-1}^{1} d\b \int_{-1 + |\b|}^{1 - |\b|} d\a \;
e^{ - i \b \ol P \cdot z + i \a \D \cdot z / 2}
\Bigg[ \frac{\ol P \cdot z}{M} G_E(\b, \a; t) \phantom{space}
\\
- \frac{i}{2 M} \varepsilon^{\mu \nu \a \b} z_\mu \D_\nu \ol P_\a \gamma_\b \gamma_5 G_M(\b,\a;t)
- \frac{\D \cdot z}{4 M} \left( 1 - \frac{t }{4 M^2} \right) G(\b,\a;t)
\Bigg]  u_{\l}(P) 
,\end{multline}
where we have defined new double distributions
\begin{align}
G_E(\b,\a;t) &= F(\b,\a;t) + \frac{t}{4 M^2} K(\b,\a;t) \\
G_M(\b,\a;t) &= F(\b,\a;t) + K(\b,\a;t)
\end{align}
in analogy with the Sachs electric and magnetic form factors. After integrating by parts and neglecting
the surface terms, we have
\begin{multline}
\langle P^\prime,\lp | 
\ol \psi_q \left( - z/2 \right) \rlap\slash z
\psi_q \left( z/2 \right) 
| P,\l \rangle 
= \frac{i}{M} \int_{-1}^{1} d\b \int_{-1 + |\b|}^{1 - |\b|} d\a \;
e^{ - i \b \ol P \cdot z + i \a \D \cdot z / 2}
\\
\times \frac{ 1 }{1 - \frac{t}{4 M^2} } \ol u_{\lp}(\Pp) 
\Bigg[ 
 N(\b, \a; t) 
- \frac{1}{2}  \varepsilon^{\mu \nu \a \b} z_\mu \D_\nu \ol P_\a \gamma_\b \gamma_5 \, G_M(\b,\a;t)
\Bigg]   u_{\l}(P) 
,\label{eqn:curl2} \end{multline}
where 
\begin{equation}
N(\b,\a;t) =  \frac{\partial}{\partial \b} G_E(\b,\a;t) + \frac{1}{2} \left( 1 - \frac{t}{4 M^2} \right) \frac{\partial}{\partial \a} G(\b,\a;t)
.\end{equation}

Now we define an arbitrary potential function $\chi(\b,\a;t)$ which has the property: $\chi(\b,\a;t) = - \chi(\b,-\a;t)$. 
The above expression Eq.~\eqref{eqn:curl2}  is invariant under the transformation
\begin{align}
G_E(\b,\a;t) & \to G_E(\b,\a;t) - \left( 1 - \frac{t}{4 M^2} \right) \frac{\partial}{\partial \a} \, \chi(\b,\a;t) \notag \\ 
G(\b,\a;t) & \to G(\b, \a; t) + 2 \frac{\partial}{\partial \b} \, \chi(\b,\a;t)
\label{eqn:transf} \end{align}
which implicitly leaves $G_M(\b,\a;t)$ unchanged. Notice the $\a$ symmetry of the distributions is preserved 
because the potential $\chi(\b,\a;t)$ is an odd function. Translating the transformation Eq.~\eqref{eqn:transf} 
back into the original DDs, we have invariance under
\begin{align} 
F(\b,\a;t) & \to F(\b,\a;t) -   \frac{\partial}{\partial \a}  \,\chi(\b,\a;t) \notag \\
K(\b,\a;t) & \to K(\b,\a;t) +    \frac{\partial}{\partial \a}  \,\chi(\b,\a;t) \notag \\
G(\b,\a;t) & \to G(\b,\a;t) + 2   \frac{\partial}{\partial \b}  \,\chi(\b,\a;t) 
.\label{eqn:trans} \end{align} 

One may try and choose the potential so that $G(\b,\a;t)$ can be completely removed. The analysis is now the same as in \cite{Teryaev:2001qm}.
To choose such a $\chi(\b,\a;t)$, we must maintain the zero boundary conditions so that surface terms produced by partial integration vanish. 
The following choice of potential preserves the boundary conditions 
\begin{equation} \label{eqn:elim}
\chi_o(\b,\a;t) = -\frac{1}{4} \left[\int_{-1 + |\a|}^{\b} d\b' \, G(\b',\a;t) - \int_{\b}^{1 - |\a|} d \b' \, G(\b',\a;t)  
- \sign (\b)  D(\a;t) \right]
,\end{equation}
where the $D$-term is given by
\begin{equation}
D(\a;t) = \int_{-1 + |\a|}^{1 - |\a|} d\b \,  G(\b,\a;t)
.\end{equation}
Under the transformation generated by $\chi_o(\b,\a;t)$, the function $G(\b,\a;t)$ is changed accordingly to
\begin{equation}
G(\b, \a;t) \to 
\begin{cases}
0, \text{ if } G(-\b,\a;t) = G(\b,\a;t),\\
\delta(\b) D(\a;t), \text{ otherwise.}
\end{cases}
\end{equation}

\chapter{Basic Identities}\label{sec:gamma}

For reference we list identities used in computing DDs above. 
To calculate numerator structures so that 
symmetric and traceless tensors can be calculated, we used generalized Gordon identities
\cite{Brown:1992db} of the form
\begin{equation} \label{eqn:GGordon}
\ol u_{\l'}(P') \Gamma u_\l(P) = \frac{1}{4M} \ol u_{\l'} (P') \left( 2 \{ \rlap\slash \ol P, \Gamma \} + [\rlap\slash \Delta, \Gamma] \right) u_{\l}(P)
,\end{equation}
where $\Gamma$ is any Dirac matrix, $\ol P {}^\mu = \frac{1}{2} ( \Pp + P)^\mu$ and $\D^\mu = ( P^{\prime} - P)^\mu$.
The usual Gordon identity is a special case of Eq.~\eqref{eqn:GGordon}, namely for $\Gamma = \gamma^\mu$
\begin{equation}
\ol u_{\l'}(P') \gamma^\mu u_{\l}(P)= \frac{1}{2 M} \ol u_{\l'}(P') \left( 2 \ol P {}^\mu + i \sigma^{\mu \nu} \D_\nu  \right)  u_{\l}(P)
.\end{equation}
We also require the following two cases of the general identity with $\gamma_5 = i \gamma^0 \gamma^1 \gamma^2 \gamma^3$
and $\varepsilon^{0123} = +1$.
\begin{align}
\ol u_{\l'}(P') \gamma^\mu \gamma_5 \, u_{\l}(P) 
& = \frac{1}{2 M}  \ol u_{\l'}(P') 
\left(  
\D^\mu \gamma_5 
 - \varepsilon^{\mu \nu \a \b} \sigma_{\a \b} \ol P_\nu  
\right)  u_{\l}(P) \\
\ol u_{\l'}(P')\sigma^{\mu \nu} u_{\l}(P) 
& = \frac{i}{2 M}  \ol u_{\l'}(P') 
\left( 
\gamma^{\nu} \D^\mu - \gamma^\mu \D^\nu 
+ 2 i  \varepsilon^{\mu \nu \a \b} \ol P_\a \gamma_\b  \gamma_5 
\right)  u_{\l}(P)
.\end{align}

Also of use is the gamma triple product
\begin{equation}
\gamma^\mu \gamma^\nu \gamma^\rho = S^{\mu \nu \rho \sigma} \gamma_\sigma  + i \varepsilon^{\mu \nu \rho \sigma} \gamma_\sigma \gamma_5
,\end{equation}
where
\begin{equation}
S^{\mu \nu \rho \sigma} = g^{\mu \nu} g^{\rho \sigma} - g^{\mu \rho} g^{\nu \sigma} + g^{\mu \sigma} g^{\nu \rho}
\notag
.\end{equation}

\chapter{Calculations on the Light-Cone}\label{sec:LC}

Here we include our conventions for projecting quantities onto the light cone. While the development and derivations
above rely exclusively on manifest Lorentz invariance, the light-cone Fock representation provides transparent 
physical intuition about our model. The light-cone wavefunctions for our model are admittedly simple and thus provide
a useful guide to understanding the DDs constructed above.

The light-cone spinor $u_{\l}(k,m)$ satisfies the Dirac equation $(\rlap\slash k - m) u_{\l}(k,m) = 0$
and is explicitly given by (e.g., see \cite{Heinzl:1998kz})
\begin{equation} \label{eqn:lcspinor}
u_{\l}(k,m) = \frac{1}{2^{1/4}\sqrt{k^+}} 
\left( \sqrt{2} k^+ + \beta m + \bm{\a}^\perp \cdot \mathbf{k}^\perp  \right) X_{\l}
,\end{equation}
where $\b = \gamma^0$ and $\bm{\a} = \gamma^0 \bm{\gamma}$ and the unit spinors $X_\l$ are given by
\begin{align}
X_\uparrow^\dagger & = \frac{1}{\sqrt{2}} \left( 1,0,1,0 \right) \notag \\
X_\downarrow^\dagger & = \frac{1}{\sqrt{2}} \left( 0,1,0,-1 \right) \notag
.\end{align}
Using Eq.~\eqref{eqn:lcspinor}, we derive the useful product of spinors of different mass and momentum
\begin{equation} \label{eqn:spinorprod}
\ol u_{\lp}(k,m) u_\l(P,M) = \frac{1}{\sqrt{k^+ P^+}} 
\left[ 
\delta_{\lp,\l} (k^+ M + P^+ m) 
- \l \, \delta_{\lp,-\l} ( k^+ P_\l - P^+ k_\l)
\right]
,\end{equation}
where the notation $a_\l = a^1 + i \l a^2$ has been employed with the understanding 
that spins correspond to the signs $\uparrow = +1$ and $\downarrow = -1$.

Using the light-cone spinors, one can find the $H_q(x,\x,t)$ and $E_q(x,\x,t)$ GPDs 
in terms of the light-cone, non-diagonal matrix element $\mathcal{M}_q^{\lp,\l}(x,\x,t)$, namely
\begin{align}
\mathcal{M}_q^{\l,\l}(x,\x,t) & = \frac{1}{\sqrt{1 - \x^2}} \left[ (1 - \x^2) H_q(x,\x,t) - \x^2 E_q(x,\x,t) \right] \\
\mathcal{M}_q^{\l,-\l}(x,\x,t) & =  - \frac{ \l \D_\l }{2 M \sqrt{1-\x^2}}  \, E_q(x,\x,t)
.\end{align} 
These expressions can alternately be used to derive GPDs for the model considered in Section \ref{sec:DDcalc}
by directly inserting the quark bilocal operator between non-diagonal proton states. Expressions are derived in 
this manner by integration over the relative light-cone energy $k^-$ at the cost of sacrificing manifest Lorentz
invariance. This description in terms of light-cone Fock components, however, is more intuitive to work with than the DD
formulation. Moreover, the light-cone energy integration obviates the positivity properties of our model GPDs. 
We shall not present complete expressions for the GPDs on the light-cone, however, the diagonal overlap
region $x>\xi$ is particularly simple to consider. 

For our model, we can find the lowest Fock component's 
light-cone wavefunction by projecting the covariant amplitude onto the 
surface $x^+ = 0$, namely
\begin{eqnarray} 
\psi_{\text{LC}} (x,\mathbf{k}^\perp_{\text{rel}}; s_i,\l) &=& \frac{1}{(2 P^+)^2} \int \frac{d k^-}{2 \pi}
\frac{\ol u_{s_1} (P-k,m)}{\sqrt{1-x}} 
\Big( \gamma^+ \gamma_5 \, i C \Big) 
\ol u_{s_2}(P-k,m)^{\text{T}} \notag \\ 
&& \times \frac{\ol u_{s_3} (k,m) }{\sqrt{x}} \gamma^+ \Psi(k,P) u_\l(P,M) 
,\label{eqn:lcwfn} \end{eqnarray}
where $x$ is the fraction of the proton's plus momentum carried by the residual quark $(x = k^+ / P^+)$, 
and the relative transverse momentum is $\mathbf{k}^\perp_{\text{rel}} = \mathbf{k}^\perp - x \mathbf{P}^\perp$. 
Above, $\l$ labels the spin of the proton, whereas the $s_i$ label spins of the three quarks.
We have omitted the color and isospin parts of the wavefunction which are trivial: $\propto \varepsilon_{c_1,c_2,c_3} (\delta_{1,u} \delta_{2,d} - 
\delta_{1,d} \delta_{2,u} ) \delta_{3,u}$ for the scalar diquark.
In the scalar channel, the quark-diquark Bethe-Salpeter wavefunction is
\begin{equation}
\Psi(k,P) = \frac{i}{\rlap \slash k  - m + i \varepsilon}  \Big ( - i g^{(s)} \Big)  D(P - k)
,\end{equation} 
and above we have chosen a single Dirac component $\Gamma^{(s)} = - i g^{(s)}$ and the coupling constant $g$ has been 
absorbed into the overall normalization above. Our choice of vertex functions corresponds to only modeling a subset 
of the possible three quark light-cone wavefunctions of the proton \cite{Ji:2002xn}. Carrying out the projection in Eq.~\eqref{eqn:lcwfn}, yields
\begin{equation} \label{eqn:psi}
\psi_{\text{LC}} (x,\mathbf{k}^\perp_{\text{rel}}; s_i,\l) \propto 
\frac{ s_1 \delta_{s_1,-s_2} [\delta_{s_3,\l} (x M + m ) + \l \, \delta_{s_3, -\l} k_{\text{rel},\l}]}
{x^a \sqrt{ 1 - x} \left[M^2 - \frac{\mathbf{k}^\perp_{\text{rel}} {}^2 }{x (1-x)} - \frac{m^2}{x} - \frac{m_D^2}{1 -x}\right]^a}
.\end{equation}
Above $a = 1$ for the valence light-cone wavefunction. The effective wavefunction $\psi_{\text{LC}}^{\text{eff}} (x,\mathbf{k}^\perp_{\text{rel}}; s_i,\l)$ 
generated by the analytic regularization is given by Eq.~\eqref{eqn:psi} with $a =2$ (or in general $a > 1$). 
There are no true higher Fock components in this model since the interaction kernel is instantaneous in light-cone 
time \cite{Tiburzi:2002sw}.
Focusing on the spin structure, the scalar diquark channel consists of two states: a state 
where the proton spin is aligned with residual quark's spin and a state where the quark and diquark are in a relative
$p$-wave. 

The quark distribution function can be obtained from 
\begin{equation}
f_u(x) = \sum_{s_i} \int d \mathbf{k}^\perp \left| \psi_{\text{LC}}^{\text{eff}} (x,\mathbf{k}^\perp; s_i,\l) \right|^2 
,\end{equation}
and agrees with the covariant calculation of $q_u(x)$ from its moments in Section \ref{sec:DDcalc}. Similarly the Dirac and Pauli
form factors can be expressed in terms of the effective wavefunction since the light-cone contour integration 
only receives contribution from the diquark pole. The expression are
\begin{eqnarray}
F_1(t) &=& \sum_{s_i} \int d  \mathbf{k}^\perp  \, dx \, \psi_{\text{LC}}^{*\text{eff}}(x, \mathbf{k}^\perp + (1  - x ) \mathbf{\D}^\perp; s_i, \l)
\psi_{\text{LC}}^{\text{eff}}(x, \mathbf{k}^\perp; s_i, \l) 
\\
- \frac{\l \D_\l} {2 M} F_2(t) &=& \sum_{s_i }\int d  \mathbf{k}^\perp  \, dx \, 
\psi_{\text{LC}}^{*\text{eff}}(x, \mathbf{k}^\perp + (1  - x ) \mathbf{\D}^\perp; s_i, -\l)
\psi_{\text{LC}}^{\text{eff}}(x, \mathbf{k}^\perp; s_i, \l) 
\end{eqnarray}
Also for this reason, we can express the GPDs as simple convolutions in the diagonal overlap region $x>\x$. Let $x_1 = \frac{x + \x}{1 + \x}$ and
$x_2 = \frac{ x - \x}{ 1 - \x}$ and then we have
\begin{multline}
\theta( x - \x) \mathcal{M}^{\l,\l}_q (x,\x,t)  = \sum_{s_i} \int d  \mathbf{k}^\perp  \, 
\psi_{\text{LC}}^{*\text{eff}} \left(x_2, \mathbf{k}^\perp + (1  - x_2 ) \frac{\mathbf{\D}^\perp}{2}; s_i, \l \right)
\\
\times
\psi_{\text{LC}}^{\text{eff}} \left(x_1, \mathbf{k}^\perp - (1  - x_1 ) \frac{\mathbf{\D}^\perp}{2}; s_i, \l \right) 
\end{multline}
with a very similar expression holding for the spin-flip amplitude. 
In the above form the positivity bound is manifest.

In the axial-vector diquark channel we have the orthogonal amplitude
\begin{eqnarray} 
\psi (x,\mathbf{k}^\perp_{\text{rel}}; s_i,\l) 
&=& 
\frac{1}{(2 P^+)^2} \int \frac{d k^-}{2 \pi}
\frac{\ol u_{s_1} (P-k,m)}{\sqrt{1-x}} 
\Big( \gamma^+ \gamma_\mu \, i C \Big) 
\ol u_{s_2}(P-k,m)^{\text{T}} \notag \\ 
&& \times \frac{\ol u_{s_3} (k,m) }{\sqrt{x}} \gamma^+ \Psi^\mu(k,P) u_\l(P,M) 
.\label{eqn:lcwfn2} \end{eqnarray}
Above, we have omitted the color and isospin parts of the wavefunction which are: $\propto \varepsilon_{c_1,c_2,c_3} [ (\delta_{1,u} \delta_{2,d} + 
\delta_{1,d} \delta_{2,u} ) \delta_{3,u} - 2 \delta_{1,u} \delta_{2,u} \delta_{3,d}]$
for the axial diquark. In the axial-vector channel, the quark-diquark Bethe-Salpeter vector wavefunction is
\begin{equation}
\Psi^\mu(k,P) = \frac{i}{\rlap \slash k  - m + i \varepsilon} \Big ( - i g^{(a)} \gamma_\nu \gamma_5  \Big)  D^{\mu \nu}(P - k) 
,\end{equation} 
where we have chosen a single Dirac component $\Gamma^{(a)} = - i g^{(a)}$ and the coupling constant $g^{(a)} = g^{(s)}$, as above. 
Carrying out the projection in Eq.~\eqref{eqn:lcwfn2} yields the axial diquark piece of the light-cone wavefunction.
Analogous formulas to the above hold for the axial diquark contribution to 
the quark distribution, form factors and GPDs for $x > \x$.

\chapter{Derivation of the Double Distributions} \label{sec:DDDD}

In this Appendix, we detail the calculation of the DDs in the scalar diquark channel and comment on the calculation in the 
axial-vector channel. The crucial steps in the derivation hinge upon reducing the numerator by factors present in the denominator
or by use of the simple identity:
\begin{equation} \label{eqn:notwogammas}
\gamma^{\{\mu_i} \gamma^{\mu_j\}} = 0
.\end{equation}

In the scalar diquark channel, let us take Eq.~\eqref{eqn:impulse} as our starting point. 
Denote the propagators simply by $\mathfrak{A} = (k - \ol P + \D /2)^2 - m_D^2 + i \varepsilon$, 
$\mathfrak{B} = (k + \D)^2 - m^2 + i \varepsilon$, and $\mathfrak{C} = k^2 - m^2 + i \varepsilon$.
The DDs in the scalar channel can be deduced without reducing factors in the numerator. We merely
introduce two Feynman parameters $\{x,y\}$ to cast the denominator specifically in the form
$[ x \mathfrak{A} + y \mathfrak{B} + (1-x-y) \mathfrak{C}]^{2 a + 1}$. One then translates $k^\mu$ to
render the integral hyperspherically symmetric via the definition $k^\mu = l^\mu + \b \ol P {}^\mu - (\a + 1) \D^\mu / 2$.
Here $\b = x$ and $\a  = x + 2 y - 1$. Carrying out this procedure on Eq.~\eqref{eqn:impulse} produces
\begin{multline} \label{eqn:impulse2}
\frac{1}{2} \int_0^1 d\b \int_{-1 + \b}^{1 - \b} d\a  \int d^4 l  \frac{[(1 - \b)^2 - \a^2]^{a -1} \Gamma(2 a + 1) }{4^{a - 1} \Gamma(a)^2}
\\ \times
\ol u_{\l'}(\Pp)
\frac{N^{\{ \mu} (l + \b \ol P - \a \D /2)^{\mu_1} \cdots (l + \b \ol P - \a \D /2)^{\mu_n \} }   }
{[l^2 - D_o(\b,\a;t)]^{2 a + 1}}
u_\l(P)
,\end{multline}   
where the numerator Dirac structure is given by
\begin{eqnarray} \label{eqn:firstN}
N^\mu &=& (\rlap \slash l + \b \rlap \slash \ol P - (\a - 1) \rlap \slash \D / 2 ) \gamma^{\mu} ( \rlap \slash l + \b \rlap \slash \ol P - (\a +1) \rlap \slash \D / 2 )
\notag \\
&& + m^2 \gamma^\mu + i m \sigma^{\mu \nu} \D_\nu + 2 m ( l + \b \ol P - \a \D/ 2)^\mu 
.\end{eqnarray}

Using Eq.~\eqref{eqn:notwogammas}, as well as the identities in Appendix \ref{sec:gamma}, we can cast Eq.~\eqref{eqn:impulse2} in 
the form
\begin{multline} \label{eqn:impulse3}
\frac{1}{2} \int_0^1 d\b \int_{-1 + \b}^{1 - \b} d\a  \int d^4 l  \frac{[(1 - \b)^2 - \a^2]^{a -1} \Gamma(2 a + 1) }{4^{a - 1} \Gamma(a)^2}
[l^2 - D_o(\b,\a;t)]^{-2 a - 1}
\\ \times
\ol u_{\l'}(\Pp)
\mathcal{N}^{\{\mu}  u_\l(P)
\sum_{k=0}^{n} \frac{n!}{ k! (n-k)!}      
\ol P {}^{\mu_1} \cdots \ol P {}^{\mu_{n-k}} 
\left( - \frac{\D}{2}\right)^{\mu_{n-k+1}} \cdots \left( - \frac{\D}{2}\right)^{\mu_{n}\}}
,\end{multline}
where
\begin{eqnarray}
 \mathcal{N}^\mu &=& \left\{ (m + \b M)^2 - \frac{l^2}{2} + \frac{t}{4} [ (1-\b)^2 - \a^2]  \right\} \gamma^\mu  
\notag \\
&& + i \sigma^{\mu \nu} \D_\nu ( 1 - \b) ( m + \b M) - \a ( m + \b M) \D^\mu 
.\end{eqnarray}   
The $l$ integration is then standard and the DDs can be read off simply by using Eqs.~\eqref{eqn:moments} -- \eqref{eqn:generated}
and we arrive at Eq.~\eqref{eqn:Fs}.

Calculation in the axial-vector diquark channel is similar, however, the numerator is more complicated. In the same units as Eq.~\eqref{eqn:impulse}, 
we have the contribution from Fig.~\ref{ftwist} for the axial diquark
\begin{equation} \label{eqn:aximpulse}
\frac{1}{9} Z(a) \int  d^4 k \, D^{\a \b}(k - P ) 
\frac{\ol u_{\l'} (\Pp) 
\gamma_5 \gamma_\a (\rlap\slash k + \rlap\slash \D + m) \Gamma^{\mu \mu_1 \ldots \mu_n} (\rlap\slash k + m ) \gamma_\b \gamma_5  u_\l(P)}
{[k^2 - m^2 + i \varepsilon]{}^a [(k+\D)^2 - m^2 + i \varepsilon]{}^a }
,\end{equation}
where the vector propagator $D^{\a \b}(k - P)$ is given in Eq.~\eqref{eqn:vector}. The terms which result
from the $g^{\a \b}$ structure in the vector propagator can be dealt with straightforwardly after evaluating the contracted gamma
matrices. The integrals encountered are then similar to those in the scalar channel. 




\chapter*{Vita}
Brian Charles Tiburzi graduated in May, 1999, from Amherst College (Amherst, Massachusetts) 
with a Bachelor of Arts degree, \emph{summa cum laude}, in Physics 
and Mathematics. Amherst College has the highest number of tennis courts per student of any institution of higher learning.
In December, 2000, he earned a Master of Science degree in Physics at the 
University of Washington in Seattle, Washington. There he received his Doctor of 
Philosophy degree in Physics in September, 2004.  Indeed it rained quite a bit during this 
period in Seattle. Throughout the grayness and despite constant pressure, however, Brian did not succumb to a coffee addiction;
although, he has been known to order a ``grande non-fat no-water chai'' on occasion.
Brian's other publications which are orthogonal to the scope of this work 
include \cite{Arndt:2003ww,Arndt:2003we,Arndt:2003vd,Arndt:2004we} and his life expectancy at birth was
roughly $74.0$ years.

\end{document}